\documentclass[longauth]{aa}
\usepackage{graphicx}
\usepackage{txfonts}
\usepackage{natbib}
\bibpunct{(}{)}{;}{a}{}{,}
\usepackage{xcolor}
\usepackage{float}
\usepackage{pifont}
\usepackage{upgreek}
\newcommand{\cmark}{\ding{51}}
\newcommand{\xmark}{\ding{55}}
\usepackage{hyperref}

\def\as {\ifmmode {\rlap.}$\,$''$\,$\! \else ${\rlap.}$\,$''$\,$\!$\fi}

\begin{document} 

\title{Clustered star formation at early evolutionary stages.}

\subtitle{Physical and chemical analysis of the young star-forming regions ISOSS\,J22478+6357 and ISOSS\,J23053+5953}

\author{C. Gieser\inst{1}\fnmsep\thanks{Fellow of the International Max Planck Research School for Astronomy and Cosmic Physics at the University of Heidelberg (IMPRS-HD).}
	\and
	H.~Beuther\inst{1}
	\and
	D.~Semenov\inst{1,2}
	\and
	S.~Suri\inst{1}
	\and
	J.D.~Soler\inst{1}
	\and
	H.~Linz\inst{1}
	\and
	J.~Syed\inst{1}
	\and
	Th.~Henning\inst{1}
	\and
	S.~Feng\inst{3}
	\and
	T.~M{\"o}ller\inst{4}
	\and
	A.~Palau\inst{5}
	\and
	J.M.~Winters\inst{6}
	\and
	M.T.~Beltr{\'a}n\inst{7}
	\and
	R.~Kuiper\inst{8}
	\and
	L.~Moscadelli\inst{7}
	\and
	P.~Klaassen\inst{9}
	\and
	J.S.~Urquhart\inst{10}
	\and
	T.~Peters\inst{11}
	\and
	S.N.~Longmore\inst{12}
	\and
	{\'A}.~S{\'a}nchez-Monge\inst{4}
	\and
	R.~Galv{\'a}n-Madrid\inst{5}
	\and
	R.E.~Pudritz\inst{13}
	\and
	K.G.~Johnston\inst{14}
	}

	\institute{Max Planck Institute for Astronomy, Königstuhl 17, 69117 Heidelberg, Germany,\\
	\email{gieser@mpia.de}
	\and
	Department of Chemistry, Ludwig Maximilian University, Butenandtstr. 5-13, 81377 Munich, Germany
	\and
	Department of Astronomy, Xiamen University, Xiamen, Fujian 361005, P. R. China
	\and
	I. Physikalisches Institut, Universit{\"a}t zu K{\"o}ln, Z{\"u}lpicher Str. 77, D-50937, K{\"o}ln, Germany
	\and
	Instituto de Radioastronom{\'i}a y Astrof{\'i}sica, Universidad Nacional Aut{\'o}noma de M{\'e}xico, P.O. Box 3-72, 58090, Morelia, Michoac{\'a}n, Mexico
	\and
	Institut de Radioastronomie Millim\'{e}trique (IRAM), 300 Rue de la Piscine, F-38406 Saint Martin d'H\`{e}res, France
	\and
	INAF, Osservatorio Astrofisico di Arcetri, Largo E. Fermi 5, I-50125 Firenze, Italy
	\and
	Zentrum f{\"u}r Astronomie der Universit{\"a}t Heidelberg, Institut f{\"u}r Theoretische Astrophysik, Albert-Ueberle-Straße 2, 69120 Heidelberg, Germany
	\and
	UK Astronomy Technology Centre, Royal Observatory Edinburgh, Blackford Hill, Edinburgh EH9 3HJ, UK
	\and
	Centre for Astrophysics and Planetary Science, University of Kent, Canterbury, CT2 7NH, UK
	\and
	Max-Planck-Institut f{\"u}r Astrophysik, Karl-Schwarzschild-Str. 1, 85748 Garching, Germany
	\and
	Astrophysics Research Institute, Liverpool John Moores University, Liverpool, L3 5RF, UK
	\and
	Department of Physics and Astronomy, McMaster University, 1280 Main St. W, Hamilton, ON L8S 4M1, Canada
	\and
	School of Physics and Astronomy, University of Leeds, Leeds LS2 9JT, United Kingdom
	}

	\date{Received...; accepted...}
 
	\abstract
	{The process of high-mass star formation during the earliest evolutionary stages and the change over time of the physical and chemical properties of individual fragmented cores are still not fully understood.}
	{We aim to characterize the physical and chemical properties of fragmented cores during the earliest evolutionary stages in the very young star-forming regions ISOSS\,J22478+6357 and ISOSS\,J23053+5953.}
	{NOrthern Extended Millimeter Array (NOEMA) 1.3\,mm data are used in combination with archival mid- and far-infrared Spitzer and Herschel telescope observations to construct and fit the spectral energy distributions of individual fragmented cores. The radial density profiles are inferred from the 1.3\,mm continuum visibility profiles, and the radial temperature profiles are estimated from H$_{2}$CO rotation temperature maps. Molecular column densities are derived with the line fitting tool \texttt{XCLASS}. The physical and chemical properties are combined by applying the physical-chemical model MUlti Stage ChemicaL codE (\texttt{MUSCLE}) in order to constrain the chemical timescales of a few line-rich cores. The morphology and spatial correlations of the molecular emission are analyzed using the histogram of oriented gradients (HOG) method.}
	{The mid-infrared data show that both regions contain a cluster of young stellar objects. Bipolar molecular outflows are observed in the CO $2-1$ transition toward the strong millimeter (mm) cores, indicating protostellar activity. We find strong molecular emission of SO, SiO, H$_{2}$CO, and CH$_{3}$OH in locations that are not associated with the mm cores. These shocked knots can be associated either with the bipolar outflows or, in the case of ISOSS\,J23053+5953, with a colliding flow that creates a large shocked region between the mm cores. The mean chemical timescale of the cores is lower ($\sim$20\,000\,yr) compared to that of the sources of the more evolved CORE sample ($\sim$60\,000\,yr). With the HOG method, we find that the spatial emission of species that trace the extended emission and of shock-tracing molecules are well correlated within transitions of these groups.}
	{Clustered star formation is observed toward both regions. Comparing the mean results of the density and temperature power-law index with the results of the original CORE sample of more evolved regions, it appears that neither change significantly from the earliest evolutionary stages to the hot molecular core stage. However, we find that the 1.3\,mm flux, kinetic temperature, H$_{2}$ column density, and core mass of the cores increase in time, which can be traced both in the $M$/$L$ ratio and the chemical timescale, $\tau_\mathrm{chem}$.}


	\maketitle

\section{Introduction}\label{sec:introduction}

	Massive stars ($M_{\star} \geq 8$\,$M_{\odot}$) are rare, compared to low-mass stars with $M_{\star} \lesssim 1$\,$M_{\odot}$, but important in shaping the physical and chemical properties of a galaxy. For example, their high bolometric luminosities dominate the stellar contribution to the spectral energy distribution (SED) of galaxies \citep[e.g.,][]{Walcher2011} and thus provide a high flux of ionizing radiation. Energetic, mechanical, and radiative feedback is provided to the interstellar medium (ISM) through outflows \citep[e.g.,][]{Beuther2002,Koelligan2018}, stellar winds \citep[e.g.,][]{Meynet1994,Gatto2017}, and supernova explosions \citep[e.g.,][]{McKee1977,Girichidis2016}.
	
	In the earliest stages of high-mass star formation (HMSF), the stellar birthplaces are still deeply embedded within their dense parental molecular cloud and not visible at optical or infrared (IR) wavelengths. Therefore, the cold gas and dust can only be studied at millimeter (mm) or sub-mm wavelengths. For comprehensive reviews of massive star formation, we refer to, for example, \citet{Beuther2007}, \citet{Bonnell2007}, \citet{Zinnecker2007}, \citet{Smith2009}, \citet{Tan2014}, \citet{Schilke2015}, \citet{Motte2018}, and \citet{Rosen2020}. 
	
	Once a protostar forms, gas accretion and bipolar outflows are commonly observed, and the young stellar objects (YSOs) become bright at mid-infrared (MIR) wavelengths. HMSF is generally clustered and observed toward giant molecular clouds with sizes of $R \gtrsim 1$\,pc \citep[e.g.,][]{RomanDuval2010}, providing a large mass reservoir for individual massive clumps with typical sizes of $R \gtrsim 0.1$\,pc. Following the nomenclature of \citet{Beuther2007} and \citet{Zhang2009}, these clumps are commonly observed to fragment into individual cores at high angular resolution, $R \approx 0.01 - 0.1$\,pc, where a single or a small system of multiple protostars forms.
	
	A large diversity of core fragmentation properties is observed. In some regions, there is a single dominant massive core \citep[e.g.,][]{Beuther2018,Maud2019}, while in other regions a large number of cores are found \citep[e.g.,][]{Palau2014, Palau2015, Beuther2018,Beuther2021}. In addition, it has been suggested that HMSF could occur at hubs of filamentary structures \citep[e.g.,][]{Myers2009,GalvanMadrid2010,Schneider2012,GalvanMadrid2013,Tige2017,Kumar2020}. It is not yet fully understood what drives the accretion of the surrounding gas and dust from cloud and clump scales onto the cores \citep[e.g.,][]{Smith2009}. Variations in magnetic field strengths \citep[e.g.,][]{Hennebelle2019} and density profiles could also be important factors that influence the multiplicity and core mass \citep[e.g.,][]{Commercon2011,Girichidis2011}.
	
	High-mass star-forming regions (HMSFRs) are sites with a rich molecular content and for dedicated reviews we refer to \citet{Herbst2009} and \citet{Jorgensen2020}. Most of the total of $\sim$200 molecules detected in the ISM are found toward HMSFRs \citep{McGuire2018}. Chemical reactions occur in the gas phase, on dust grain surfaces, and in the icy mantle layers \citep[e.g.,][]{Garrod2008}. A current challenge is to understand, for example, the chemical segregation between nitrogen(N)- and oxygen(O)-bearing species \citep[e.g.,][]{Wyrowski1999, JimenezSerra2012, Feng2015, Allen2017, Gieser2019}.
	
	In order to study HMSF at core scales, we carried out the survey ``CORE - Fragmentation and disk formation during high-mass star formation,'' which is a large program with the NOrthern Extended Millimeter Array (NOEMA) targeting 18 HMSFRs in the northern hemisphere with the currently highest possible angular resolution at 1.3\,mm ($\sim$0\as4). An overview of the CORE project and analysis of the 1.3\,mm continuum data are presented in \citet{Beuther2018}. The molecular line data, revealing a high degree of chemical complexity on scales $<$10\,000\,au, are analyzed by \citet{Gieser2021}. 
	
	The observations of the CORE sample were carried out between $217.2 - 220.8$\,GHz ($\sim$1.3\,mm) with the WideX correlator and a single pointing toward each region. Two additional pilot regions were previously observed in a similar setup \citep{Feng2016}.
	
	In the physical and chemical analysis of the full CORE sample \citep{Gieser2021}, 22 objects are analyzed, defined as ``cores'', that have a radially decreasing temperature profile and associated 1.3\,mm continuum emission. Molecular column densities are derived from the 1.3\,mm spectra (e.g., C$^{18}$O, SO, OCS, HNCO, HC$_{3}$N, CH$_{3}$OH, CH$_{3}$CN), and each core is modeled with the physical-chemical model MUlti Stage ChemicaL codE (\texttt{MUSCLE}) to estimate chemical timescales, $\tau_\mathrm{chem}$. The luminosities toward these regions are high \citep[$L > 10^{4}$\,$L_{\odot}$;][]{Beuther2018}, suggesting that these regions contain massive YSOs and that the evolutionary stages of most of the regions are between the high-mass protostellar object (HMPO) and hot molecular core (HMC) phase. Some regions have already formed hyper- or ultra-compact H{\sc ii} (HCH{\sc ii} or UCH{\sc ii}) regions. For a more detailed description of the physical properties of the IRDC, HMPO, HMC, and UCH{\sc ii} evolutionary phases, we refer to, for example, \citet{Gerner2014} and \citet{Gieser2021}. With \texttt{MUSCLE}, gradients of $\tau_\mathrm{chem}$ are found for cores located within the same region if the separations are large, while nearby cores have similar chemical timescales.
	
	In this paper, we study the core properties in the two young and cold intermediate- to high-mass star-forming regions ISOSS\,J22478+6357 and ISOSS\,J23053+5953. The data were obtained as an extension of the CORE project (``CORE-extension'') with both regions observed in multiple pointings creating large mosaics and the new PolyFiX correlator increasing the spectral bandwidth by a factor of four. An overview and the analysis of the fragmentation and kinematic properties is presented in \citet{Beuther2021}. These two additional regions are at very early evolutionary stages, similar to typical infrared dark clouds (IRDCs), with $T_\mathrm{clump} \approx 20$\,K \citep{Ragan2012}. Thus, these two sources complement the CORE sample with regions that tend to occupy an earlier phase in the evolutionary sequence of HMSF \citep{Beuther2007}. Both regions were selected based on targets of the ISOPHOT Serendipity Survey (ISOSS) observing the sky during slew times at 170\,$\upmu$m \citep{Krause2004}.
	
\begin{table*}
\caption{Overview of the observed regions.}
\label{tab:regions}
\centering
\begin{tabular}{lcccccccc}
\hline\hline
Region & \multicolumn{2}{c}{Coordinates} & Distance & Galactocentric distance & Galactic height & Velocity & \multicolumn{2}{c}{Isotopic ratios}\\
\cline{2-3} \cline{8-9}
 & $\alpha$ & $\delta$ & $d$ & $d_{\mathrm{gal}}$ & $z_{\mathrm{gal}}$ & $\varv_{\mathrm{LSR}}$ & $^{12}$C/$^{13}$C & $^{16}$O/$^{18}$O\\
 & (J2000) & (J2000) & (kpc) & (kpc) & (pc) & (km\,s$^{-1}$) & & \\
\hline
ISOSS\,J22478+6357 & 22:47:49.23 & +63:56:45.3 & $3.23$ & $9.7$ & $270$ & $-39.7$ & $80$ & $607$\\ 
ISOSS\,J23053+5953 & 23:05:22.47 & +59:53:52.6 & $4.31$ & $10.4$ & $11$ & $-51.7$ & $86$ & $649$\\ 
\hline 
\end{tabular}
\tablefoot{The distance, $d$, and systemic velocity, $\varv_{\mathrm{LSR}}$, are taken from \citet{Ragan2012}. The calculation of the isotopic ratios is based on \citet{Wilson1994}, including the dependence on the galactocentric distance, $d_{\mathrm{gal}}$.}
\end{table*}

	The ISOSS\,J22478+6357 region was first identified with the Infrared Astronomical Satellite (IRAS) as IRAS\,22460+6341 \citep{Beichman1988}, and is also known as J224749.9+635647 \citep{DiFrancesco2008}. We note that the associated IRAS source \citep{Kerton2003} is spatially offset by $\sim$41$''$ from the studied IR and mm counterparts in this and other more recent studies \citep{Hennemann2008,Ragan2012,Bihr2015,Beuther2021}. Wide-field Infrared Survey Explorer (WISE) images at 12 and 22\,$\upmu$m show extended emission around the IRAS source, which can be interpreted as a signpost of intermediate-mass star formation \citep{Lundquist2014}. 
	
	With Herschel observations using the Photodetector Array Camera and Spectrometer \citep[PACS,][]{Poglitsch2010} instrument as part of the Earliest Phases of Star Formation (EPoS) program, \citet{Ragan2012} determine a kinematic distance of $d = 3.23^{+0.61}_{-0.60}$\,kpc \citep[using the model by][]{Reid2009} and a total gas mass of 104\,$M_\odot$ within a total of seven resolved clumps. The systemic velocity $\varv_\mathrm{LSR}$ is $-39.7$\,km\,s$^{-1}$. The region is located in the outer galaxy, $d_{\mathrm{gal}} = 9.7$\,kpc away from the Galactic center. 
	
	A detailed study of a few ISOSS regions, including ISOSS\,J22478+6357, is presented in \citet{Hennemann2008}. These authors detect two submm clumps with the Submillimetre Common User Bolometer Array (SCUBA) at the James Clerk Maxwell Telescope (JCMT) at 450\,$\upmu$m and 850\,$\upmu$m, ``SMM1\,E'' and ``SMM1\,W,'' with sizes of 0.14\,pc and 0.24\,pc, gas masses of 64\,$M_{\odot}$ and 116\,$M_{\odot}$, and dust temperatures of 15\,K and 14\,K, respectively. By fitting the SED of a bright MIR source, one of the more evolved YSOs in the SMM1\,E clump is classified as an intermediate-mass star with central mass $M_{\star} = 6 - 8.5$\,$M_{\odot}$ and a system age of $10^{6}-6\times10^{6}$\,yr. Using the 850\,$\upmu$m JCMT SCUBA observations, \citet{Bihr2015} estimate a total gas mass of 140\,$M_\odot$ in ISOSS\,J22478+6357.
	
	The region ISOSS\,J23053+5953 was originally identified as a point source by IRAS under the identifier IRAS\,23032+5937 \citep{Beichman1988}. In the literature it is also known as J230523.6+595356 \citep{DiFrancesco2008} or G109.995-00.282 \citep{Rosolowsky2010}. With the Bolocam Galactic Plane Survey (BGPS), a 1.1\,mm flux of 0.826\,Jy is derived within a radius of $R = 38\as32$ \citep{Rosolowsky2010}. \citet{Schlingman2011} estimate the following properties using the BGPS data: $M = 130$\,$M_\odot$, beam average density $n = 3000$\,cm$^{-3}$, volume averaged density $n = 840$\,cm$^{-3}$, free-fall timescale $\tau_{\mathrm{ff}} = 1.4 \times 10^{6}$\,yr, and crossing timescale $\tau_{\mathrm{cross}} = 4.0 \times 10^{5}$\,yr.
		
	Single-dish observations with a linear resolution of $\approx 0.3 - 1$\,pc reveal molecular line emission toward the region with line widths of $1.5 - 5$\,km\,s$^{-1}$ \citep[e.g., CO, CS, HCO$^{+}$, N$_{2}$H$^{+}$, CH$_{3}$CCH, CH$_{3}$CHO;][]{Harju1993, Wouterloot1993, Bronfman1996, Alakoz2002, Shirley2013, Vasyunina2014}. Based on NH$_{3}$ observations with the Effelsberg 100m telescope, the kinetic temperature was estimated to be $\sim$15\,K \citep{Wouterloot1988,Harju1993}. \citet{Vasyunina2014} estimate a kinetic temperature of $31 \pm 1$\,K based on CH$_{3}$CCH line emission. Early studies found that there are two velocity components in molecular line emission, for example, for NH$_{3}$ \citep[at $-52.3$\,km\,s$^{-1}$ and $-50.8$\,km\,s$^{-1}$,][]{Wouterloot1988}, H$_{2}$O \citep[at $-52.3$\,km\,s$^{-1}$ and $-49.5$\,km\,s$^{-1}$,][]{Wouterloot1993}, and CS \citep[at $-51.8$\,km\,s$^{-1}$ and $-51.2$\,km\,s$^{-1}$,][]{Larionov1999}. 
	
	Very Large Array (VLA) and Effelsberg 100m observations of the NH$_{3}$ (1,1) and (2,2) lines at an angular resolution of $4''$ are analyzed by \citet{Bihr2015}. These authors find a steep velocity gradient of $>$30\,km\,s$^{-1}$\,pc$^{-1}$ toward the region suggesting a dynamical collapse and/or converging gas flow. With JCMT SCUBA 850\,$\upmu$m observations, a total gas mass of 610\,$M_\odot$ is estimated by the authors in ISOSS\,J23053+5953. The kinematic data of the CORE-extension project further resolve this steep velocity gradient in DCO$^{+}$ ($3-2$) being higher than 50\,km\,s$^{-1}$\,pc$^{-1}$ \citep{Beuther2021}.
	
	\citet{Wouterloot1986} and \citet{Wouterloot1993} detected H$_{2}$O maser emission. But no H$_{2}$O maser emission was detected in follow-up studies \citep{Comoretto1990,Palagi1993,Palla1993, Slysh1999, Valdettaro2001,Sunada2007}, so the presence or potential variability of H$_{2}$O maser emission remains unclear. No CH$_{3}$OH maser emission was detected \citep{Wouterloot1993, Kalenskii1994}.
	
	ISOSS\,J23053+5953 is a target of the EPoS survey as well \citep{Ragan2012}. The authors derive a gas mass of 488\,$M_\odot$ within three resolved clumps and a kinematic distance of $d = 4.31^{+0.64}_{-0.62}$\,kpc at a systemic velocity of $-52.5$\,km\,s$^{-1}$ \citep[using the model by][]{Reid2009}. This is in agreement with other distance estimates, for example, by \citet[][$d = 4.82$\,kpc]{Yang2002}, \citet[][$d = 4.253$\,kpc]{Schlingman2011}, \citet[][$d = 4.44_{-0.70}^{+0.74}$\,kpc]{Ellsworth-Bowers2015a}, and the \citet[][$d = 5.07$\,kpc]{Planck2015}. With a galactocentric distance of $d_\mathrm{gal} \approx 10$\,kpc \citep{Wouterloot1989b, Ellsworth-Bowers2015a} the region is located in the outer galaxy as well.
	
	\citet{Birkmann2007} find a young and accreting massive protostar toward one of the clumps with an associated outflow. The line profile of the optically thick HCO$^{+}$ $3-2$ transition indicates infalling material with self-absorption in the red-shifted line wing compared the optically thin H$^{13}$CO$^{+}$ $3-2$ isotopologue.

	\citet{Pitann2011} studied the region using Spitzer observations with the InfraRed Array Camera \citep[IRAC,][]{Fazio2004} and the Multiband Imaging Photometer for Spitzer \citep[MIPS,][]{Rieke2004} instruments. These authors find that the region, harboring a cluster of sources, has extended polycyclic aromatic hydrocarbons (PAH) emission and warm dust components. In addition, the emission of forbidden transitions of Fe{\sc ii}, Si{\sc ii}, and S{\sc i} imply post-shocked gas that occurred by a J-shock. This is confirmed by the detection of high-energy H$_{2}$ transitions from $S$(0) to $S$(7) with excitation energies ranging between $510 - 7200$\,K. These authors do not find an indication of a photodissociation region (PDR).

	An overview of the 1.3\,mm observations with NOEMA, the core fragmentation properties, and a detailed analysis of the kinematic properties of ISOSS\,J22478+6357 and ISOSS\,J23053+5953 are presented in \citet{Beuther2021}. We expand the analysis with a detailed investigation of the molecular lines detected at 1.3\,mm. In addition, we used archival MIR and far-infrared (FIR) data to study the clustered nature of both regions. The MIR observations reveal emission of more evolved protostars while the FIR emission traces the cold dust emission.

	The paper is organized as follows: The NOEMA and Institut de RadioAstronomie Millim\'{e}trique (IRAM) 30m telescope observations and data calibration are described in Sect. \ref{sec:observations}. The analysis of the continuum data is given in Sect. \ref{sec:continuum}. The molecular line data are analyzed in Sect. \ref{sec:spectralline}. In Sect. \ref{sec:MUSCLE} we apply a physical-chemical model to a few line-rich cores. We discuss our results in Sect. \ref{sec:discussion} and a summary and conclusions are given in Sect. \ref{sec:conclusions}.
	
\section{Observations}\label{sec:observations}

\begin{figure*}
\includegraphics[]{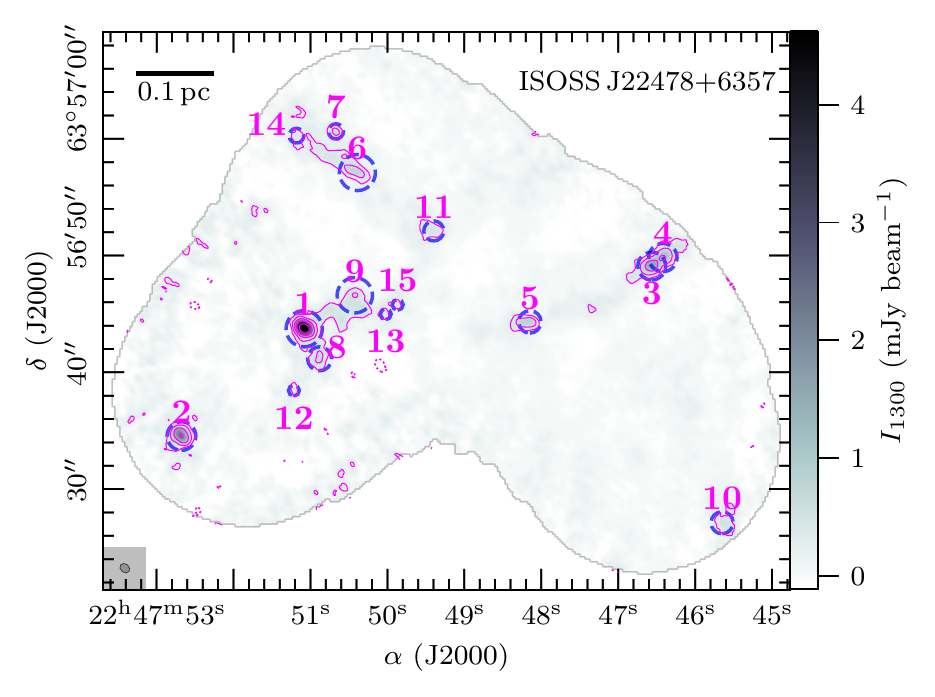}
\includegraphics[]{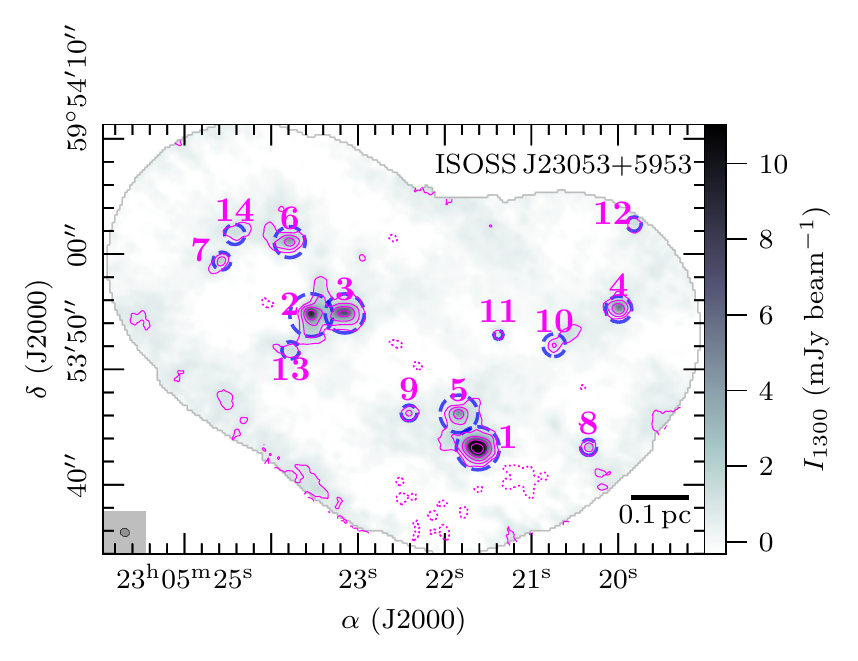}
\caption{NOEMA 1.3\,mm continuum emission of ISOSS\,J22478+6357 (\textit{left panel}) and ISOSS\,J23053+5953 (\textit{right panel}). The NOEMA continuum data are presented in color scale and pink contours. The dotted pink contour marks the $-5\sigma_\mathrm{cont}$ level. Solid pink contour levels are 5, 10, 20, 40, and 80$\sigma_\mathrm{cont}$. All mm cores identified in \citet{Beuther2021} are labeled in pink. The core outer radius $r_\mathrm{out}$ is indicated by a dashed blue circle. The synthesized beam is shown in the bottom left corner. A linear spatial scale of 0.1\,pc is indicated by a black scale bar.}
\label{fig:continuum}
\end{figure*}

\begin{table}
\caption{Detected emission lines.}
\label{tab:lineprops}
\centering
\begin{tabular}{lcccl}
\hline\hline
Molecule (Transition) & Rest & Einstein & Upper & Refer- \\
 & frequency & coefficient & energy & ence \\
 & & & level & \\
 & $\nu_{0}$ & log $A_{\mathrm{12}}$ & $E_{\mathrm{u}}$/$k_{\mathrm{B}}$ & \\
 & (GHz) & (log s$^{-1}$) & (K) & \\
\hline 
SO ($5_{5}-4_{4}$) & 215.221 & $-3.9$ & 44 & JPL\\ 
DCO$^{+}$ ($3-2$) & 216.113 & $-2.6$ & 21 & CDMS\\ 
H$_{2}$S ($2_{2,0}-2_{1,1}$) & 216.710 & $-4.3$ & 84 & CDMS\\ 
CH$_{3}$OH ($5_{1,4}-4_{2,3}E$) & 216.946 & $-4.9$ & 56 & CDMS\\ 
SiO ($5-4$) & 217.105 & $-3.3$ & 31 & JPL\\ 
c-C$_{3}$H$_{2}$ ($6_{1,6}-5_{0,5}$) & 217.822 & $-3.2$ & 39 & JPL\\ 
H$_{2}$CO ($3_{0,3}-2_{0,2}$) & 218.222 & $-3.6$ & 21 & JPL\\ 
HC$_{3}$N ($24-23$) & 218.325 & $-3.1$ & 131 & JPL\\ 
CH$_{3}$OH ($4_{2,3}-3_{1,2}E$) & 218.440 & $-4.3$ & 45 & CDMS\\ 
H$_{2}$CO ($3_{2,2}-2_{2,1}$) & 218.476 & $-3.8$ & 68 & JPL\\ 
H$_{2}$CO ($3_{2,1}-2_{2,0}$) & 218.760 & $-3.8$ & 68 & JPL\\ 
OCS ($18-17$) & 218.903 & $-4.5$ & 100 & JPL\\ 
C$^{18}$O ($2-1$) & 219.560 & $-6.2$ & 16 & JPL\\ 
HNCO ($10_{0,10}-9_{0,9}$) & 219.798 & $-3.8$ & 58 & CDMS\\ 
H$_{2}^{13}$CO ($3_{1,2}-2_{1,1}$) & 219.909 & $-3.6$ & 33 & JPL\\ 
SO ($6_{5}-5_{4}$) & 219.949 & $-3.9$ & 35 & JPL\\ 
CH$_{3}$OH ($8_{0,8}-7_{1,6}E$) & 220.079 & $-4.6$ & 97 & CDMS\\ 
$^{13}$CO ($2-1$) & 220.399 & $-6.2$ & 16 & JPL\\ 
CH$_{3}$CN ($12_{0}-11_{0}$) & 220.747 & $-3.2$ & 69 & JPL\\ 
CH$_{3}$OH ($8_{1,8}-7_{0,7}E$) & 229.759 & $-4.4$ & 89 & CDMS\\ 
CO ($2-1$) & 230.538 & $-6.2$ & 17 & JPL\\ 
OCS ($19-18$) & 231.061 & $-4.4$ & 111 & JPL\\ 
$^{13}$CS ($5-4$) & 231.221 & $-3.6$ & 33 & JPL\\ 
N$_{2}$D$^{+}$ ($3-2$) & 231.322 & $-2.7$ & 22 & JPL\\ 
SO$_{2}$ ($4_{2,2}-3_{1,3}$) & 235.152 & $-4.1$ & 19 & JPL\\ 
HC$_{3}$N ($26-25$) & 236.513 & $-3.0$ & 153 & JPL\\ 
\hline
\end{tabular}
\tablefoot{Line properties are taken from either the CDMS \citep{CDMS} or JPL \citep{JPL} databases. The CH$_{3}$CN $12_{0}-11_{0}$ transition is partially blended with CH$_{3}$CN $12_{1}-11_{1}$ at 220.743\,GHz.}
\end{table}

\begin{table*}
\caption{Overview of the merged (combined NOEMA + IRAM\,30m) spectral line data products.}
\label{tab:lineobs}
\centering
\begin{tabular}{llc|cccc|cccc}
\hline\hline
 & & & \multicolumn{4}{c|}{ISOSS\,J22478+6357} & \multicolumn{4}{c}{ISOSS\,J23053+5953}\\
\cline{4-7} \cline{7-11}
Molecule (Transition) & \texttt{CLEAN} & Spectral & \multicolumn{2}{c}{Synthesized} & Line noise & Detected? & \multicolumn{2}{c}{Synthesized} & Line noise & Detected? \\
 & algorithm & resolution & \multicolumn{2}{c}{beam} & & & \multicolumn{2}{c}{beam} & \\
\cline{4-5} \cline{8-9}
 & & $\delta \varv$ & $\theta_\mathrm{maj} \times \theta_\mathrm{min}$ & PA & $\sigma_\mathrm{line}$ & & $\theta_\mathrm{maj} \times \theta_\mathrm{min}$ & PA & $\sigma_\mathrm{line}$ & \\
 & & (km\,s$^{-1}$) & ($''\times''$) & ($^\circ$) & (K\,channel$^{-1}$) & & ($''\times''$) & ($^\circ$) & (K\,channel$^{-1}$) & \\
\hline 
SO ($5_{5}-4_{4}$) & SDI & 3.0 & 0.98$\times$0.77 & 50 & 0.072 & \cmark & 0.91$\times$0.78 & 58 & 0.061 & \cmark\\ 
DCO$^{+}$ ($3-2$) & Clark & 0.5 & 0.97$\times$0.77 & 50 & 0.17 & \cmark & 0.90$\times$0.78 & 60 & 0.17 & \cmark\\ 
H$_{2}$S ($2_{2,0}-2_{1,1}$) & Clark & 3.0 & 0.97$\times$0.77 & 51 & 0.067 & \xmark & 0.90$\times$0.78 & 59 & 0.065 & \cmark\\ 
CH$_{3}$OH ($5_{1,4}-4_{2,3}E$) & Clark & 3.0 & 0.96$\times$0.78 & 63 & 0.069 & \xmark & 0.90$\times$0.78 & 59 & 0.067 & \cmark\\ 
SiO ($5-4$) & SDI & 0.5 & 0.97$\times$0.77 & 50 & 0.21 & \cmark & 0.90$\times$0.78 & 59 & 0.19 & \cmark\\ 
c-C$_{3}$H$_{2}$ ($6_{1,6}-5_{0,5}$) & Clark & 3.0 & 0.97$\times$0.77 & 50 & 0.086 & \cmark & 0.89$\times$0.77 & 59 & 0.072 & \cmark\\ 
H$_{2}$CO ($3_{0,3}-2_{0,2}$) & SDI & 0.5 & 0.97$\times$0.78 & 50 & 0.23 & \cmark & 0.89$\times$0.77 & 59 & 0.17 & \cmark\\ 
HC$_{3}$N ($24-23$) & Clark & 0.5 & 0.98$\times$0.77 & 50 & 0.17 & \xmark & 0.89$\times$0.77 & 60 & 0.16 & \cmark\\ 
CH$_{3}$OH ($4_{2,3}-3_{1,2}E$) & Clark & 0.5 & 0.97$\times$0.77 & 51 & 0.17 & \cmark & 0.89$\times$0.77 & 59 & 0.16 & \cmark\\ 
H$_{2}$CO ($3_{2,2}-2_{2,1}$) & SDI & 0.5 & 0.97$\times$0.77 & 49 & 0.17 & \cmark & 0.89$\times$0.77 & 59 & 0.17 & \cmark\\ 
H$_{2}$CO ($3_{2,1}-2_{2,0}$) & SDI & 0.5 & 0.96$\times$0.80 & 68 & 0.17 & \cmark & 0.89$\times$0.77 & 59 & 0.19 & \cmark\\ 
OCS ($18-17$) & Clark & 0.5 & 0.97$\times$0.77 & 50 & 0.15 & \xmark & 0.89$\times$0.77 & 59 & 0.15 & \cmark\\ 
C$^{18}$O ($2-1$) & SDI & 0.5 & 0.92$\times$0.82 & 81 & 0.19 & \cmark & 0.89$\times$0.77 & 59 & 0.18 & \cmark\\ 
HNCO ($10_{0,10}-9_{0,9}$) & Clark & 3.0 & 0.96$\times$0.77 & 50 & 0.072 & \xmark & 0.89$\times$0.77 & 59 & 0.059 & \cmark\\ 
H$_{2}^{13}$CO ($3_{1,2}-2_{1,1}$) & Clark & 3.0 & 0.97$\times$0.77 & 49 & 0.078 & \xmark & 0.88$\times$0.77 & 60 & 0.056 & \cmark\\ 
SO ($6_{5}-5_{4}$) & SDI & 3.0 & 0.96$\times$0.77 & 50 & 0.085 & \cmark & 0.88$\times$0.77 & 61 & 0.062 & \cmark\\ 
CH$_{3}$OH ($8_{0,8}-7_{1,6}E$) & Clark & 3.0 & 0.96$\times$0.76 & 50 & 0.076 & \xmark & 0.88$\times$0.77 & 59 & 0.058 & \cmark\\ 
$^{13}$CO ($2-1$) & SDI & 0.5 & 0.96$\times$0.77 & 50 & 0.20 & \cmark & 0.88$\times$0.76 & 60 & 0.19 & \cmark\\ 
CH$_{3}$CN ($12_{0}-11_{0}$) & Clark & 0.5 & 0.96$\times$0.76 & 50 & 0.20 & \xmark & 0.88$\times$0.76 & 59 & 0.17 & \cmark\\ 
CH$_{3}$OH ($8_{1,8}-7_{0,7}E$) & Clark & 0.5 & 0.94$\times$0.74 & 49 & 0.17 & \cmark & 0.84$\times$0.74 & 63 & 0.18 & \cmark\\ 
CO ($2-1$) & SDI & 0.5 & 0.93$\times$0.74 & 48 & 0.25 & \cmark & 0.83$\times$0.73 & 62 & 0.29 & \cmark\\ 
OCS ($19-18$) & Clark & 0.5 & 0.93$\times$0.74 & 49 & 0.21 & \xmark & 0.83$\times$0.73 & 62 & 0.17 & \cmark\\ 
$^{13}$CS ($5-4$) & Clark & 0.5 & 0.93$\times$0.74 & 48 & 0.21 & \xmark & 0.83$\times$0.73 & 61 & 0.21 & \cmark\\ 
N$_{2}$D$^{+}$ ($3-2$) & Clark & 0.5 & 0.93$\times$0.74 & 48 & 0.21 & \cmark & 0.83$\times$0.73 & 61 & 0.19 & \cmark\\ 
SO$_{2}$ ($4_{2,2}-3_{1,3}$) & Clark & 0.5 & 0.92$\times$0.72 & 48 & 0.20 & \xmark & 0.82$\times$0.72 & 61 & 0.18 & \cmark\\ 
HC$_{3}$N ($26-25$) & Clark & 0.5 & 0.91$\times$0.72 & 49 & 0.24 & \xmark & 0.82$\times$0.71 & 61 & 0.23 & \cmark\\ 
\hline
\end{tabular}
\tablefoot{The line properties are summarized in Table \ref{tab:lineprops}. Emission lines with extended emission are \texttt{CLEANed} with the SDI algorithm \citep{Steer1984}, while species with compact emission are \texttt{CLEANed} with the Clark algorithm \citep{Clark1980}. Transitions covered by the high-resolution units are re-binned to a velocity resolution of 0.5\,km\,s$^{-1}$ and transitions only covered by the low-resolution units of PolyFiX have a velocity resolution of 3.0\,km\,s$^{-1}$. For each region, it is marked if the transition is detected (\cmark) or not (\xmark) in the line integrated intensity maps (Sect. \ref{sec:moment0}).}
\end{table*}

\begin{figure*}
\centering
\includegraphics[]{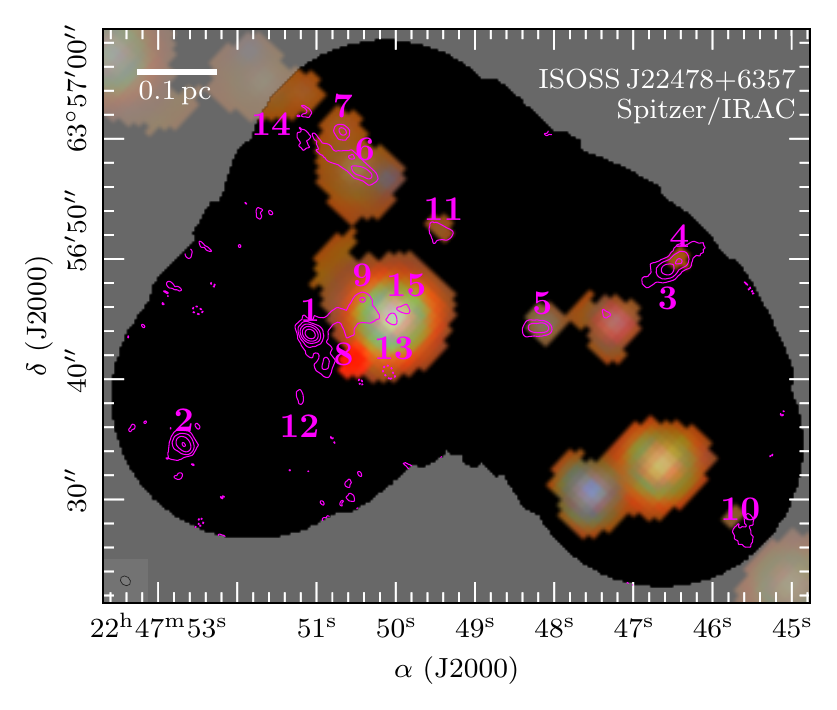}
\includegraphics[]{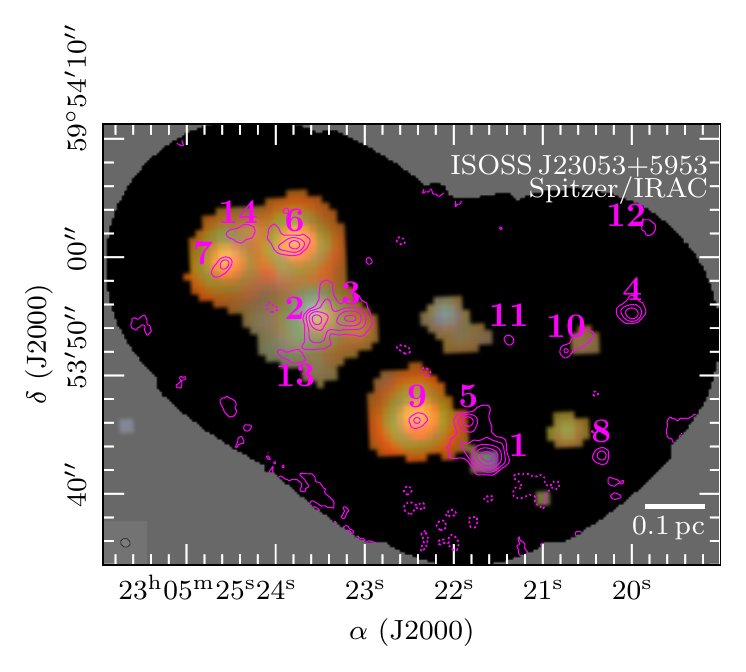}\\
\includegraphics[]{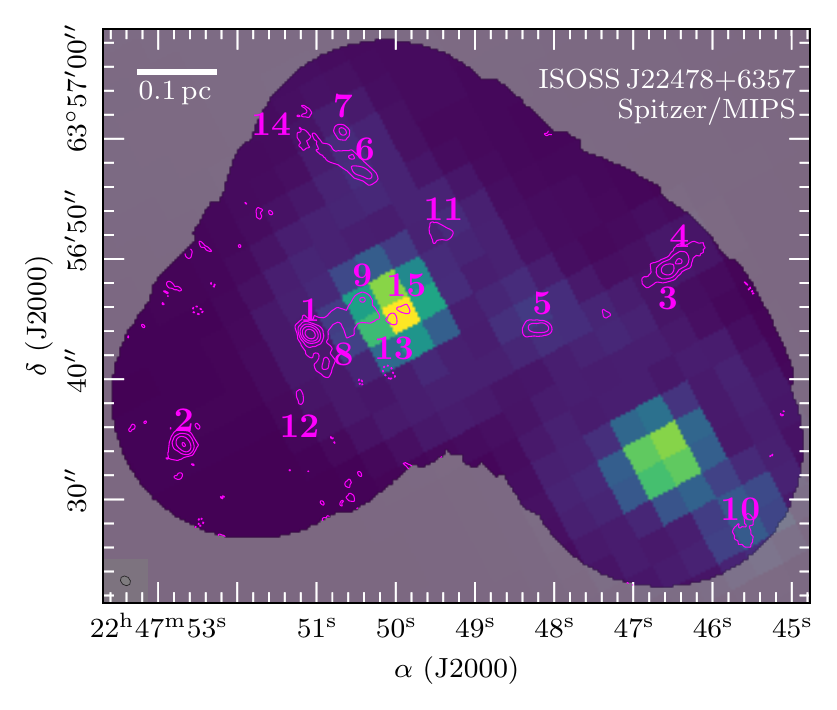}
\includegraphics[]{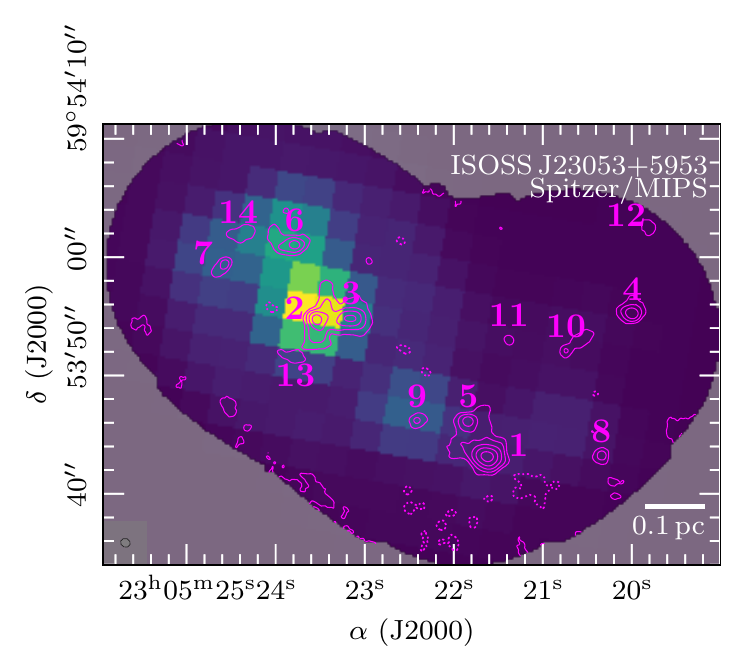}\\
\includegraphics[]{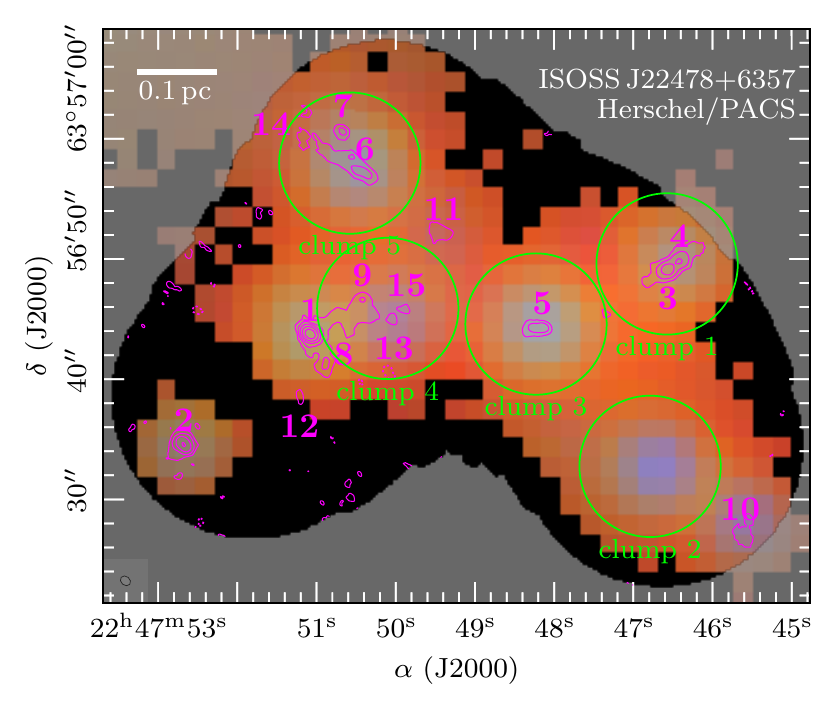}
\includegraphics[]{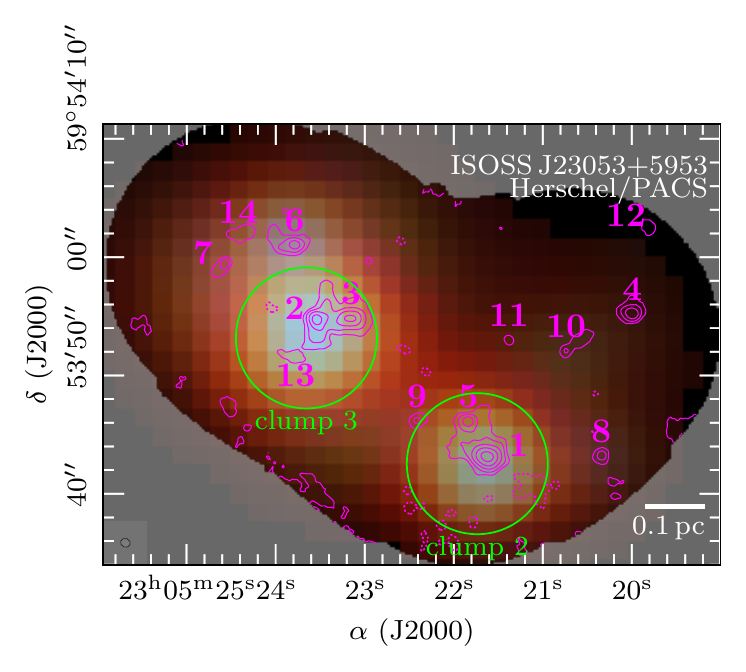}
\caption{Multi-wavelength overview of ISOSS\,J22478+6357 (\textit{left panel}) and ISOSS\,J23053+5953 (\textit{right panel}). The \textit{top panels} show a Spitzer IRAC composite RGB image at 4.5 (blue), 5.8 (green), and 8.0\,$\upmu$m (red). Emission with S/N $< 5$ above the sky background is clipped and shown in black. The \textit{middle panels} show Spitzer MIPS data at 24\,$\upmu$m. The \textit{bottom panels} show a composite RGB image of Herschel PACS observations at 70 (blue), 100 (green), and 160\,$\upmu$m (red). Emission with S/N $< 5$ above the sky background is clipped and shown in black. The Herschel clumps studied in \citet{Ragan2012} are indicated by green circles and labeled in green. In all panels, the NOEMA continuum data are shown in pink contours. The dotted pink contour marks the $-5\sigma_\mathrm{cont}$ level. Solid pink contour levels are 5, 10, 20, 40, and 80$\sigma_\mathrm{cont}$. The area outside of the NOEMA field-of-view is gray-shaded. All mm cores identified in \citet{Beuther2021} are labeled in pink. The synthesized beam of the NOEMA continuum data is shown in the bottom left corner. A linear spatial scale of 0.1\,pc is indicated by a white scale bar.}
\label{fig:overview}
\end{figure*}	
	
\subsection{CORE-extension data}
	The NOEMA mosaic observations were taken in February and March 2019 with ten antennas in the A, C, and D array configurations at 1.3\,mm (Band 3) using the PolyFiX correlator. ISOSS\,J22478+6357 and ISOSS\,J23053+5953 were observed with six and four NOEMA pointings, respectively. A summary of both regions, for example, the coordinates and velocities, is shown in Table \ref{tab:regions}. Complementary IRAM 30m observations in the same frequency range, to include short-spacing information of the spectral line data, were obtained using the Eight MIxer Receiver \citep[EMIR,][]{EMIR} in June 2019.
	
\subsubsection{NOEMA data}

	The NOEMA data were calibrated using the \texttt{CLIC} package in \texttt{GILDAS}\footnote{\url{https://www.iram.fr/IRAMFR/GILDAS/}}. The PolyFiX correlator simultaneously covers $\sim$8\,GHz in two sidebands (lower sideband, LSB, and upper sideband, USB) and in the two orthogonal linear polarizations (horizontal and vertical) with a fixed channel spacing of 2\,MHz ($\sim$2.7\,km\,s$^{-1}$ at 1.3\,mm). Rest frequencies from 213.3 GHz to 221.3 GHz in the LSB and from 228.7 GHz to 236.7 GHz in the USB were covered by the observations. High-resolution units with a channel spacing of 62.5\,kHz ($\sim$0.084\,km\,s$^{-1}$ at 1.3\,mm) were placed within the broadband correlator units.
	
	The NOEMA 1\,mm data of the original CORE sample consisted of single pointing observations toward each region, which allowed us to self-calibrate these data sets with the \texttt{GILDAS} \texttt{selfcal} task \citep[a detailed description of the method is presented in][]{Gieser2021}. Since self-calibration of mosaic observations is currently not possible, we use the standard calibrated NOEMA data for the CORE-extension data presented in \citet{Beuther2021} and in this work.
	
\subsubsection{IRAM 30m data}
	
	The EMIR data were calibrated using the \texttt{CLASS} package in \texttt{GILDAS}. The EMIR instrument consists of four basebands with a width of $\sim$4\,GHz in each baseband: the lower outer (LO, $213.5- 217.5$\,GHz), lower inner (LI, $217.3 - 221.3$\,GHz), upper inner (UI, $229.2 - 233.2$\,GHz), and upper outer (UO, $233.0 - 237.0$\,GHz) baseband. The chosen Fast Fourier Transform Spectrometer (FTS) backend delivers a channel separation of 200\,kHz ($\sim$0.27\,km\,s$^{-1}$ at 1.3\,mm). The half power beam width (HPBW) is 11\as8 in the lower basebands and 11\as2 in the upper basebands. After an initial inspection of the data, spiky channels caused by noise artifacts were filled with Gaussian noise. The antenna temperature $T_{\mathrm{A}}^{*}$ was converted to main beam temperature $T_{\mathrm{mb}}$ using $T_{\mathrm{mb}}=\frac{F_{\mathrm{eff}}}{B_{\mathrm{eff}}} \times T_{\mathrm{A}}^{*}$ with $F_{\mathrm{eff}}=0.915$ and $B_{\mathrm{eff}}=0.57$.

\subsubsection{Imaging}\label{sec:imaging}

	The NOEMA continuum and combined NOEMA + IRAM 30m (``merged'') spectral line data were imaged using the \texttt{MAPPING} package in \texttt{GILDAS}. Primary beam correction was applied to the final continuum and spectral line data products.
	
	 In a first inspection of the NOEMA spectral line data of both regions, we identified all detected emission lines. An overview of all lines and their properties is shown in Table \ref{tab:lineprops}. All of these emission lines were covered by the IRAM 30m EMIR observations, and hence the interferometric and single-dish data can be combined in order to recover missing flux filtered out by the interferometer. If available, we used the data obtained with the high-resolution units smoothed to a spectral resolution of 0.5\,km\,s$^{-1}$ to increase the signal-to-noise ratio (S/N) and otherwise with the low-resolution basebands smoothed to a spectral resolution of 3.0\,km\,s$^{-1}$ to increase the S/N.

	The continuum is extracted from the low-resolution spectral line data by masking out all channels with line emission using the \texttt{uv\_filter} and \texttt{uv\_continuum} tasks. The continuum data in the LSB and USB were merged with the task \texttt{uv\_merge}. The continuum data were \texttt{CLEANed} with natural weighting using the Clark algorithm \citep{Clark1980}. The synthesized beam ($\theta_{\mathrm{maj}}\times\theta_{\mathrm{min}}$, PA) and noise $\sigma_\mathrm{cont}$ of the continuum image are $0\as92\times0\as73$, $50^{\circ}$ and 0.057\,mJy\,beam$^{-1}$ for ISOSS\,J22478+6357 and $0\as84\times0\as74$, $ 67^{\circ}$ and 0.16\,mJy\,beam$^{-1}$ for ISOSS\,J23053+5953. Short-spacing information is only available for the spectral line data, so spatial filtering affects the continuum data. In \citet{Beuther2021} it is estimated that only $10-20\%$ of the total flux is recovered by the NOEMA observations.

	The continuum was subtracted from the spectral line data with the task \texttt{uv\_baseline}. For each visibility, a baseline is fitted and subtracted with all channels of line emission masked out (listed in Table \ref{tab:lineobs}). The NOEMA observations were combined with the IRAM 30m data using the \texttt{uvshort} task. 
	
	The continuum-subtracted spectral line data were \texttt{CLEANed} with natural weighting using the SDI algorithm \citep{Steer1984} for lines with extended emission within the field-of-view (all CO isotopologues, SO, SiO, and H$_{2}$CO) and with the Clark algorithm \citep{Clark1980} for lines with less extended emission. A detailed comparison of these two \texttt{CLEAN} algorithms applied to the CORE region W3\,IRS4 is presented in \citet{Mottram2020}. The properties of the spectral line data products are summarized in Table \ref{tab:lineobs}. The line noise $\sigma_{\mathrm{line}}$, computed in emission-free channels, is $\sim$0.2\,K\,channel$^{-1}$ and $\sim$0.07\,K\,channel$^{-1}$ in the high-resolution and low-resolution line data, respectively.
	
	\subsection{Archival data}\label{sec:archivaldata}
	
	In addition to the high angular resolution NOEMA data at 1.3\,mm, we used archival MIR to FIR observations obtained with the Spitzer and Herschel space telescopes to study the clustered nature of the regions and constrain the SEDs. For both regions, archival data covering the full field-of-view (FOV) of the NOEMA mosaic exists.

	In the FIR, both regions were targets in the Herschel key program EPoS and the properties of the cold dust are analyzed in \citet{Ragan2012}. Three photometric bands of the PACS instrument at 70, 100, and 160\,$\upmu$m, with angular resolutions of $\sim$5$''$, $\sim$7$''$, and $\sim$12$''$, respectively, cover the peak of the SED of cold dust emission. Science-ready data products were taken from the Herschel Science Archive.

	The Spitzer IRAC instrument has four photometric bands at 3.6, 4.5, 5.8, and 8.0\,$\upmu$m. The angular resolution is $1''-2''$, so comparable to the 1.3\,mm observations with NOEMA. The Spitzer MIPS observations at 24\,$\upmu$m have an angular resolution of $7''$. The 70\,$\upmu$m MIPS data were not used, as the Herschel PACS 70\,$\upmu$m data have a higher angular resolution. Science-ready data products were obtained from the Spitzer Heritage Archive. For a detailed description of the Spitzer observations, we refer to \citet{Hennemann2008} for ISOSS\,J22478+6357 and to \citet{Birkmann2007} and \citet{Pitann2011} for ISOSS\,J23053+5953.

\section{Continuum}\label{sec:continuum}

\begin{table*}
\caption{Overview of the analyzed mm cores.}
\label{tab:positions}
\centering
\begin{tabular}{lcccccccccccc}
\hline\hline
Core & \multicolumn{2}{c}{Coordinates} & & & & & & & & & CO & Herschel\\
\cline{2-3}
 & $\alpha$ & $\delta$ & $\varv_{\mathrm{LSR}}$ & $I_\mathrm{peak}$ & $F_{1300}$ & $r_\mathrm{out}$ & $T_\mathrm{kin}$ & $M_\mathrm{core}$ & $N$(H$_{2}$) & $\tau_{\mathrm{cont}}$ & outflow & clump\\
 & (J2000) & (J2000) & (km\,s$^{-1}$) & (mJy & (mJy) & (au) & (K) & ($M_\odot$) & ($10^{22}$ & ($10^{-3}$) & & \\
 & & & & beam$^{-1}$) & & & & & cm$^{-2}$) & & & \\
\hline
ISOSS\,J22478+6357 1 & 22:47:51.08 & +63:56:43.7 & $-39.8$ & 7.69 & 12.59 & 5007 & 42 & 1.84 & 27.4 & 7.0 & \cmark & $\ldots$\\ 
ISOSS\,J22478+6357 2 & 22:47:52.68 & +63:56:34.5 & $-39.7$ & 2.38 & 5.14 & 3967 & 13 & 3.33 & 37.5 & 9.8 & \cmark & $\ldots$\\ 
ISOSS\,J22478+6357 3 & 22:47:46.56 & +63:56:49.1 & $-39.0$ & 1.80 & 4.02 & 3841 & 33 & 0.78 & 8.5 & 2.2 & \cmark & $\ldots$\\ 
ISOSS\,J22478+6357 4 & 22:47:46.42 & +63:56:49.8 & $-39.1$ & 1.25 & 3.49 & 3915 & 22 & 1.1 & 9.6 & 2.5 & \cmark & clump 1\\ 
ISOSS\,J22478+6357 5 & 22:47:48.15 & +63:56:44.3 & $-40.1$ & 0.89 & 2.02 & 3108 & 37 & 0.34 & 3.7 & 0.9 & \cmark & clump 3\\ 
ISOSS\,J22478+6357 6 & 22:47:50.39 & +63:56:57.1 & $-40.5$ & 0.78 & 4.08 & 4986 & 27 & 0.97 & 4.5 & 1.2 & \xmark & clump 5\\ 
ISOSS\,J22478+6357 7 & 22:47:50.67 & +63:57:00.6 & $-41.0$ & 0.73 & 0.79 & 2096 & 10 & 0.76 & 17.1 & 4.5 & \xmark & $\ldots$\\ 
ISOSS\,J22478+6357 8 & 22:47:50.88 & +63:56:41.2 & $-39.0$ & 0.68 & 1.90 & 3326 & 27 & 0.47 & 4.1 & 1.0 & \xmark & $\ldots$\\ 
ISOSS\,J22478+6357 9 & 22:47:50.42 & +63:56:46.6 & $-40.7$ & 0.62 & 3.91 & 4891 & 10 & 3.76 & 14.5 & 3.8 & \xmark & $\ldots$\\ 
ISOSS\,J22478+6357 10 & 22:47:45.65 & +63:56:27.1 & $-39.5$ & 0.56 & 1.38 & 3060 & 10 & 1.32 & 13.1 & 3.5 & \xmark & $\ldots$\\ 
ISOSS\,J22478+6357 11 & 22:47:49.40 & +63:56:52.1 & $-39.3$ & 0.47 & 1.07 & 2768 & 18 & 0.45 & 4.8 & 1.2 & \xmark & $\ldots$\\ 
ISOSS\,J22478+6357 12 & 22:47:51.22 & +63:56:38.4 & $-38.8$ & 0.40 & 0.40 & 1510 & 57 & 0.04 & 1.0 & 0.3 & \xmark & $\ldots$\\ 
ISOSS\,J22478+6357 13 & 22:47:50.02 & +63:56:45.0 & $-40.4$ & 0.37 & 0.37 & 1510 & 15 & 0.19 & 4.7 & 1.2 & \xmark & clump 4\\ 
ISOSS\,J22478+6357 14 & 22:47:51.18 & +63:57:00.3 & $-40.9$ & 0.36 & 0.52 & 2039 & 27 & 0.13 & 2.2 & 0.6 & \xmark & $\ldots$\\ 
ISOSS\,J22478+6357 15 & 22:47:49.87 & +63:56:45.8 & $-39.9$ & 0.35 & 0.35 & 1510 & 20 & 0.12 & 3.0 & 0.8 & \xmark & $\ldots$\\ 
\hline 
ISOSS\,J23053+5953 1 & 23:05:21.62 & +59:53:43.2 & $-52.2$ & 18.38 & 62.67 & 8063 & 186 & 3.29 & 13.2 & 4.1 & \cmark & clump 2\\ 
ISOSS\,J23053+5953 2 & 23:05:23.54 & +59:53:54.7 & $-52.0$ & 9.82 & 35.39 & 8018 & 139 & 2.51 & 9.5 & 3.0 & \xmark & clump 3\\ 
ISOSS\,J23053+5953 3 & 23:05:23.15 & +59:53:54.9 & $-51.3$ & 7.55 & 30.09 & 7324 & 67 & 4.61 & 15.8 & 5.0 & \xmark & $\ldots$\\ 
ISOSS\,J23053+5953 4 & 23:05:19.99 & +59:53:55.2 & $-51.5$ & 5.88 & 11.76 & 4916 & 33 & 3.95 & 27.0 & 8.6 & \xmark & $\ldots$\\ 
ISOSS\,J23053+5953 5 & 23:05:21.84 & +59:53:46.1 & $-51.0$ & 4.92 & 18.08 & 7055 & 112 & 1.61 & 6.0 & 1.9 & \xmark & $\ldots$\\ 
ISOSS\,J23053+5953 6 & 23:05:23.79 & +59:54:01.0 & $-52.2$ & 4.57 & 13.02 & 5793 & 88 & 1.5 & 7.2 & 2.2 & \xmark & $\ldots$\\ 
ISOSS\,J23053+5953 7 & 23:05:24.57 & +59:53:59.4 & $-52.4$ & 2.32 & 3.53 & 3349 & 54 & 0.69 & 6.2 & 1.9 & \xmark & $\ldots$\\ 
ISOSS\,J23053+5953 8 & 23:05:20.34 & +59:53:43.2 & $-51.9$ & 2.29 & 2.94 & 2993 & 38 & 0.84 & 9.0 & 2.8 & \xmark & $\ldots$\\ 
ISOSS\,J23053+5953 9 & 23:05:22.41 & +59:53:46.2 & $-51.8$ & 1.88 & 2.52 & 2970 & 92 & 0.28 & 2.8 & 0.9 & \cmark & $\ldots$\\ 
ISOSS\,J23053+5953 10 & 23:05:20.74 & +59:53:52.1 & $-51.8$ & 1.73 & 4.76 & 4264 & 30 & 1.8 & 9.0 & 2.8 & \cmark & $\ldots$\\ 
ISOSS\,J23053+5953 11 & 23:05:21.38 & +59:53:53.0 & $-51.8$ & 1.50 & 1.50 & 1737 & 26 & 0.67 & 9.1 & 2.9 & \xmark & $\ldots$\\ 
ISOSS\,J23053+5953 12 & 23:05:19.82 & +59:54:02.6 & $-51.7$ & 1.46 & 1.89 & 2681 & 39 & 0.53 & 5.6 & 1.8 & \xmark & $\ldots$\\ 
ISOSS\,J23053+5953 13 & 23:05:23.78 & +59:53:51.7 & $-52.4$ & 1.39 & 2.56 & 3262 & 61 & 0.44 & 3.2 & 1.0 & \xmark & $\ldots$\\ 
ISOSS\,J23053+5953 14 & 23:05:24.42 & +59:54:01.7 & $-52.3$ & 1.29 & 3.51 & 3808 & 116 & 0.3 & 1.5 & 0.5 & \xmark & $\ldots$\\ 
\hline 
\end{tabular}
\tablefoot{The coordinates, peak intensity $I_\mathrm{peak}$, integrated flux $F_{1300}$, outer radius $r_\mathrm{out}$, $T_\mathrm{kin}$ obtained from the H$_{2}$CO rotation temperature map, core mass $M_\mathrm{core}$, and molecular hydrogen column density $N$(H$_{2}$) are taken from \citet{Beuther2021}. The systemic velocity $\varv_{\mathrm{LSR}}$ of each core is estimated from the optically thin C$^{18}$O $2-1$ line (Sect. \ref{sec:molecularcolumndensities}). The continuum optical depth $\tau_{\mathrm{cont}}$ is calculated according to Eq. \eqref{eq:opticaldepth} in Sect. \ref{sec:fragmentation}. The presence of protostellar outflows are investigated in Sect. \ref{sec:outflows} using CO $2-1$ line wing emission. Cores with an associated bipolar outflow are marked with a \cmark, whereas cores with no or no clear bipolar outflow signature are marked with a \xmark. The last column lists Herschel clumps studied by \citet{Ragan2012} associated with the mm cores (Fig. \ref{fig:overview}). We note that due to poor angular resolutions at FIR wavelengths, multiple mm cores can be embedded within a Herschel clump. Here we list for the mm cores the closest clump considering the clump peak intensity.}
\end{table*}

\begin{table*}
\caption{Overview of the physical and chemical properties of cores with S/N $> 20$ in the continuum data.}
\label{tab:cores}
\centering
\begin{tabular}{lccccccc}
\hline\hline
Core & $r_\mathrm{out}$ & $M_\mathrm{core}$ & $T_{500}$ & $q$ & $\alpha$ & $p$ & $\tau_\mathrm{chem}$\\
 & (au) & ($M_\odot$) & (K) & & & & (yr)\\
\hline
ISOSS\,J22478+6357 1 & 5007 & 1.84 & $105.8\pm9.4$ & $0.95\pm0.05$ & $-0.32\pm0.02$ & $1.73\pm0.05$ & 3.1$\times$10$^{4}$\\ 
ISOSS\,J22478+6357 2 & 3967 & 3.33 & $$\ldots$$ & $$\ldots$$ & $-0.34\pm0.10$ & $$\ldots$$ & $\ldots$\\ 
ISOSS\,J22478+6357 3 & 3841 & 0.78 & $57.5\pm20.1$ & $0.72\pm0.26$ & $$\ldots$$ & $$\ldots$$ & $\ldots$\\ 
ISOSS\,J22478+6357 4 & 3915 & 1.10 & $$\ldots$$ & $$\ldots$$ & $$\ldots$$ & $$\ldots$$ & $\ldots$\\ 
ISOSS\,J23053+5953 1 & 8063 & 3.29 & $247.0\pm22.8$ & $0.38\pm0.05$ & $-1.62\pm0.05$ & $1.00\pm0.07$ & 1.9$\times$10$^{4}$\\ 
ISOSS\,J23053+5953 2 & 8018 & 2.51 & $181.7\pm14.1$ & $0.14\pm0.03$ & $-0.78\pm0.05$ & $2.08\pm0.06$ & 1.9$\times$10$^{4}$\\ 
ISOSS\,J23053+5953 3 & 7324 & 4.61 & $$\ldots$$ & $$\ldots$$ & $-1.20\pm0.05$ & $$\ldots$$ & $\ldots$\\ 
ISOSS\,J23053+5953 4 & 4916 & 3.95 & $$\ldots$$ & $$\ldots$$ & $-0.64\pm0.09$ & $$\ldots$$ & $\ldots$\\ 
ISOSS\,J23053+5953 5 & 7055 & 1.61 & $99.8\pm17.1$ & $0.14\pm0.10$ & $-0.99\pm0.06$ & $1.87\pm0.12$ & $\ldots$\\ 
ISOSS\,J23053+5953 6 & 5793 & 1.50 & $153.2\pm5.5$ & $0.29\pm0.02$ & $-1.18\pm0.07$ & $1.53\pm0.07$ & 1.8$\times$10$^{4}$\\ 
\hline 
\end{tabular}
\tablefoot{The outer radius $r_\mathrm{out}$ and core mass $M_\mathrm{core}$ are taken from \citet{Beuther2021}. The radial temperature profile (Eq. \ref{eq:temperatureprofile}), described by $T_{500}$, the temperature at a radius of 500\,au, and the power-law index $q$, are derived from the H$_{2}$CO temperature maps in Sect. \ref{sec:temperatureprofile}. The density power-law index $p$ is derived from the 1.3\,mm continuum visibilities in Sect. \ref{sec:density}. The chemical age $\tau_\mathrm{chem}$ of cores, with eight or more detected species in the spectral line data and for which both density and temperature profiles could be derived, is estimated in Sect. \ref{sec:chemicaltimescales}.}
\end{table*}

\begin{table*}
\caption{Photometric data points of mm cores with associated Spitzer IRAC sources in at least three bands (Fig. \ref{fig:overview}) and SED fit results (Fig. \ref{fig:SED}).}
\label{tab:photometry}
\centering
\begin{tabular}{lccccc|ccc||ccccccc||c}
\hline\hline
Core & \multicolumn{5}{c|}{Spitzer} & \multicolumn{3}{c||}{Herschel} & \multicolumn{7}{|c||}{SED fit} &\\
\cline{2-6}\cline{7-9}\cline{10-16}
 & \multicolumn{4}{c}{IRAC} & MIPS & \multicolumn{3}{c||}{PACS} & \multicolumn{7}{|c||}{} & \\
 & $F_{3.6}$ & $F_{4.5}$ & $F_{5.8}$ & $F_{8.0}$ & $F_{24}$ & $F_{70}$ & $F_{100}$ & $F_{160}$ & $T_{1}$ & $L_{1}$ & $T_{2}$ & $L_{2}$ & $T_{3}$ & $L_{3}$ & $L$ & $M$/$L$\\
 & (mJy) & (mJy) & (mJy) & (mJy) & (mJy) & (Jy) & (Jy) & (Jy) & (K) & ($L_{\odot}$) & (K) & ($L_{\odot}$) & (K) & ($L_{\odot}$) & ($L_{\odot}$) & ($M_{\odot}$/$L_{\odot}$)\\
\hline
ISOSS\,J22478+6357 4 & $\ldots$ & 0.3 & 0.8 & 1.2 & $\ldots$ & 0.2 & 0.5 & 2.1 & 375 & 0.3 & 20 & 7.0 & $\ldots$ & $\ldots$ & 7.4 & $1.5\times 10^{-1}$\\ 
ISOSS\,J22478+6357 5 & 1.5 & 2.5 & 3.3 & 3.8 & $\ldots$ & 0.3 & 0.8 & 1.4 & 548 & 1.0 & 24 & 7.9 & $\ldots$ & $\ldots$ & 8.9 & $3.8\times 10^{-2}$\\ 
ISOSS\,J22478+6357 6 & $\ldots$ & 2.8 & 4.6 & 6.3 & $\ldots$ & 0.3 & 0.6 & 1.2 & 464 & 1.5 & 21 & 7.5 & $\ldots$ & $\ldots$ & 9.0 & $1.1\times 10^{-1}$\\ 
ISOSS\,J22478+6357 13 & 44.7 & 67.0 & 95.0 & 105.4 & 234.0 & $\ldots$ & $\ldots$ & $\ldots$ & 552 & 27.3 & 54 & 47.0 & $\ldots$ & $\ldots$ & 74.3 & $2.6\times 10^{-3}$\\ 
ISOSS\,J23053+5953 1 & 0.7 & 1.6 & 1.6 & 1.1 & $\ldots$ & 7.6 & 22.5 & 35.9 & 620 & 0.7 & 23 & 379.3 & $\ldots$ & $\ldots$ & 380.1 & $8.7\times 10^{-3}$\\ 
ISOSS\,J23053+5953 2 & 2.7 & 8.1 & 11.5 & 12.4 & 1190.0 & 18.2 & 44.1 & 67.0 & 479 & 5.7 & 109 & 83.0 & 28 & 706.6 & 795.3 & $3.2\times 10^{-3}$\\ 
ISOSS\,J23053+5953 9 & 1.5 & 6.6 & 14.7 & 24.3 & 391.0 & $\ldots$ & $\ldots$ & $\ldots$ & 374 & 11.4 & 48 & 333.7 & $\ldots$ & $\ldots$ & 345.1 & $8.1\times 10^{-4}$\\ 
\hline 
\end{tabular}
\tablefoot{The photometry of the Spitzer IRAC and MIPS data is described in Sect. \ref{sec:SED}. The Herschel PACS fluxes are taken from \citet{Ragan2012} by cross-matching the position of the clumps and the mm cores (Table \ref{tab:positions}). Due to poor angular resolutions at FIR wavelengths, core 13 in ISOSS\,J22478+6357 and core 9 in ISOSS\,J23053+5953 cannot be assigned to a Herschel clump and therefore their SEDs are not fitted in the FIR regime. In order to fit the full SED of the cores, two or three components are required. The temperature $T_{i}$ and luminosity $L_{i}$ of the $i$-th component, as well as the total bolometric luminosity $L = \sum_{i} L_{i}$, are presented. The $M$/$L$ ratio is calculated from the core mass $M_\mathrm{core}$ divided by the total bolometric luminosity $L$.}
\end{table*}

\begin{figure*}[!htb]
\centering
\includegraphics[]{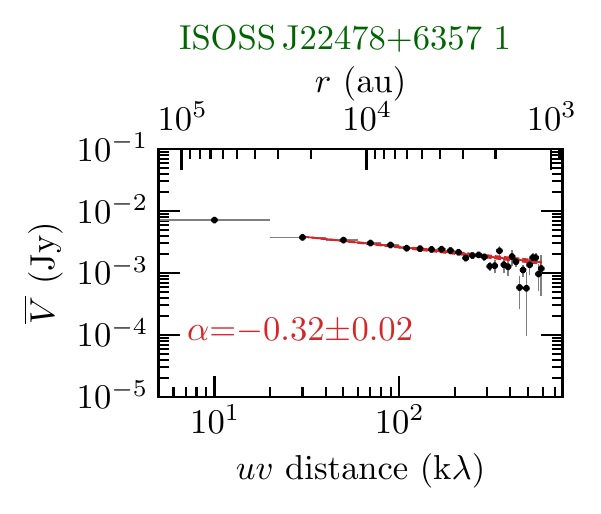}
\includegraphics[]{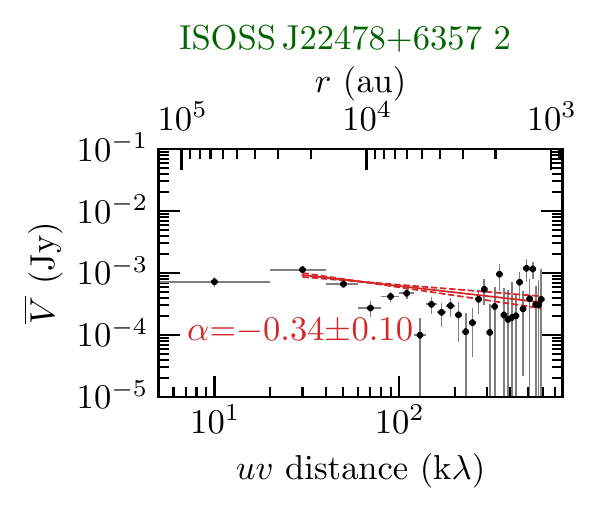}
\includegraphics[]{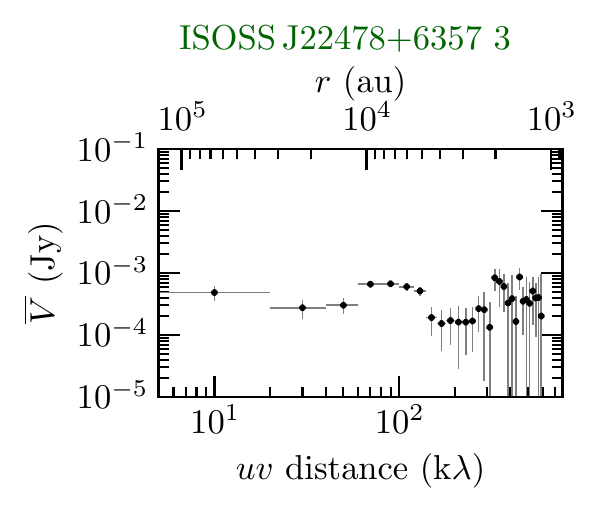}
\includegraphics[]{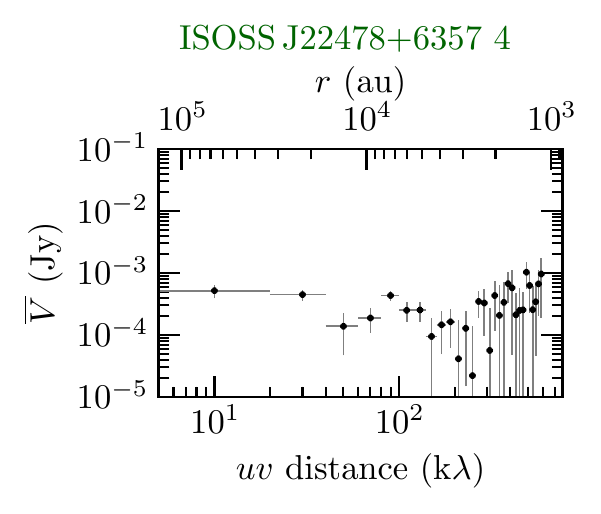}
\includegraphics[]{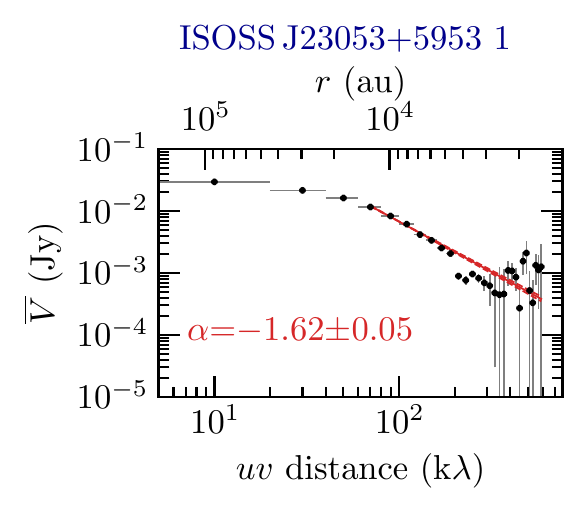}
\includegraphics[]{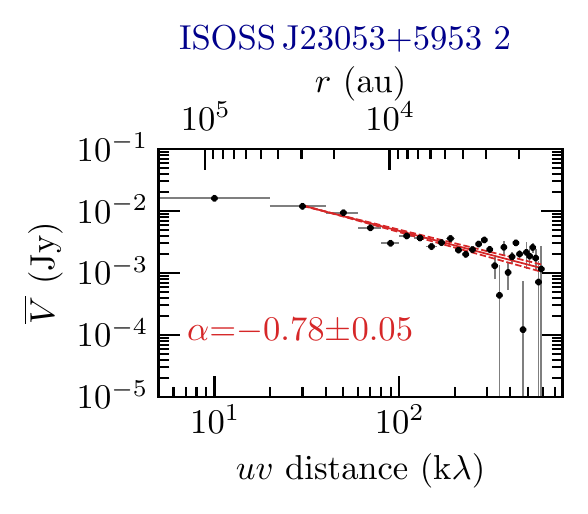}
\includegraphics[]{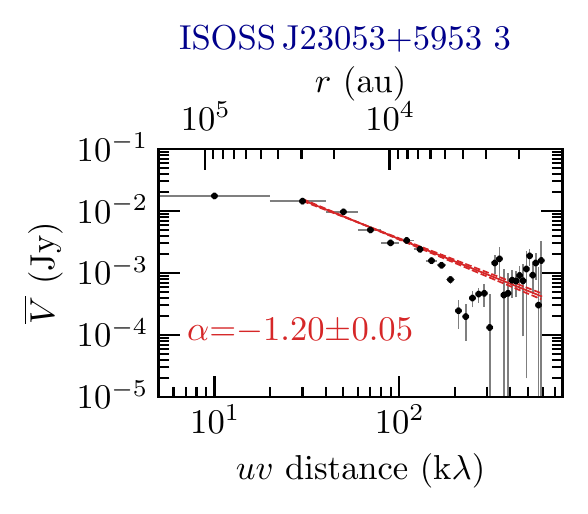}
\includegraphics[]{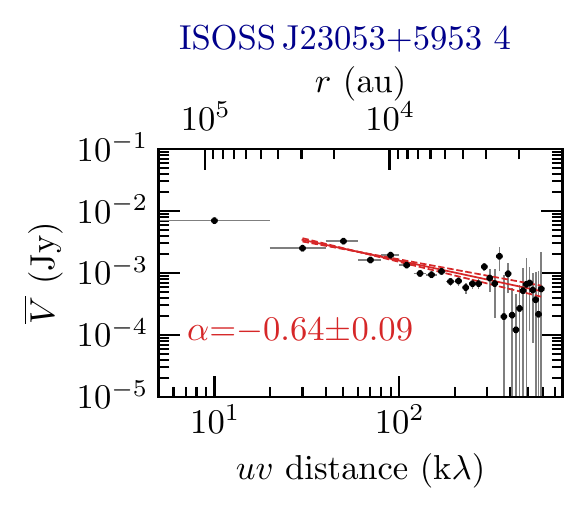}
\includegraphics[]{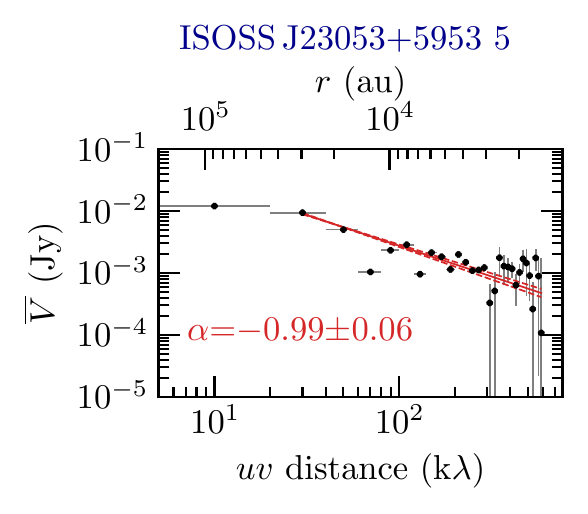}
\includegraphics[]{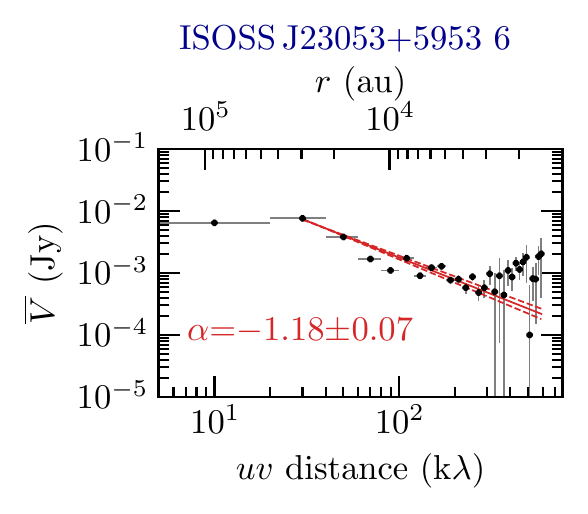}
\caption{Radial visibility profiles of cores $1-4$ in ISOSS\,J22478+6357 and cores $1-6$ in ISOSS\,J23053+5953. The azimuthally averaged visibility amplitudes $\overline V$, considering the real and imaginary components, are shown in black. The bottom x-axis shows the $uv$ distance and the top x-axis is converted to the corresponding linear spatial scale. A fit to the data and its uncertainties ($\pm$1$\sigma$) are shown by the red solid and dashed lines, respectively.}
\label{fig:visibilityprofile}
\end{figure*}

	The NOEMA 1.3\,mm continuum data reveal the compact dust emission in the two star-forming regions. The subarcsecond resolution achieved with NOEMA allows us to study individual fragmented millimeter (mm) cores. In this section, the core properties are analyzed using the NOEMA 1.3\,mm continuum emission in combination with archival MIR and FIR data.

\subsection{Fragmentation properties}\label{sec:fragmentation}

	The NOEMA 1.3\,mm continuum data are shown in Fig. \ref{fig:continuum} for ISOSS\,J22478+6357 and ISOSS\,J23053+5953. In both regions, the continuum data reveal a large number of fragmented mm cores. In total, ten bright mm cores with S/N $> 20$ are detected, four in ISOSS\,J22478+6357 and six in ISOSS\,J23053+5953.
	
	The fragmentation properties are studied in detail in \citet{Beuther2021}. By applying the \texttt{clumpfind} algorithm \citep{Williams1994} to the continuum data, 15 and 14 individual mm cores can be identified for ISOSS\,J22478+6357 and ISOSS\,J23053+5953, respectively. The cores are labeled in Fig. \ref{fig:continuum}. In this study, we investigate the physical and chemical properties of these 29 cores. An overview and summary of the results from \citet{Beuther2021}, including the core position, 1.3\,mm peak intensity $I_\mathrm{peak}$, 1.3\,mm integrated flux $F_{1300}$, and outer radius $r_\mathrm{out}$ derived with \texttt{clumpfind}, is shown in Table \ref{tab:positions}. 
	
	A comparison of the NOEMA 1.3\,mm continuum and archival MIR and FIR observations obtained with the Spitzer and Herschel space telescopes is presented in Fig. \ref{fig:overview}. The Spitzer IRAC and MIPS observations highlight evolved protostars. The Herschel PACS observations at 70, 100, and 160\,$\upmu$m trace the cold dust emission on clump scales. The mm cores have counterparts not only at FIR wavelengths, such that they are embedded in the cold clumps, but there are also a few cores that are MIR-bright (e.g., mm core 13 in ISOSS\,J22478+6357 and mm core 2 in ISOSS\,J23053+5953). The SED of mm cores with associated Spitzer IRAC sources are constructed and modeled in Sect. \ref{sec:SED}.
	
	Based on the 1.3\,mm continuum peak intensity $I_\mathrm{peak}$ and integrated flux $F_{1300}$ of each core, \citet{Beuther2021} calculate the molecular hydrogen column density $N$(H$_{2}$) and core mass $M_\mathrm{core}$ assuming optically thin dust emission. For the kinetic temperature $T_\mathrm{kin}$, the H$_{2}$CO rotation temperature $T$(H$_{2}$CO) was used assuming local thermal equilibrium (LTE) conditions (Sect. \ref{sec:XCLASSMapFitting}). In order to check if the assumption of optically thin dust emission is valid, we computed the 1.3\,mm continuum optical depth $\tau_{\mathrm{cont}}$ at the position of the peak intensity of each core with
	
	\begin{equation}
	\label{eq:opticaldepth}
	\tau_{\mathrm{cont}} = -\mathrm{ln}\bigg( 1 - \frac{I_\mathrm{peak}}{ B_{\nu}(T_\mathrm{kin})} \bigg),
	\end{equation}
	
	where $B_{\nu}$ is the Planck function.	
	The continuum optical depth, $\tau_{\mathrm{cont}}$, as well as $N$(H$_{2}$), $M_\mathrm{core}$, and $T_\mathrm{kin}$ for each core are summarized in Table \ref{tab:positions}. The optical depth is $< $0.01 for all 29 cores; therefore, the assumption of optically thin dust emission is valid.
	
	The core masses $M_\mathrm{core}$, derived from the integrated 1.3\,mm flux, vary between 0.04 and 4.61\,$M_\odot$ (Table \ref{tab:positions}). This indicates that most of the mm cores form low- and intermediate-mass stars. The sum of all core masses is 17\,$M_\odot$ and 23\,$M_\odot$ in ISOSS\,J22478+6357 and ISOSS\,J23053+5953, respectively. The mass estimates using single-dish observations \citep{Ragan2012,Bihr2015} are a factor of $10-30$ higher than estimated from the 1.3\,mm NOEMA data. The interferometric observations of the dust continuum suffer from spatially filtering the extended emission, the estimated core masses are therefore only lower limits. The single-dish observations at FIR wavelengths from Herschel PACS and JCMT SCUBA reveal that both regions have a large gas mass reservoir of a few 100\,$M_\odot$ \citep{Ragan2012,Bihr2015}. This provides a mass reservoir for further growth and indicates that mass assembly is not complete. Bright MIR sources suggest the presence of at least intermediate-mass protostars in both regions. 
		
	\citet{Birkmann2007} derive core masses of 26, 4.4, and 4.4\,$M_\odot$ for cores 1, 2, and 3, respectively, in ISOSS\,J23053+5953 using 1.3\,mm observations. These values are higher than the estimates by \citet{Beuther2021}. As in the calculation by \citet{Birkmann2007} the assumed kinetic temperature was derived from lower dust temperatures based on the single-dish observations. The estimated core mass of mm core 13 in ISOSS\,J22478+6357 is low (0.44\,$M_\odot$), but \citet{Hennemann2008} infers that this MIR source (their ``source 3'') has a central mass of $6-8.5$\,$M_\odot$ using the YSO SED models by \citet{Robitaille2007}. By applying the same SED models to mm core 2 (their ``SMM East'') in ISOSS\,J23053+5953, \citet{Pitann2011} estimate a central mass of $4.1-6.6$\,$M_\odot$, whereas \citet{Beuther2021} derive a core mass of 2.51\,$M_\odot$.
	
	The calculation of the core mass depends strongly on the assumed temperature. In \citet{Beuther2021}, core masses are also derived assuming a uniform temperature of 20\,K. In cases when the kinetic temperature is actually higher, derived from the H$_{2}$CO rotation temperature maps (Sect. \ref{sec:XCLASSMapFitting}), the core masses are overestimated by a factor of a few. While the assumption of a low temperature at clump scales is valid, the high angular resolution data reveals that some mm cores already harbor protostars that heat up the surrounding envelope. Differences in the mass estimates are also caused by a higher assumed gas-to-dust mass ratio \citep[150,][]{Beuther2021} instead of the canonical value of 100 used in the literature. Given the mass reservoir and that the regions are in an early evolutionary stage and thus gas accretion will continue for a few $10^{5}-10^{6}$\,yr, some of the low-mass cores might form intermediate- to high-mass protostars in the future.

\subsection{Radial density profiles}\label{sec:density}

	Observations of individual high-mass cores suggest that the density profiles of the gas and dust envelope around the forming protostars can be described by power-law profiles with power-law indices $p$ typically ranging from 1.5 to 2.0 \citep[e.g.,][]{Beuther2007B,Zhang2009,Palau2014,Palau2021,Gieser2021,Gomez2021}. The density profiles are described as
	
	\begin{equation}
	n(r) = n_{\mathrm{in}} \times \bigg( \frac{r}{r_{\mathrm{in}}} \bigg) ^{-p},
	\label{eq:densityprofile}
	\end{equation}
	
	with density $n_{\mathrm{in}}$ and arbitrary radius $r_{\mathrm{in}}$.
	The density profiles can be inferred, for example, from the radial intensity profiles of the dust continuum emission \citep[][]{vanderTak2000,Beuther2002B,Palau2014}. While the aforementioned studies are based on single-dish observations, the interferometric continuum data suffer from significantly spatially filtering the extended emission \citep{Beuther2021}. This effect can cause intensity profiles in the image domain to be much steeper than in reality and result in too steep ($p > 3$) density profiles. In order to overcome this problem, we therefore analyze the continuum data in the $uv$ plane - excluding the regime of missing baselines - in order to estimate the power-law index $p$ of each core. The same method to infer the core density profiles as presented in \citet{Gieser2021} is applied to the mm cores. Assuming optically thin dust emission (which is valid for all cores, Table \ref{tab:positions}), the power-law index $p$ can be calculated from the power-law index of the azimuthally averaged visibility amplitudes $\overline V \sim s^{\alpha}$ (taking into account both the real and imaginary components), with $uv$ distance $s$ \citep[e.g.,][]{Adams1991, Looney2003}. The density power-law profile then equals to
	
	\begin{equation}
	p = \alpha - q + 3,
	\label{eq:p}
	\end{equation}
	
	where $q$ is the power-law index of the temperature profile. The temperature profile $T$($r$) of the cores is also assumed to have a spherically symmetric distribution and is described as
	
	\begin{equation}
	T(r) = T_{\mathrm{in}} \times \left(\frac{r}{r_\mathrm{in}}\right)^{-q},
	\label{eq:temperatureprofile}
	\end{equation}
	with temperature $T_\mathrm{in} = T(r_\mathrm{in})$ and arbitrary radius $r_\mathrm{in}$. The temperature profiles are estimated from the H$_{2}$CO temperature maps in Sect. \ref{sec:temperatureprofile}.
	
	Many of the fainter mm cores are only marginally or not resolved. In order to reliably derive the radial density and temperature profiles, we restrict the analysis to mm cores that are detected with S/N $> 20$ in the 1.3\,mm continuum data. This is the case for cores $1-4$ in ISOSS\,J22478+6357 and cores $1-6$ in ISOSS\,J23053+5953.
	
	The azimuthally averaged visibility amplitude $\overline V$ and its standard deviation were computed for each core with the \texttt{uvamp} task in \texttt{MIRIAD} \citep{miriad} considering the real and imaginary components. The phase center was shifted to the location of the core (Table \ref{tab:positions}). The visibilities were binned in a bin size of 20\,k$\lambda$ up to 600\,k$\lambda$, which corresponds to the longest NOEMA baseline (774\,m).
	
	The computation of the azimuthally averaged visibility amplitudes of a certain core can be influenced by other mm-bright cores in the field (e.g., in ISOSS\,J23053+5953 mm core 2 is close to mm core 3, Fig. \ref{fig:continuum}). Thus, before the radial visibility amplitudes of a certain core were computed, we first subtracted the emission of the remaining four brightest cores in the region. Using the \texttt{uv\_fit} task in \texttt{GILDAS MAPPING}, we found that the emission of the mm cores is modeled best by fitting two circular Gaussian functions with a full width half maximum (FWHM) of 0\as5, and 2\as0, respectively, in order to take into account both compact and extended emission. The flux of the two components was varied for each core separately in order to take into account that the cores have a varying morphology (e.g., compact core or a compact core embedded in a more extended envelope). The model emission of the four remaining brightest mm cores, described each by the two Gaussian functions, was then subtracted from the original $uv$ data set. We carefully checked, for each core, that the emission of the four brightest remaining cores was subtracted correctly by imaging the cores-subtracted $uv$ data set.
	
	As an example, after core 3 in ISOSS\,J23053+5953, the four remaining brightest cores within the FOV are cores 1, 2, 4, and 5 (Table \ref{tab:positions}). We first fitted and subtracted the model emission of cores 1, 2, 4, and 5 with the \texttt{uv\_fit} task. We imaged this cores-subtracted $uv$ data set and validated that the emission of cores 1, 2, 4, and 5 was removed without affecting the emission of core 3. Then we shifted the phase center in the cores-subtracted $uv$ data to the location of core 3 and computed the azimuthally averaged visibility amplitudes using the \texttt{uvamp} task as described above.
	
	The azimuthally averaged visibility profiles for cores $1-4$ in ISOSS\,J22478+6357 and cores $1-6$ in ISOSS\,J23053+5953 are shown in Fig. \ref{fig:visibilityprofile}. The radial visibility profiles of most of the cores follow a single power-law profile. Only at the shortest baselines, at $0-20$\,k$\lambda$, the profiles flatten for most of the resolved cores. The flattening at short $uv$ distances is caused by spatial filtering since the shortest baseline is $\sim$18\,m. In general, at large $uv$ distances ($> 300$\,k$\lambda$), there is an increased scatter of the data points due to the fact that the number of long baselines of the NOEMA interferometer is smaller compared to the number of short baselines. This also causes an increase in the uncertainties of the binned data points. Cores 3 and 4 in ISOSS\,J22478+6357 have a flat profile, since these cores are only marginally resolved (Fig. \ref{fig:continuum}). The visibility profile of core 1 in ISOSS\,J23053+5953 flattens at $uv$ distances $<$60\,k$\lambda$. A large bipolar molecular outflow is observed toward this core (Sect. \ref{sec:outflows}) that might impact the radial profile of the dust envelope.
	
	 Cores 3 and 4 in ISOSS\,J22478+6357 and cores 2 and 3 in ISOSS\,J23053+5953 are very close, 1\as2 (3\,900\,au) and 2\as9 (12\,000\,au), respectively (Fig. \ref{fig:continuum}). The subtraction of the emission of the nearby cores is difficult in these cases, since the cores are embedded within a common envelope.
	
	In order to estimate the density profile according to Eq. \eqref{eq:p}, we fitted the observed visibility profiles with a power-law profile for $uv$ distances ranging between $20 - 600$\,k$\lambda$ for all cores except for core 1 in ISOSS\,J23053+5953. Here, we fitted the visibility profile between $60 - 600$\,k$\lambda$ not considering the flat distribution toward the smaller baselines. Since the visibility profiles of cores 3 and 4 ISOSS\,J22478+6357 are flat, we do not fit their profiles. The temperature power-law index $q$, required for the calculation of $p$, is derived in Sect. \ref{sec:temperatureprofile} and varies between 0.1 and 1 (Table \ref{tab:cores}). 
	
	The results for $\alpha$ and $p$ are summarized in Table \ref{tab:cores}. The density power-law index ranges between $1 - 2$. A detailed discussion and comparison with the density profiles of 22 cores in the more evolved CORE sample \citep{Gieser2021} is given in Sect. \ref{sec:discore}.

\subsection{Spectral energy distribution}\label{sec:SED}

\begin{figure*}
\centering
\includegraphics[]{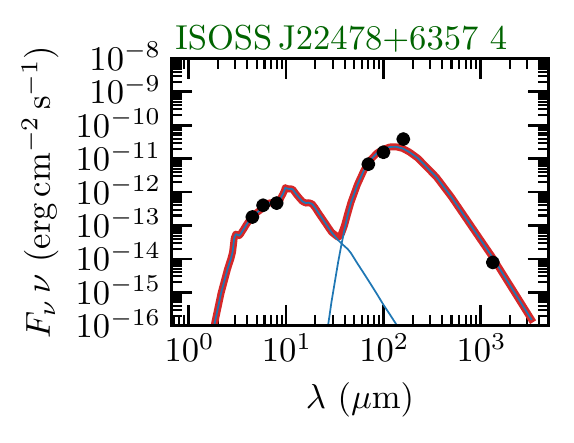}
\includegraphics[]{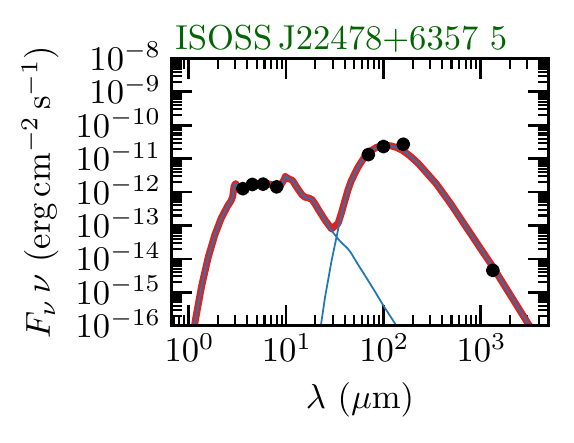}
\includegraphics[]{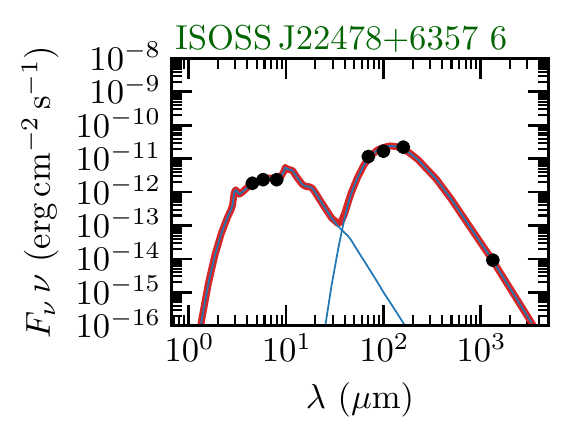}
\includegraphics[]{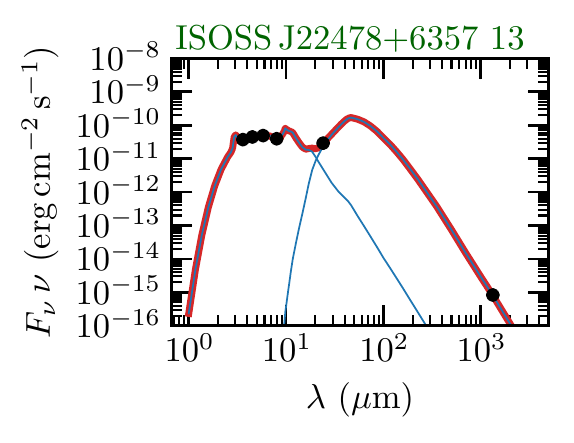}
\includegraphics[]{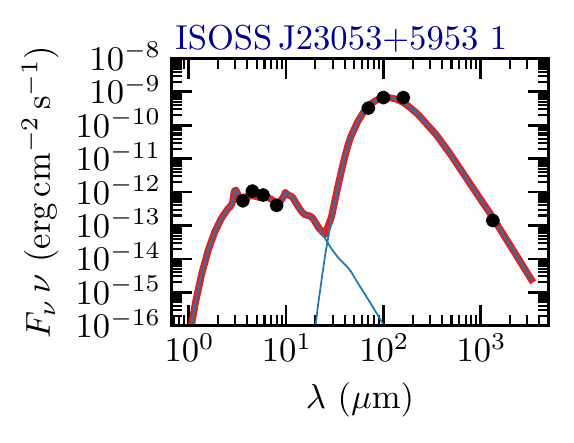}
\includegraphics[]{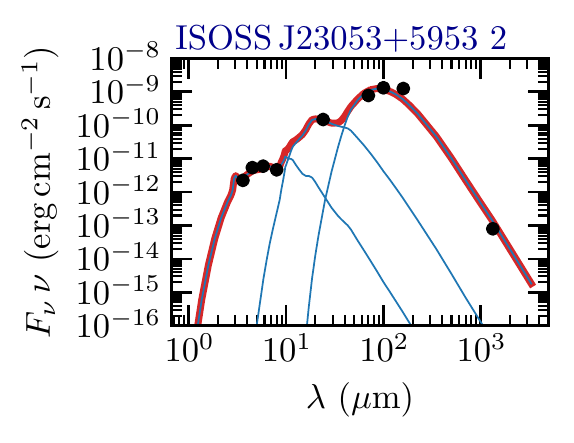}
\includegraphics[]{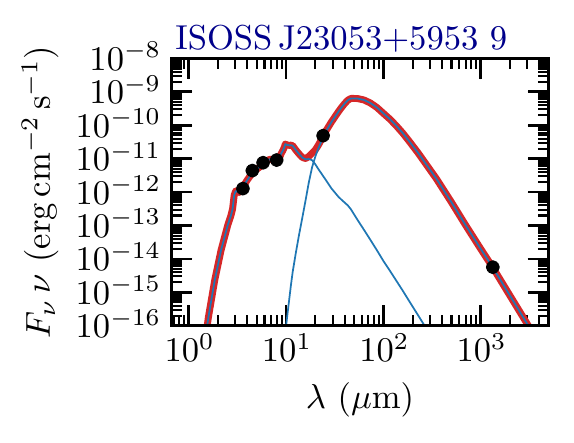}
\caption{SED of mm cores with associated Spitzer IRAC emission in at least three bands. The fluxes (black points) are obtained from Spitzer IRAC (3.6, 4.5, 5.8, and 8.0\,$\upmu$m), Spitzer MIPS (24\,$\upmu$m), Herschel PACS (70, 100, and 160\,$\upmu$m), and NOEMA 1.3\,mm observations (Table \ref{tab:positions} and \ref{tab:photometry}). The total SED fit is shown by a red line and individual components are indicated by blue lines.}
\label{fig:SED}
\end{figure*}

	\begin{figure*}
\centering
\includegraphics[]{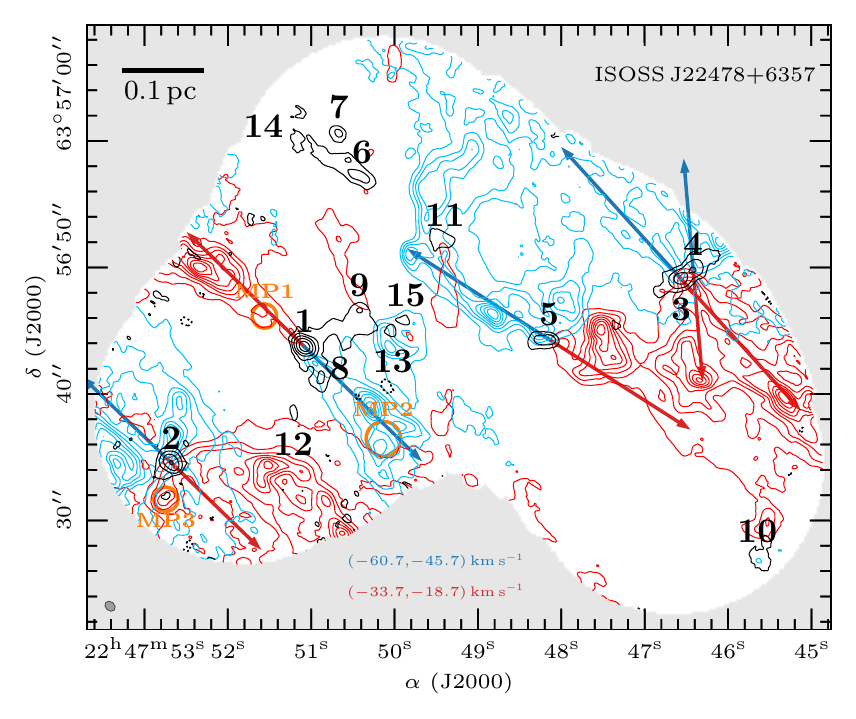}
\includegraphics[]{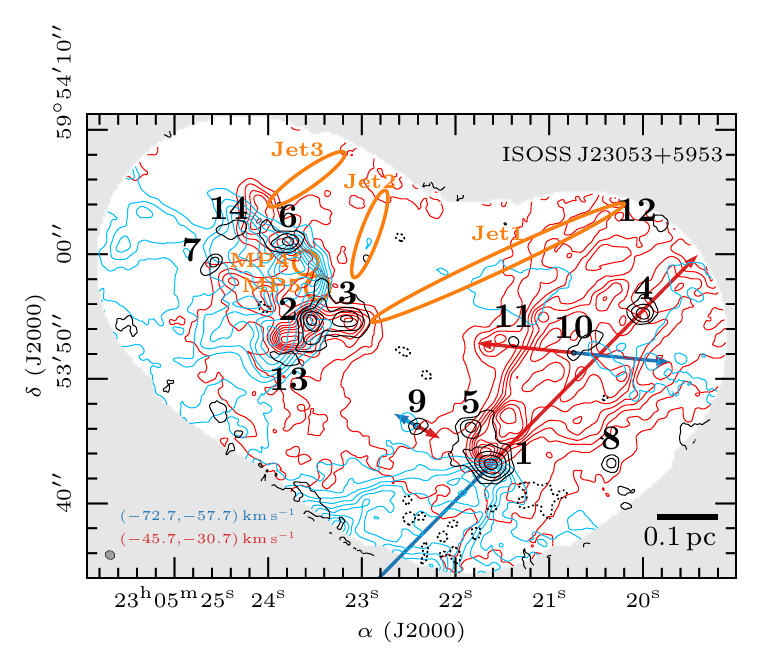}
\caption{CO $2-1$ outflows in ISOSS\,J22478+6357 (\textit{left panel}) and ISOSS\,J23053+5953 (\textit{right panel}). The blue and red contours show the integrated
intensity of the blue- and red-shifted line wings, respectively. The intensity is integrated over $\varv_\mathrm{LSR} - 21$ to $\varv_\mathrm{LSR} - 6$ and over $\varv_\mathrm{LSR} + 6$ to $\varv_\mathrm{LSR} + 21$\,km\,s$^{-1}$ for the blue- and red-shifted integrated intensity, respectively. The corresponding velocity ranges are shown in each panel. Contour levels range from $10\%-90\%$ of the peak integrated intensity with steps of 10\%. In ISOSS J22478+6357 the red- and blue-shifted peak integrated intensity is 45\,K\,km\,s$^{-1}$ and 47\,K\,km\,s$^{-1}$ and in ISOSS J23053+5953 it is 112\,K\,km\,s$^{-1}$ and 140\,K\,km\,s$^{-1}$, respectively. Red- and blue-shifted outflow directions are indicated by red and blue arrows, respectively. The NOEMA 1.3\,mm continuum data are shown in black contours. The dotted black contour marks the $-5\sigma_\mathrm{cont}$ level. Solid black contour levels are 5, 10, 20, 40, and 80$\sigma_\mathrm{cont}$. All mm cores identified in \citet{Beuther2021} are labeled in black. Positions with a peak in molecular emission, but no associated continuum (``molecular peaks,'' MP) are indicated by orange circles (Sect. \ref{sec:moment0}). Jet-like structures, seen in SiO $5-4$ emission, are indicated by orange ellipses (Fig. \ref{fig:moment0_SiO_5_4}). The synthesized beam of the continuum data is shown in the bottom left corner. The synthesized beam of the spectral line data is similar. A linear spatial scale of 0.1\,pc is indicated by a black scale bar.}
\label{fig:outflows}
\end{figure*}

	The cold clumps toward both regions are studied by \citet{Ragan2012} using Herschel PACS observations at 70, 100, and 160\,$\upmu$m. Both regions were also observed with the Spitzer space telescope at 3.6, 4.5, 5.8, and 8.0\,$\upmu$m using the IRAC instrument and at 24\,$\upmu$m using the MIPS instrument. An overview of these archival data sets in comparison with the NOEMA 1.3\,mm emission is shown in Fig. \ref{fig:overview}.
	
	In the ISOSS\,J22478+6357 region, all mm cores are embedded within the $R \approx 0.1$\,pc sized Herschel clumps. A clump with strong 70\,$\upmu$m emission is detected toward the southwest with no associated 1.3\,mm continuum emission (``clump 2''). This source also shows strong emission in the Spitzer IRAC and MIPS data. This suggests a more evolved source where the cold dust envelope was already disrupted by the protostar. The strongest MIR source is around the mm cores 13 and 15. These two mm cores, with a projected separation of $\sim$4\,200\,au, might in reality be the remaining dust envelope that is currently being disrupted by the protostar. \citet{Hennemann2008} find that this YSO is an intermediate-mass protostar. Cores 4, 5, and 6 also have bright counterparts in the Spitzer IRAC data.
	
	The spatial extent of the two Herschel clumps in ISOSS\,J23053+5953 is larger, $R \approx 0.2 - 0.3$\,pc, compared to the clumps in ISOSS\,J22478+6357. While cores 2, 3, 6, 7, 13, 14 are embedded in the northeastern clump peaking toward core 2, cores 1, 4, 5, 8, 9, 10, and 11 are associated with the southwestern clump peaking toward cores 1 and 5. MIR emission is detected around the following cores in ISOSS\,J23053+5953: cores 1, 2, 6, 7, 9, and 14. The southwestern clump is elongated toward core 12, but no significant FIR emission is detected there. There is also no corresponding MIR counterpart. Located at the edge of the primary beam with enhanced noise, core 12 might be an artifact.
	
	The temperature and bolometric luminosity of the cold clumps are derived by \citet{Ragan2012} with Herschel PACS observations by fitting the SED. Following the nomenclature of \citet{Ragan2012}, clumps $1-5$ toward ISOSS\,J22478+6357 and clumps 2 and 3 toward ISOSS\,J23053+5953 were covered by our observations (Fig. \ref{fig:overview}). The five clumps in ISOSS\,J22478+6357 have a temperature ranging between $18-21$\,K and bolometric luminosity between $7-10$\,$L_\odot$. In the northeastern (clump 3) and southwestern clump (clump 2) in ISOSS\,J23053+5953 the dust temperature is 21\, and 22\,K and the bolometric luminosity is 441\,$L_\odot$ and 869\,$L_\odot$, respectively. However, the Spitzer data show that there is an additional warmer component. 
	
	We therefore refitted the SED of the cores with clear MIR counterparts in order to derive a more reliable estimate of the bolometric luminosity $L$. There are sources detected in the Spitzer data with no mm counterpart; however, in the following analysis we only focused on the sources with mm counterparts. It should be noted that while the Spitzer IRAC and NOEMA observations have a comparable angular resolution, the angular resolution of the Spitzer MIPS and Herschel PACS data is lower (Sect. \ref{sec:archivaldata}). Smoothing the data to a common resolution would smear out all core features; instead, we performed a conservative cross-matching of the Spitzer MIPS sources and Herschel PACS clumps with the mm cores, as explained below.
	
	In order to derive fluxes of the sources in the Spitzer IRAC and MIPS data, we performed aperture photometry using the \texttt{photutils} package. Background subtraction was performed by clipping sources with emission $>$3$\sigma$ and then the median was computed and subtracted from the data. The \texttt{DAOFIND} algorithm \citep{Stetson1987} was used to identify sources with emission $>$5$\sigma$. A circular aperture with a radius of 2\as4 and 7\as6 was used to calculate the flux of the extracted sources in the IRAC and MIPS data, respectively. We did not subtract the background with a background annulus due to the crowded fields in the IRAC data. Instead we subtracted the background using the median value in the region as described above. The Spitzer sources are cross-matched to the closest mm core, if the projected distance is $<$1$''$ and $<$4$''$ for the IRAC and MIPS data, respectively. Aperture correction was applied to the derived fluxes, with a correction factor of 1.215, 1.233, 1.366, 1.568, and 1.61 for the fluxes at 3.6, 4.5, 5.8, 8.0, and 24\,$\upmu$m, respectively \citep{Reach2005,Engelbracht2007}. 
	
	In order to reliably fit the SED, we require that the mm cores are detected in three or four IRAC bands and therefore fitted the SED of evolved protostars. As a second further constraint, the core must either be detected at 24\,$\upmu$m or be associated with a Herschel clump peaking toward the position of the mm core (Table \ref{tab:positions}). This ensures that the mm flux can be accurately fitted with a cold component. For example, in the case of cores 6 and 7 in ISOSS\,J23053+5953, the mm cores have strong Spitzer IRAC counterparts; however, due to the poor angular resolution in the FIR, the Herschel clump (``clump 3'' in Fig. \ref{fig:overview}) is not resolved toward these cores and peaks toward core 2 instead. While the warm component can be estimated with the Spitzer IRAC fluxes, a second cold component based on the mm flux cannot be constrained with only one data point. On the other hand, around core 13 in ISOSS\,J22478+6357, the Herschel clump (``clump 4'' in Fig. \ref{fig:overview}) is not resolved also showing significant emission toward core 1. In this case, the Herschel PACS data points are not included in the fitting, but the strong 24\,$\upmu$m MIPS flux provided an additional data point peaking toward core 13. In summary, we can construct the SED from MIR to mm wavelengths for cores 4, 5, 6, and 13 in ISOSS\,J22478+6357 and for cores 1, 2, and 9 in ISOSS\,J23053+5953.
	
	The fluxes of the Herschel PACS clumps are taken from \citet{Ragan2012}, for which we cross-match the studied clumps to the closest mm core if one is detected (Table \ref{tab:positions}). The FIR observations resolve linear scales of $0.1 - 0.2$\,pc (Fig. \ref{fig:overview}), so the fluxes are integrated over a much larger area compared to the NOEMA 1.3\,mm and Spitzer IRAC data. The 1.3\,mm fluxes are taken from \citet{Beuther2021} and are listed in Table \ref{tab:positions}. The MIR and FIR photometric data points are listed in Table \ref{tab:photometry} and the core SEDs are shown in Fig. \ref{fig:SED}. 
	
	For each core, we fit the SED with two or three components of a modified black body, 
	\begin{equation}
	B_\nu(T) = \frac{F_\nu \eta d^2}{\kappa_\nu M},
	\end{equation}
in order to estimate the temperature and bolometric luminosity \citep[see also][]{Beuther2007C,Beuther2010,Linz2010,Ragan2012}. The wavelength-dependent dust opacities, $\kappa_\nu$, are taken from \citet{Ossenkopf1994} for densities of 10$^{6}$\,cm$^{-3}$ and 10$^{5}$\,yr of coagulation and with thin ice mantles. In consistency with all previous CORE and CORE-extension studies, we assume a gas-to-dust mass ratio of $\eta = 150$ \citep[$\eta = \frac{M_{\mathrm{gas}}}{M_{\mathrm{H}}}$/$\frac{M_{\mathrm{dust}}}{M_{\mathrm{H}}}$, with $\frac{M_{\mathrm{gas}}}{M_{\mathrm{H}}} = 1.4$ and $\frac{M_{\mathrm{dust}}}{M_{\mathrm{H}}} = 0.0091$; Table 1.4 and 23.1 in][respectively]{Draine2011}. This is higher than the typically assumed ratio of 100; however, recent observations suggest that $\eta$ increases with galactocentric distance $d_\mathrm{gal}$ \citep{Giannetti2017}. These authors find a relation, where $\eta = 130 - 320$ at $d_\mathrm{gal} = 10$\,kpc. For most cores, two components were sufficient to model their SED; however, for core 2 in ISOSS\,J23053+5953, three components were required to reliably model all photometric data points.
	
	The results of the temperature $T_{i}$ and luminosity $L_{i}$ of each $i$-th component, and total bolometric luminosity $L$ are summarized in Table \ref{tab:photometry}. The fitted total SED as well as each component are shown in Fig. \ref{fig:SED}. The contribution of cold dust with $T \approx 20 - 40$\,K is not sufficient to explain the observed fluxes at MIR wavelengths. In addition to the cold component, one or two warmer components with $T = 400 - 600$\,K are required to reliably fit the SED (Table \ref{tab:photometry}). This gives further evidence that some of the mm cores contain YSOs in a more evolved stage. 
	
	As an example for mm core 1 in ISOSS\,J23053+5953, two components with $T = 23$\,K and $T = 620$\,K are required to properly reproduce the FIR+mm and MIR fluxes, respectively. A strong 24\,$\upmu$m source is detected toward mm core 2 in ISOSS\,J23053+5953, for which three components with $T = 28$\,K, $T = 109$\,K, and $T = 479$\,K are needed to describe the full SED.

	One uncertainty arises from the fact that, even though the MIR and FIR sources are associated with the mm cores, the angular resolution of the observations from MIR to mm wavelengths vary and thus the fluxes are integrated over different angular sizes. This effect is particularly strong for the 160\,$\upmu$m fluxes whose observed fluxes are usually higher than the fluxes derived from the SED fit (Fig. \ref{fig:SED}).
	
\section{Spectral line emission}\label{sec:spectralline}

	In this section, we study the composition of the molecular gas by analyzing the 1.3\,mm spectral line data. In the analysis of the original CORE data set \citep{Gieser2021}, the low-resolution data were used, which provided a continuous spectrum. However, as ISOSS\,J22478+6357 and ISOSS\,J23053+5953 are much colder and younger with typical line widths $<$3\,km\,s$^{-1}$ \citep[compared to a mean line width of 6\,km\,s$^{-1}$ in the original CORE sample,][]{Gieser2021}, we used the high-resolution ($\sim$0.5\,km\,s$^{-1}$) data if a high-resolution unit was placed toward the line. The properties of the spectral line data products (such as velocity resolution, synthesized beam, line noise $\sigma_\mathrm{line}$) are summarized in Table \ref{tab:lineobs}.
	
\subsection{Molecular outflows}\label{sec:outflows}

\begin{figure*}
\centering
\includegraphics[]{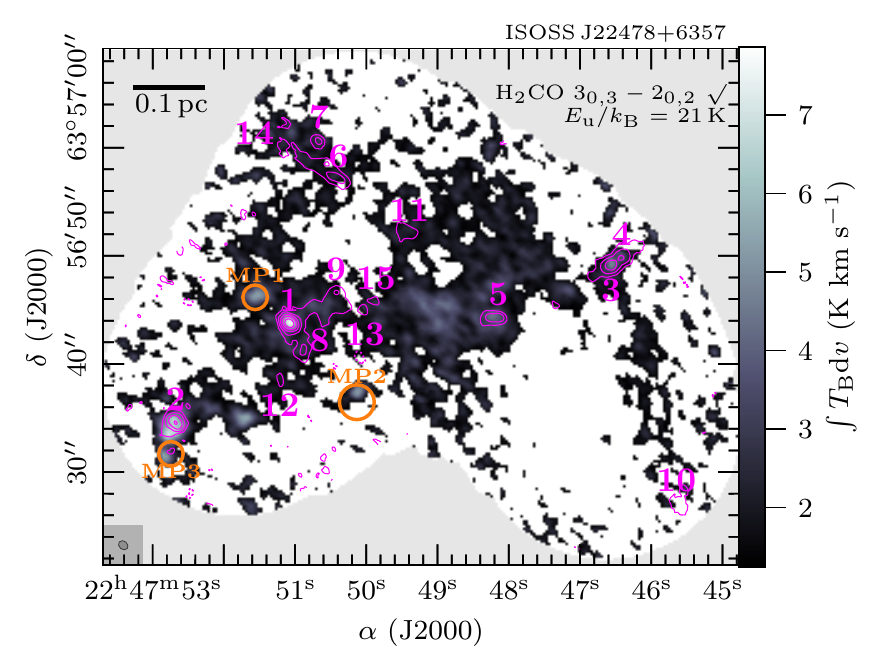}
\includegraphics[]{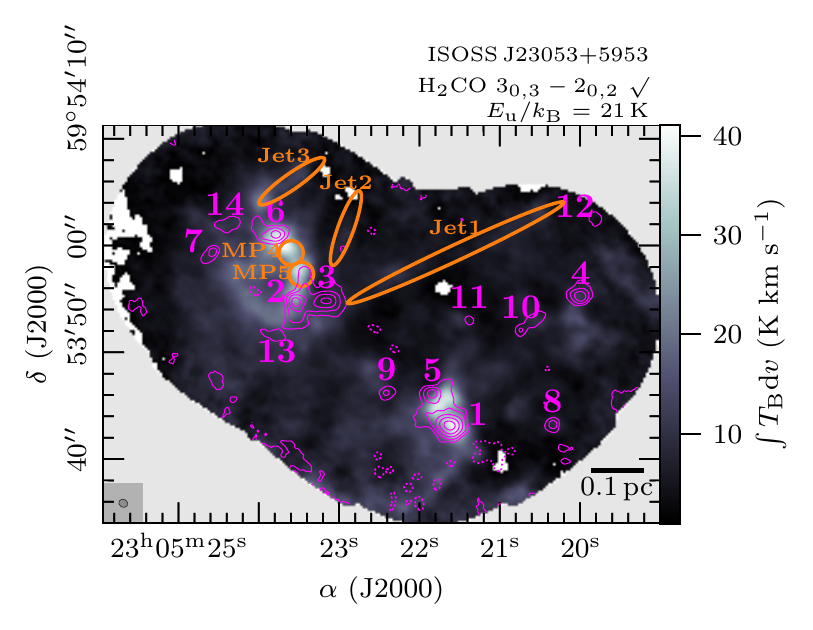}
\caption{Line integrated intensity map of H$_{2}$CO $3_{0,3}-2_{0,2}$ of ISOSS\,J22478+6357 (\textit{left panel}) and ISOSS\,J23053+5953 (\textit{right panel}). The integrated intensity of the transition with a threshold of S/N $\geq 3$ is presented in color scale. The line intensity is integrated from a velocity of $\varv_\mathrm{LSR} - 3$\,km\,s$^{-1}$ to $\varv_\mathrm{LSR} + 3$\,km\,s$^{-1}$. It is marked if the transition is detected ($\surd$) or not ($\times$), further explained in Sect. \ref{sec:moment0}. The NOEMA 1.3\,mm continuum data are shown in pink contours. The dotted pink contour marks the $-5\sigma_\mathrm{cont}$ level. Solid pink contour levels are 5, 10, 20, 40, and 80$\sigma_\mathrm{cont}$. All mm cores identified in \citet{Beuther2021} are labeled in pink. Positions with a peak in molecular emission, but no associated continuum (``molecular peaks,'' MP) are indicated by orange circles (Sect. \ref{sec:moment0}). Jet-like structures, seen in SiO $5-4$ emission, are indicated by orange ellipses (Fig. \ref{fig:moment0_SiO_5_4}). The synthesized beam of the continuum data is shown in the bottom left corner. The synthesized beam of the spectral line data is similar. A linear spatial scale of 0.1\,pc is indicated by a black scale bar.}
\label{fig:moment0_H2CO_3_03_2_02}
\end{figure*}

	Molecular outflows are ubiquitous in low- and high-mass star-forming regions indicating indirectly the presence of gas accretion toward protostars \citep[e.g.,][]{Beuther2002,Wu2004,Zhang2005,Koelligan2018}. In low-mass star-forming regions, both the disk and outflow are commonly observed around protostars. Toward high-mass protostars, large bipolar outflows are found \citep[e.g.,][]{Beuther2002, Arce2007, Frank2014}, but disk structures remain challenging to observe \citep[e.g.,][]{Cesaroni2017, Ahmadi2019, Maud2019, Beltran2020,Johnston2020}.
	
	We used the integrated CO $2-1$ emission in the blue- and red-shifted line wings to search for molecular outflows in ISOSS\,J22478+6357 and ISOSS\,J23053+5953. The blue- and red-shifted integrated intensity was computed from $\varv_\mathrm{LSR} - 21$\,km\,s$^{-1}$ to $\varv_\mathrm{LSR} - 6$\,km\,s$^{-1}$ and from $\varv_\mathrm{LSR} + 6$\,km\,s$^{-1}$ to $\varv_\mathrm{LSR} + 21$\,km\,s$^{-1}$, respectively. The results are shown in Fig. \ref{fig:outflows} for ISOSS\,J22478+6357 and ISOSS\,J23053+5953.
	
	Both regions show multiple bipolar outflow signatures, indicated by arrows in Fig. \ref{fig:outflows}. In Table \ref{tab:positions} we list all mm cores with molecular outflows seen in CO $2-1$. We find outflow signatures toward cores $1-5$ in ISOSS\,J22478+6357. The outflow toward core 1 is collimated. The outflow around core 2 has a quadrupolar morphology, similar to the outflow of the intermediate-mass YSO IRAS\,22198+6336 \citep{SanchezMonge2010}. Either the quadrupolar shape arises from the cavity walls of a wide-angle outflow or two distinct outflows of unresolved multiple protostars are present. Toward the northwest of core 5 there could be a bipolar outflow with no associated mm continuum core. All outflow directions show a preferred orientation from the northeast to the southwest and are almost parallel with respect to each other. The orientation of the outflows are perpendicular to the filamentary gas traced by DCO$^{+}$ for ISOSS\,J22478+6357 (Fig. \ref{fig:moment0_DCO+_3_2}). Core 11 might have a north-south bipolar outflow, but the signature is not clear.
	
	In ISOSS\,J23053+5953 a large-scale bipolar outflow is seen toward core 1 with the red- and blue-shifted lobes directed toward the northwest and southeast, respectively. A small projected outflow is observed toward core 9, which might be either a very young outflow just being launched or the result of an inclination effect. The CO $2-1$ line profile of core 9 is not significantly broader than the one of core 1, so an inclination effect is unlikely. Core 10 also hosts an outflow that is partially overlapping with the red-shifted outflow of core 1. Toward the northeast of ISOSS\,J23053+5953, red- and blue-shifted emission is detected around the location of cores 2, 3, 6, 7, 13, and 14; however, no clear outflow directions can be identified. For example, toward both cores 7 and 6 there is a weak signature of northeast-southwest bipolar outflows. We discuss in Sect. \ref{sec:kinematics} that the region shows a steep velocity gradient toward these cores caused by a colliding flow, which is also seen in the CO line wing emission. It is therefore difficult to determine if these cores host outflows or not.

	Clear detections of line wings from $-60$\,km\,s$^{-1}$ to $-40$\,km\,s$^{-1}$ were found toward ISOSS\,J23053+5953 using the CO $1-0$ line observed with the IRAM 30m telescope at an angular resolution of 21$''$ \citep{Wouterloot1989a}. These authors derived the outflow properties and estimated a total outflow mass of $M_\mathrm{out} = 14.2 - 34.2$\,$M_{\odot}$, momentum of $P_\mathrm{out} = 71.8 - 172.4$\,$M_{\odot}$\,km\,s$^{-1}$, energy of $E_\mathrm{out} = (40.5 - 97.4) \times 10^{44}$\,ergs, a size of $R_\mathrm{out} = 1$\,pc for both blue- and red-shifted lobes, and an outflow timescale of $\tau_\mathrm{out} = 10^{5}$\,yr. Their outflow can be assigned to the large-scale outflow from mm core 1 with a red-shifted lobe directed toward the northwest and the blue-shifted lobe directed toward the southeast \citep[Fig. 6 in][]{Wouterloot1989a}. 
	
	Comparing the mass outflow rate of core 1, $\dot M_\mathrm{out} = \frac{M_\mathrm{out}}{\tau_\mathrm{out}} \approx 2\times10^{-4}$\,$M_\odot$\,yr$^{-1}$ with the mass outflow rates of low- to high-mass YSOs \citep[Fig. 11 in][]{Henning2000}, suggests that core 1 will form an intermediate- to high-mass star \citep[see also][for a comparison of the mass outflow rate and bolometric luminosity]{Beuther2002,Wu2004,Wu2005,LopezSepulcre2009,Maud2015B}. The outflow has also been tentatively detected by \citet{Birkmann2007} using Plateau de Bure Interferometer (PdBI) observations of CO $2-1$ (we note that the red- and blue-shifted outflow directions are swapped in their Fig. 3).
	
	The molecular outflows observed with NOEMA in both regions can be assigned to mm cores; however, their extent is larger than the observed mosaic, so a larger FOV would be required to properly derive the outflow properties (such as mass, momentum, energy, and timescale) of individual outflows. The multiplicity of bipolar outflows suggests that ongoing clustered star formation is occurring in both regions. A recent study of the properties of the molecular outflow of core 1 in ISOSS\,J23053+5953 is presented in \citet{Rodriguez2021b} using observations with the Submillimeter Array (SMA) and the VLA. These authors detect compact centimeter (cm) emission toward the mm core, suggesting the presence of an ionized jet. The larger FOV of the SMA primary beam allowed them to properly derive the outflow properties, with $M_\mathrm{out} = 45.2\pm12.6$\,$M_{\odot}$ and $\tau_\mathrm{out} = 1.5-7.2\times10^{4}$\,yr. The resulting outflow rate $\dot M_\mathrm{out} \approx 10^{-3}$\,$M_\odot$\,yr$^{-1}$ is even higher than previously determined by single-dish observations \citep{Wouterloot1989a}. A further discussion of the protostellar outflows in ISOSS\,J22478+6357 and ISOSS\,J23053+5953 is given in Sect. \ref{sec:discore}.
	
\subsection{Spatial morphology and correlations}\label{sec:morphology}

\begin{figure*}
\centering
\includegraphics[]{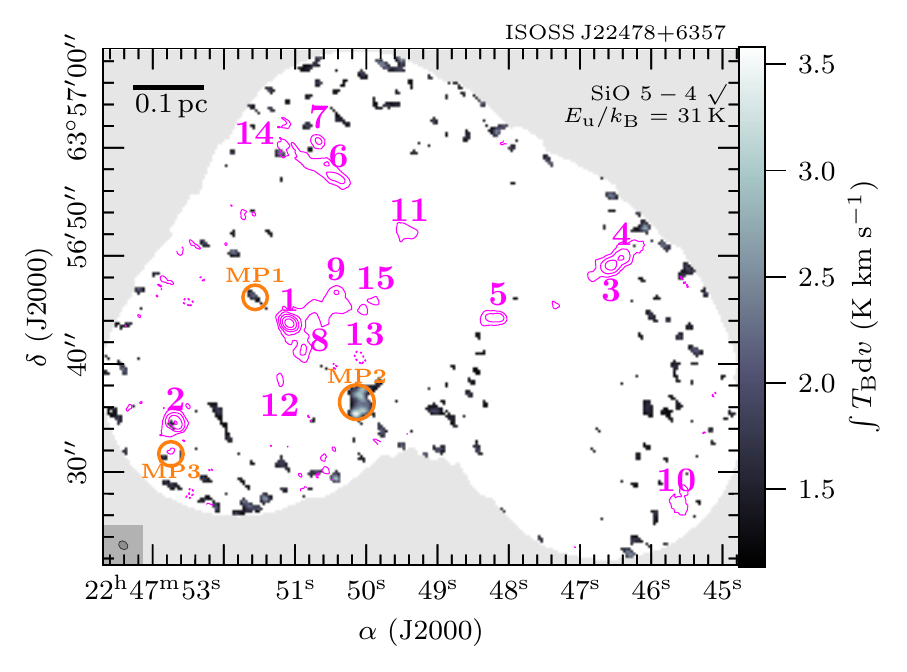}
\includegraphics[]{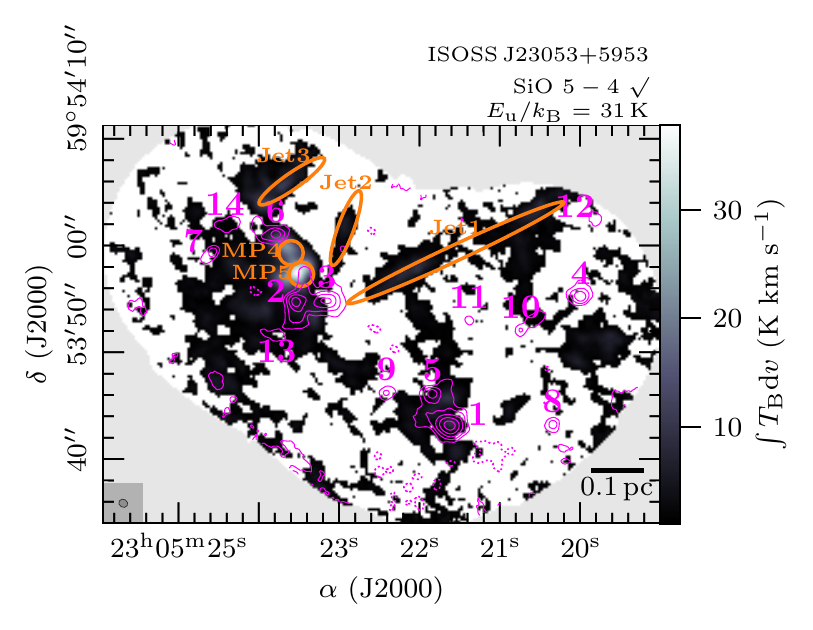}
\caption{The same as Fig. \ref{fig:moment0_H2CO_3_03_2_02}, but for SiO $5-4$.}
\label{fig:moment0_SiO_5_4}
\end{figure*}

\begin{figure*}
\centering
\includegraphics[]{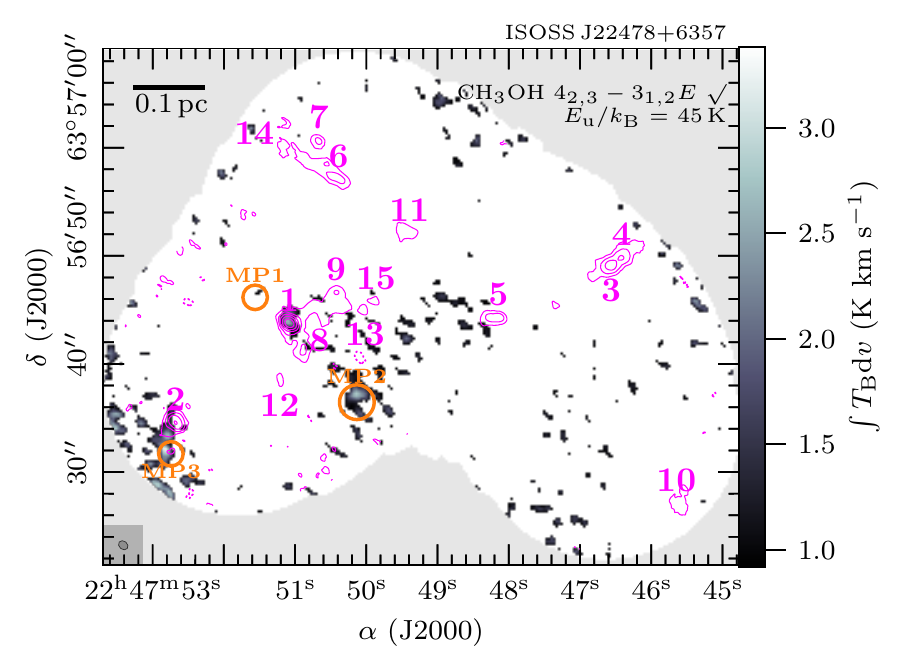}
\includegraphics[]{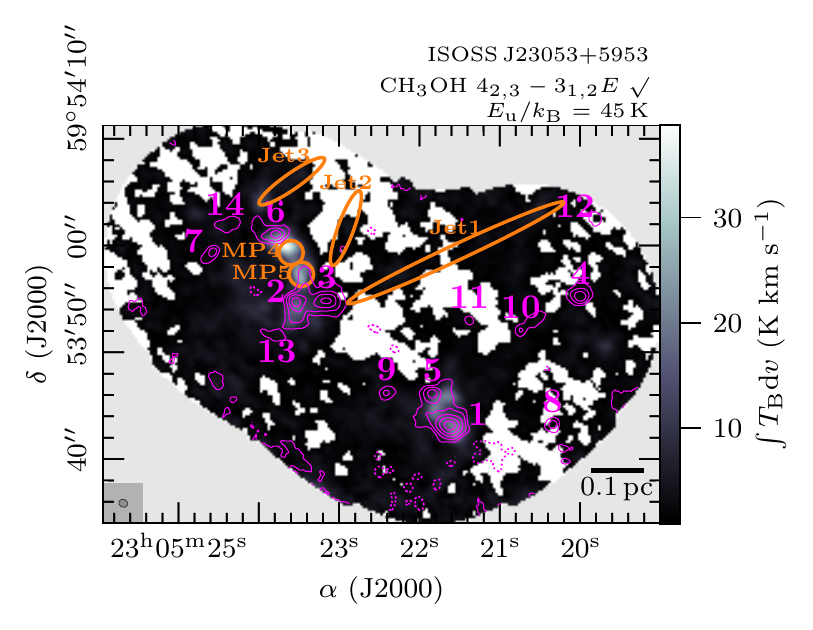}
\caption{The same as Fig. \ref{fig:moment0_H2CO_3_03_2_02}, but for CH$_{3}$OH $4_{2,3}-3_{1,2}E$.}
\label{fig:moment0_CH3OH_4_2_3_1}
\end{figure*}

	The spectral setup covers in total 26 emission lines detected in at least one of the two regions and an overview of the transition properties is shown in Table \ref{tab:lineprops}. We detect simple species consisting of $2 - 5$ atoms (CO, $^{13}$CO, C$^{18}$O, $^{13}$CS, SO, H$_{2}$S, OCS, SO$_{2}$, H$_{2}$CO, H$_{2}^{13}$CO, HNCO, HC$_{3}$N); deuterated ions (N$_{2}$D$^{+}$, DCO$^{+}$); the shock tracer SiO; the cyclic molecule c-C$_{3}$H$_{2}$; and two complex organic molecules (CH$_{3}$OH, CH$_{3}$CN).
	
	\subsubsection{Line integrated intensity}\label{sec:moment0}
	
	In order to investigate the spatial morphology of each emission line, we computed the integrated intensity around the region systemic velocity from $\varv_\mathrm{LSR} - 3$\,km\,s$^{-1}$ to $\varv_\mathrm{LSR} + 3$\,km\,s$^{-1}$. The systemic velocity $\varv_\mathrm{LSR}$ for both regions is listed in Table \ref{tab:regions}. This velocity range covers 3 and 13 channels in the low- and high-resolution data, respectively. The integrated intensity (``moment 0'') maps for strong transitions of three key species, H$_{2}$CO ($3_{0,3}-2_{0,2}$), SiO ($5-4$), and CH$_{3}$OH ($4_{2,3}-3_{1,2}E$) are shown in Figs. \ref{fig:moment0_H2CO_3_03_2_02}, \ref{fig:moment0_SiO_5_4}, and \ref{fig:moment0_CH3OH_4_2_3_1}, respectively. The moment 0 maps of the remaining lines are shown in Appendix \ref{app:moment0maps}. The noise in the integrated intensity maps $\sigma_\mathrm{int.intensity}$ was calculated by considering the velocity resolution ($\delta \varv$), line noise ($\sigma_\mathrm{line}$), and number of channels ($\# \mathrm{channels}$): $\sigma_\mathrm{int.intensity} = \delta \varv \times \sigma_\mathrm{line} \times (\# \mathrm{channels})^{0.5}$. In all integrated intensity maps, only locations with an integrated intensity $> 3\sigma_\mathrm{int.intensity}$ are presented.
	
	The properties of the line data products, such as synthesized beam and line noise, are summarized in Table \ref{tab:lineobs}. It can be clearly seen in the integrated intensity maps that the noise is not uniform throughout the mosaic increasing toward the edge. For a comparison between the two regions, regardless of whether a transition shows significant emission throughout the FOV, we carefully investigated each line integrated intensity map, especially for the fainter lines, and searched for spatially resolved emission with S/N $> 3$ that is not caused by noise artifacts. In Table \ref{tab:lineobs} and in the integrated intensity maps we indicate for each region if the line is considered as detected (\cmark) or not (\xmark) in the integrated intensity map. For example, the OCS $18-17$ line (Fig. \ref{fig:moment0_OCS_18_17}) is clearly detected in ISOSS\,J23053+5953 toward core 1, 2 and 6, whereas in ISOSS\,J22478+6357 the line emission is irregular and caused by noise artifacts and therefore marked as not detected (\xmark). In Sect. \ref{sec:molecularcolumndensities}, molecular column densities are derived from the core spectrum and there it was checked separately, if the transition has emission $> 3 \sigma_\mathrm{line}$.
	 	
	While all transitions are detected in ISOSS\,J23053+5953, ISOSS\,J22478+6357 is more line-poor, with many transitions at higher upper energy levels ($E_\mathrm{u}$/$k_\mathrm{B} \gtrsim 50$\,K) not detected (Table \ref{tab:lineobs}). In both regions, the emission of the three CO isotopologues (CO, $^{13}$CO, C$^{18}$O, Figs. \ref{fig:moment0_CO_2_1}, \ref{fig:moment0_13CO_2_1}, and \ref{fig:moment0_C18O_2_1}, respectively) is widespread across the FOV. The optically thick CO and $^{13}$CO lines trace the outer parts of the cloud and/or clump structure, while even the optically thin C$^{18}$O emission is detected everywhere in the FOV. The detected H$_{2}$CO transitions (Figs. \ref{fig:moment0_H2CO_3_03_2_02}, \ref{fig:moment0_H2CO_3_22_2_21}, and \ref{fig:moment0_H2CO_3_21_2_20}) have extended emission in both regions. H$_{2}$CO is a good thermometer at temperatures $<$100\,K and therefore its emission is used in Sect. \ref{sec:XCLASSMapFitting} to create temperature maps of the regions.
	
	 In ISOSS\,J22478+6357, the line-richest object is core 1 with emission peaks of H$_{2}$CO, CH$_{3}$OH, SO, c-C$_{3}$H$_{2}$ (Figs. \ref{fig:moment0_H2CO_3_03_2_02}, \ref{fig:moment0_CH3OH_4_2_3_1}, \ref{fig:moment0_SO_5_5_4_4}, \ref{fig:moment0_c-C3H2_6_16_5_05}, \ref{fig:moment0_H2CO_3_22_2_21}, \ref{fig:moment0_H2CO_3_21_2_20}, \ref{fig:moment0_SO_6_5_5_4}, and \ref{fig:moment0_CH3OH_8_-1_7_0}). The DCO$^{+}$ and N$_{2}$D$^{+}$ emission (Figs. \ref{fig:moment0_DCO+_3_2} and \ref{fig:moment0_N2D+_3_2}) peak toward core 2, but both molecules also have large-scale filamentary emission that is connecting the mm cores. H$_{2}$CO (Fig. \ref{fig:moment0_H2CO_3_03_2_02}) and SO (Figs. \ref{fig:moment0_SO_5_5_4_4} and \ref{fig:moment0_SO_6_5_5_4}) also show emission peaks toward core 2. The cyclic molecule c-C$_{3}$H$_{2}$ has distinct emission peaks toward cores 1, 3, and 4 (Fig. \ref{fig:moment0_c-C3H2_6_16_5_05}) tracing UV irradiated gas \citep[e.g.,][]{Pety2005,Fontani2012,Mottram2020}. There is no known PDR or UCH{\sc ii} region nearby and the emission peaks are co-spatial with the continuum peak of the cores. This suggests that the irradiation stems from the central protostar. Complementary observations at cm wavelengths would draw a clearer picture about the presence of any UCH{\sc ii} region.

	We find locations in ISOSS\,J22478+6357 with strong molecular emission, but no 1.3\,mm continuum counterpart. Toward the northeast and southwest of core 1 there are two molecular peaks (MPs), MP1 and MP2, seen clearly in H$_{2}$CO $3_{0,3}-2_{0,2}$ (Fig. \ref{fig:moment0_H2CO_3_03_2_02}) and SO $6_{5}-5_{4}$ emission (Fig. \ref{fig:moment0_SO_6_5_5_4}). The fact that these MPs are located at both sides of core 1 suggests that they are most likely shocked regions caused by the bipolar outflow of core 1 (Fig. \ref{fig:outflows}). The dust emission toward the MPs can be too faint to be detected at our sensitivity limit, but due to shocks dust grain destruction of the mm -sized grains into undetectable $\upmu$m-sized fragments might play a role as well. Toward the south of core 2, we also find a molecular emission peak only associated with faint 1.3\,mm continuum emission (MP3). This is clearly seen in H$_{2}$CO $3_{0,3}-2_{0,2}$ (Fig. \ref{fig:moment0_H2CO_3_03_2_02}), CH$_{3}$OH $4_{2,3}-3_{1,2}E$ (Fig. \ref{fig:moment0_CH3OH_4_2_3_1}), and both SO transitions (Figs. \ref{fig:moment0_SO_5_5_4_4} and \ref{fig:moment0_SO_6_5_5_4}). MP3 is connected to the red-shifted outflow cavity of core 2 (Fig. \ref{fig:outflows}). ISOSS\,J22478+6357 is generally line-poor with simple species tracing the envelope, in which the cores are embedded. Molecular emission peaks with no mm continuum counterpart can be linked to molecular outflows, thus tracing shocked gas. Here, molecules, such as SO, H$_{2}$CO, and CH$_{3}$OH, which were initially frozen on the dust grains, are evaporated into the gas phase by the shock. Toward most of the cores, no distinct molecular emission is detected. This suggests that the cores are still too cold to have high molecular abundances in the gas phase.
	 
	ISOSS\,J23053+5953 is richer in line emission compared to ISOSS\,J22478+6357 (Table \ref{tab:lineobs}). Most of the molecular emission peaks at core 1, with some species peaking at cores 2 and 6. Core 1 has spatial emission peaks of SO (Figs. \ref{fig:moment0_SO_5_5_4_4} and \ref{fig:moment0_SO_6_5_5_4}), $^{13}$CS (Fig. \ref{fig:moment0_13CS_5_4}), H$_{2}$S (Fig. \ref{fig:moment0_H2S_2_20_2_11}), OCS (Figs. \ref{fig:moment0_OCS_18_17} and \ref{fig:moment0_OCS_19_18}), H$_{2}$CO (Figs. \ref{fig:moment0_H2CO_3_03_2_02}, \ref{fig:moment0_H2CO_3_22_2_21}, and \ref{fig:moment0_H2CO_3_21_2_20}), HC$_{3}$N (Figs. \ref{fig:moment0_HC3N_24_23} and \ref{fig:moment0_HC3N_26_25}), and CH$_{3}$OH (Figs. \ref{fig:moment0_CH3OH_4_2_3_1}, \ref{fig:moment0_CH3OH_5_1_4_2}, \ref{fig:moment0_CH3OH_8_0_7_1}, and \ref{fig:moment0_CH3OH_8_-1_7_0}).
	
	Core 2 has a prominent emission peak in the CH$_{3}$OH $5_{1,4}-4_{2,3}$ line (Fig. \ref{fig:moment0_CH3OH_5_1_4_2}), but also in OCS (Figs. \ref{fig:moment0_OCS_18_17} and \ref{fig:moment0_OCS_19_18}), HNCO (Fig. \ref{fig:moment0_HNCO_10_010_9_09}), and CH$_{3}$CN (\ref{fig:moment0_CH3CN_12_0_11_0}) emission. Core 6 also shows many emission peaks and in comparison to core 1, strong emission of HNCO (Fig. \ref{fig:moment0_HNCO_10_010_9_09}), c-C$_{3}$H$_{2}$ (Fig. \ref{fig:moment0_c-C3H2_6_16_5_05}), and CH$_{3}$CN (Fig. \ref{fig:moment0_CH3CN_12_0_11_0}) is detected here, but also SO (Figs. \ref{fig:moment0_SO_5_5_4_4} and \ref{fig:moment0_SO_6_5_5_4}), OCS (Figs. \ref{fig:moment0_OCS_18_17} and \ref{fig:moment0_OCS_19_18}), and H$_{2}$S (Fig. \ref{fig:moment0_H2S_2_20_2_11}). 
	
	Similar to ISOSS\,J22478+6357, DCO$^{+}$ is distributed throughout the FOV (Fig. \ref{fig:moment0_DCO+_3_2}). Toward core 3, DCO$^{+}$ emission peaks toward the north, while N$_{2}$D$^{+}$ emission (Fig. \ref{fig:moment0_N2D+_3_2}) peaks toward the west. A possible scenario could be that there is a severe depletion of CO in the western position resulting in a low DCO$^{+}$ abundance. This is further discussed in Sect. \ref{sec:dis23053}.
	
	The cavity of the red-shifted outflow lobe of core 1 (Sect. \ref{sec:outflows}) is seen between cores 1 and 5 in H$_{2}$CO (Fig. \ref{fig:moment0_H2CO_3_03_2_02}), CH$_{3}$OH (Figs. \ref{fig:moment0_CH3OH_4_2_3_1}, \ref{fig:moment0_CH3OH_5_1_4_2}, \ref{fig:moment0_CH3OH_8_0_7_1}, and \ref{fig:moment0_CH3OH_8_-1_7_0}), SiO (Fig. \ref{fig:moment0_SiO_5_4}), and SO (Figs. \ref{fig:moment0_SO_5_5_4_4} and \ref{fig:moment0_SO_6_5_5_4}), and HNCO (Fig. \ref{fig:moment0_HNCO_10_010_9_09}) emission. The bipolar outflow is also seen as a dark lane in CO (Fig. \ref{fig:moment0_CO_2_1}). In SiO (Fig. \ref{fig:moment0_SiO_5_4}) and SO (Figs. \ref{fig:moment0_SO_5_5_4_4} and \ref{fig:moment0_SO_6_5_5_4}), three jets can be identified toward the north of the region (labeled as Jet1, Jet2, and Jet3 in the integrated intensity maps). Jet1 is also seen in H$_{2}$CO emission (Fig. \ref{fig:moment0_H2CO_3_03_2_02}). While we do not find clear outflow signatures of the cores in this location, these jet features might indeed be caused by protostellar outflows (Sect. \ref{sec:outflows}).
	
	Toward the south of core 6 and toward the north of core 2, two emission peaks can be identified (MP4 and MP5) seen in H$_{2}$CO (Figs. \ref{fig:moment0_H2CO_3_03_2_02}, \ref{fig:moment0_H2CO_3_22_2_21}, and \ref{fig:moment0_H2CO_3_21_2_20}), H$_{2}^{13}$CO (Fig. \ref{fig:moment0_H213CO_3_12_2_11}), CH$_{3}$OH (Figs. \ref{fig:moment0_CH3OH_4_2_3_1}, \ref{fig:moment0_CH3OH_5_1_4_2}, \ref{fig:moment0_CH3OH_8_0_7_1}, and \ref{fig:moment0_CH3OH_8_-1_7_0}), SiO (Fig. \ref{fig:moment0_SiO_5_4}), SO (Figs. \ref{fig:moment0_SO_5_5_4_4} and \ref{fig:moment0_SO_6_5_5_4}), SO$_{2}$ (Fig. \ref{fig:moment0_SO2_4_22_3_13}), HC$_{3}$N (Figs. \ref{fig:moment0_HC3N_24_23} and \ref{fig:moment0_HC3N_26_25}), OCS (Figs. \ref{fig:moment0_OCS_18_17} and \ref{fig:moment0_OCS_19_18}), and HNCO (Fig. \ref{fig:moment0_HNCO_10_010_9_09}) emission. DCO$^{+}$ emission is only enhanced toward MP5 (Fig. \ref{fig:moment0_DCO+_3_2}). It could be that MP4 and MP5 are connected to potential molecular outflows as it is the case for MP1, MP2, and MP3 seen in SO, SiO, and H$_{2}$CO emission toward ISOSS\,J22478+6357. Unfortunately, the CO $2-1$ line wing emission toward the location around cores 2, 3, and 6 is very complex and it is not possible to identify clear bipolar outflow signatures (Fig. \ref{fig:outflows}). While we cannot rule out the presence of protostellar outflows causing these shocked regions, it coincides with a steep large-scale velocity gradient, for example, seen in NH$_{3}$ \citep{Bihr2015} and DCO$^{+}$ \citep{Beuther2021}, which hints at the presence of a colliding flow. This velocity gradient is further investigated in Sect. \ref{sec:kinematics}.
	
	The emission of MP4 is elongated to the northeast toward mm core 6. All species having an enhanced emission toward MP4 and MP5 are known to trace shocked regions caused by protostellar outflows \citep{Leurini2011,Benedettini2013,Moscadelli2013,Shimajiri2015,Palau2017, Tychoniec2019,Okoda2020,Taquet2020}. There is enhanced c-C$_{3}$H$_{2}$ emission toward the north and south of core 2 (Fig. \ref{fig:moment0_c-C3H2_6_16_5_05}) that could potentially trace, in addition to the colliding flow, a bipolar outflow.
	
	A ring-like structure is seen in H$_{2}$CO emission in (Fig. \ref{fig:moment0_H2CO_3_03_2_02}) and in CO absorption (Fig. \ref{fig:moment0_CO_2_1}) connecting the core 6, MP4, MP5, and core 2. This dynamically complex region might be an overlap of a colliding flow and the presence of protostellar outflows.
	
	\subsubsection{Spatial correlations}\label{sec:HOG}
	
\begin{figure*}
\centering
\includegraphics[]{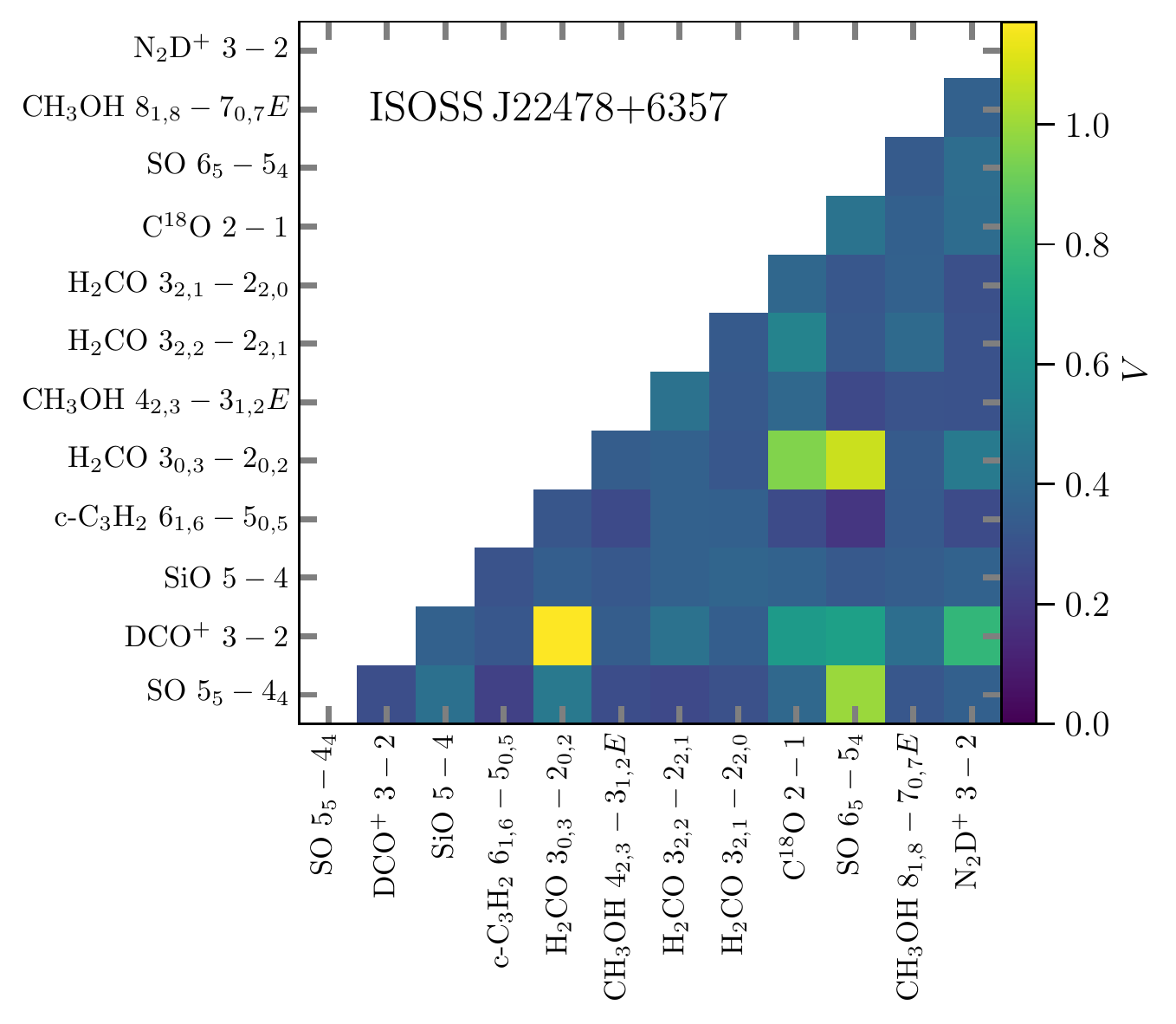}
\includegraphics[]{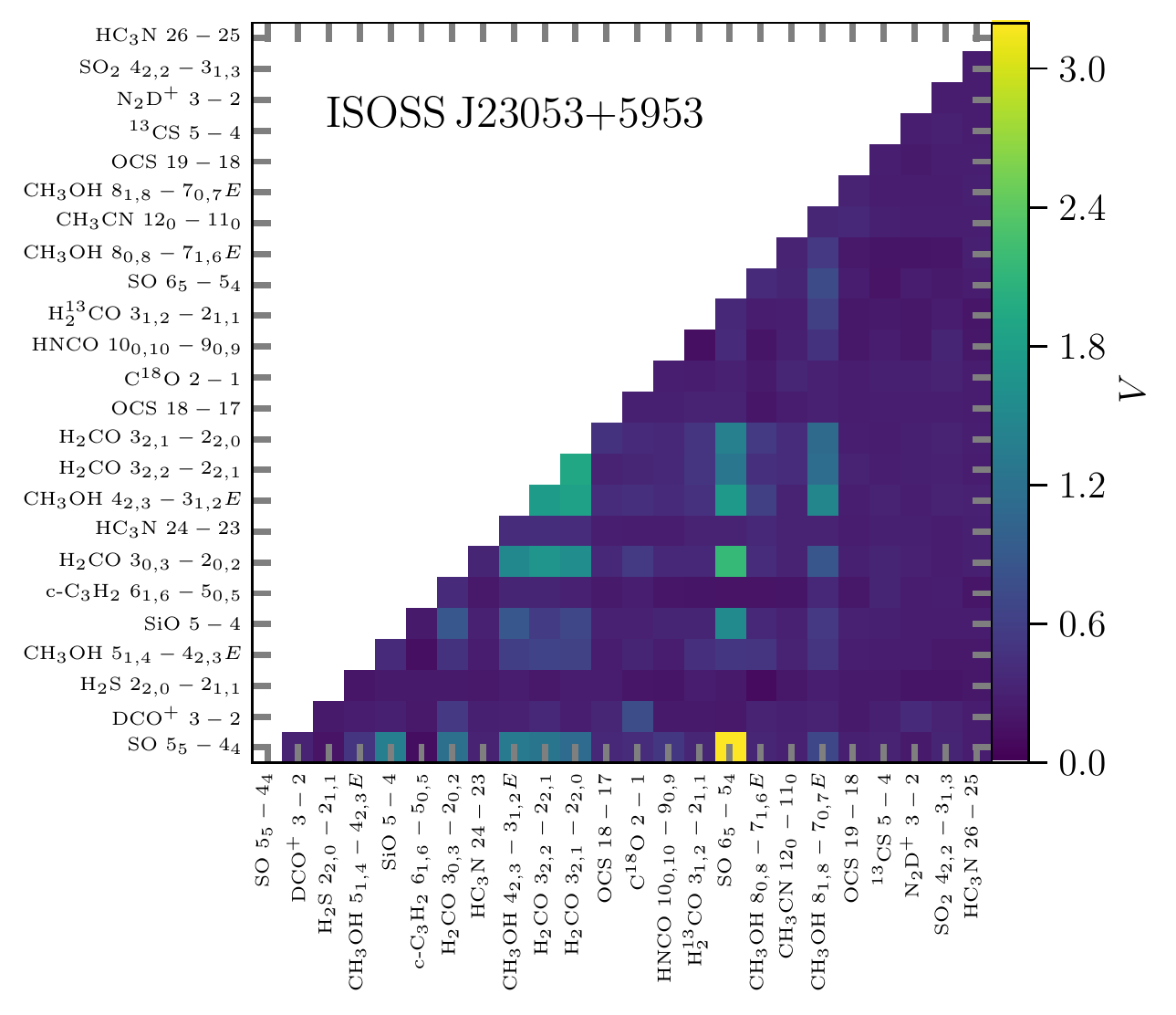}
\caption{HOG results for ISOSS\,J22478+6357 (\textit{top panel}) and ISOSS\,J23053+5953 (\textit{bottom panel}). The peak projected Rayleigh statistic $V$ is shown for each transition pair (Sect. \ref{sec:HOG}). Examples of high spatial correlations in velocity space are shown in Fig. \ref{fig:HOG_example}.}
\label{fig:HOG_summary}
\end{figure*}

\begin{figure*}
\centering
\includegraphics[]{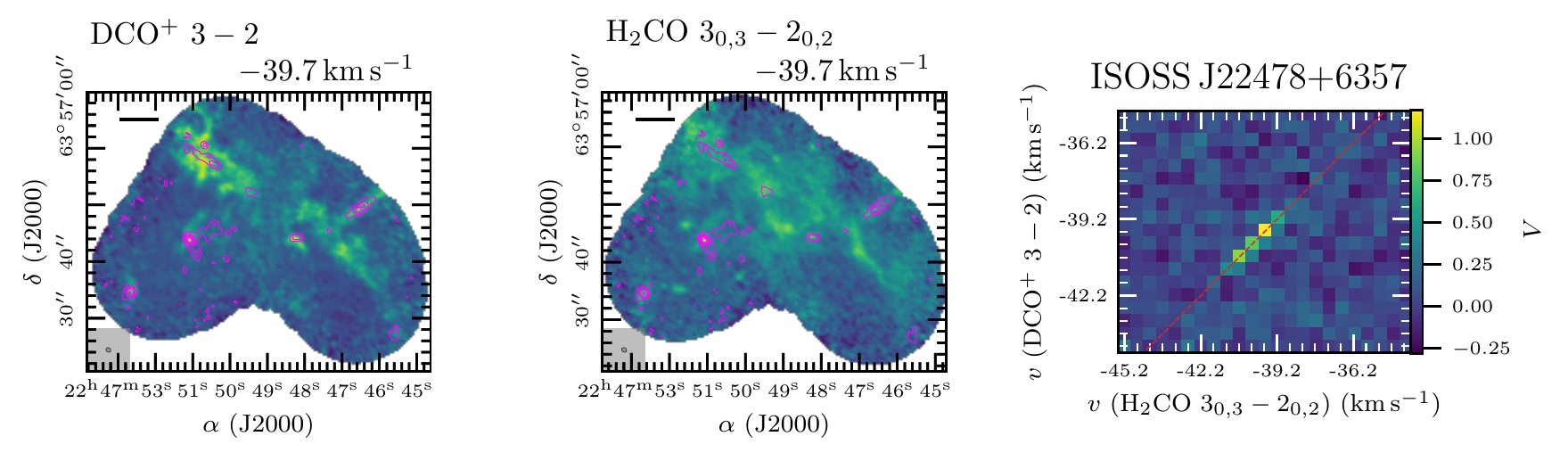}
\includegraphics[]{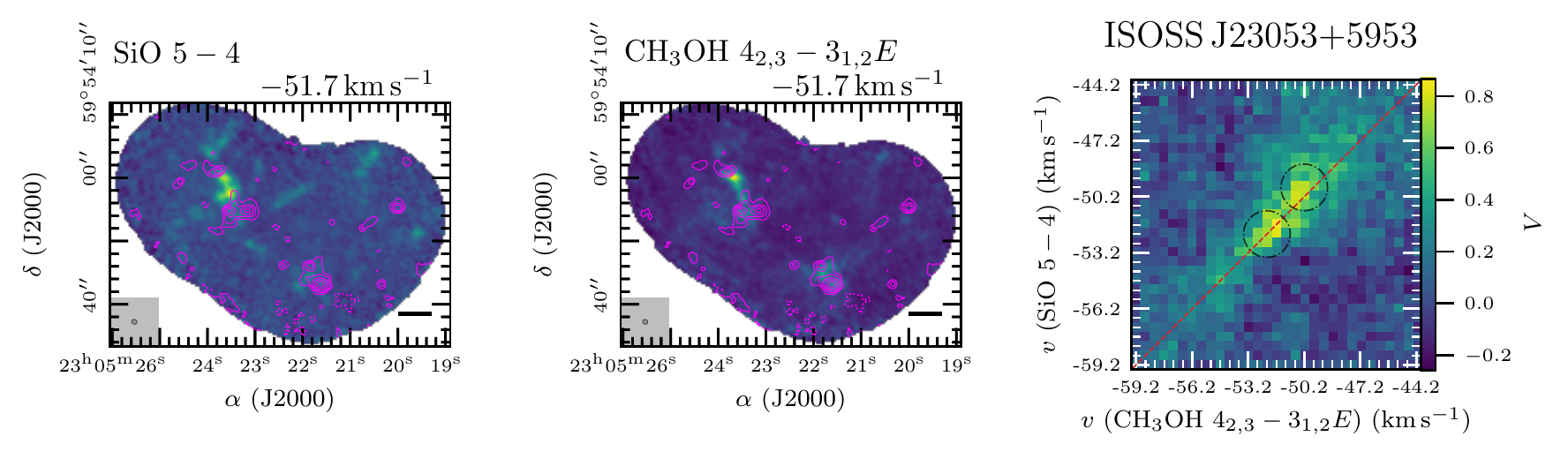}
\caption{HOG correlation results for DCO$^{+}$ ($3-2$) and H$_{2}$CO ($3_{0,3}-2_{0,2}$) toward ISOSS\,J22478+6357 (\textit{top panel}) and for SiO ($5-4$) and CH$_{3}$OH ($4_{2,3}-3_{1,2}E$) toward ISOSS\,J23053+5953 (\textit{bottom panel}). Emission maps of the channels with the highest correlation are shown in the \textit{left} and \textit{middle} panels. The NOEMA 1.3\,mm continuum data are shown in pink contours. The dotted pink contour marks the $-5\sigma_\mathrm{cont}$ level. Solid pink contour levels are 5, 10, 20, 40, and 80$\sigma_\mathrm{cont}$. The synthesized beam of the continuum data is shown in the bottom left corner. The synthesized beam of the spectral line data is similar. A linear spatial scale of 0.1\,pc is indicated by a black scale bar. The projected Rayleigh statistic $V$, computed for all velocity pairs, is shown in the \textit{right panel}. The red dashed line indicates equal velocity in both transitions. For ISOSS\,J23053+5953, the velocity regime around the two velocity components seen in line emission at $\varv \approx -52$\,km\,s$^{-1}$ and $-50$\,km\,s$^{-1}$ (Sect. \ref{sec:kinematics}) are highlighted by black dash-dotted circles.}
\label{fig:HOG_example}
\end{figure*}

\begin{figure*}
\centering
\includegraphics[]{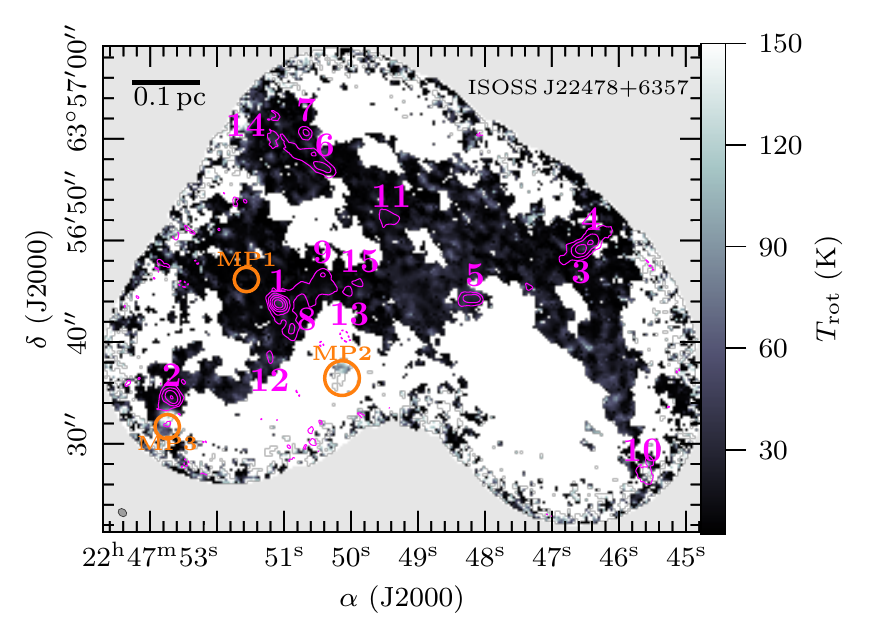}
\includegraphics[]{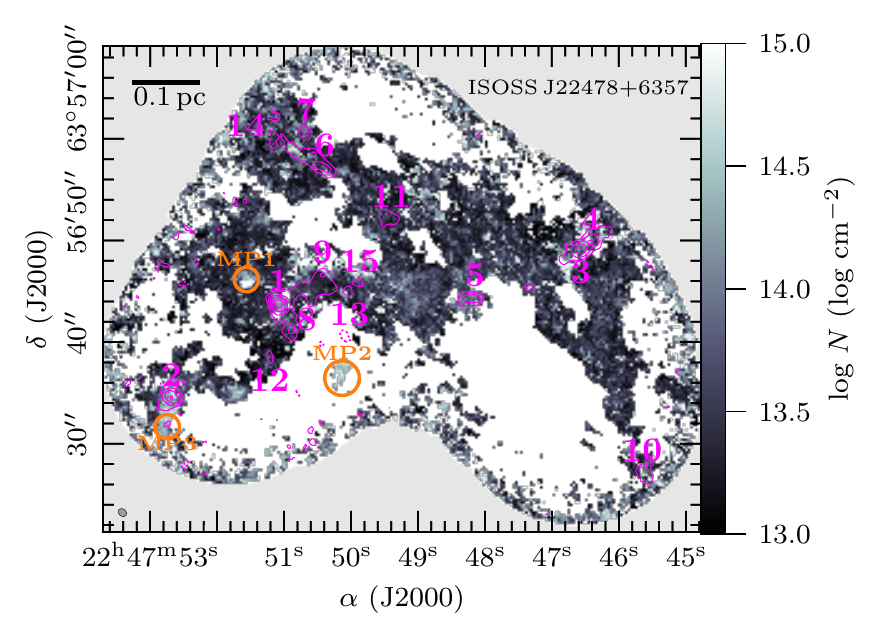}
\includegraphics[]{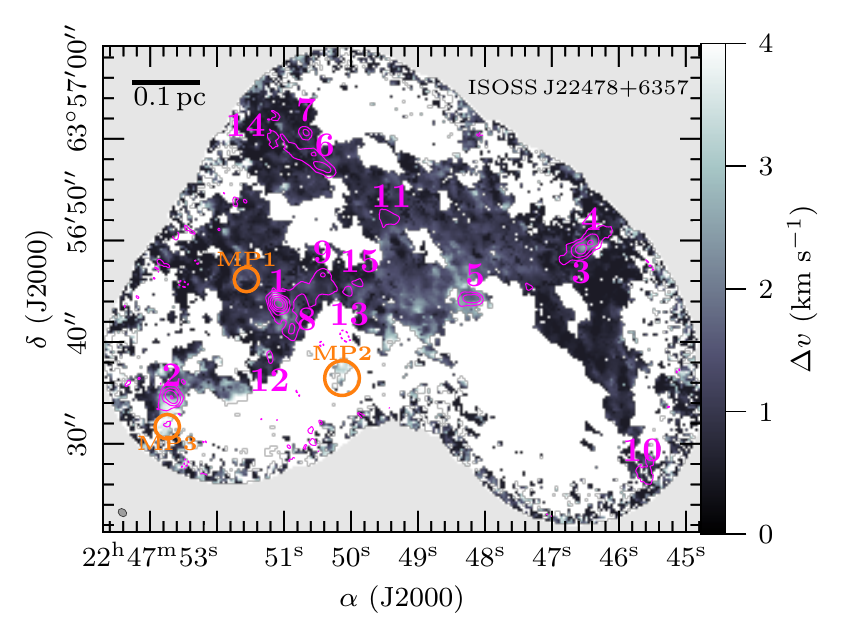}
\includegraphics[]{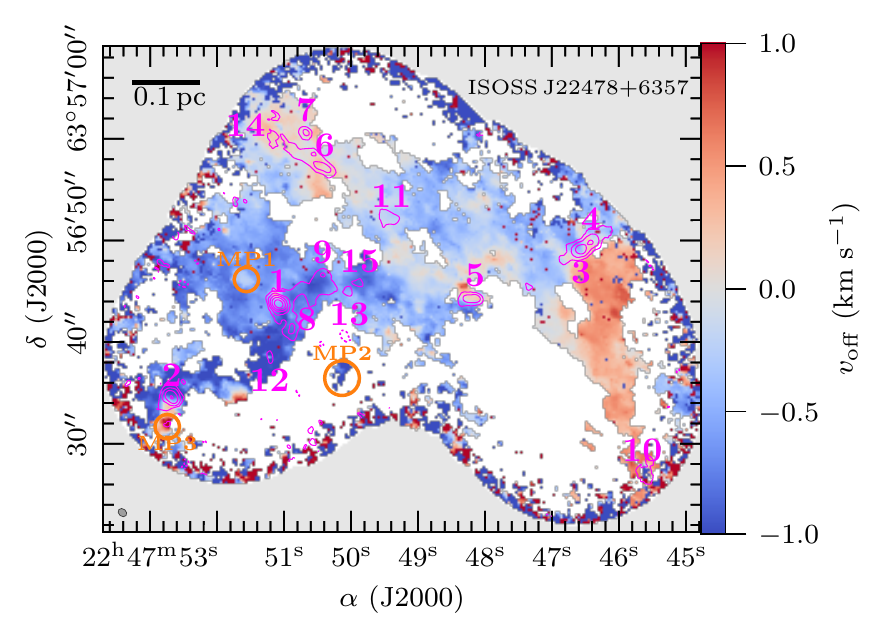}
\caption{H$_{2}$CO parameter maps of ISOSS\,J22478+6357 derived with \texttt{XCLASS}. The fit parameter maps ($T_\mathrm{rot}$, $N$, $\Delta \varv$, $\varv_\mathrm{off}$) for H$_{2}$CO are presented in color scale. The NOEMA 1.3\,mm continuum data are shown in pink contours. The dotted pink contour marks the $-5\sigma_\mathrm{cont}$ level. Solid pink contour levels are 5, 10, 20, 40, and 80$\sigma_\mathrm{cont}$. All mm cores identified in \citet{Beuther2021} are labeled in pink. Positions with a peak in molecular emission, but no associated continuum (``molecular peaks,'' MP) are indicated by orange circles (Sect. \ref{sec:moment0}). The synthesized beam of the continuum data is shown in the bottom left corner. The synthesized beam of the spectral line data is similar. A linear spatial scale of 0.1\,pc is indicated by a black scale bar.}
\label{fig:map_22478_H2CO}
\end{figure*}

\begin{figure*}
\centering
\includegraphics[]{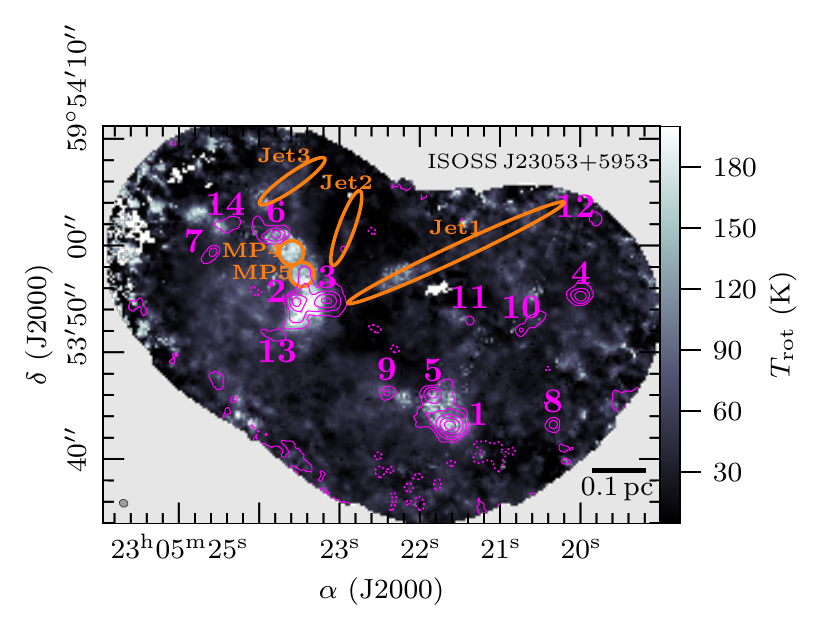}
\includegraphics[]{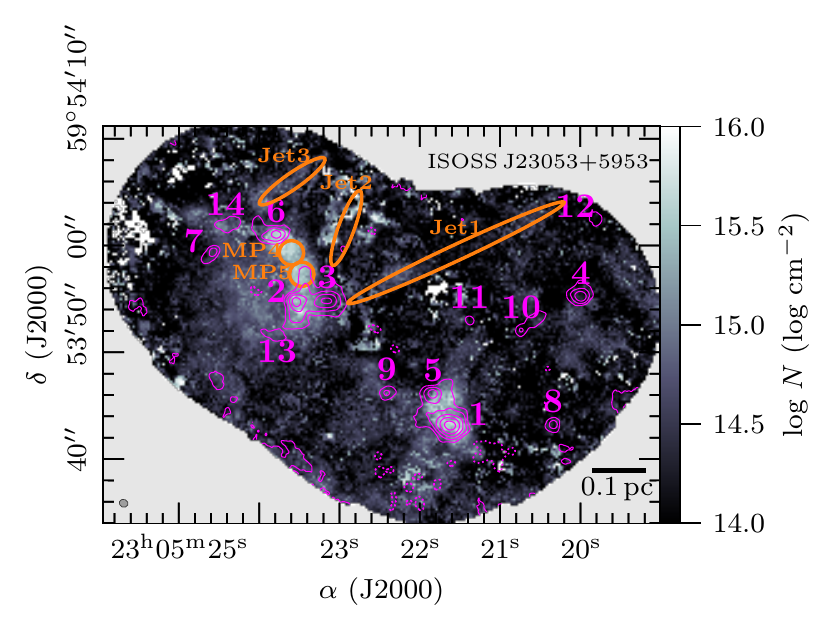}
\includegraphics[]{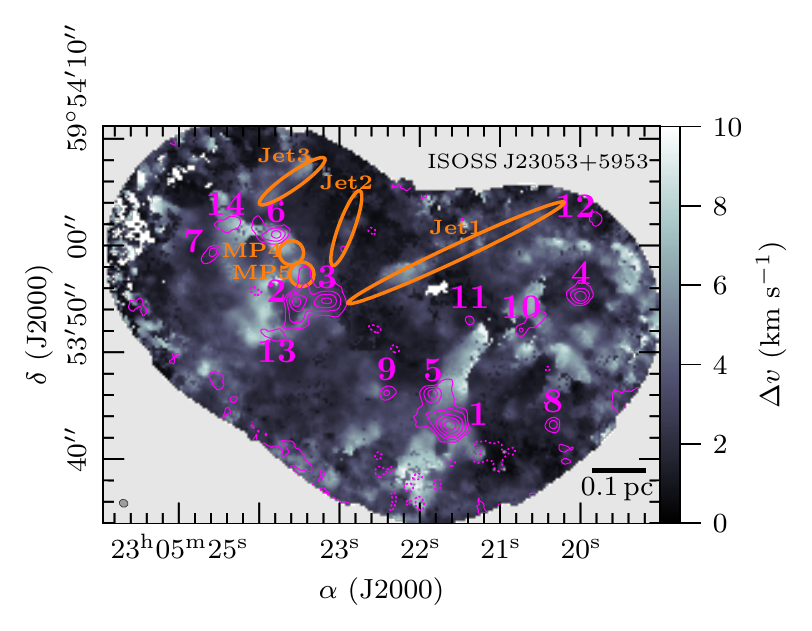}
\includegraphics[]{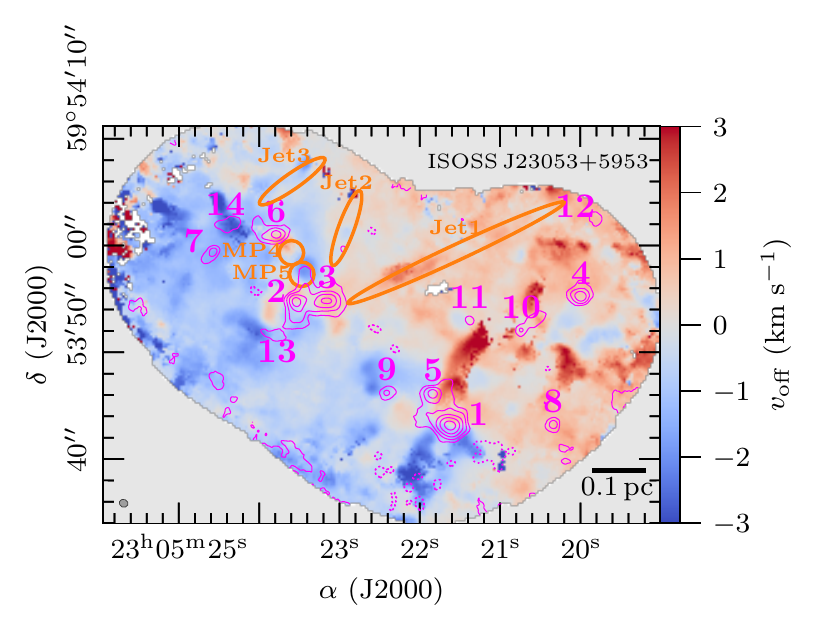}
\caption{The same as Fig. \ref{fig:map_22478_H2CO}, but for ISOSS\,J23053+5953. Jet-like structures, seen in SiO $5-4$ emission, are indicated by orange ellipses (Fig. \ref{fig:moment0_SiO_5_4}).}
\label{fig:map_23053_H2CO}
\end{figure*}

	To quantify the spatial correlation of the detected molecular emission lines, we applied the histogram of oriented gradients (HOG) method, for which a detailed description is given in \citet{Soler2019}. In summary, HOG computes the relative orientation of the local intensity gradients of two position-position-velocity (PPV) cubes $I^{\mathrm{A}}_{ijl}$ and $I^{\mathrm{B}}_{ijm}$, where $i$ and $j$ run over the spatial axes and $l$ and $m$ over the spectral axes. 
	
	With the increasing bandwidth of correlators, sensitivity, and number of observed regions, it has become challenging to compare the spatial distribution of molecular emission and it is basically impossible to do that by eye on a channel-by-channel basis. The correlation function can also be used to study spatial correlations of the integrated intensity \citep[e.g.,][]{Guzman2018,Law2021}. The HOG method also allows us to find a similar spatial morphology between two transitions. As the comparison is carried out in each velocity channel, potential velocity offsets between two molecules can be identified for example \citep{Soler2019}.
	
	For the two regions in this study, we are able to compare the results obtained with HOG with the morphology of the integrated intensity maps (Sect. \ref{sec:moment0}). However, for future line surveys toward star-forming regions and the analysis of the spatial morphology of the original CORE sample, HOG provides a convenient method to compare the line emission at high angular resolution and to find and study spatial correlations of molecular emission.
	
	The projected Rayleigh statistic, ($V$), is a statistical test to determine whether the distribution of angles $\phi_{ijlm}$ between the gradients is nonuniform and peaked at a particular angle \citep[see, e.g.,][]{Durand1958,Batschelet1972,Jow2018}. In this application, the angle of interest is $0$ degrees, which corresponds to the alignment of the iso-intensity contours in the PPV cubes.
	
	In our application, we accounted for the statistical correlation brought in by the beam by introducing the statistical weights $w_{ij}$. Relative orientation angles of local intensity gradients $\phi_{ijlm}$ are computed after applying a Gaussian filter with kernel size $\Delta$. The weighting is either
$w_{ijlm} = \frac{\delta}{\Delta}$ with pixel size $\delta$ or $w_{ijlm} = 0$ in noisy regions. The projected Rayleigh statistic $V_{lm}$ quantifies the amount of spatial correlation between the velocity channels $l$ and $m$ of two PPV cubes and is calculated from

\begin{equation}
V_{lm} = \frac{\sum_{ij}w_{ijlm}\mathrm{cos}(2\phi_{ijlm})}{\sqrt{\sum_{ij}\frac{w_{ijlm}^{2}}{2}}}.
\end{equation}

	Hence high values of $V_{lm}$ correspond to a high spatial correlation, while low values of $V_{lm}$ correspond to low or no spatial correlation, for example, when comparing two velocity channels dominated by noise. To recover the smallest spatial scales, we adopted a kernel size $\Delta$ equal to the synthesized beam. For both regions, we compare all pairs of detected emission lines listed in Table \ref{tab:lineobs}, except for the optically thick CO and $^{13}$CO transitions. For all transitions, the emission was compared in channels within a velocity range between $\varv_\mathrm{LSR}-7.5$\,km\,s$^{-1}$ and $\varv_\mathrm{LSR}+7.5$\,km\,s$^{-1}$. The only exception is the H$_{2}$CO $3_{0,3}-2_{0,2}$ line observed toward ISOSS\,J22478+6357, which lies at the edge of the high-resolution unit. Here we only considered a velocity range between $\varv_\mathrm{LSR}-5.5$\,km\,s$^{-1}$ and $\varv_\mathrm{LSR}+5.5$\,km\,s$^{-1}$. The peak projected Rayleigh statistic $V$ for all transitions is shown in Fig. \ref{fig:HOG_summary} for ISOSS\,J22478+6357 and ISOSS\,J23053+5953. A detailed example of two transitions with a high spatial correlation for each region is shown in Fig. \ref{fig:HOG_example}.
	
	In ISOSS\,J22478+6357, many transitions are not detected and thus excluded from the analysis (detections and non-detections for both regions are listed in Table \ref{tab:lineobs}). The low-energy transitions of H$_{2}$CO and SO have a high correlation. C$^{18}$O, DCO$^{+}$, and H$_{2}$CO trace large-scale emission. As expected, high correlations of transitions of the same species are found (H$_{2}$CO and SO). The high correlation between DCO$^{+}$ and N$_{2}$D$^{+}$ is originating from the strong emission around mm core 2.
	
	For ISOSS\,J23053+5953 we also find high correlations among combinations of transitions of the same species, which is the case for SO, H$_{2}$CO, and CH$_{3}$OH. While transitions with higher upper energy levels are generally less extended, the emission stems from the same location. The observed SiO transition shows a high correlation with both SO transitions toward the outflow of core 1, Jet1, and the shocked region caused by the putative colliding flow where MP4 and MP5 are located. With the exception of cores 1, 2, and 6 in ISOSS\,J23053+5953, the molecular emission does not peak toward the cores, which would be expected if the gas temperatures are already high enough to evaporate species frozen on the dust grains or allow efficient gas phase chemistry reactions. On the contrary, the emission of SO, SiO, H$_{2}$CO, and CH$_{3}$OH transitions peak toward the shocked region (MP4 and MP5).
	
	In summary, with HOG we are able to find molecular species that show a high spatial correlation. This can be the case for chemically related species (e.g., H$_{2}$CO and CH$_{3}$OH) and for species tracing physical conditions such as a shock (e.g., SiO and SO). As expected, multiple transitions of the same molecule also have a high spatial correlation (e.g., for transitions of SO, H$_{2}$CO, and CH$_{3}$OH). In addition, the kinematic features of the molecular line emission can be studied in detail. We do not find significant velocity offsets for species with a high spatial correlation, but the two velocity components in ISOSS\,J23053+5953 can be clearly identified in Fig. \ref{fig:HOG_example} (right panel) at $-52$\,km\,s$^{-1}$ and at $-50$\,km\,s$^{-1}$ and are indicated by black dash-dotted circles in the figure.
	
\subsection{Kinematic properties}\label{sec:kinematics}

	A detailed study of the kinematic properties using the NOEMA + IRAM 30m observations of ISOSS\,J22478+6357 and ISOSS\,J23053+5953 is presented in \citet{Beuther2021}. DCO$^{+}$ is a good tracer of the early stages of HMSF \citep[e.g.,][]{Gerner2015}. 
	
	Using the DCO$^{+}$ $3-2$ transition, both regions show distinct emission features (Fig. \ref{fig:moment0_DCO+_3_2}). The integrated line intensity shows that the 15 cores in ISOSS\,J22478+6357 are connected by filamentary structures (Fig. \ref{fig:moment0_DCO+_3_2}). Multiple velocity components are resolved toward substructures within the region and the line widths (FWHM) are small, on the order of $\sim$1\,km\,s$^{-1}$ \citep{Beuther2021}. The DCO$^{+}$ line integrated intensity of ISOSS\,J23053+5953 does not show filamentary, but extended emission with many emission peaks close, but slightly offset from the core positions (Fig. \ref{fig:moment0_DCO+_3_2}). There are two distinct velocity components, $\varv \approx -53$\,km\,s$^{-1}$ in the southeast direction and $\varv \approx -51$\,km\,s$^{-1}$ in the northwest direction \citep{Beuther2021}. This velocity gradient has already been reported by \citet{Bihr2015} using NH$_{3}$ emission at lower angular resolution and can be explained by a colliding gas flow triggering star formation. The line widths of individual DCO$^{+}$ components are also small within the region, $\Delta \varv \leqslant 1.5$\,km\,s$^{-1}$, the only exceptions being cores 1 and 2 with $\Delta \varv > 2$\,km\,s$^{-1}$ \citep{Beuther2021}. The thermal line width is 0.2\,km\,s$^{-1}$ at 20\,K \citep{Beuther2021}.
	
	The presence of two velocity components is seen in various molecular tracers using single-dish \citep{Wouterloot1988,Wouterloot1993,Larionov1999} and interferometric observations \citep{Bihr2015,Beuther2021}. In \citet{Beuther2021} the DCO$^{+}$ ($3-2$) intensity-weighted peak velocity (``moment 1'') map is shown to highlight the velocity gradients. For a more complete picture, we show in Fig. \ref{fig:moment1} moment 1 maps of all observed lines with extended emission ($^{13}$CO, C$^{18}$O, SO, SiO, DCO$^{+}$, H$_{2}$CO, and CH$_{3}$OH) with the exception of the optically thick CO $2-1$ transition. The velocity gradient is clearly seen in all transitions. This velocity gradient is suggested to be caused by a colliding flow \citep{Bihr2015,Beuther2021}.
	
	Another tracer of the dynamical processes in the regions is SiO. SiO is produced in shocked regions, for example, due to outflows, disk winds or converging gas flows. In a shock, silicon is sputtered off the grains and subsequently forms SiO in the gas phase \citep{Schilke1997}. Using the SiO $5-4$ transition, significant emission is only detected toward the south in ISOSS\,J22478+6357 with no nearby continuum source. In Sect. \ref{sec:outflows}, we find that this emission peak is shocked gas that is directed along the blue-shifted side of the outflow of core 1. In ISOSS\,J23053+5953, the spatial extent of the line integrated intensity of the SiO $5-4$ is much larger peaking between cores 2, 3, and 6. Jet-like features (Jet1 and Jet2), which might be protostellar outflows originating from core 3, are seen toward the north of the region. SiO emission is also seen between cores 1 and 5 and likely caused by the outflow of core 1 (Sect. \ref{sec:outflows}).
	 
\subsection{Formaldehyde distribution}\label{sec:XCLASSMapFitting}

	The spectral line data can be used to derive molecular properties such as the rotation temperature $T_\mathrm{rot}$ and column density $N$ in each pixel and to create parameter maps within the full FOV. It is computationally expensive to apply this method for all pixels and all detected molecules (Table \ref{tab:lineobs}); therefore, we only applied this pixel-by-pixel analysis to formaldehyde (H$_{2}$CO), for which we detect three transitions. For the regions of the original CORE sample, temperature maps are also derived using the high-density tracer CH$_{3}$CN \citep{Gieser2021}, but since it is not detected in ISOSS\,J22478+6357 and the emission is not extended in ISOSS\,J23053+5953, only H$_{2}$CO can be used to probe the gas temperature.
	
	We employ the eXtended CASA Line Analysis Software Suite \citep[\texttt{XCLASS},][]{XCLASS} to derive the following parameter set for a molecule: source size $\theta_{\mathrm{source}}$, rotation temperature $T_{\mathrm{rot}}$ (``excitation temperature''), column density $N$, line width $\Delta \varv$, and offset velocity $\varv_{\mathrm{off}}$ with respect to the local standard of rest $\varv_{\mathrm{LSR}}$. A detailed description of the used \texttt{XCLASS} setup is given in Appendix \ref{app:XCLASSfits}.
	
	The H$_{2}$CO rotation temperature maps are already presented in \citet{Beuther2021}. Assuming $T_{\mathrm{kin}} \approx T_{\mathrm{rot}}$, the H$_{2}$CO rotation temperature was used as input to estimate the H$_{2}$ column density $N$(H$_{2}$) and core mass $M_\mathrm{core}$ from the 1.3\,mm continuum data (Table \ref{tab:positions}). Pixels with $T_{\mathrm{rot}} < 3$\,K are only found toward the edge of the FOV where the noise is high and the fits are unreliable.

	The H$_{2}$CO parameter maps of ISOSS\,J22478+6357 are shown in Fig. \ref{fig:map_22478_H2CO}. The rotation temperature $T_{\mathrm{rot}}$ is generally low varying between $10 - 50$\,K. Toward MP2 there is a $\approx 3'' \times 3''$ region with a high rotation temperature, $T_{\mathrm{rot}} > 150$\,K. As discussed in Sect. \ref{sec:outflows} and \ref{sec:moment0}, this location is directed along the blue-shifted outflow of core 1 and is a shocked region with an enhanced H$_{2}$CO abundance and temperature increase. This can be seen in the column density map where $N$(H$_{2}$CO) is highest toward the core positions with strong mm continuum emission, but also toward this shocked region. The line widths are small, on the order of 1\,km\,s$^{-1}$. The line width significantly increases at the positions of MP2 and MP3 with $\Delta \varv > 3$\,km\,s$^{-1}$. A small east-west velocity gradient within the region is observed in the H$_{2}$CO envelope with $\varv_{\mathrm{off}}$ varying between $\pm 1$\,km\,s$^{-1}$.
	
	The H$_{2}$CO parameter maps of ISOSS\,J23053+5953 are shown in Fig. \ref{fig:map_23053_H2CO}. The average rotation temperature is slightly higher with an average of $T_{\mathrm{rot}} \approx 70$\,K compared to ISOSS\,J22478+6357. A high increase with rotation temperatures $> 150$\,K is observed toward MP4 and MP5, in cores 2 and 13, and toward cores 1 and 5. We note that the highest temperatures are not located toward the continuum peak positions of the cores, but are offset and/or between the cores. This implies that the cores themselves are still cold, in agreement with temperature estimates using NH$_{3}$ \citep{Bihr2015}, but shocks increase the local gas temperature and enhance the H$_{2}$CO abundance by a factor of two, and increase the line width. The shocked region in the northeast is associated with a colliding flow and potential outflows, while the shocked region between cores 1 and 5 is most likely associated with the strong outflow of core 1, which is also seen by strong velocity gradients in $\varv_{\mathrm{off}}$ (see Sect. \ref{sec:outflows} for a discussion of the outflows in the regions). Surrounding core 4 a ring-like structure of increase in $T_{\mathrm{rot}}$, $\Delta \varv$, and $\varv_{\mathrm{off}}$ can be observed, which could be caused by the red-shifted outflow of core 1.
	
	In summary, the H$_{2}$CO parameter maps reveal the large-scale gas distribution in both regions. We find that the temperature is enhanced toward the MPs that are caused by shocks. At this high angular resolution, we find that the temperature is higher in some locations compared to estimates using the Herschel PACS observations \citep[$\sim$20\,K,][]{Ragan2012}. Toward some mm cores and the shocked locations, the H$_{2}$CO column density and line width are enhanced as well. A small east-west velocity gradient is observed in ISOSS\,J22478+6357, while in ISOSS\,J22478+6357 there is a steep velocity gradient from the northeast to the southwest.
	
\subsection{Radial temperature profiles}\label{sec:temperatureprofile}
\begin{figure*}
\centering
\includegraphics[]{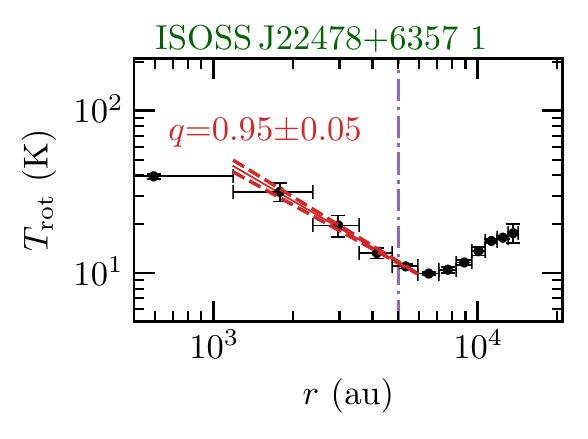}
\includegraphics[]{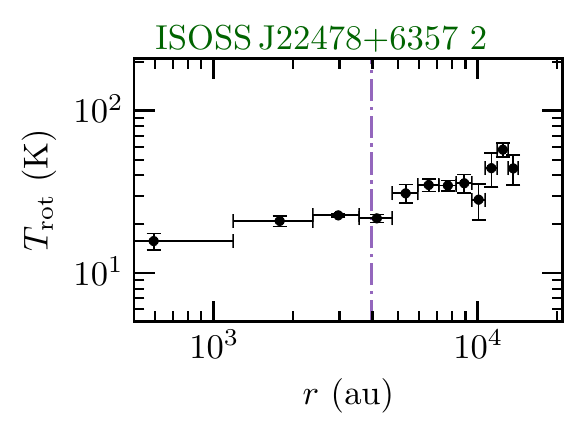}
\includegraphics[]{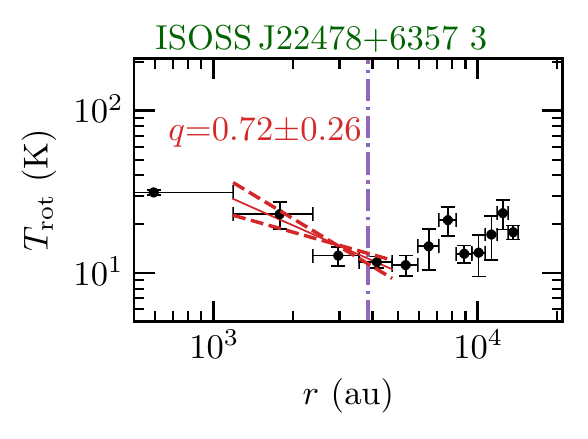}
\includegraphics[]{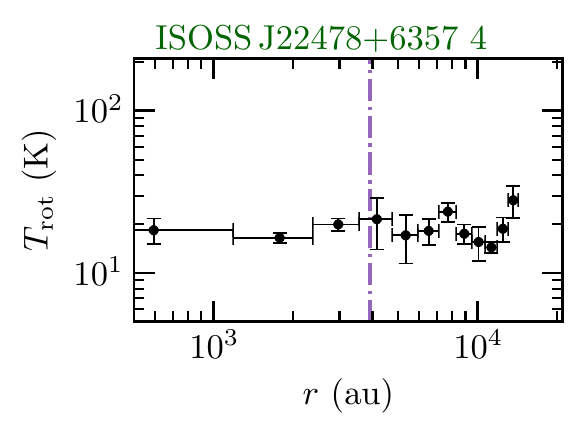}
\includegraphics[]{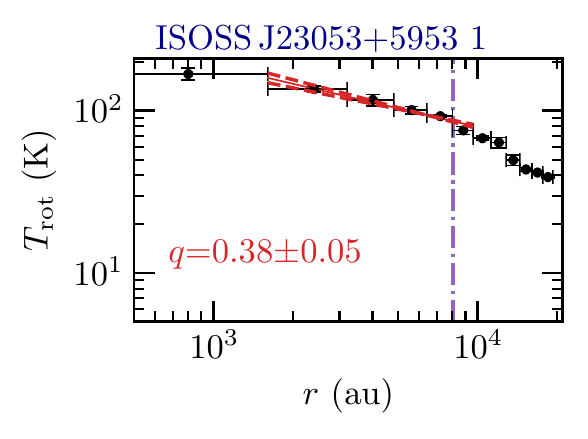}
\includegraphics[]{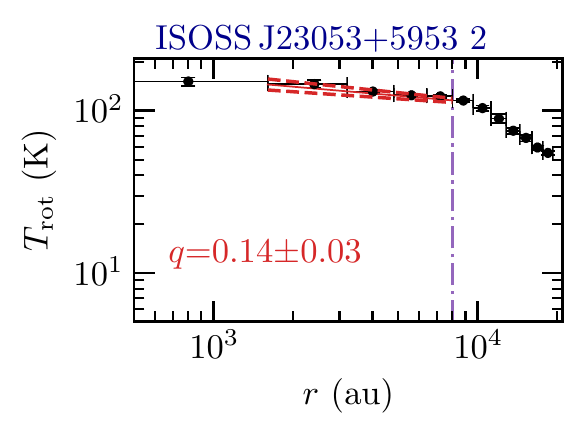}
\includegraphics[]{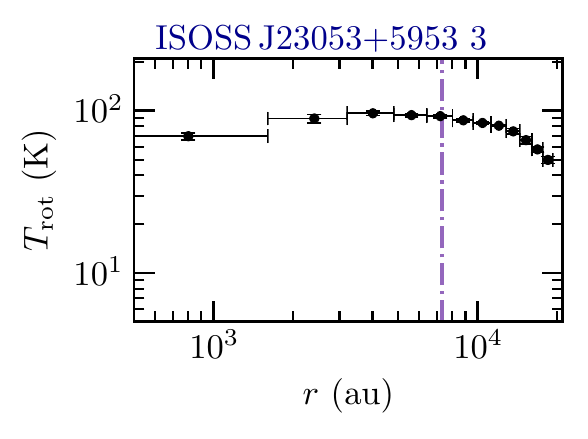}
\includegraphics[]{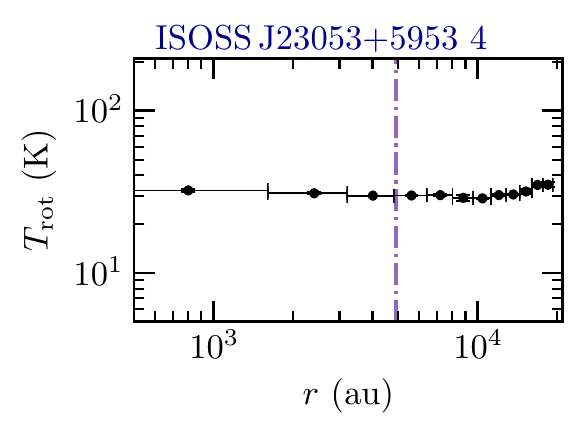}
\includegraphics[]{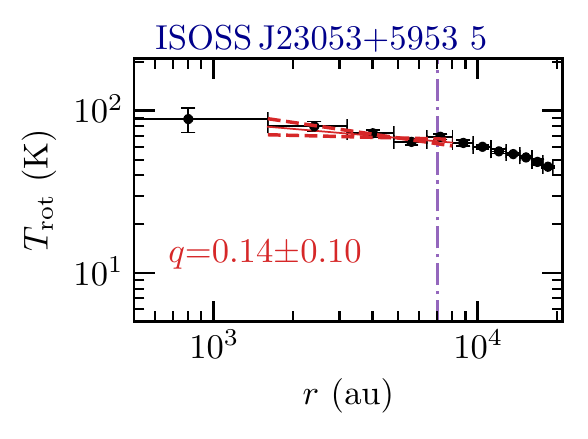}
\includegraphics[]{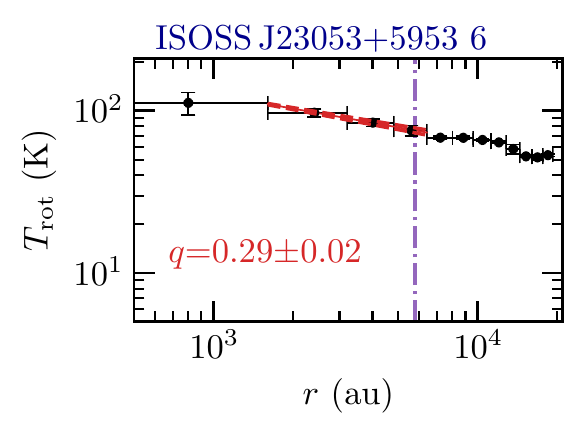}
\caption{Radial temperature profile for cores $1-4$ in ISOSS\,J22478+6357 and cores $1-6$ in ISOSS\,J23053+5953. The radial H$_{2}$CO temperature profile is shown by the black data points. The outer radius $r_\mathrm{out}$, estimated with the \texttt{clumpfind} algorithm in the continuum data (Table \ref{tab:positions}), is indicated by the dash-dotted purple vertical line. A fit to the data and its uncertainties ($\pm$1$\sigma$) are shown by the red solid and dashed lines, respectively.}
\label{fig:temperatureprofile}
\end{figure*}
	
	The derived H$_{2}$CO temperature maps can be used to create radial temperature profiles of the mm cores. To be consistent with the analysis of the density profiles (Sect. \ref{sec:density}), we only applied the following analysis to cores with S/N $> 20$. Radial temperature profiles were extracted from the H$_{2}$CO temperature map for cores $1-4$ in ISOSS\,J22478+6357 and for cores $1-6$ in ISOSS\,J23053+5953. We binned the data with a bin size $0.5 \times \Delta r$ corresponding to half of the average beam size ($\Delta r = \frac{\theta_{\mathrm{maj}}+\theta_{\mathrm{min}}}{2}$). The radial temperature profile for cores $1-4$ in ISOSS\,J22478+6357 and cores $1-6$ in ISOSS\,J23053+5953 are shown in Fig. \ref{fig:temperatureprofile}. Cores 3 and 4 in ISOSS\,J22478+6357 are close so their temperature profiles overlap.
	
	We fitted the data assuming a power-law profile according to Eq. \eqref{eq:temperatureprofile}. We excluded the inner-most data point, which is diluted by the beam, and only fitted the data up to the bin that contains the outer radius $r_\mathrm{out}$ (Table \ref{tab:positions}). The outer radius $r_\mathrm{out}$ of each core is taken from \citep{Beuther2021} and estimated by applying the \texttt{clumpfind} algorithm \citep{Williams1994} to the 1.3\,mm continuum data. For the characteristic radius and temperature we used $r_{500} = 500$\,au and $T_{500} = T(r_{500})$. Flat temperature profiles (core 4 in ISOSS\,J22478+6357 and core 4 in ISOSS\,J23053+5953) or profiles that have an increasing temperature at increasing radii (core 2 in ISOSS\,J22478+6357 and core 3 in ISOSS\,J23053+5953) were not fitted.
	
	The results for $T(r_{500})$, and $q$ are summarized in Table \ref{tab:cores}. The temperature $T(r_{500})$ varies between 60 and 250\,K indicating that some protostars have already heated up their surrounding envelope. The temperature power-law index $q$ varies between 0.1 and 1.0. The results are consistent with results obtained in low- and high-mass cores \citep[e.g.,][]{Shirley2000,Palau2014,Gieser2021}. A comparison of the temperature profile with the cores of the CORE sample is discussed in Sect. \ref{sec:discore}.
	
	For cores 1 and 2 in ISOSS\,J22478+6357, there is a significant temperature increase for $r > r_\mathrm{out}$. The H$_{2}$CO temperature map (Fig. \ref{fig:map_22478_H2CO}) reveals that toward the shocked gas of MP2 and MP3 the obtained rotation temperature is higher ($>$100\,K) compared to the ambient gas ($\approx 20 - 30$\,K). Thus, the radial temperature increase for $r > r_\mathrm{out}$ is related to the bipolar outflows of cores 1 and 2 (Sect. \ref{sec:outflows}). The radial temperature increase of core 2 is also connected to the fact that it is located at the edge of the temperature map, where the temperature fits are unreliably due to an increase in noise (Fig. \ref{fig:map_22478_H2CO}).
	
	Flat temperature profiles are observed if the temperature gradient is not resolved. In addition, a flat profile can be observed if protostellar heating is not yet high enough to heat up the envelope. A high optical depth of the H$_{2}$CO lines could have an impact toward the densest regions of the cores, but since we can clearly derive temperature gradients for the densest cores (e.g., core 1 in ISOSS\,J23053+5953), line optical depth effects causing unreliable temperature estimates should not be an issue for the less dense cores. In \citet{Gieser2021} it was found that only in evolved HMSFRs containing HMCs the H$_{2}$CO temperature maps are not reliable in the densest central regions. In these cases, CH$_{3}$CN was used as a thermometer instead.
	
\subsection{Molecular column densities}\label{sec:molecularcolumndensities}

\begin{figure*}
\centering
\includegraphics[]{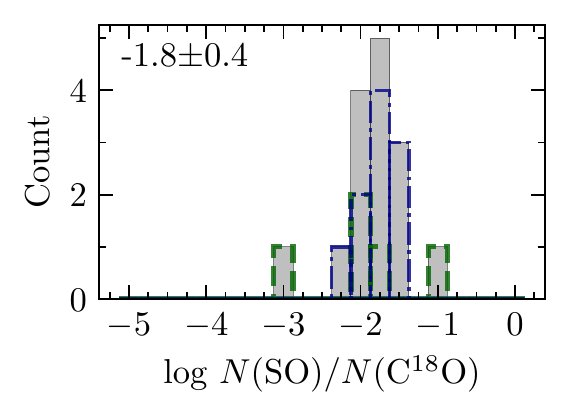}
\includegraphics[]{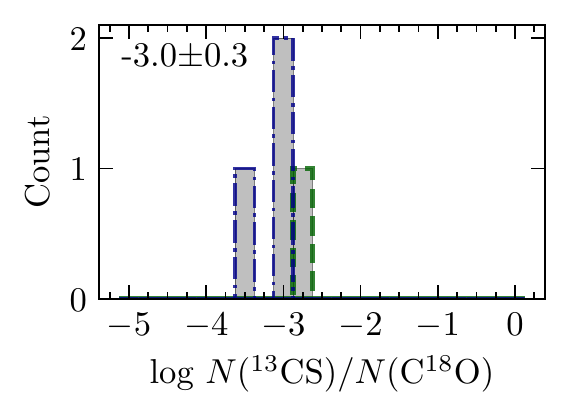}
\includegraphics[]{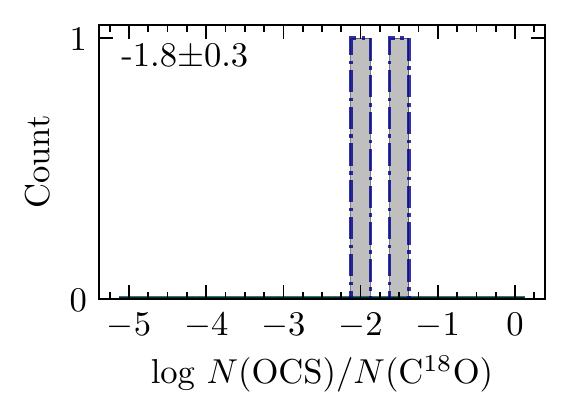}
\includegraphics[]{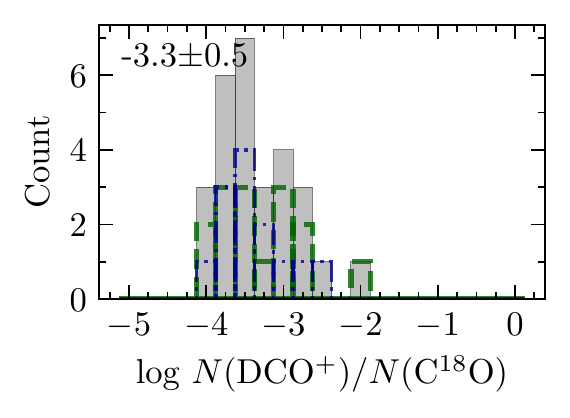}
\includegraphics[]{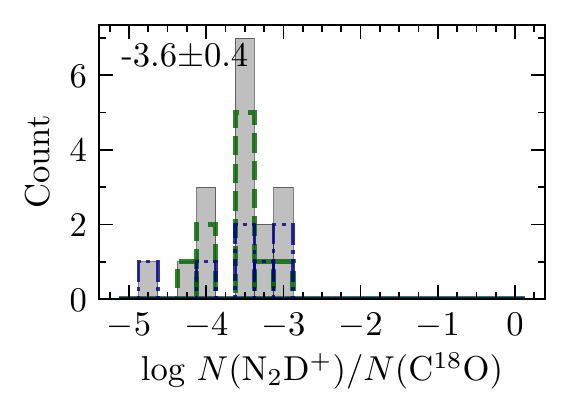}
\includegraphics[]{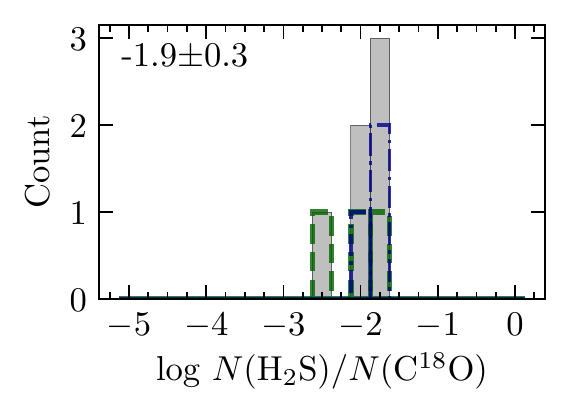}
\includegraphics[]{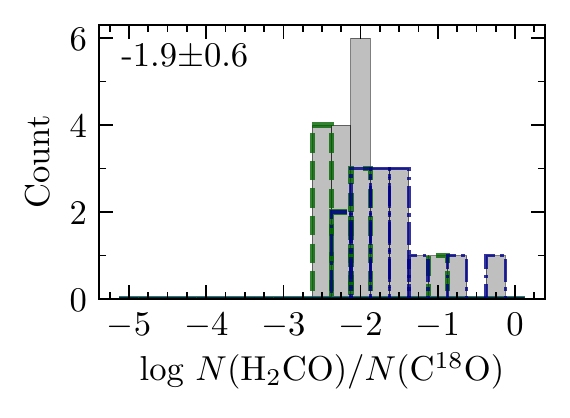}
\includegraphics[]{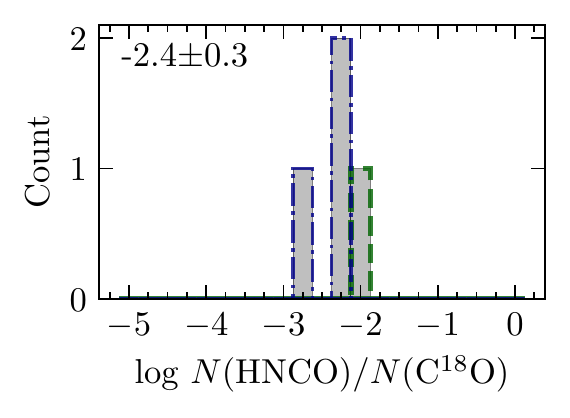}
\includegraphics[]{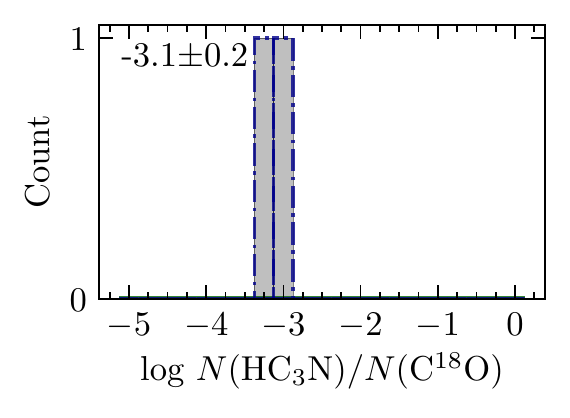}
\includegraphics[]{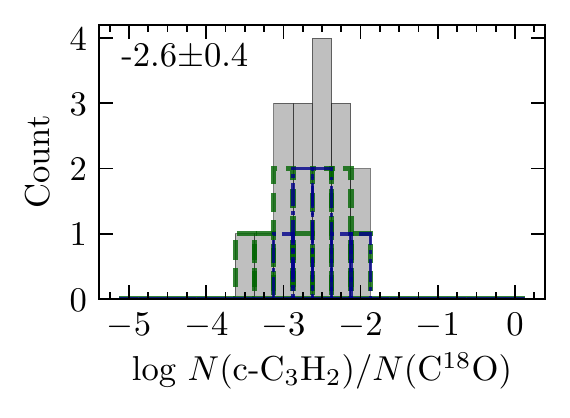}
\includegraphics[]{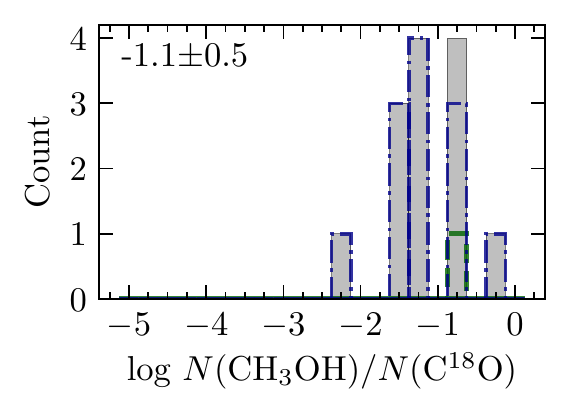}
\includegraphics[]{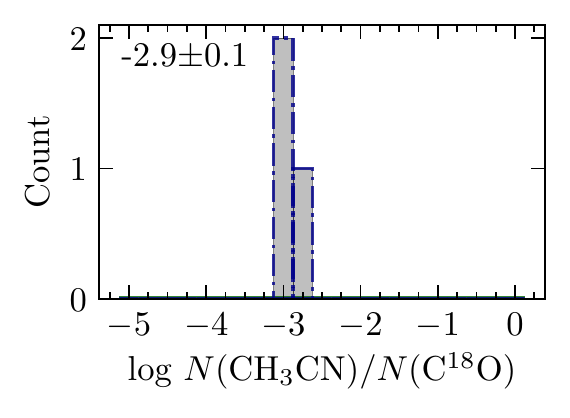}
\caption{Abundance histograms (relative to C$^{18}$O) of all species fitted with \texttt{XCLASS}. The gray histogram shows the combined results of all 29 cores. The mean and standard deviation are shown in the top left corner. The green dashed and blue dash-dotted histograms show the results for the cores in ISOSS\,J22478+6357 and ISOSS\,J23053+5953, respectively.}
\label{fig:histo_abund}
\end{figure*}

	We quantify the molecular content by analyzing the molecular column densities $N$ with \texttt{XCLASS} detected toward spectra of the 29 cores. The general setup of our \texttt{XCLASS} method is described in Appendix \ref{app:XCLASSfits}. Here, we use the \texttt{myXCLASSFit} function to estimate the molecular column densities of detected molecules (Table \ref{tab:lineobs}). For each core, all species were fitted with one emission component. The column densities of the species are an input for the physical-chemical modeling of the cores (Sect. \ref{sec:MUSCLE}). We excluded CO and $^{13}$CO from the \texttt{XCLASS} fitting because of the high optical depth, and only fitted the optically thin C$^{18}$O line. H$_{2}^{13}$CO and SO$_{2}$ were also excluded, as there is no strong emission toward the 29 cores. Our physical-chemical model does not include shock chemistry; therefore, the column density of SiO was not derived either. It should be noted that the abundances of other species are also affected by shock chemistry \citep[e.g.,][]{Palau2017}, in the case of our observations we see shocked regions in enhanced SO, H$_{2}$CO, and CH$_{3}$OH emission (see Sect. \ref{sec:moment0}).
	
	Tables \ref{tab:XCLASSresults1} and \ref{tab:XCLASSresults2} summarize the best-fit column density for all 29 cores and species fitted with \texttt{XCLASS}. Upper limits are given, if the model peak intensity is $<$3$\sigma_\mathrm{line}$, considering all transitions of a molecule. Upper limits are also given in cases where the upper column density error is larger than a factor of ten and when the lower column density error is zero. With these two additional constraints, unreliable fits can be further discarded.
	
	As an example, we show the observed and \texttt{XCLASS} modeled spectra of core 1 in ISOSS\,J23053+5953 for all fitted transitions in Fig. \ref{fig:spectrum_fit_23053}. Most of the transitions show a Gaussian-shaped line profile. The H$_{2}$CO $3_{0,3}-2_{0,2}$ line is affected by self-absorption toward the systemic velocity and has broad line wings due to the bipolar outflow. The blue-shifted peak has a lower intensity than the red-shifted peak. This asymmetry arises from the outflowing gas being partially absorbed along the line-of-sight. The outflow is also seen in OCS $18-17$ and $19-18$ transitions having a weak double peaked line profile. The intensity of the CH$_{3}$OH lines that were only observed at a spectral resolution of 3\,km\,s$^{-1}$ is overproduced by a factor of two in the \texttt{XCLASS} best-fit spectrum. The upper energy levels of these transitions are not significantly different from the remaining transitions observed with the high-resolution units. This mismatch is most likely due to the differences in spectral resolution, as the remaining two transitions are fitted well. In the low-resolution spectra, the emission of a few lines, for example for SO, is broader than the frequency range considered in the fit. This enabled that, in \texttt{XCLASS}, the non-Gaussian line wings caused by the outflows do not have a strong impact, which might cause broad features in the fitted line profile. For the remaining cores, the line widths are smaller and therefore the fit frequency range is sufficient to consider the emission of the full line. An example of the observed and \texttt{XCLASS} modeled spectra of core 1 in ISOSS\,J22478+6357 is presented in Fig. \ref{fig:spectrum_fit_22478}, for which clearly fewer emission lines are detected, and hence column densities for less species can be estimated.
	
	For all species, we computed the abundance relative to $N$(C$^{18}$O) and abundance histograms are shown in Fig. \ref{fig:histo_abund}. As the H$_{2}$ column density is derived from the NOEMA 1.3\,mm continuum emission \citep{Beuther2021} and suffers from spatial filtering, we use $N$(C$^{18}$O) for a reliable comparison instead \citep[see also][]{Gieser2021}. Each panel shows the results considering all cores, and separate histograms for ISOSS\,J22478+6357 and ISOSS\,J23053+5953 are presented as well. The logarithmic mean and standard variation of the abundance relative to $N$(C$^{18}$O) considering both regions are shown in each panel. Some species are only detected toward ISOSS\,J23053+5953 (OCS, HC$_{3}$N, and CH$_{3}$CN). For SO, $^{13}$CS, DCO$^{+}$, N$_{2}$D$^{+}$, H$_{2}$S, c-C$_{3}$H$_{2}$, and CH$_{3}$OH both regions have a similar abundance. However, the H$_{2}$CO abundance is about one order of magnitude higher in ISOSS\,J23053+5953 cores compared to the cores in ISOSS\,J22478+6357. A comparison of the mean abundance found toward the ISOSS regions compared to the CORE regions is given in Sect. \ref{sec:discore}.

\section{Physical-chemical modeling}\label{sec:MUSCLE}

	By analyzing the molecular line emission in Sect. \ref{sec:molecularcolumndensities}, we find that the molecular content varies between the mm cores. Chemical timescales $\tau_{\mathrm{chem}}$ of individual cores can be estimated with the physical-chemical model \texttt{MUSCLE}. The model was originally developed to study the evolutionary timescales of the IRDC, HMPO, HMC, and UCH{\sc ii} stages using single-dish observations of 59 HMSFRs \citep{Gerner2014, Gerner2015}. The model consists of a static 1D physical structure coupled to the time-dependent gas-grain chemical network \texttt{ALCHEMIC} \citep{Semenov2010}. A detailed description of the physical-chemical model setup is given in Sect. \ref{sec:MUSCLEsetup}. By definition, the chemical timescale of the initial IRDC stage is $\tau_{\mathrm{chem}} = 0$\,yr, when the density reaches $n = 10^{4}$\,cm$^{-3}$ \citep{Gerner2014}.
	
	\texttt{MUSCLE} was applied to high angular resolution observations of the CORE pilot regions, NGC\,7538S and NGC\,7538IRS1 \citep{Feng2016}. In both regions, the cores have short chemical timescales of 11\,000-12\,000\,yr. The model parameters were investigated in detail with CORE observations of the well-studied HMC AFGL\,2591\,VLA3 \citep{Gieser2019}, for which the derived chemical timescale is in agreement with estimates derived in the literature. A complete physical-chemical analysis of 22 cores in the CORE sample is presented in \citet{Gieser2021}. The estimated chemical timescales varies from $20\,000 - 100\,000$\,yr. The chemical composition of the molecular gas of the cores varies even within a single region. An explanation for this variety is subsequent star formation. A few CORE targets contain UCH{\sc ii} regions, such as S106 or G139.9091, which both already show emission at cm wavelengths. Toward these regions, compact mm cores are found in the 1.3\,mm continuum emission, but not many emission lines are detected, with the exception of CO isotopologues, SO, and H$_{2}$CO. In these cases, molecules are destroyed by the protostellar radiation and thus the spectra are line-poor compared to spectra toward HMCs of the CORE sample.

\subsection{Model setup}\label{sec:MUSCLEsetup}

	\texttt{MUSCLE} consists of a static spherically symmetric model core described by 40 radial grid points coupled the time-dependent gas-grain chemical network \texttt{ALCHEMIC} \citep{Semenov2010}. The density $n$($r$) and temperature $T$($r$) profiles are described by power-law profiles up to a radius $r_{\mathrm{out}}$, according to Eq. \eqref{eq:densityprofile} and \eqref{eq:temperatureprofile}, respectively. From an inner radius $r_{\mathrm{in}}$ and inward, both density and temperature reach a constant plateau, $n$($r \leq r_{\mathrm{in}}$) = $n_{\mathrm{in}}$ and $T$($r \leq r_{\mathrm{in}}$) = $T_{\mathrm{in}}$.
	
	We used the same model parameters for the dust properties, interstellar radiation field, and initial chemical conditions used in the study of the 22 cores of the CORE sample, summarized in Table 5 in \citet{Gieser2021}. Instead of running the model with only 100 logarithmic steps up to $\tau_\mathrm{model} = 100\,000$\,yr \citep{Gieser2021}, we used a finer grid of 999 logarithmic steps up to $\tau_\mathrm{model} = 10^{7}$\,yr. For the inner radius $r_{\mathrm{in}}$ we chose 500\,au (Table \ref{tab:cores}). In \citet{Gieser2021}, the outer radius $r_\mathrm{out}$ was determined from the temperature profiles, but for the ISOSS, we used the outer radius derived from the \texttt{clumpfind} analysis presented in \citet{Beuther2021} and summarized in Table \ref{tab:positions}.
	
	We used the molecular hydrogen column density $N$(H$_{2}$) derived from the 1.3\,mm continuum emission (Table \ref{tab:positions}). Short-spacing information is available for the spectral line data; however, the continuum data are obtained with NOEMA only, so missing flux can be an issue. Column densities derived with \texttt{XCLASS} of the following molecules were included in the model: C$^{18}$O, $^{13}$CS, SO, OCS, DCO$^{+}$, N$_{2}$D$^{+}$, H$_{2}$S, H$_{2}$CO, HNCO, HC$_{3}$N, c-C$_{3}$H$_{2}$, CH$_{3}$OH, and CH$_{3}$CN. The chemical network \texttt{ALCHEMIC} does not consider isotopologues except for a sophisticated deuterium (D) network, thus the column densities of C$^{18}$O and $^{13}$CS were converted to the column density of their main isotopologue (CO and CS) using the isotopic ratio calculations from \citet{Wilson1994} depending on the galactocentric distance $d_\mathrm{gal}$. The $^{12}$C/$^{13}$C isotopic ratio is 80 and 86 and the $^{16}$O/$^{18}$O isotopic ratio is 607 and 649 for ISOSS J22478+6357 and ISOSS J23053+5953, respectively (Table \ref{tab:regions}). The $^{12}$C/$^{13}$C isotopic ratios are in agreement with a study by \citet{Giannetti2014}. These authors predict a ratio between $40 - 90$ and between $45 -95$ at the corresponding galactocentric distance for ISOSS\,J22478+6357 and ISOSS\,J23053+5953, respectively.
	
	A static spherically symmetric physical model is not sufficient to describe the protostellar evolution where the temperature and density are gradually increasing. However, coupling a chemical network with thousands of reactions to a hydrodynamical model is computationally expensive. As complex molecules, such as CH$_{3}$OH, are observed toward both regions, a simple chemical network would not be sufficient to describe the formation of all observed molecular species. Instead, we approximate the warm-up and density increase as step functions of several evolutionary stages with different temperature and density profiles. Using a sample of 59 HMFRS, \citet{Gerner2014, Gerner2015} created template initial chemical conditions for the IRDC, HMPO, HMC and UCH{\sc ii} stage. For each core, we ran \texttt{MUSCLE} with IRDC, HMPO and HMC initial chemical conditions, referred to as the HMPO, HMC, and UCH{\sc ii} model, respectively, to derive the best-fit chemical timescale $\tau_\mathrm{chem}$. While we follow this nomenclature, it is not our goal to classify the cores into these evolutionary stages \citep[see also][]{Gieser2021}.
	
	For each initial condition model, the chemical age $\tau_\mathrm{chem}$ was estimated from the best match of the modeled and observed column densities. The physical properties of each core were fixed and set to the observed physical properties ($p$, $q$, $T_{500}$, listed in Table \ref{tab:cores}). Therefore, the observed molecular column densities (Tables \ref{tab:XCLASSresults1} and \ref{tab:XCLASSresults2}) constrained the best-fit model. As both regions are in general line-poor, only cores, for which at least eight of the in total 14 column density points are considered as detections and good fits and not upper limits, are modeled with \texttt{MUSCLE}. This includes four cores: core 1 in ISOSS\,J22478+6357 and cores 1, 2, and 6 in ISOSS\,J23053+5953.
	
	Not all three initial condition models were considered for each core. The cores gradually warm up due to the protostellar radiation and a temperature decrease with increasing time would be unphysical. We therefore required that the observed temperature of the core at $r_{\mathrm{in}}$ is equal or higher than the temperature at $r_{\mathrm{in}}$ of the initial condition model, $T_{\mathrm{in}} \geq T_{\mathrm{in,init}}$. 
	
	In each time step, the modeled radial abundance profiles are converted to beam convolved column densities with the same beam size as in the observations assumed. The modeled and observed column densities of the input molecules are compared by a least $\chi^2$ analysis. We used a weighted $\chi^2$, taking into account the percentage of well modeled molecules $\Upsilon$: $\bar \chi^2 = \frac{\chi^2}{\Upsilon^2}$. The model with the lowest $\bar \chi^2$ was then considered as the best-fit model. The column density of a molecule is well modeled if the modeled and observed column densities agree within a factor of ten. The chemical timescale is the sum of the timescale of the inital condition model $\tau_\mathrm{init}$ and of the core model $\tau_\mathrm{model}$: $\tau_\mathrm{chem} = \tau_\mathrm{init} + \tau_\mathrm{model}$.

\subsection{Chemical timescales}\label{sec:chemicaltimescales}

\begin{figure*}
\centering
\includegraphics[]{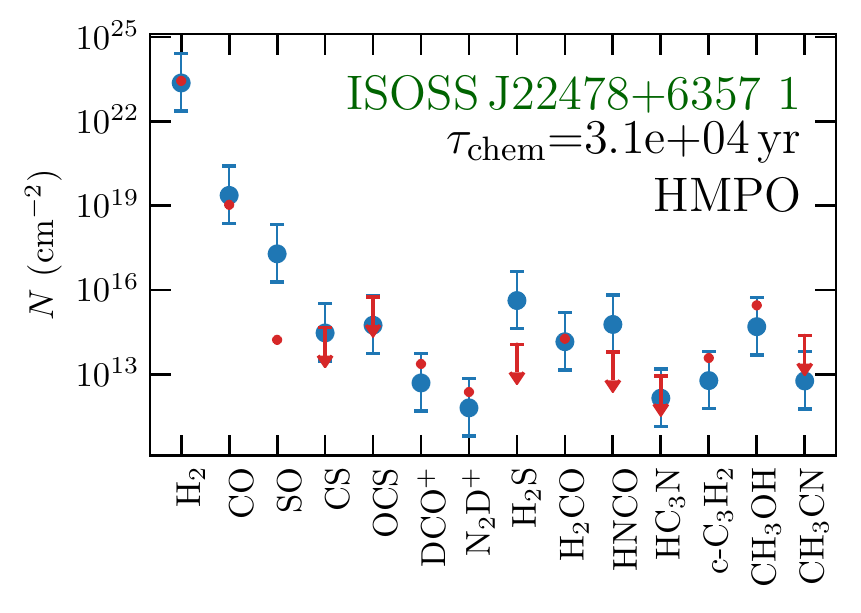}
\includegraphics[]{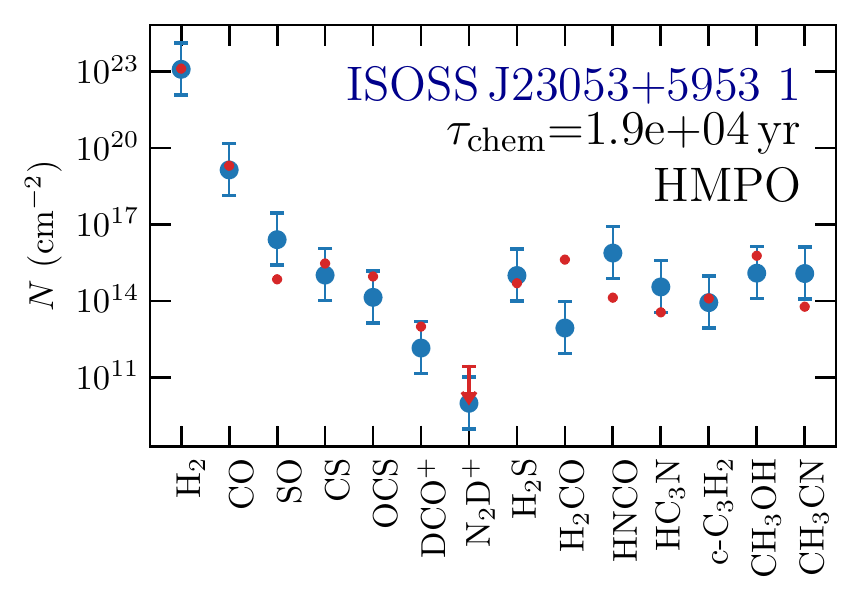}
\includegraphics[]{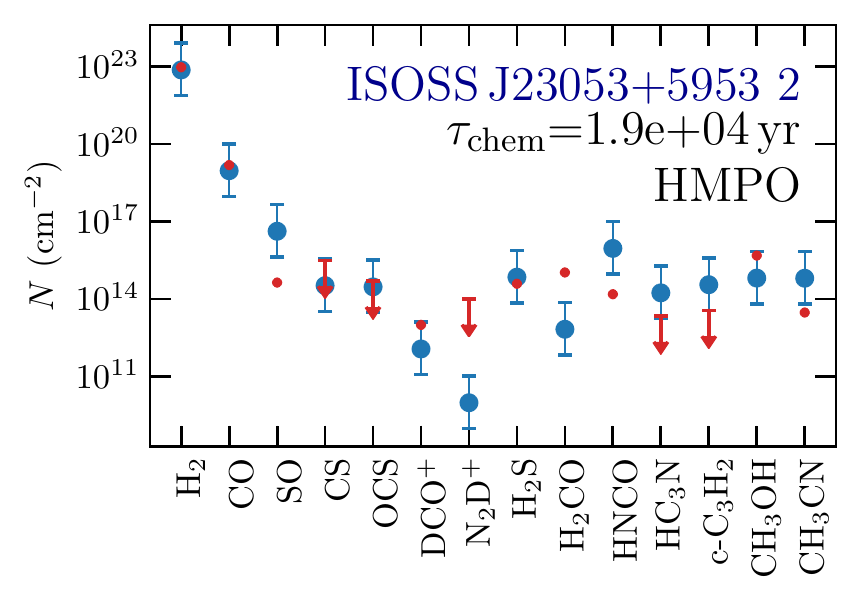}
\includegraphics[]{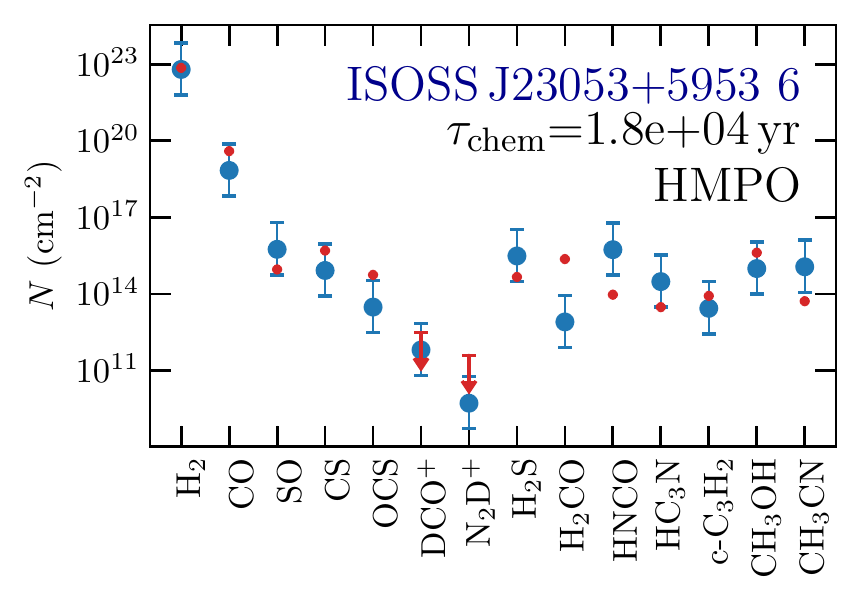}
\caption{Comparison of the observed and modeled column densities of all cores modeled with \texttt{MUSCLE}. The observed column density is shown in red and upper limits are indicated by arrows (Tables \ref{tab:XCLASSresults1} and \ref{tab:XCLASSresults2}). The column densities of the \texttt{MUSCLE} best-fit model are shown by the blue data points assuming a factor of ten uncertainty.}
\label{fig:MUSCLE_columndensity}
\end{figure*}

	The best-fit chemical timescales are summarized in Table \ref{tab:cores} for each modeled core. Detailed model results for each initial condition model are presented in Table \ref{tab:MUSCLEresults}. Following the analysis of the mm cores of the original CORE sample \citep{Gieser2021}, we assume that the chemical timescales are uncertain by a factor of two \citep[see also][]{Gerner2014,Gerner2015}. A comparison of the observed and modeled column density is shown in Fig. \ref{fig:MUSCLE_columndensity} for all four modeled cores. Since in \texttt{MUSCLE} a factor of ten uncertainty for the modeled column densities is assumed and none for the observed column densities when deciding if a molecule is well modeled or not, we only show error bars for the modeled column density data points.
	
	In the ISOSS\,J22478+6357 region, only core 1 can be modeled with \texttt{MUSCLE} and the estimated chemical timescale is 31\,000\,yr. This is in agreement with the fact that ISOSS\,J22478+6357 is young and still in a cold stage. With most molecules still frozen on the dust grains or not formed yet, the spectra appear line-poor. The modeled SO and H$_{2}$S column densities are both higher than a factor of ten compared to the observed column densities. As discussed in \citet{Gerner2014} and in \citet{Gieser2021}, the sulfur chemistry, especially SO, is poorly understood and is also affected by the presence of shocks. The elemental sulfur abundance had to be increased in the \citet{Gerner2014,Gerner2015} models in order to properly explain the observed column densities of S-bearing species, but even in their case, the SO abundances could not be explained with \texttt{MUSCLE}. The same issue was found when modeling the 22 cores of the original CORE sample \citep{Gieser2021}. It is still under debate where and in what form sulfur is locked, for example, on the grains as OCS or H$_{2}$S or in the gas phase as neutral S \citep[e.g.,][and references within]{Goicoechea2021}.
	
	For core 1 in ISOSS\,J23053+5953, we estimate a chemical timescale of $19\,000$\,yr. Both HMPO and UCH{\sc ii} model have a similar $\chi^{2}$ of 0.458 and 0.457, respectively. However, the number of well-modeled molecules is higher for the HMPO model (71.4\%) than for the UCH{\sc ii} model (64.3\%), so the weighted $\bar \chi^2$ is lower for the HMPO model. In the physical-chemical analysis of the CORE sample, cores with $\tau_\mathrm{chem} > 80\,000$\,yr show emission from vibrationally excited lines that are not detected in the two ISOSS regions. In addition, the observed spectra toward the CORE cores with $\tau_\mathrm{chem} > 80\,000$ show typical hot-core type spectra and with more emission lines detected compared to core 1 in ISOSS\,J23053+5953. This suggests that the HMPO model is indeed the more realistic model. The best-fit chemical age for cores 2 and 6 are similar to the chemical timescale of core 1, 19\,000\,yr and 18\,000\,yr, respectively.
	
	The \texttt{MUSCLE} model overestimates the SO column density for all cores except for core 6 in ISOSS\,J23053+5953. The H$_{2}$CO column density is underestimated by \texttt{MUSCLE} for all cores except core 1 in ISOSS\,J22478+6357. These species are difficult to model with \texttt{MUSCLE} since the sulfur chemistry was already an issue in the initial abundance models by \citet{Gerner2014,Gerner2015} and shock chemistry is not included, which is important to consider, for example, for outflow sources having enhances SO and H$_{2}$CO emission (see also Sect. \ref{sec:moment0}). A comparison of the chemical timescales $\tau_\mathrm{chem}$ of the four CORE-extension cores with the 22 cores of the CORE sample is discussed in Sect. \ref{sec:discore}.

\section{Discussion}\label{sec:discussion}

	The core properties of both regions are discussed and compared with the results of the 22 cores of the original more evolved CORE sample \citep{Gieser2021} in Sect. \ref{sec:discore}. A more detailed discussion of the complex dynamics in ISOSS\,J23053+5953 is given in Sect. \ref{sec:dis23053}.

\subsection{Core properties}\label{sec:discore}

	Both regions, located in the outer galaxy ($d_\mathrm{gal} \approx 10$\,kpc), fragment into a total of 29 mm cores (with S/N $\geq 5$) as revealed by the high angular resolution observations with NOEMA at 1.3\,mm \citep{Beuther2021}. Herschel observations at FIR wavelengths show that the mm cores are embedded in cold clumps \citep{Ragan2012}. MIR observations with the IRAC and MIPS instruments on the Spitzer space telescope reveal that multiple mm cores have MIR counterparts (Fig. \ref{fig:overview}). This suggests that in both regions star formation is occurring subsequently with some more evolved sources already becoming bright at MIR wavelengths. The strong mm cores with no or a weak associated MIR sources are likely younger. In this case, the luminosity of the protostars is still too faint and/or the protostars are still too embedded.
	
	\subsubsection{Clustered star formation}

	The total clump masses of the cold gas are 104\,$M_\odot$ and 488\,$M_\odot$ for ISOSS\,J22478+6357 and ISOSS\,J23053+5953 \citep{Ragan2012}. \citet{Tackenberg2012} estimate that a gas reservoir of 1000\,$M_\odot$ is required to form a 20\,$M_\odot$ star on clump scales at a size of $R \approx 0.3 - 1.0$\,pc. This suggests that in ISOSS\,J22478+6357 low- and intermediate-mass stars can form. This is in agreement with the results of \citet{Hennemann2008} who estimate that the bright MIR source toward mm core 13 is forming an intermediate-mass star. However, toward ISOSS\,J23053+5953, core 1 with its strong bipolar outflow \citep{Wouterloot1989a,Rodriguez2021b} and being embedded within a large gas reservoir \citep{Birkmann2007} is an accreting intermediate-mass protostar, destined to become a high-mass star \citep[see, e.g.,][]{Beuther2007}.
	
	The SED fitting (Sect. \ref{sec:SED}) shows that for individual cores the bolometric luminosity is not only dominated by the FIR emission of the cold dust, but a significant contribution can arise at MIR wavelengths. In ISOSS\,J22478+6357, the strong mm cores 1 and 2 have no associated MIR source, while in ISOSS\,J23053+5953 mm cores 1 and 2 can be associated with MIR emission. Some of the faint mm cores show MIR emission, for example, core 13 in ISOSS\,J22478+6357 is the brightest in the MIR. As both regions harbor young intermediate- to high-mass protostars (mm-bright) surrounded by more evolved low-mass YSOs (MIR-bright) suggests that clustered subsequent star formation occurs. It seems that first a generation of low-mass stars form, with bright emission in the near-infrared (NIR) and MIR, and then higher-mass protostars, bright at mm wavelengths, form afterward. As an example, in ISOSS\,J22478+6357 the MIR-bright source toward mm core 13 has an estimated age of $\sim 1-6$\,Myr \citep{Hennemann2008} and for the nearby mm core 1 we derive a shorter chemical timescale of 31\,000\,yr (Table \ref{tab:cores}). While the angular resolution of the Spitzer IRAC and NOEMA observations is comparable and resolve the core features, the lower angular resolution of the Spitzer MIPS and Herschel PACS data trace the clump scales, in which multiple cores can be embedded (Fig. \ref{fig:overview}). Unfortunately, high angular resolution observations at FIR wavelengths, which would significantly improve the SED fit and bolometric luminosity estimate of cold components toward individual mm cores, are currently not possible \citep[e.g.,][]{Linz2020}.
	
	In both regions a small cluster is undergoing active star formation (Fig. \ref{fig:overview}). \citet{Kumar2003} studied young star-forming regions that are classified as precursors to UCH{\sc ii} regions with no free-free emission detected at radio wavelengths. Using observations in the NIR and FIR, these authors found that the regions form stars in a clustered mode. Toward the young massive protocluster IRAS\,22134+5834, a massive protostar shows strong emission in the FIR and has a weak radio continuum, while the surrounding protostars on the other hand are bright in the NIR \citep{Kumar2003b,Palau2013,Wang2016}. This is an indication for an early phase of massive star formation with protostars at a range of evolutionary stages. This is also observed toward both ISOSS regions in our study where strong mm cores and bright MIR sources coexist (Fig. \ref{fig:overview}). The formation and evolution of massive stars is much faster, but it is also commonly observed that high-mass protostars are surrounded by a cluster of more evolved low-mass protostars \citep[see also e.g.,][]{Kumar2006}. 
	
	 This is in agreement with the scenario proposed by \citet{Kumar2020} that low-mass star formation occurs in filaments, while HMSF occurs in hubs, which take longer to assemble. Thus, high-mass protostars are observed to be surrounded by more evolved low-mass protostars. We find that in ISOSS\,J22478+6357, the mm cores are embedded within filamentary structures, whereas in ISOSS\,J23053+5953 the system consist of two bigger clumps resembling more of a hub (Sect. \ref{sec:moment0}). This also supports the theory that star formation occurs at the intersection of filaments and gas hubs, and that the growing massive cores could be fed by fresh matter from and/or via these filaments \citep[e.g.,][]{Myers2009,Peretto2013,Tige2017,Kumar2020}. However, it is difficult from an observational point of view to observe this possible gas accretion from large to small scales. Both ISOSS\,J22478+6357 and ISOSS\,J23053+5953 seem to be embedded in a filament-hub system as observed with the Herschel Spectral and Photometric Imaging REceiver (SPIRE). But currently no velocity information of these structures is available and thus follow-up line observations of a larger FOV are required in order to investigate the clustered star-formation properties in both regions.
	
	Another explanation could be that additional external feedback triggers a further generation of stars. In ISOSS\,J23053+5953, the colliding flow could have triggered further star formation at the colliding front toward cores 2, 3, and 6. In ISOSS\,J22478+6357 the nearby IRAS source, which could be an expanding H{\sc ii} region, but has not been studied, could have an impact on the cluster.

	\subsubsection{Physical properties}
		
	In both regions, clear signatures of protostellar outflows are observed in line wings of the CO $2-1$ transition (Fig. \ref{fig:outflows}). For comparison, \citet{Zhang2007} find multiple collimated outflows toward the high-mass protocluster AFGL\,5142 with CO, SO, and SiO observations. In AFGL\,5142, the bipolar outflows are randomly oriented and the outflows are associated with mm cores. In the filamentary IRDC G28.34+0.06 P1, \citet{Wang2011} detect mm cores aligned in a filamentary structure. Even though the cores show no line emission except for CO, all cores harbor bipolar jet-like CO outflows. The outflows are oriented almost parallel to each other and perpendicular to the filament structure. The molecular outflows are observed in SO, SiO, H$_{2}$CO, and CH$_{3}$OH emission as well \citep{Zhang2015}. 
	
	This is similar to the line-poor mm cores toward ISOSS\,J22478+6357, for which the outflows are aligned and perpendicular to the filamentary gas as well (Fig. \ref{fig:outflows}). This suggests that for protostars forming along filaments, there is a preferred orientation of the outflows and that the filaments can provide gas that can be accreted. The magnetic field morphology is a topic for future investigations since the magnetic field is typically also oriented perpendicular to the high-density filaments \citep[e.g.,][]{Soler2017}.
	
	 On the other hand, the outflows that can clearly be identified in ISOSS\,J23053+5953 do not have a preferred orientation. There is no clear signature of a filament, but rather the presence of two massive clumps (Fig. \ref{fig:overview}) and a colliding flow toward the location of the mm cores (Fig. \ref{fig:moment1}). Similar to AFGL\,5142, outflows toward more massive star-forming regions are influenced by the gravitational interaction between the protostars, which might cause the outflows to precess and change their orientation with time \citep[for detailed simulations of high-mass outflows, see, e.g.,][]{Peters2014}.
	
	The physical structure, specifically the radial temperature and density profile, is analyzed for the ten strongest cores within both regions. The 1.3\,mm visibility profiles (Fig. \ref{fig:visibilityprofile}) were used to estimate the density power-law index $p$ of five cores (Table \ref{tab:cores}). We derive a mean density power-law index of $p = 1.6\pm 0.4$. The same analysis was applied to 22 cores of the original CORE sample \citep{Gieser2021}, with a mean of $p = 2.0 \pm 0.2$. This suggests that from the IRDC to HMC stage, the density profile does not change significantly within these short evolutionary timescales. In \citet{Gieser2021} the mean density power-law index was compared to results at various spatial scales in the literature. It is found that $p$ does not change from clump to core scales. A mean value of $p \approx 2$ is also found toward low-mass star-forming regions \citep{Motte2001}. Hence a density profile around two appears to be present over a broad range in evolutionary timescale as well as protostellar mass. The density profile of an initial (high-mass) starless core and in a more evolved UCH{\sc ii} region phase might differ from the intermediate IRDC, HMPO, and HMC stage, though.
	
	The radial temperature profile of the cores (Fig. \ref{fig:temperatureprofile}) is derived from the H$_{2}$CO rotation temperature maps (Figs. \ref{fig:map_22478_H2CO} and \ref{fig:map_23053_H2CO}). The mean temperature power-law index is $q = 0.4 \pm 0.3$ ranging between $q = 0.1 - 1$ within the six mm cores with a radially decreasing temperature profile (Table \ref{tab:cores}). In some cores, we do not observe a radial temperature gradient, which might not be resolved by our observations. For the cores of the original CORE sample, H$_{2}$CO and, in addition, CH$_{3}$CN temperature maps were used to estimate the temperature profiles of individual cores. The mean temperature power-law index of the 22 cores of the original CORE sample is $q = 0.4 \pm 0.1$, which is predicted from theoretical calculations \citep{Emerson1988,Osorio1999,vanderTak2000} and similar to the value found toward the ISOSS cores. 
	
	It is expected that the initially cold cores develop a radially decreasing temperature profile due to protostellar heating. The fact that the density and temperature power-law index are similar from the IRDC to HMC phase suggests that the density and temperature gradient form very early on and stay roughly constant. Since observational constraints of the temperature and density profiles at core scales are still scarce, our findings provide useful constraints for physical models and simulations, but in addition, the density and temperature structure are crucial for chemical models as well (Sect. \ref{sec:MUSCLE}).
	
	Comparing results of the H$_{2}$CO temperature maps (Table \ref{tab:cores}) and the temperature of individual components in the SED fit (Table \ref{tab:photometry}) reveals a temperature gradient of decreasing gas temperature at increasing spatial scale. The Herschel observations tracing the larger scale clumps at $0.1-0.3$\,pc reveal a cold component of $\sim$20\,K. Thus, the mm cores form within a cold clump. The CORE-extension observations show that at a spatial scale of $\approx 3\,000 - 4\,000$\,au the gas temperature increases toward the cores, with H$_{2}$CO rotation temperatures up to $\sim$200\,K. The MIR sources appear as point sources in the Spitzer IRAC observations. The strong MIR emission arises from the close vicinity of the protostar and reveals a hot component of $400 - 600$\,K toward the mm cores with MIR counterparts.
	
	\subsubsection{Molecular gas}
	
\begin{figure}
\centering
\includegraphics[]{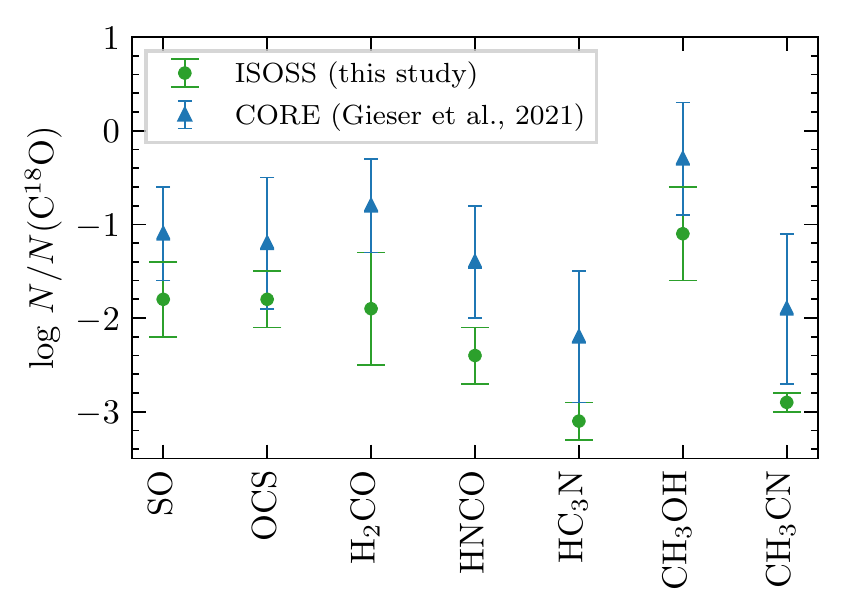}
\caption{Mean abundances (relative to C$^{18}$O) and standard deviation of the cores in both ISOSS regions (green circles) in comparison with the cores in the CORE sample (blue triangles) taken from \citet{Gieser2021}. Only species that were studied both in this work and in \citet{Gieser2021} are shown.}
\label{fig:abund_comparison}
\end{figure}

	Both ISOSS\,J22478+6357 and ISOSS\,J23053+5953 appear line-poor compared to many regions in the original CORE sample. In total, 26 transitions can be identified (Table \ref{tab:lineprops}), all of them toward ISOSS\,J23053+5953, while toward ISOSS\,J22478+6357 only 14 transitions are detected (Table \ref{tab:lineobs}). Toward the regions of the CORE sample, strong line emission of SO$_{2}$, vibrationally excited HC$_{3}$N and torsionally excited CH$_{3}$OH and less abundant isotopologues, such as O$^{13}$CS and $^{34}$SO$_{2}$ are commonly detected in the spectra. In addition, more transitions of HNCO and CH$_{3}$CN with higher upper energy levels are detected. This contrast is in agreement with the fact that the gas temperature and density in ISOSS\,J22478+6357 and ISOSS\,J23053+5953 are lower suggesting that these regions are less evolved.
	
	The mm cores are embedded in an envelope seen in H$_{2}$CO and DCO$^{+}$ emission (Figs. \ref{fig:moment0_H2CO_3_03_2_02} and \ref{fig:moment0_DCO+_3_2}). While both regions have similar clump properties \citep{Ragan2012}, the high angular resolution data reveal complex dynamics that affect the molecular gas of the regions (Sects. \ref{sec:moment0} and \ref{sec:kinematics}). 
	
	While ISOSS\,J22478+6357 is more quiescent and line-poor, a colliding flow in ISOSS\,J23053+5953 is creating a large shocked region that is seen in line-rich emission knots (MP4 and MP5), which are not associated with the mm cores and are in a steep velocity gradient (Fig. \ref{fig:moment1}). In both sources, the temperatures are low throughout the region, so most of the molecules around individual cores are either still frozen onto the dust grains or have not yet formed. Shocked regions are caused either by bipolar outflows (MP1, MP2, and MP3 in ISOSS\,J22478+6357, Fig. \ref{fig:outflows}) or by a large-scale colliding flow (MP4 and MP5 in ISOSS\,J23053+5953, Fig. \ref{fig:moment1}). These shocks are traced by emission of SO, SiO, H$_{2}$CO, and CH$_{3}$OH (e.g., Figs. \ref{fig:moment0_SO_5_5_4_4}, \ref{fig:moment0_SiO_5_4}, \ref{fig:moment0_H2CO_3_03_2_02}, and \ref{fig:moment0_CH3OH_5_1_4_2}). The presence of molecular outflows, which might also create shocks, toward mm cores surrounding MP4 an MP5 cannot be ruled out.
	
	The mean abundance, relative to $N$(C$^{18}$O) and standard deviation of the 29 cores in both ISOSS regions in comparison with the mean abundance of the 22 cores in the CORE sample is shown in Fig. \ref{fig:abund_comparison}. Only abundances of species that were analyzed in both studies are shown. While there is clearly a scatter for the ISOSS and more evolved CORE cores, which is most likely caused by individual cores being at slightly different evolutionary stages, the mean abundances of all species of the ISOSS cores are about an order of magnitude lower than the mean abundances of the CORE cores. This can be interpreted in an evolutionary sense that most of the species are still depleted early on and the gas phase abundances and gas density increase during the evolution of the cores.
	
	The molecular emission in the large mosaics allows us to investigate different physical processes. The larger scale molecular cloud is traced by CO isotopologues (CO, $^{13}$CO, and C$^{18}$O, Figs. \ref{fig:moment0_CO_2_1}, \ref{fig:moment0_13CO_2_1}, and \ref{fig:moment0_C18O_2_1}), with even C$^{18}$O being detected everywhere in the FOV. We find that DCO$^{+}$ and H$_{2}$CO are also good tracers of the larger scale emission, that is, the filamentary structure in ISOSS\,J22478+6357 and the clumps in ISOSS \,J23053+5953 (Figs. \ref{fig:moment0_DCO+_3_2} and \ref{fig:moment0_H2CO_3_03_2_02}). Distinct molecular emission peaking toward the mm cores (SO, $^{13}$CS, OCS, DCO$^{+}$, N$_{2}$D$^{+}$, H$_{2}$S, H$_{2}$CO, H$_{2}^{13}$CO, HNCO, HC$_{3}$N, c-C$_{3}$H$_{2}$, CH$_{3}$OH and CH$_{3}$CN) is seen for cores 1 and 2 in ISOSS\,J22478+6357 and for cores 1, 2, and 6 in ISOSS \,J23053+5953. Bipolar molecular outflows are traced by broad CO line wing emission (Fig. \ref{fig:moment0_CO_2_1}), SO (Figs. \ref{fig:moment0_SO_5_5_4_4} and \ref{fig:moment0_SO_6_5_5_4}), SiO (Fig. \ref{fig:moment0_SiO_5_4}), H$_{2}$CO (Fig. \ref{fig:moment0_H2CO_3_03_2_02}), H$_{2}^{13}$CO (Fig. \ref{fig:moment0_H213CO_3_12_2_11}), and CH$_{3}$OH (Fig. \ref{fig:moment0_CH3OH_4_2_3_1}). 

	Multiple molecular emission peaks in molecular emission are detected in both regions (MPs $1 - 5$). \citet{Qiu2009} detect a MP in the HMSFR HH\,80-81 also with no mm continuum counterpart. In their case, a CO outflow arises from this molecular core, while in our case, the molecular emission, which also is not associated with any mm continuum, stems from shocks caused by outflows and a colliding flow.

	\subsubsection{Physical-chemical modeling results}
	
	In \citet{Gieser2021}, the physical-chemical model \texttt{MUSCLE} is applied to 22 cores in order to estimate the chemical timescale, $\tau_\mathrm{chem}$, and a mean of 60\,000\,yr, with a spread of 20\,000 - 100\,000\,yr, is estimated. It is not possible to apply \texttt{MUSCLE} to all cores in ISOSS\,J22478+6357 and ISOSS \,J23053+5953 as the spectra are in general line-poor. The mean chemical timescale of the ISOSS cores is 20\,000\,yr. The 1.3\,mm spectral setup is ideal to study the molecular emission around cores in the HMPO and HMC stage \citep{Gieser2021}, but in the IRDC stage spectra appear line-poor (Figs. \ref{fig:spectrum_fit_22478} and \ref{fig:spectrum_fit_23053}). Complementary observations at 3\,mm with many simple species, including their deuterated counterparts, in the ground state levels would help to better constrain the chemical timescales of the mm cores in ISOSS\,J22478+6357, but also of the line-poor mm cores in the original CORE sample.
	
	For ISOSS\,J22478+6357 we are only able to model core 1 that shows the most emission lines compared to the remaining cores in the region. As expected from the low temperature in the region, the estimated chemical timescale is low ($\sim$31\,000\,yr). The remaining strong mm cores hosting bipolar outflows as well, are most likely in an early evolutionary stage, where more species are still frozen onto the grains, which is also seen in low kinetic temperatures of the cores (Table \ref{tab:positions}). The estimated chemical timescale of core 1 in ISOSS\,J22478+6357 is $\approx 10\,000$\,yr higher compared to the mm cores in ISOSS\,J23053+5953. Since $M_\mathrm{core}$ and $T_{500}$ are low, this low-mass protostar cannot reach densities as high as in core 1 in ISOSS \,J23053+5953 in order to have a more complex molecular content. The ISOSS\,J22478+6357 cluster seems slightly older, with a MIR-bright YSO around mm core 13 at an estimated system age of $10^{6}$\,yr \citep{Hennemann2008} already present.
	
	Core 1 in ISOSS \,J23053+5953 is most likely a young intermediate- to high-mass protostar with a strong bipolar outflow at an estimated chemical timescale of $\approx 20\,000$\,yr. The single-dish observations reported in \citet{Vasyunina2014}, centered on core 1, show the presence of a few more complex organic molecules, such as CH$_{3}$CCH and CH$_{3}$CHO, detected at 2 and 3\,mm. This indicates the onset of some more complex chemistry toward this particular core. Unresolved CH$_{3}$CN emission in our NOEMA data toward the continuum peak already suggests that the inner parts already reach high temperatures and densities (Fig. \ref{fig:moment0_CH3CN_12_0_11_0}). For cores 2 and 6 in ISOSS \,J23053+5953, we also estimate a comparable chemical timescale. Both can be associated with MIR sources as well (Fig. \ref{fig:overview}), both significantly brighter in the MIR than core 1, and also have distinct MPs in the line integrated intensity maps (Sect. \ref{sec:moment0}).
	
	\subsubsection{Correlations}
	
\begin{figure*}[!htb]
\centering
\includegraphics[]{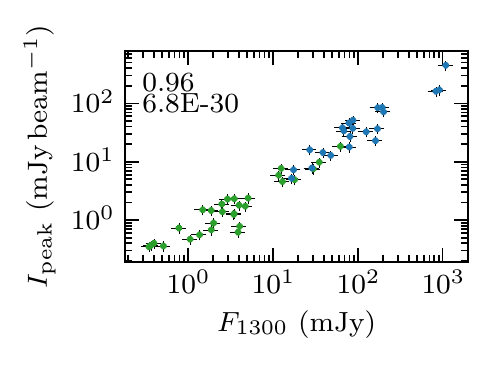}
\includegraphics[]{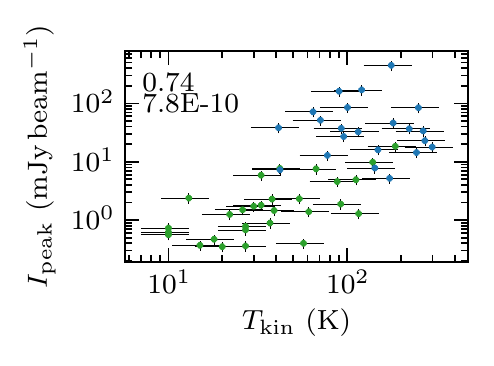}
\includegraphics[]{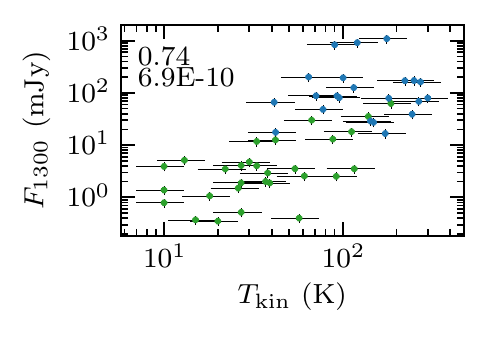}
\includegraphics[]{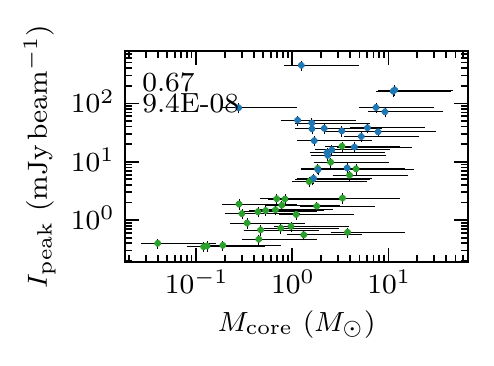}
\includegraphics[]{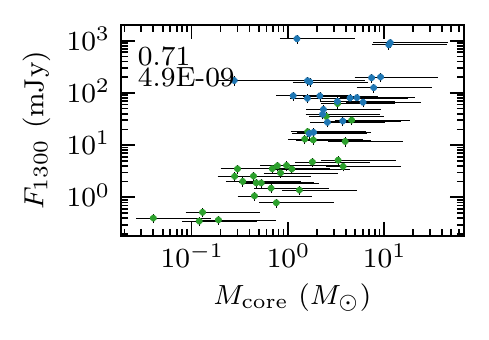}
\includegraphics[]{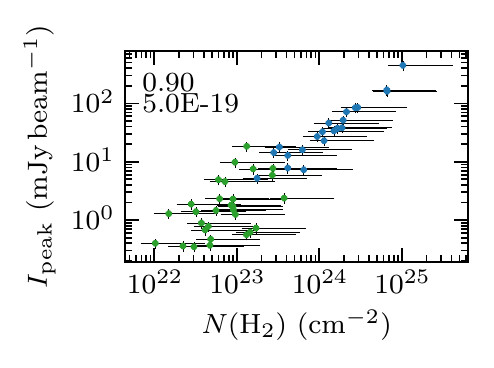}
\includegraphics[]{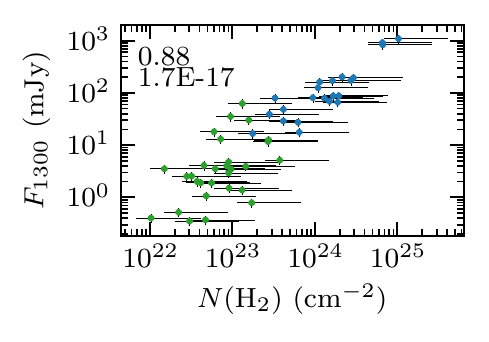}
\includegraphics[]{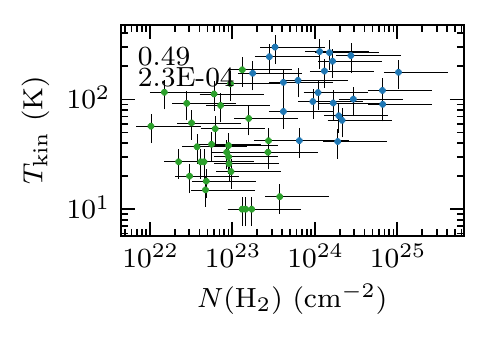}
\includegraphics[]{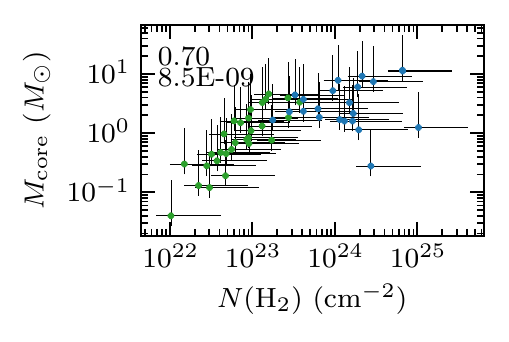}
\includegraphics[]{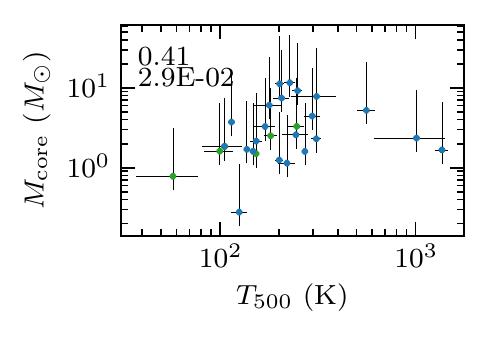}
\includegraphics[]{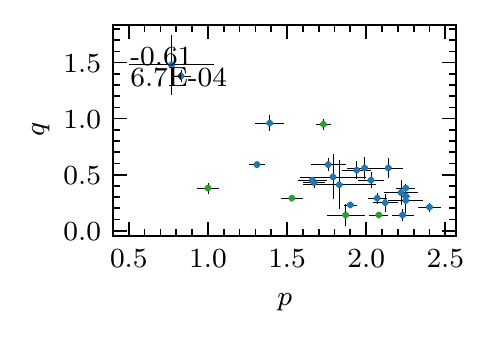}
\includegraphics[]{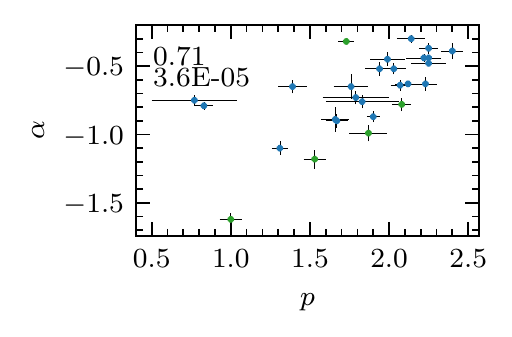}
\includegraphics[]{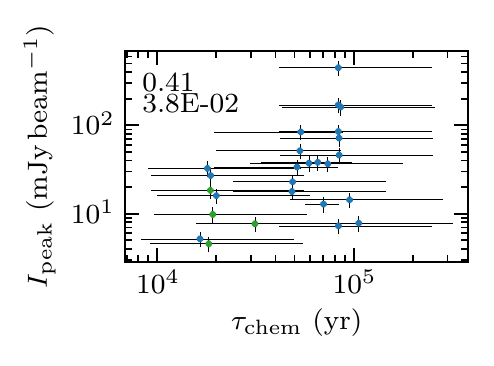}
\includegraphics[]{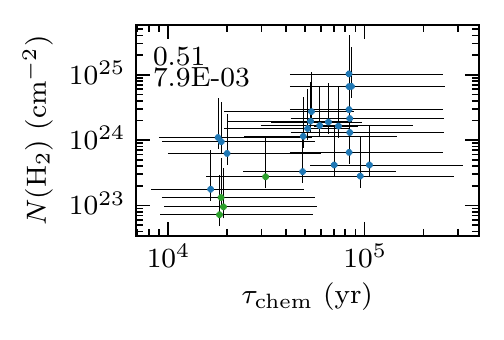}
\includegraphics[]{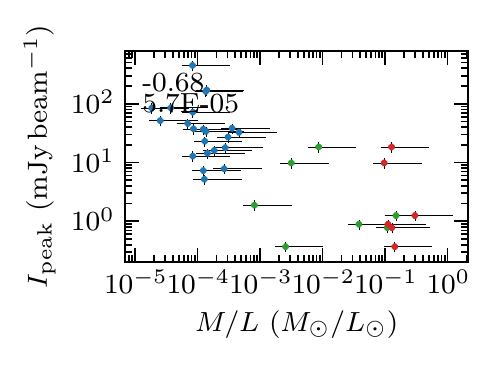}
\includegraphics[]{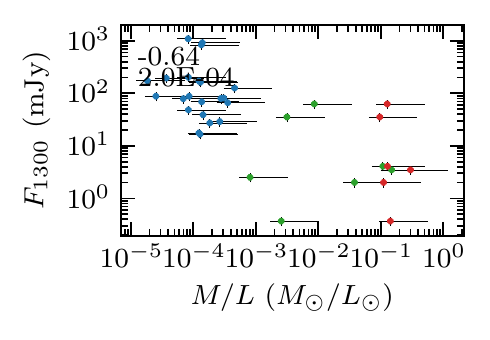}
\includegraphics[]{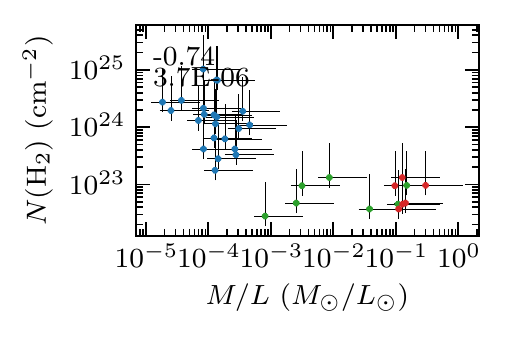}
\includegraphics[]{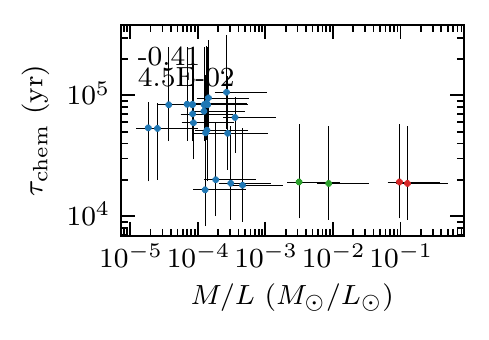}
\caption{Core parameters with a Spearman correlation coefficient of $|r_\mathrm{S}| \geq 0.4$. The results for the cores in ISOSS\,J22478+6357 and ISOSS\,J23053+5953 are shown in green (this work, Tables \ref{tab:positions}, \ref{tab:cores}, and \ref{tab:photometry}) and the results for the cores of the original CORE sample are presented in blue \citep{Gieser2021}. In each panel, the correlation coefficient $r_\mathrm{S}$ and $p$-value are shown in the top left corner. For completeness, the clump-average $M$/$L$ ratio taken from \citet{Ragan2012} of the corresponding ISOSS cores (Table \ref{tab:positions}) is shown in red.}
\label{fig:correlations}
\end{figure*}
	
	Using the CORE and CORE-extension observations, the physical and chemical properties of 32 fragmented mm cores were studied in detail \citep[ten cores in this work and 22 cores in][]{Gieser2021}. The following derived properties are investigated for correlations: the 1.3\,mm peak intensity and integrated flux ($I_\mathrm{peak}$ and $F_{1300}$), the kinetic temperature estimated from the H$_{2}$CO temperature map ($T_\mathrm{kin}$), core mass $M_\mathrm{core}$, molecular hydrogen column density $N$(H$_{2}$), kinetic temperature at $r = 500$\,au obtained from the radial temperature profile $T_{500}$, visibility, density, and temperature power-law index ($\alpha$, $p$, and $q$), chemical timescale ($\tau_\mathrm{chem}$), and $M$/$L$ ratio.
	
	The absolute flux calibration of the NOEMA data is expected to be 20\% or better. We therefore assume a 20\% uncertainty for $I_\mathrm{peak}$ and $F_{1300}$. Using the MCMC error estimation algorithm, the uncertainty of the H$_{2}$CO rotation temperature estimated with \texttt{XCLASS} is $\sim$30\% \citep{Gieser2021}. Since $T_\mathrm{kin}$ is approximated by $T_\mathrm{rot}$(H$_{2}$CO), we assume a 30\% uncertainty for $T_\mathrm{kin}$. The core mass and H$_{2}$ column density are uncertain by a factor of three \citep[taking into account further uncertainties of, for example, the distance, gas-to-dust mass ratio, and dust absorption coefficient,][]{Beuther2021}. The uncertainties for $T_{500}$, $\alpha$, $p$, and $q$ are taken from the uncertainty of the fit (Table \ref{tab:cores}). The chemical timescales are estimated to be uncertain by a factor of two (Sect. \ref{sec:chemicaltimescales}). The $M$/$L$ ratio ($M$/$L = \frac{M_\mathrm{core}}{L}$, Table \ref{tab:photometry}) is assumed to be uncertain by a factor of three. The $M$/$L$ ratio of the CORE cores, based on single-dish observations and taken from \citet{Beuther2018}, is an average of the whole region. But for the ISOSS cores we are able to constrain the $M$/$L$ ratio on core scales (Sect. \ref{sec:SED}). For completeness, we also show the clump-averaged $M$/$L$ ratio taken from \citet{Ragan2012} of the ISOSS cores that have an associated Herschel clump peaking toward the core (core-clump associations are listed in Table \ref{tab:positions}). The clump-averaged $M$/$L$ ratio of the ISOSS cores is shown in red in Fig. \ref{fig:correlations}, but not considered in the calculation of the correlation coefficient.
	
	In order to find correlations or anticorrelations, we computed the Spearman correlation coefficient $r_{S}$ \citep{Cohen1988} for all parameter combinations. If $r_{S} = 1$, a positive correlation exists and if $r_{S} = -1$ an anticorrelation exists between the parameters. For uncorrelated parameters, $r_{S} = 0$ holds. All parameter pairs with $|r_{S}| \geq 0.4$ are shown in Fig. \ref{fig:correlations}. The ISOSS and CORE cores are shown in green and blue, respectively.
	
	A high correlation exists between $I_\mathrm{peak}$ and $F_{1300}$, but the figure reveal that the cores of the ISOSS regions are fainter than the cores of the CORE regions with a small overlap in between. This correlation spans over four orders of magnitude highlighting that the CORE-extension cores are a complement of the CORE cores. Brighter mm cores also have a higher kinetic temperature $T_\mathrm{kin}$. As $N$(H$_{2}$) and $M_\mathrm{core}$ are computed from $I_\mathrm{peak}$ and $F_{1300}$, respectively, we find a high correlation here as well. Thus, $M_\mathrm{core}$ and $N$(H$_{2}$) also have a high correlation. $T_\mathrm{kin}$ and $T_{500}$ correlate with $N$(H$_{2}$) and $M_\mathrm{core}$, respectively.
	
	The temperature index $	q$ is anticorrelated with the density index $p$, such that cores with flatter temperature profiles have a steeper density profile. A strong correlation between $\alpha$ and $p$ can be attributed to Eq. \eqref{eq:p}. We do not find correlations for $\alpha$, $p$, and $q$ with other physical parameters. The mean indices for the density and temperature power-law indices are similar in both studies, implying that these parameters do not change significantly from the IRDC to HMC stage.
	
	Both the peak intensity and H$_{2}$ column density of the mm cores correlate with $\tau_\mathrm{chem}$ such that brighter mm cores with higher H$_{2}$ column densities have a higher estimated chemical timescale. The $M$/$L$ ratio is a tracer of the evolutionary trend such that more evolved protostars have a lower $M$/$L$ ratio \citep[e.g.,][]{Sridharan2002,Molinari2008,Molinari2010,Maud2015,Molinari2016, Urquhart2018,Molinari2019}. The peak intensity, integrated flux, and H$_{2}$ column density of the mm cores are anticorrelated with the $M$/$L$ ratio. As would be expected as a tracer of the evolutionary trend, the chemical timescale $\tau_\mathrm{chem}$ is also anticorrelated with the $M$/$L$ ratio.
	
	The correlation of the physical parameters of the mm cores in the CORE and CORE-extension sample reveal that some of the studied physical properties change with time: cores with a high mm brightness also have a higher temperature, core mass, H$_{2}$ column density and lower $M$/$L$ ratio. This suggests that the initial low-mass cores gain mass with time while also increasing the gas temperature and density. Both the $M$/$L$ ratio and chemical timescales $\tau_\mathrm{chem}$ are tracers of the evolutionary sequence. It is difficult to determine the bolometric luminosity $L$ of cores, which are clearly resolved in MIR and mm observations, but not at FIR wavelengths (Fig. \ref{fig:overview}). Since it is currently impossible to reach subarcsecond resolutions at FIR wavelengths \citep[e.g.,][]{Linz2020}, more reliable SED fits remain challenging, especially in crowded regions (Sect. \ref{sec:SED}). However, chemical timescales estimated with \texttt{MUSCLE} are determined by the observed molecular column densities, for which subarcsecond resolution observations are possible with current interferometers. We do not find that the indices of the temperature and density power-law profiles have a high correlation with the remaining parameters suggesting that they are set very early on and do not change significantly from the IRDC to HMC stage.

\subsection{Dynamics in ISOSS J23053+5953}\label{sec:dis23053}
	
	In ISOSS\,J23053+5953 a colliding gas flow creates enhanced abundances of shock tracers such as SO, $^{13}$CS, SiO, OCS, SO$_{2}$, H$_{2}$CO, H$_{2}^{13}$CO, HC$_{3}$N, CH$_{3}$OH, and CH$_{3}$CN (indicated as MP4 and MP5 in the integrated intensity maps, Sect. \ref{sec:moment0}). These shocked locations reveal molecules that were initially depleted. Therefore, it allows us to investigate which molecules already resided on the grains. In addition, SiO is created by Si sputtered off the grains and forming SiO in the gas phase \citep{Schilke1997}. These emission peaks are spatially located at a velocity gradient seen clearly for all species with extended emission (Fig. \ref{fig:moment1}). However, since both cores 2 and 6 are young protostars, molecular outflows are expected to be present, but difficult to identify (Fig. \ref{fig:outflows}).
	
	Core 2, which is embedded within a common dust envelope with core 3 (Fig. \ref{fig:continuum}), shows emission of c-C$_{3}$H$_{2}$ (Fig. \ref{fig:moment0_c-C3H2_6_16_5_05}) toward the north and south of the mm core. This might trace an outflow cavity, but a clear outflow direction cannot be inferred due to the complex dynamics. We therefore conclude that a colliding flow creates a shocked region between cores 2, 3, and 6, but protostellar outflows are most likely present as well.
	
	Colliding flows are found in other regions of the Galaxy. For example, in the well-known giant molecular cloud complex NGC\,6334, colliding flows are reported in the source known as ``V'' by \citet{Juarez2017}, where two velocity components separated about 2.4\,km\,s$^{-1}$ are seen in H$^{13}$CO$^{+}$, with SiO emission at the intermediate velocities and also at the interface of the two flows \citep[Fig. 7 of ][]{Juarez2017}. Multiple IR, cm, and mm sources are found at the flow interface as well \citep[Figs. 1 and 4 of ][]{Juarez2017}. Overall, the similarities of NGC\,6334\,V and ISOSS\,J23053+5953 are remarkable, despite clearly being located in different parts of the Galaxy. Converging flows are also reported in DR21(OH) \citep{Csengeri2011}, Orion \citep{Lee2013}, G035.39-00.33 \citep{JimenezSerra2014}, G0.253+0.016 \citep{Johnston2014}, the Galactic Center \citep{Schwoerer2019}, the Musca filament \citep{Bonne2020}, and G+0.693-0.03 \citep{Zeng2020}.
	
	Toward core 3 in ISOSS\,J23053+5953, a chemical differentiation between DCO$^{+}$ peaking toward the north (Fig. \ref{fig:moment0_DCO+_3_2}) and N$_{2}$D$^{+}$ peaking toward the west (Fig. \ref{fig:moment0_N2D+_3_2}) of the mm core is observed. In general, N$_{2}$D$^{+}$ is less depleted than DCO$^{+}$ due to N$_{2}$ residing longer in the gas phase compared to CO and CO leading to an enhancement of deuterated species \citep{Emprechtinger2009}. Due to the fact that all CO isotopologues show wide spread emission, the difference between the two species might not be due to CO freeze-out. \citet{Murillo2018} also find spatially offset emission of these two species in the cold envelope of the low-mass protostar IRAS\,16293-2422. In their models, the different locations of the molecular emission are due to a temperature effect, where DCO$^{+}$ is located in a colder region at the disk-envelope interface. The colliding flow might have caused a temperature change toward core 3 or the spatial difference might be caused by potential molecular outflows. Two jet-like features are observed in SiO (Jet1 and Jet2), both pointing to core 3 (Fig. \ref{fig:moment0_SiO_5_4}).

\section{Conclusions}\label{sec:conclusions}

	We used high angular resolution observations at 1.3\,mm of the young star-forming regions ISOSS\,J22478+6357 and ISOSS\,J23053+5953 and MIR and FIR archival data from the Spitzer and Herschel telescopes to study the physical and chemical properties of fragmented mm cores. This study (CORE-extension) is a follow-up of the CORE large program, for which the fragmentation and physical-chemical properties of fragmented cores toward HMSFRs classified as HMPOs, HMCs, and UCH{\sc ii} regions are presented in \citet{Beuther2018} and \citet{Gieser2021}, respectively. Here, we focused on the physical (temperature and density profiles, outflows) and chemical properties of younger mm cores.
	
	Our main results are summarized as follows:

\begin{enumerate}
	\item Comparing the 1.3\,mm continuum with Spitzer IRAC and MIPS observations, we find that some of the fainter mm cores already have strong MIR counterparts suggesting clustered subsequent star formation occurring in both regions. A generation of more evolved low-mass protostars surrounds the younger intermediate- and high-mass protostars. The bolometric luminosity of the cores, which can be resolved in the MIR and FIR, was estimated from fitting their SEDs from MIR to mm wavelengths. Two or three temperature components were necessary to properly fit the full SEDs.
	\item The radial density profile was inferred from the 1.3\,mm continuum visibilities and the radial temperature profile was inferred from H$_{2}$CO rotation temperature maps. We assumed that both profiles follow a power-law profile, and we derive a mean density and temperature power-law index of $p = 1.6 \pm 0.4$ and $q = 0.4 \pm 0.3$, respectively. We find similar values toward the cores of the original more evolved CORE sample \citep{Gieser2021}, suggesting that both the radial density and temperature power-law index do not change significantly from the IRDC to HMC stage.
	\item The 1.3\,mm spectra are generally line-poor, compared to spectra observed in the original CORE sample, and mostly simple species with low upper energy levels are detected. Molecular column densities were inferred with \texttt{XCLASS}. The abundances (relative to C$^{18}$O) of the cores compared to the cores of the original CORE sample \citep{Gieser2021} are about one order of magnitude lower suggesting that most species are still depleted in the younger ISOSS sources studied here.
	\item The CO ($2-1$) line wing emission reveals multiple bipolar molecular outflows in both regions. In ISOSS\,J22478+6357, we find five mm cores with outflows that are aligned almost parallel in the plane of the sky. These mm cores are connected by a filamentary structure seen in DCO$^{+}$ and the outflows are perpendicular to this filament. A large-scale bipolar outflow is seen toward core 1 in ISOSS\,J23053+5953, which is a high-mass protostar.
	\item Most of the mm cores do not show distinct molecular emission peaks (except for cores 1 and 2 in ISOSS\,J22478+6357, and for cores 1, 2, and 6 in ISOSS\,J23053+5953). Multiple molecular emission peaks are detected within the FOV in both regions with no associated mm continuum cores for transitions of SO, SiO, H$_{2}$CO, and CH$_{3}$OH being well-known shock tracers. These regions can be linked to bipolar outflows ISOSS\,J22478+6357 and a combination of a colliding flow and outflows in ISOSS\,J23053+5953 creating shocks in the gas surrounding the mm cores.
	\item We applied the statistical tool HOG to the line data in both regions to find velocity channels with a high spatial correlation between two molecular transitions. As expected, transitions of the same species have a high correlation, but also the shock-tracing species and the molecules tracing the larger-scale emission are spatially correlated with no significant velocity offsets. With HOG it will be possible for future programs to easily identify spatial correlations in a large sample of regions and many detected transitions on a channel-by-channel comparison.
	\item A large-scale velocity gradient is seen in ISOSS\,J23053+5953 toward the cores 2, 3 and 6. Between these cores, a colliding flow is creating a shock enhancing the abundances of many detected species. This colliding flow might have triggered further star formation.
	\item With the physical-chemical model \texttt{MUSCLE}, we estimate chemical timescales of four cores with the highest number of detected species and for which we were able to derive density and temperature profiles. We find a mean chemical timescale of 20\,000\,yr, while for the cores of the original more evolved CORE sample a mean of 60\,000\,yr is derived \citep{Gieser2021}. Most cores in the ISOSS regions are in an early evolutionary phase, which is supported by the shorter chemical timescales, smaller molecular abundances, and lower temperatures. However, we also find some more evolved cores in the region with higher temperatures, signposts of complex chemistry and MIR signatures that indicates subsequent on-going star formation (cores 1 and 6 in ISOSS\,J23053+5953).
	\item Comparing the mm core properties studied in the ISOSS and CORE regions, we find that some properties clearly change with time. The 1.3\,mm peak intensity and integrated flux, core mass, H$_{2}$ column density, and kinetic temperature of a mm core increase with time, since these properties are anticorrelated with the $M$/$L$ ratio and correlated with the chemical timescale $\tau_\mathrm{chem}$.
\end{enumerate}

	The CORE and CORE-extension studies reveal that subarcsecond angular resolution observations are required in order to investigate the properties of fragmented cores that show complex dynamical and chemical properties on these spatial scales ($<$10\,000\,au). The number of observed and analyzed mm cores is still not high; however, this analysis demonstrates a pathway for ongoing and future large programs targeting hundreds or thousands of embedded young (high-mass) protostars. The 1.3\,mm setup of the CORE project is not ideal for studying the molecular composition in young and cold regions in the IRDC phase, as well as in evolved UCH{\sc ii} regions. High angular resolution observations at 3\,mm covering the ground-state transitions of many simple species, including many deuterated molecules, would provide a better insight on the earliest evolutionary stages. An intriguing result of this study is that both regions harboring young intermediate- and high-mass protostars are already embedded in a cluster of more evolved low-mass MIR-bright YSOs.

\begin{acknowledgements}
	The authors would like to thank the anonymous referee whose comments helped to improve the clarity of this paper. This work is based on observations carried out under project number L14AB with the IRAM NOEMA Interferometer and the IRAM\,30\,m telescope. IRAM is supported by INSU/CNRS (France), MPG (Germany), and IGN (Spain). C.G. and H.B. acknowledge support from the European Research Council under the Horizon 2020 Framework Programme via the ERC Consolidator Grant CSF-648505. D.S. acknowledges support by the Deutsche Forschungsgemeinschaft through SPP 1833: ``Building a Habitable Earth (SE 1962/6-1).'' R.K. acknowledges financial support via the Emmy Noether and Heisenberg Research Grants funded by the German Research Foundation (DFG) under grant no. KU 2849/3 and 2849/9. A.P. acknowledges financial support from the UNAM-PAPIIT IN111421 grant and the Sistema Nacional de Investigadores of CONACyT. R.G.-M. acknowledges support from UNAM-PAPIIT project IN104319. A. P. and R.G.-M. also acknowledge support from CONACyT Ciencia de Frontera project number 86372. R.E.P. is supported by an NSERC Discovery grant. This research made use of Astropy\footnote{\url{http://www.astropy.org}}, a community-developed core Python package for Astronomy \citep{Astropy2013, Astropy2018}. This research made use of \texttt{photutils}, an Astropy package for detection and photometry of astronomical sources \citep{Bradley2021}.
\end{acknowledgements}

\bibliographystyle{aa} 
\bibliography{bibliography_ISOSS} 

\begin{thebibliography}{183}
\expandafter\ifx\csname natexlab\endcsname\relax\def\natexlab#1{#1}\fi

\bibitem[{Bei(1988)}]{Beichman1988}
 1988, {Infrared Astronomical Satellite (IRAS) Catalogs and Atlases.Volume 1:
  Explanatory Supplement.}, Vol.~1

\bibitem[{{Adams}(1991)}]{Adams1991}
{Adams}, F.~C. 1991, \apj, 382, 544

\bibitem[{{Ahmadi} {et~al.}(2019){Ahmadi}, {Kuiper}, \& {Beuther}}]{Ahmadi2019}
{Ahmadi}, A., {Kuiper}, R., \& {Beuther}, H. 2019, \aap, 632, A50

\bibitem[{{Alakoz} {et~al.}(2002){Alakoz}, {Kalenskii}, {Promislov},
  {Johansson}, \& {Winnberg}}]{Alakoz2002}
{Alakoz}, A.~V., {Kalenskii}, S.~V., {Promislov}, V.~G., {Johansson}, L.~E.~B.,
  \& {Winnberg}, A. 2002, Astronomy Reports, 46, 551

\bibitem[{{Allen} {et~al.}(2017){Allen}, {van der Tak}, {S{\'a}nchez-Monge},
  {Cesaroni}, \& {Beltr{\'a}n}}]{Allen2017}
{Allen}, V., {van der Tak}, F.~F.~S., {S{\'a}nchez-Monge}, {\'A}., {Cesaroni},
  R., \& {Beltr{\'a}n}, M.~T. 2017, \aap, 603, A133

\bibitem[{{Arce} {et~al.}(2007){Arce}, {Shepherd}, {Gueth}, {Lee}, {Bachiller},
  {Rosen}, \& {Beuther}}]{Arce2007}
{Arce}, H.~G., {Shepherd}, D., {Gueth}, F., {et~al.} 2007, in Protostars and
  Planets V, ed. B.~{Reipurth}, D.~{Jewitt}, \& K.~{Keil}, 245

\bibitem[{{Astropy Collaboration} {et~al.}(2018){Astropy Collaboration},
  {Price-Whelan}, {Sip{\H o}cz}, {G{\"u}nther}, {Lim}, {Crawford}, {Conseil},
  {Shupe}, {Craig}, {Dencheva}, {Ginsburg}, {VanderPlas}, {Bradley},
  {P{\'e}rez-Su{\'a}rez}, {de Val-Borro}, {Aldcroft}, {Cruz}, {Robitaille},
  {Tollerud}, {Ardelean}, {Babej}, {Bach}, {Bachetti}, {Bakanov}, {Bamford},
  {Barentsen}, {Barmby}, {Baumbach}, {Berry}, {Biscani}, {Boquien}, {Bostroem},
  {Bouma}, {Brammer}, {Bray}, {Breytenbach}, {Buddelmeijer}, {Burke},
  {Calderone}, {Cano Rodr{\'{\i}}guez}, {Cara}, {Cardoso}, {Cheedella},
  {Copin}, {Corrales}, {Crichton}, {D'Avella}, {Deil}, {Depagne}, {Dietrich},
  {Donath}, {Droettboom}, {Earl}, {Erben}, {Fabbro}, {Ferreira}, {Finethy},
  {Fox}, {Garrison}, {Gibbons}, {Goldstein}, {Gommers}, {Greco}, {Greenfield},
  {Groener}, {Grollier}, {Hagen}, {Hirst}, {Homeier}, {Horton}, {Hosseinzadeh},
  {Hu}, {Hunkeler}, {Ivezi{\'c}}, {Jain}, {Jenness}, {Kanarek}, {Kendrew},
  {Kern}, {Kerzendorf}, {Khvalko}, {King}, {Kirkby}, {Kulkarni}, {Kumar},
  {Lee}, {Lenz}, {Littlefair}, {Ma}, {Macleod}, {Mastropietro}, {McCully},
  {Montagnac}, {Morris}, {Mueller}, {Mumford}, {Muna}, {Murphy}, {Nelson},
  {Nguyen}, {Ninan}, {N{\"o}the}, {Ogaz}, {Oh}, {Parejko}, {Parley}, {Pascual},
  {Patil}, {Patil}, {Plunkett}, {Prochaska}, {Rastogi}, {Reddy Janga},
  {Sabater}, {Sakurikar}, {Seifert}, {Sherbert}, {Sherwood-Taylor}, {Shih},
  {Sick}, {Silbiger}, {Singanamalla}, {Singer}, {Sladen}, {Sooley},
  {Sornarajah}, {Streicher}, {Teuben}, {Thomas}, {Tremblay}, {Turner},
  {Terr{\'o}n}, {van Kerkwijk}, {de la Vega}, {Watkins}, {Weaver}, {Whitmore},
  {Woillez}, {Zabalza}, \& {Astropy Contributors}}]{Astropy2018}
{Astropy Collaboration}, {Price-Whelan}, A.~M., {Sip{\H o}cz}, B.~M., {et~al.}
  2018, \aj, 156, 123

\bibitem[{{Astropy Collaboration} {et~al.}(2013){Astropy Collaboration},
  {Robitaille}, {Tollerud}, {Greenfield}, {Droettboom}, {Bray}, {Aldcroft},
  {Davis}, {Ginsburg}, {Price-Whelan}, {Kerzendorf}, {Conley}, {Crighton},
  {Barbary}, {Muna}, {Ferguson}, {Grollier}, {Parikh}, {Nair}, {Unther},
  {Deil}, {Woillez}, {Conseil}, {Kramer}, {Turner}, {Singer}, {Fox}, {Weaver},
  {Zabalza}, {Edwards}, {Azalee Bostroem}, {Burke}, {Casey}, {Crawford},
  {Dencheva}, {Ely}, {Jenness}, {Labrie}, {Lim}, {Pierfederici}, {Pontzen},
  {Ptak}, {Refsdal}, {Servillat}, \& {Streicher}}]{Astropy2013}
{Astropy Collaboration}, {Robitaille}, T.~P., {Tollerud}, E.~J., {et~al.} 2013,
  \aap, 558, A33

\bibitem[{{Batschelet}(1972)}]{Batschelet1972}
{Batschelet}, E. 1972, {Recent Statistical Methods for Orientation Data}, Vol.
  262, 61

\bibitem[{{Beltr{\'a}n}(2020)}]{Beltran2020}
{Beltr{\'a}n}, M. 2020, arXiv e-prints, arXiv:2005.06912

\bibitem[{{Benedettini} {et~al.}(2013){Benedettini}, {Viti}, {Codella},
  {Gueth}, {G{\'o}mez-Ruiz}, {Bachiller}, {Beltr{\'a}n}, {Busquet},
  {Ceccarelli}, \& {Lefloch}}]{Benedettini2013}
{Benedettini}, M., {Viti}, S., {Codella}, C., {et~al.} 2013, \mnras, 436, 179

\bibitem[{{Beuther} {et~al.}(2007{\natexlab{a}}){Beuther}, {Churchwell},
  {McKee}, \& {Tan}}]{Beuther2007}
{Beuther}, H., {Churchwell}, E.~B., {McKee}, C.~F., \& {Tan}, J.~C.
  2007{\natexlab{a}}, Protostars and Planets V, 165

\bibitem[{{Beuther} {et~al.}(2021){Beuther}, {Gieser}, {Suri}, {Linz},
  {Klaassen}, {Semenov}, {Winters}, {Henning}, {Soler}, {Urquhart}, {Syed},
  {Feng}, {M{\"o}ller}, {Beltr{\'a}n}, {S{\'a}nchez-Monge}, {Longmore},
  {Peters}, {Ballesteros-Paredes}, {Schilke}, {Moscadelli}, {Palau},
  {Cesaroni}, {Lumsden}, {Pudritz}, {Wyrowski}, {Kuiper}, \&
  {Ahmadi}}]{Beuther2021}
{Beuther}, H., {Gieser}, C., {Suri}, S., {et~al.} 2021, \aap, 649, A113

\bibitem[{{Beuther} {et~al.}(2010){Beuther}, {Henning}, {Linz}, {Krause},
  {Nielbock}, \& {Steinacker}}]{Beuther2010}
{Beuther}, H., {Henning}, T., {Linz}, H., {et~al.} 2010, \aap, 518, L78

\bibitem[{{Beuther} {et~al.}(2007{\natexlab{b}}){Beuther}, {Leurini},
  {Schilke}, {Wyrowski}, {Menten}, \& {Zhang}}]{Beuther2007B}
{Beuther}, H., {Leurini}, S., {Schilke}, P., {et~al.} 2007{\natexlab{b}}, \aap,
  466, 1065

\bibitem[{{Beuther} {et~al.}(2018){Beuther}, {Mottram}, {Ahmadi}, {Bosco},
  {Linz}, {Henning}, {Klaassen}, {Winters}, {Maud}, {Kuiper}, {Semenov},
  {Gieser}, {Peters}, {Urquhart}, {Pudritz}, {Ragan}, {Feng}, {Keto},
  {Leurini}, {Cesaroni}, {Beltran}, {Palau}, {S{\'a}nchez-Monge},
  {Galvan-Madrid}, {Zhang}, {Schilke}, {Wyrowski}, {Johnston}, {Longmore},
  {Lumsden}, {Hoare}, {Menten}, \& {Csengeri}}]{Beuther2018}
{Beuther}, H., {Mottram}, J.~C., {Ahmadi}, A., {et~al.} 2018, \aap, 617, A100

\bibitem[{{Beuther} {et~al.}(2002{\natexlab{a}}){Beuther}, {Schilke}, {Menten},
  {Motte}, {Sridharan}, \& {Wyrowski}}]{Beuther2002B}
{Beuther}, H., {Schilke}, P., {Menten}, K.~M., {et~al.} 2002{\natexlab{a}},
  \apj, 566, 945

\bibitem[{{Beuther} {et~al.}(2002{\natexlab{b}}){Beuther}, {Schilke},
  {Sridharan}, {Menten}, {Walmsley}, \& {Wyrowski}}]{Beuther2002}
{Beuther}, H., {Schilke}, P., {Sridharan}, T.~K., {et~al.} 2002{\natexlab{b}},
  \aap, 383, 892

\bibitem[{{Beuther} \& {Steinacker}(2007)}]{Beuther2007C}
{Beuther}, H. \& {Steinacker}, J. 2007, \apjl, 656, L85

\bibitem[{{Bihr} {et~al.}(2015){Bihr}, {Beuther}, {Linz}, {Ragan}, {Hennemann},
  {Tackenberg}, {Smith}, {Krause}, \& {Henning}}]{Bihr2015}
{Bihr}, S., {Beuther}, H., {Linz}, H., {et~al.} 2015, \aap, 579, A51

\bibitem[{{Birkmann} {et~al.}(2007){Birkmann}, {Krause}, {Hennemann},
  {Henning}, {Steinacker}, \& {Lemke}}]{Birkmann2007}
{Birkmann}, S.~M., {Krause}, O., {Hennemann}, M., {et~al.} 2007, \aap, 474, 883

\bibitem[{{Bonne} {et~al.}(2020){Bonne}, {Bontemps}, {Schneider}, {Clarke},
  {Arzoumanian}, {Fukui}, {Tachihara}, {Csengeri}, {Guesten}, {Ohama},
  {Okamoto}, {Simon}, {Yahia}, \& {Yamamoto}}]{Bonne2020}
{Bonne}, L., {Bontemps}, S., {Schneider}, N., {et~al.} 2020, \aap, 644, A27

\bibitem[{{Bonnell}(2007)}]{Bonnell2007}
{Bonnell}, I.~A. 2007, in Astronomical Society of the Pacific Conference
  Series, Vol. 367, Massive Stars in Interactive Binaries, ed. N.~{St.-Louis}
  \& A.~F.~J. {Moffat}, 303

\bibitem[{{Bradley} {et~al.}(2021){Bradley}, {Sip{\H{o}}cz}, {Robitaille},
  {Tollerud}, {Vin{\'\i}cius}, {Deil}, {Barbary}, {Wilson}, {Busko},
  {G{\"u}nther}, {Cara}, {Conseil}, {Bostroem}, {Droettboom}, {Bray}, {Andersen
  Bratholm}, {Lim}, {Barentsen}, {Craig}, {Rathi}, {Pascual}, {Perren},
  {Donath}, {Georgiev}, {De Val-Borro}, {Kerzendorf}, {Bach}, {Quint},
  {Souchereau}, \& {Weaver}}]{Bradley2021}
{Bradley}, L., {Sip{\H{o}}cz}, B., {Robitaille}, T., {et~al.} 2021,
  {astropy/photutils: 1.0.2}

\bibitem[{{Bronfman} {et~al.}(1996){Bronfman}, {Nyman}, \&
  {May}}]{Bronfman1996}
{Bronfman}, L., {Nyman}, L.~A., \& {May}, J. 1996, \aaps, 115, 81

\bibitem[{{Carter} {et~al.}(2012){Carter}, {Lazareff}, {Maier}, {Chenu},
  {Fontana}, {Bortolotti}, {Boucher}, {Navarrini}, {Blanchet}, {Greve}, {John},
  {Kramer}, {Morel}, {Navarro}, {Pe{\~n}alver}, {Schuster}, \& {Thum}}]{EMIR}
{Carter}, M., {Lazareff}, B., {Maier}, D., {et~al.} 2012, \aap, 538, A89

\bibitem[{{Cesaroni} {et~al.}(2017){Cesaroni}, {S{\'a}nchez-Monge},
  {Beltr{\'a}n}, {Johnston}, {Maud}, {Moscadelli}, {Mottram}, {Ahmadi},
  {Allen}, {Beuther}, {Csengeri}, {Etoka}, {Fuller}, {Galli},
  {Galv{\'a}n-Madrid}, {Goddi}, {Henning}, {Hoare}, {Klaassen}, {Kuiper},
  {Kumar}, {Lumsden}, {Peters}, {Rivilla}, {Schilke}, {Testi}, {van der Tak},
  {Vig}, {Walmsley}, \& {Zinnecker}}]{Cesaroni2017}
{Cesaroni}, R., {S{\'a}nchez-Monge}, {\'A}., {Beltr{\'a}n}, M.~T., {et~al.}
  2017, \aap, 602, A59

\bibitem[{{Clark}(1980)}]{Clark1980}
{Clark}, B.~G. 1980, \aap, 89, 377

\bibitem[{Cohen(1988)}]{Cohen1988}
Cohen, J. 1988, Statistical power analysis for the behavioral sciences
  (Routledge)

\bibitem[{{Commer{\c{c}}on} {et~al.}(2011){Commer{\c{c}}on}, {Hennebelle}, \&
  {Henning}}]{Commercon2011}
{Commer{\c{c}}on}, B., {Hennebelle}, P., \& {Henning}, T. 2011, \apjl, 742, L9

\bibitem[{{Comoretto} {et~al.}(1990){Comoretto}, {Palagi}, {Cesaroni}, {Felli},
  {Bettarini}, {Catarzi}, {Curioni}, {Curioni}, {Di Franco}, {Giovanardi},
  {Massi}, {Palla}, {Panella}, {Rossi}, {Speroni}, \& {Tofani}}]{Comoretto1990}
{Comoretto}, G., {Palagi}, F., {Cesaroni}, R., {et~al.} 1990, \aaps, 84, 179

\bibitem[{{Csengeri} {et~al.}(2011){Csengeri}, {Bontemps}, {Schneider},
  {Motte}, {Gueth}, \& {Hora}}]{Csengeri2011}
{Csengeri}, T., {Bontemps}, S., {Schneider}, N., {et~al.} 2011, \apjl, 740, L5

\bibitem[{{Di Francesco} {et~al.}(2008){Di Francesco}, {Johnstone}, {Kirk},
  {MacKenzie}, \& {Ledwosinska}}]{DiFrancesco2008}
{Di Francesco}, J., {Johnstone}, D., {Kirk}, H., {MacKenzie}, T., \&
  {Ledwosinska}, E. 2008, \apjs, 175, 277

\bibitem[{{Draine}(2011)}]{Draine2011}
{Draine}, B.~T. 2011, {Physics of the Interstellar and Intergalactic Medium}
  (Princeton: Princeton University Press)

\bibitem[{Durand \& Greenwood(1958)}]{Durand1958}
Durand, D. \& Greenwood, J.~A. 1958, The Journal of Geology, 66, 229

\bibitem[{{Ellsworth-Bowers} {et~al.}(2015){Ellsworth-Bowers}, {Rosolowsky},
  {Glenn}, {Ginsburg}, {Evans}, {Battersby}, {Shirley}, \&
  {Svoboda}}]{Ellsworth-Bowers2015a}
{Ellsworth-Bowers}, T.~P., {Rosolowsky}, E., {Glenn}, J., {et~al.} 2015, \apj,
  799, 29

\bibitem[{{Emerson}(1988)}]{Emerson1988}
{Emerson}, J.~P. 1988, in NATO Advanced Science Institutes (ASI) Series C, Vol.
  241, NATO Advanced Science Institutes (ASI) Series C, ed. A.~K. {Dupree} \&
  M.~T.~V.~T. {Lago}, 21

\bibitem[{{Emprechtinger} {et~al.}(2009){Emprechtinger}, {Caselli}, {Volgenau},
  {Stutzki}, \& {Wiedner}}]{Emprechtinger2009}
{Emprechtinger}, M., {Caselli}, P., {Volgenau}, N.~H., {Stutzki}, J., \&
  {Wiedner}, M.~C. 2009, \aap, 493, 89

\bibitem[{{Endres} {et~al.}(2016){Endres}, {Schlemmer}, {Schilke}, {Stutzki},
  \& {M{\"u}ller}}]{Endres2016}
{Endres}, C.~P., {Schlemmer}, S., {Schilke}, P., {Stutzki}, J., \&
  {M{\"u}ller}, H.~S.~P. 2016, Journal of Molecular Spectroscopy, 327, 95

\bibitem[{{Engelbracht} {et~al.}(2007){Engelbracht}, {Blaylock}, {Su}, {Rho},
  {Rieke}, {Muzerolle}, {Padgett}, {Hines}, {Gordon}, {Fadda},
  {Noriega-Crespo}, {Kelly}, {Latter}, {Hinz}, {Misselt}, {Morrison},
  {Stansberry}, {Shupe}, {Stolovy}, {Wheaton}, {Young}, {Neugebauer},
  {Wachter}, {P{\'e}rez-Gonz{\'a}lez}, {Frayer}, \&
  {Marleau}}]{Engelbracht2007}
{Engelbracht}, C.~W., {Blaylock}, M., {Su}, K.~Y.~L., {et~al.} 2007, \pasp,
  119, 994

\bibitem[{{Fazio} {et~al.}(2004){Fazio}, {Hora}, {Allen}, {Ashby}, {Barmby},
  {Deutsch}, {Huang}, {Kleiner}, {Marengo}, {Megeath}, {Melnick}, {Pahre},
  {Patten}, {Polizotti}, {Smith}, {Taylor}, {Wang}, {Willner}, {Hoffmann},
  {Pipher}, {Forrest}, {McMurty}, {McCreight}, {McKelvey}, {McMurray}, {Koch},
  {Moseley}, {Arendt}, {Mentzell}, {Marx}, {Losch}, {Mayman}, {Eichhorn},
  {Krebs}, {Jhabvala}, {Gezari}, {Fixsen}, {Flores}, {Shakoorzadeh}, {Jungo},
  {Hakun}, {Workman}, {Karpati}, {Kichak}, {Whitley}, {Mann}, {Tollestrup},
  {Eisenhardt}, {Stern}, {Gorjian}, {Bhattacharya}, {Carey}, {Nelson},
  {Glaccum}, {Lacy}, {Lowrance}, {Laine}, {Reach}, {Stauffer}, {Surace},
  {Wilson}, {Wright}, {Hoffman}, {Domingo}, \& {Cohen}}]{Fazio2004}
{Fazio}, G.~G., {Hora}, J.~L., {Allen}, L.~E., {et~al.} 2004, \apjs, 154, 10

\bibitem[{{Feng} {et~al.}(2015){Feng}, {Beuther}, {Henning}, {Semenov},
  {Palau}, \& {Mills}}]{Feng2015}
{Feng}, S., {Beuther}, H., {Henning}, T., {et~al.} 2015, \aap, 581, A71

\bibitem[{{Feng} {et~al.}(2016){Feng}, {Beuther}, {Semenov}, {Henning}, {Linz},
  {Mills}, \& {Teague}}]{Feng2016}
{Feng}, S., {Beuther}, H., {Semenov}, D., {et~al.} 2016, \aap, 593, A46

\bibitem[{{Fontani} {et~al.}(2012){Fontani}, {Palau}, {Busquet}, {Isella},
  {Estalella}, {Sanchez-Monge}, {Caselli}, \& {Zhang}}]{Fontani2012}
{Fontani}, F., {Palau}, A., {Busquet}, G., {et~al.} 2012, \mnras, 423, 1691

\bibitem[{{Frank} {et~al.}(2014){Frank}, {Ray}, {Cabrit}, {Hartigan}, {Arce},
  {Bacciotti}, {Bally}, {Benisty}, {Eisl{\"o}ffel}, {G{\"u}del}, {Lebedev},
  {Nisini}, \& {Raga}}]{Frank2014}
{Frank}, A., {Ray}, T.~P., {Cabrit}, S., {et~al.} 2014, in Protostars and
  Planets VI, ed. H.~{Beuther}, R.~S. {Klessen}, C.~P. {Dullemond}, \&
  T.~{Henning}, 451

\bibitem[{{Galv{\'a}n-Madrid} {et~al.}(2013){Galv{\'a}n-Madrid}, {Liu},
  {Zhang}, {Pineda}, {Peng}, {Zhang}, {Keto}, {Ho}, {Rodr{\'\i}guez}, {Zapata},
  {Peters}, \& {De Pree}}]{GalvanMadrid2013}
{Galv{\'a}n-Madrid}, R., {Liu}, H.~B., {Zhang}, Z.~Y., {et~al.} 2013, \apj,
  779, 121

\bibitem[{{Galv{\'a}n-Madrid} {et~al.}(2010){Galv{\'a}n-Madrid}, {Zhang},
  {Keto}, {Ho}, {Zapata}, {Rodr{\'\i}guez}, {Pineda}, \&
  {V{\'a}zquez-Semadeni}}]{GalvanMadrid2010}
{Galv{\'a}n-Madrid}, R., {Zhang}, Q., {Keto}, E., {et~al.} 2010, \apj, 725, 17

\bibitem[{{Garrod} {et~al.}(2008){Garrod}, {Widicus Weaver}, \&
  {Herbst}}]{Garrod2008}
{Garrod}, R.~T., {Widicus Weaver}, S.~L., \& {Herbst}, E. 2008, \apj, 682, 283

\bibitem[{{Gatto} {et~al.}(2017){Gatto}, {Walch}, {Naab}, {Girichidis},
  {W{\"u}nsch}, {Glover}, {Klessen}, {Clark}, {Peters}, {Derigs}, {Baczynski},
  \& {Puls}}]{Gatto2017}
{Gatto}, A., {Walch}, S., {Naab}, T., {et~al.} 2017, \mnras, 466, 1903

\bibitem[{{Gerner} {et~al.}(2014){Gerner}, {Beuther}, {Semenov}, {Linz},
  {Vasyunina}, {Bihr}, {Shirley}, \& {Henning}}]{Gerner2014}
{Gerner}, T., {Beuther}, H., {Semenov}, D., {et~al.} 2014, \aap, 563, A97

\bibitem[{{Gerner} {et~al.}(2015){Gerner}, {Shirley}, {Beuther}, {Semenov},
  {Linz}, {Albertsson}, \& {Henning}}]{Gerner2015}
{Gerner}, T., {Shirley}, Y.~L., {Beuther}, H., {et~al.} 2015, \aap, 579, A80

\bibitem[{{Giannetti} {et~al.}(2017){Giannetti}, {Leurini}, {K{\"o}nig},
  {Urquhart}, {Pillai}, {Brand}, {Kauffmann}, {Wyrowski}, \&
  {Menten}}]{Giannetti2017}
{Giannetti}, A., {Leurini}, S., {K{\"o}nig}, C., {et~al.} 2017, \aap, 606, L12

\bibitem[{{Giannetti} {et~al.}(2014){Giannetti}, {Wyrowski}, {Brand},
  {Csengeri}, {Fontani}, {Walmsley}, {Nguyen Luong}, {Beuther}, {Schuller},
  {G{\"u}sten}, \& {Menten}}]{Giannetti2014}
{Giannetti}, A., {Wyrowski}, F., {Brand}, J., {et~al.} 2014, \aap, 570, A65

\bibitem[{{Gieser} {et~al.}(2021){Gieser}, {Beuther}, {Semenov}, {Ahmadi},
  {Suri}, {M{\"o}ller}, {Beltr{\'a}n}, {Klaassen}, {Zhang}, {Urquhart},
  {Henning}, {Feng}, {Galv{\'a}n-Madrid}, {de Souza Magalh{\~a}es},
  {Moscadelli}, {Longmore}, {Leurini}, {Kuiper}, {Peters}, {Menten},
  {Csengeri}, {Fuller}, {Wyrowski}, {Lumsden}, {S{\'a}nchez-Monge}, {Maud},
  {Linz}, {Palau}, {Schilke}, {Pety}, {Pudritz}, {Winters}, \&
  {Pi{\'e}tu}}]{Gieser2021}
{Gieser}, C., {Beuther}, H., {Semenov}, D., {et~al.} 2021, \aap, 648, A66

\bibitem[{{Gieser} {et~al.}(2019){Gieser}, {Semenov}, {Beuther}, {Ahmadi},
  {Mottram}, {Henning}, {Beltran}, {Maud}, {Bosco}, {Leurini}, {Peters},
  {Klaassen}, {Kuiper}, {Feng}, {Urquhart}, {Moscadelli}, {Csengeri},
  {Lumsden}, {Winters}, {Suri}, {Zhang}, {Pudritz}, {Palau}, {Menten},
  {Galvan-Madrid}, {Wyrowski}, {Schilke}, {S{\'a}nchez-Monge}, {Linz},
  {Johnston}, {Jim{\'e}nez-Serra}, {Longmore}, \& {M{\"o}ller}}]{Gieser2019}
{Gieser}, C., {Semenov}, D., {Beuther}, H., {et~al.} 2019, \aap, 631, A142

\bibitem[{{Girichidis} {et~al.}(2011){Girichidis}, {Federrath}, {Banerjee}, \&
  {Klessen}}]{Girichidis2011}
{Girichidis}, P., {Federrath}, C., {Banerjee}, R., \& {Klessen}, R.~S. 2011,
  \mnras, 413, 2741

\bibitem[{{Girichidis} {et~al.}(2016){Girichidis}, {Walch}, {Naab}, {Gatto},
  {W{\"u}nsch}, {Glover}, {Klessen}, {Clark}, {Peters}, {Derigs}, \&
  {Baczynski}}]{Girichidis2016}
{Girichidis}, P., {Walch}, S., {Naab}, T., {et~al.} 2016, \mnras, 456, 3432

\bibitem[{{Goicoechea} {et~al.}(2021){Goicoechea}, {Aguado}, {Cuadrado},
  {Roncero}, {Pety}, {Bron}, {Fuente}, {Riquelme}, {Chapillon}, {Herrera}, \&
  {Duran}}]{Goicoechea2021}
{Goicoechea}, J.~R., {Aguado}, A., {Cuadrado}, S., {et~al.} 2021, \aap, 647,
  A10

\bibitem[{{G{\'o}mez} {et~al.}(2021){G{\'o}mez}, {V{\'a}zquez-Semadeni}, \&
  {Palau}}]{Gomez2021}
{G{\'o}mez}, G.~C., {V{\'a}zquez-Semadeni}, E., \& {Palau}, A. 2021, \mnras,
  502, 4963

\bibitem[{{Guzm{\'a}n} {et~al.}(2018){Guzm{\'a}n}, {Guzm{\'a}n}, {Garay},
  {Bronfman}, \& {Hechenleitner}}]{Guzman2018}
{Guzm{\'a}n}, A.~E., {Guzm{\'a}n}, V.~V., {Garay}, G., {Bronfman}, L., \&
  {Hechenleitner}, F. 2018, \apjs, 236, 45

\bibitem[{{Harju} {et~al.}(1993){Harju}, {Walmsley}, \&
  {Wouterloot}}]{Harju1993}
{Harju}, J., {Walmsley}, C.~M., \& {Wouterloot}, J.~G.~A. 1993, \aaps, 98, 51

\bibitem[{{Hennebelle} \& {Inutsuka}(2019)}]{Hennebelle2019}
{Hennebelle}, P. \& {Inutsuka}, S.-i. 2019, Frontiers in Astronomy and Space
  Sciences, 6, 5

\bibitem[{{Hennemann} {et~al.}(2008){Hennemann}, {Birkmann}, {Krause}, \&
  {Lemke}}]{Hennemann2008}
{Hennemann}, M., {Birkmann}, S.~M., {Krause}, O., \& {Lemke}, D. 2008, \aap,
  485, 753

\bibitem[{{Henning} {et~al.}(2000){Henning}, {Schreyer}, {Launhardt}, \&
  {Burkert}}]{Henning2000}
{Henning}, T., {Schreyer}, K., {Launhardt}, R., \& {Burkert}, A. 2000, \aap,
  353, 211

\bibitem[{{Herbst} \& {van Dishoeck}(2009)}]{Herbst2009}
{Herbst}, E. \& {van Dishoeck}, E.~F. 2009, \araa, 47, 427

\bibitem[{{Jim{\'e}nez-Serra} {et~al.}(2014){Jim{\'e}nez-Serra}, {Caselli},
  {Fontani}, {Tan}, {Henshaw}, {Kainulainen}, \&
  {Hernandez}}]{JimenezSerra2014}
{Jim{\'e}nez-Serra}, I., {Caselli}, P., {Fontani}, F., {et~al.} 2014, \mnras,
  439, 1996

\bibitem[{{Jim{\'e}nez-Serra} {et~al.}(2012){Jim{\'e}nez-Serra}, {Zhang},
  {Viti}, {Mart{\'\i}n-Pintado}, \& {de Wit}}]{JimenezSerra2012}
{Jim{\'e}nez-Serra}, I., {Zhang}, Q., {Viti}, S., {Mart{\'\i}n-Pintado}, J., \&
  {de Wit}, W.~J. 2012, \apj, 753, 34

\bibitem[{{Johnston} {et~al.}(2014){Johnston}, {Beuther}, {Linz}, {Schmiedeke},
  {Ragan}, \& {Henning}}]{Johnston2014}
{Johnston}, K.~G., {Beuther}, H., {Linz}, H., {et~al.} 2014, \aap, 568, A56

\bibitem[{{Johnston} {et~al.}(2020){Johnston}, {Hoare}, {Beuther}, {Kuiper},
  {Kee}, {Linz}, {Boley}, {Maud}, {Ahmadi}, \& {Robitaille}}]{Johnston2020}
{Johnston}, K.~G., {Hoare}, M.~G., {Beuther}, H., {et~al.} 2020, \aap, 634, L11

\bibitem[{{J{\o}rgensen} {et~al.}(2020){J{\o}rgensen}, {Belloche}, \&
  {Garrod}}]{Jorgensen2020}
{J{\o}rgensen}, J.~K., {Belloche}, A., \& {Garrod}, R.~T. 2020, \araa, 58, 727

\bibitem[{{Jow} {et~al.}(2018){Jow}, {Hill}, {Scott}, {Soler}, {Martin},
  {Devlin}, {Fissel}, \& {Poidevin}}]{Jow2018}
{Jow}, D.~L., {Hill}, R., {Scott}, D., {et~al.} 2018, \mnras, 474, 1018

\bibitem[{{Ju{\'a}rez} {et~al.}(2017){Ju{\'a}rez}, {Girart},
  {Zamora-Avil{\'e}s}, {Tang}, {Koch}, {Liu}, {Palau}, {Ballesteros-Paredes},
  {Zhang}, \& {Qiu}}]{Juarez2017}
{Ju{\'a}rez}, C., {Girart}, J.~M., {Zamora-Avil{\'e}s}, M., {et~al.} 2017,
  \apj, 844, 44

\bibitem[{{Kalenskii} \& {Sobolev}(1994)}]{Kalenskii1994}
{Kalenskii}, S.~V. \& {Sobolev}, A.~M. 1994, Astronomy Letters, 20, 91

\bibitem[{{Kerton} \& {Brunt}(2003)}]{Kerton2003}
{Kerton}, C.~R. \& {Brunt}, C.~M. 2003, \aap, 399, 1083

\bibitem[{{K{\"o}lligan} \& {Kuiper}(2018)}]{Koelligan2018}
{K{\"o}lligan}, A. \& {Kuiper}, R. 2018, \aap, 620, A182

\bibitem[{{Krause} {et~al.}(2004){Krause}, {Vavrek}, {Birkmann}, {Klaas},
  {Stickel}, {T{\'o}th}, \& {Lemke}}]{Krause2004}
{Krause}, O., {Vavrek}, R., {Birkmann}, S., {et~al.} 2004, Baltic Astronomy,
  13, 407

\bibitem[{{Kumar} {et~al.}(2003{\natexlab{a}}){Kumar}, {Davis}, \&
  {Bachiller}}]{Kumar2003}
{Kumar}, M.~S.~N., {Davis}, C.~J., \& {Bachiller}, R. 2003{\natexlab{a}},
  \apss, 287, 191

\bibitem[{{Kumar} {et~al.}(2006){Kumar}, {Keto}, \& {Clerkin}}]{Kumar2006}
{Kumar}, M.~S.~N., {Keto}, E., \& {Clerkin}, E. 2006, \aap, 449, 1033

\bibitem[{{Kumar} {et~al.}(2003{\natexlab{b}}){Kumar}, {Ojha}, \&
  {Davis}}]{Kumar2003b}
{Kumar}, M.~S.~N., {Ojha}, D.~K., \& {Davis}, C.~J. 2003{\natexlab{b}}, \apj,
  598, 1107

\bibitem[{{Kumar} {et~al.}(2020){Kumar}, {Palmeirim}, {Arzoumanian}, \&
  {Inutsuka}}]{Kumar2020}
{Kumar}, M.~S.~N., {Palmeirim}, P., {Arzoumanian}, D., \& {Inutsuka}, S.~I.
  2020, \aap, 642, A87

\bibitem[{{Larionov} {et~al.}(1999){Larionov}, {Val'tts}, {Winnberg},
  {Johansson}, {Booth}, \& {Golubev}}]{Larionov1999}
{Larionov}, G.~M., {Val'tts}, I.~E., {Winnberg}, A., {et~al.} 1999, \aaps, 139,
  257

\bibitem[{{Law} {et~al.}(2021){Law}, {Zhang}, {{\"O}berg}, {Galv{\'a}n-Madrid},
  {Keto}, {Liu}, \& {Ho}}]{Law2021}
{Law}, C.~J., {Zhang}, Q., {{\"O}berg}, K.~I., {et~al.} 2021, \apj, 909, 214

\bibitem[{{Lee} {et~al.}(2013){Lee}, {Looney}, {Schnee}, \& {Li}}]{Lee2013}
{Lee}, K., {Looney}, L.~W., {Schnee}, S., \& {Li}, Z.-Y. 2013, \apj, 772, 100

\bibitem[{{Leurini} {et~al.}(2011){Leurini}, {Codella}, {Zapata},
  {Beltr{\'a}n}, {Schilke}, \& {Cesaroni}}]{Leurini2011}
{Leurini}, S., {Codella}, C., {Zapata}, L., {et~al.} 2011, \aap, 530, A12

\bibitem[{{Linz} {et~al.}(2021){Linz}, {Beuther}, {Gerin}, {Goicoechea},
  {Helmich}, {Krause}, {Liu}, {Molinari}, {Ossenkopf-Okada}, {Pineda},
  {Sauvage}, {Schinnerer}, {van der Tak}, {Wiedner}, {Amiaux}, {Bhatia},
  {Buinhas}, {Durand}, {F{\"o}rstner}, {Graf}, \& {Lezius}}]{Linz2020}
{Linz}, H., {Beuther}, H., {Gerin}, M., {et~al.} 2021, Experimental Astronomy

\bibitem[{{Linz} {et~al.}(2010){Linz}, {Krause}, {Beuther}, {Henning}, {Klein},
  {Nielbock}, {Stecklum}, {Steinacker}, \& {Stutz}}]{Linz2010}
{Linz}, H., {Krause}, O., {Beuther}, H., {et~al.} 2010, \aap, 518, L123

\bibitem[{{Looney} {et~al.}(2003){Looney}, {Mundy}, \& {Welch}}]{Looney2003}
{Looney}, L.~W., {Mundy}, L.~G., \& {Welch}, W.~J. 2003, \apj, 592, 255

\bibitem[{{L{\'o}pez-Sepulcre} {et~al.}(2009){L{\'o}pez-Sepulcre}, {Codella},
  {Cesaroni}, {Marcelino}, \& {Walmsley}}]{LopezSepulcre2009}
{L{\'o}pez-Sepulcre}, A., {Codella}, C., {Cesaroni}, R., {Marcelino}, N., \&
  {Walmsley}, C.~M. 2009, \aap, 499, 811

\bibitem[{{Lundquist} {et~al.}(2014){Lundquist}, {Kobulnicky}, {Alexand er},
  {Kerton}, \& {Arvidsson}}]{Lundquist2014}
{Lundquist}, M.~J., {Kobulnicky}, H.~A., {Alexand er}, M.~J., {Kerton}, C.~R.,
  \& {Arvidsson}, K. 2014, \apj, 784, 111

\bibitem[{{Maud} {et~al.}(2019){Maud}, {Cesaroni}, {Kumar}, {Rivilla},
  {Ginsburg}, {Klaassen}, {Harsono}, {S{\'a}nchez-Monge}, {Ahmadi}, {Allen},
  {Beltr{\'a}n}, {Beuther}, {Galv{\'a}n-Madrid}, {Goddi}, {Hoare},
  {Hogerheijde}, {Johnston}, {Kuiper}, {Moscadelli}, {Peters}, {Testi}, {van
  der Tak}, \& {de Wit}}]{Maud2019}
{Maud}, L.~T., {Cesaroni}, R., {Kumar}, M.~S.~N., {et~al.} 2019, \aap, 627, L6

\bibitem[{{Maud} {et~al.}(2015{\natexlab{a}}){Maud}, {Lumsden}, {Moore},
  {Mottram}, {Urquhart}, \& {Cicchini}}]{Maud2015}
{Maud}, L.~T., {Lumsden}, S.~L., {Moore}, T.~J.~T., {et~al.}
  2015{\natexlab{a}}, \mnras, 452, 637

\bibitem[{{Maud} {et~al.}(2015{\natexlab{b}}){Maud}, {Moore}, {Lumsden},
  {Mottram}, {Urquhart}, \& {Hoare}}]{Maud2015B}
{Maud}, L.~T., {Moore}, T.~J.~T., {Lumsden}, S.~L., {et~al.}
  2015{\natexlab{b}}, \mnras, 453, 645

\bibitem[{{McGuire}(2018)}]{McGuire2018}
{McGuire}, B.~A. 2018, \apjs, 239, 17

\bibitem[{{McKee} \& {Ostriker}(1977)}]{McKee1977}
{McKee}, C.~F. \& {Ostriker}, J.~P. 1977, \apj, 218, 148

\bibitem[{{Meynet} {et~al.}(1994){Meynet}, {Maeder}, {Schaller}, {Schaerer}, \&
  {Charbonnel}}]{Meynet1994}
{Meynet}, G., {Maeder}, A., {Schaller}, G., {Schaerer}, D., \& {Charbonnel}, C.
  1994, \aaps, 103, 97

\bibitem[{{Molinari} {et~al.}(2019){Molinari}, {Baldeschi}, {Robitaille},
  {Morales}, {Schisano}, {Traficante}, {Merello}, {Molinaro}, {Vitello},
  {Sciacca}, \& {Liu}}]{Molinari2019}
{Molinari}, S., {Baldeschi}, A., {Robitaille}, T.~P., {et~al.} 2019, \mnras,
  486, 4508

\bibitem[{{Molinari} {et~al.}(2016){Molinari}, {Merello}, {Elia}, {Cesaroni},
  {Testi}, \& {Robitaille}}]{Molinari2016}
{Molinari}, S., {Merello}, M., {Elia}, D., {et~al.} 2016, \apjl, 826, L8

\bibitem[{{Molinari} {et~al.}(2008){Molinari}, {Pezzuto}, {Cesaroni}, {Brand},
  {Faustini}, \& {Testi}}]{Molinari2008}
{Molinari}, S., {Pezzuto}, S., {Cesaroni}, R., {et~al.} 2008, \aap, 481, 345

\bibitem[{{Molinari} {et~al.}(2010){Molinari}, {Swinyard}, {Bally}, {Barlow},
  {Bernard}, {Martin}, {Moore}, {Noriega-Crespo}, {Plume}, {Testi}, {Zavagno},
  {Abergel}, {Ali}, {Andr{\'e}}, {Baluteau}, {Benedettini}, {Bern{\'e}},
  {Billot}, {Blommaert}, {Bontemps}, {Boulanger}, {Brand}, {Brunt}, {Burton},
  {Campeggio}, {Carey}, {Caselli}, {Cesaroni}, {Cernicharo}, {Chakrabarti},
  {Chrysostomou}, {Codella}, {Cohen}, {Compiegne}, {Davis}, {de Bernardis}, {de
  Gasperis}, {Di Francesco}, {di Giorgio}, {Elia}, {Faustini}, {Fischera},
  {Fukui}, {Fuller}, {Ganga}, {Garcia-Lario}, {Giard}, {Giardino}, {Glenn},
  {Goldsmith}, {Griffin}, {Hoare}, {Huang}, {Jiang}, {Joblin}, {Joncas},
  {Juvela}, {Kirk}, {Lagache}, {Li}, {Lim}, {Lord}, {Lucas}, {Maiolo},
  {Marengo}, {Marshall}, {Masi}, {Massi}, {Matsuura}, {Meny}, {Minier},
  {Miville-Desch{\^e}nes}, {Montier}, {Motte}, {M{\"u}ller}, {Natoli}, {Neves},
  {Olmi}, {Paladini}, {Paradis}, {Pestalozzi}, {Pezzuto}, {Piacentini},
  {Pomar{\`e}s}, {Popescu}, {Reach}, {Richer}, {Ristorcelli}, {Roy}, {Royer},
  {Russeil}, {Saraceno}, {Sauvage}, {Schilke}, {Schneider-Bontemps},
  {Schuller}, {Schultz}, {Shepherd}, {Sibthorpe}, {Smith}, {Smith},
  {Spinoglio}, {Stamatellos}, {Strafella}, {Stringfellow}, {Sturm}, {Taylor},
  {Thompson}, {Tuffs}, {Umana}, {Valenziano}, {Vavrek}, {Viti}, {Waelkens},
  {Ward-Thompson}, {White}, {Wyrowski}, {Yorke}, \& {Zhang}}]{Molinari2010}
{Molinari}, S., {Swinyard}, B., {Bally}, J., {et~al.} 2010, \pasp, 122, 314

\bibitem[{{M{\"o}ller} {et~al.}(2013){M{\"o}ller}, {Bernst}, {Panoglou},
  {Muders}, {Ossenkopf}, {R{\"o}llig}, \& {Schilke}}]{MAGIX}
{M{\"o}ller}, T., {Bernst}, I., {Panoglou}, D., {et~al.} 2013, \aap, 549, A21

\bibitem[{{M{\"o}ller} {et~al.}(2017){M{\"o}ller}, {Endres}, \&
  {Schilke}}]{XCLASS}
{M{\"o}ller}, T., {Endres}, C., \& {Schilke}, P. 2017, \aap, 598, A7

\bibitem[{{Moscadelli} {et~al.}(2013){Moscadelli}, {Li}, {Cesaroni}, {Sanna},
  {Xu}, \& {Zhang}}]{Moscadelli2013}
{Moscadelli}, L., {Li}, J.~J., {Cesaroni}, R., {et~al.} 2013, \aap, 549, A122

\bibitem[{{Motte} \& {Andr{\'e}}(2001)}]{Motte2001}
{Motte}, F. \& {Andr{\'e}}, P. 2001, \aap, 365, 440

\bibitem[{Motte {et~al.}(2018)Motte, Bontemps, \& Louvet}]{Motte2018}
Motte, F., Bontemps, S., \& Louvet, F. 2018, Annual Review of Astronomy and
  Astrophysics, 56, 41

\bibitem[{{Mottram} {et~al.}(2020){Mottram}, {Beuther}, {Ahmadi}, {Klaassen},
  {Beltr{\'a}n}, {Csengeri}, {Feng}, {Gieser}, {Henning}, {Johnston}, {Kuiper},
  {Leurini}, {Linz}, {Longmore}, {Lumsden}, {Maud}, {Moscadelli}, {Palau},
  {Peters}, {Pudritz}, {Ragan}, {S{\'a}nchez-Monge}, {Semenov}, {Urquhart},
  {Winters}, \& {Zinnecker}}]{Mottram2020}
{Mottram}, J.~C., {Beuther}, H., {Ahmadi}, A., {et~al.} 2020, \aap, 636, A118

\bibitem[{{M{\"u}ller} {et~al.}(2005){M{\"u}ller}, {Schl{\"o}der}, {Stutzki},
  \& {Winnewisser}}]{CDMS}
{M{\"u}ller}, H.~S.~P., {Schl{\"o}der}, F., {Stutzki}, J., \& {Winnewisser}, G.
  2005, Journal of Molecular Structure, 742, 215

\bibitem[{{Murillo} {et~al.}(2018){Murillo}, {van Dishoeck}, {van der Wiel},
  {J{\o}rgensen}, {Drozdovskaya}, {Calcutt}, \& {Harsono}}]{Murillo2018}
{Murillo}, N.~M., {van Dishoeck}, E.~F., {van der Wiel}, M.~H.~D., {et~al.}
  2018, \aap, 617, A120

\bibitem[{{Myers}(2009)}]{Myers2009}
{Myers}, P.~C. 2009, \apj, 700, 1609

\bibitem[{{Okoda} {et~al.}(2020){Okoda}, {Oya}, {Sakai}, {Watanabe}, \&
  {Yamamoto}}]{Okoda2020}
{Okoda}, Y., {Oya}, Y., {Sakai}, N., {Watanabe}, Y., \& {Yamamoto}, S. 2020,
  \apj, 900, 40

\bibitem[{{Osorio} {et~al.}(1999){Osorio}, {Lizano}, \&
  {D'Alessio}}]{Osorio1999}
{Osorio}, M., {Lizano}, S., \& {D'Alessio}, P. 1999, \apj, 525, 808

\bibitem[{{Ossenkopf} \& {Henning}(1994)}]{Ossenkopf1994}
{Ossenkopf}, V. \& {Henning}, T. 1994, \aap, 291, 943

\bibitem[{{Palagi} {et~al.}(1993){Palagi}, {Cesaroni}, {Comoretto}, {Felli}, \&
  {Natale}}]{Palagi1993}
{Palagi}, F., {Cesaroni}, R., {Comoretto}, G., {Felli}, M., \& {Natale}, V.
  1993, \aaps, 101, 153

\bibitem[{{Palau} {et~al.}(2015){Palau}, {Ballesteros-Paredes},
  {V{\'a}zquez-Semadeni}, {S{\'a}nchez-Monge}, {Estalella}, {Fall}, {Zapata},
  {Camacho}, {G{\'o}mez}, {Naranjo-Romero}, {Busquet}, \&
  {Fontani}}]{Palau2015}
{Palau}, A., {Ballesteros-Paredes}, J., {V{\'a}zquez-Semadeni}, E., {et~al.}
  2015, \mnras, 453, 3785

\bibitem[{{Palau} {et~al.}(2014){Palau}, {Estalella}, {Girart}, {Fuente},
  {Fontani}, {Commer{\c{c}}on}, {Busquet}, {Bontemps}, {S{\'a}nchez-Monge},
  {Zapata}, {Zhang}, {Hennebelle}, \& {di Francesco}}]{Palau2014}
{Palau}, A., {Estalella}, R., {Girart}, J.~M., {et~al.} 2014, \apj, 785, 42

\bibitem[{{Palau} {et~al.}(2013){Palau}, {Fuente}, {Girart}, {Estalella}, {Ho},
  {S{\'a}nchez-Monge}, {Fontani}, {Busquet}, {Commer{\c{c}}on}, {Hennebelle},
  {Boissier}, {Zhang}, {Cesaroni}, \& {Zapata}}]{Palau2013}
{Palau}, A., {Fuente}, A., {Girart}, J.~M., {et~al.} 2013, \apj, 762, 120

\bibitem[{{Palau} {et~al.}(2017){Palau}, {Walsh}, {S{\'a}nchez-Monge},
  {Girart}, {Cesaroni}, {Jim{\'e}nez-Serra}, {Fuente}, {Zapata}, \&
  {Neri}}]{Palau2017}
{Palau}, A., {Walsh}, C., {S{\'a}nchez-Monge}, {\'A}., {et~al.} 2017, \mnras,
  467, 2723

\bibitem[{{Palau} {et~al.}(2021){Palau}, {Zhang}, {Girart}, {Liu}, {Rao},
  {Koch}, {Estalella}, {Chen}, {Baobab Liu}, {Qiu}, {Li}, {Zapata}, {Bontemps},
  {Ho}, {Beuther}, {Ching}, {Shinnaga}, \& {Ahmadi}}]{Palau2021}
{Palau}, A., {Zhang}, Q., {Girart}, J.~M., {et~al.} 2021, \apj, 912, 159

\bibitem[{{Palla} {et~al.}(1993){Palla}, {Cesaroni}, {Brand}, {Caselli},
  {Comoretto}, \& {Felli}}]{Palla1993}
{Palla}, F., {Cesaroni}, R., {Brand}, J., {et~al.} 1993, \aap, 280, 599

\bibitem[{{Peretto} {et~al.}(2013){Peretto}, {Fuller}, {Duarte-Cabral},
  {Avison}, {Hennebelle}, {Pineda}, {Andr{\'e}}, {Bontemps}, {Motte},
  {Schneider}, \& {Molinari}}]{Peretto2013}
{Peretto}, N., {Fuller}, G.~A., {Duarte-Cabral}, A., {et~al.} 2013, \aap, 555,
  A112

\bibitem[{{Peters} {et~al.}(2014){Peters}, {Klaassen}, {Mac Low}, {Schr{\"o}n},
  {Federrath}, {Smith}, \& {Klessen}}]{Peters2014}
{Peters}, T., {Klaassen}, P.~D., {Mac Low}, M.-M., {et~al.} 2014, \apj, 788, 14

\bibitem[{{Pety} {et~al.}(2005){Pety}, {Teyssier}, {Foss{\'e}}, {Gerin},
  {Roueff}, {Abergel}, {Habart}, \& {Cernicharo}}]{Pety2005}
{Pety}, J., {Teyssier}, D., {Foss{\'e}}, D., {et~al.} 2005, \aap, 435, 885

\bibitem[{{Pickett} {et~al.}(1998){Pickett}, {Poynter}, {Cohen}, {Delitsky},
  {Pearson}, \& {M{\"u}ller}}]{JPL}
{Pickett}, H.~M., {Poynter}, R.~L., {Cohen}, E.~A., {et~al.} 1998, \jqsrt, 60,
  883

\bibitem[{{Pitann} {et~al.}(2011){Pitann}, {Hennemann}, {Birkmann}, {Bouwman},
  {Krause}, \& {Henning}}]{Pitann2011}
{Pitann}, J., {Hennemann}, M., {Birkmann}, S., {et~al.} 2011, \apj, 743, 93

\bibitem[{{Planck Collaboration} {et~al.}(2016){Planck Collaboration}, {Ade},
  {Aghanim}, {Arnaud}, {Ashdown}, {Aumont}, {Baccigalupi}, {Banday},
  {Barreiro}, {Bartolo}, {Battaner}, {Benabed}, {Beno{\^\i}t},
  {Benoit-L{\'e}vy}, {Bernard}, {Bersanelli}, {Bielewicz}, {Bonaldi},
  {Bonavera}, {Bond}, {Borrill}, {Bouchet}, {Boulanger}, {Bucher}, {Burigana},
  {Butler}, {Calabrese}, {Catalano}, {Chamballu}, {Chiang}, {Christensen},
  {Clements}, {Colombi}, {Colombo}, {Combet}, {Couchot}, {Coulais}, {Crill},
  {Curto}, {Cuttaia}, {Danese}, {Davies}, {Davis}, {de Bernardis}, {de Rosa},
  {de Zotti}, {Delabrouille}, {D{\'e}sert}, {Dickinson}, {Diego}, {Dole},
  {Donzelli}, {Dor{\'e}}, {Douspis}, {Ducout}, {Dupac}, {Efstathiou}, {Elsner},
  {En{\ss}lin}, {Eriksen}, {Falgarone}, {Fergusson}, {Finelli}, {Forni},
  {Frailis}, {Fraisse}, {Franceschi}, {Frejsel}, {Galeotta}, {Galli}, {Ganga},
  {Giard}, {Giraud-H{\'e}raud}, {Gjerl{\o}w}, {Gonz{\'a}lez-Nuevo},
  {G{\'o}rski}, {Gratton}, {Gregorio}, {Gruppuso}, {Gudmundsson}, {Hansen},
  {Hanson}, {Harrison}, {Helou}, {Henrot-Versill{\'e}},
  {Hern{\'a}ndez-Monteagudo}, {Herranz}, {Hildebrandt}, {Hivon}, {Hobson},
  {Holmes}, {Hornstrup}, {Hovest}, {Huffenberger}, {Hurier}, {Jaffe}, {Jaffe},
  {Jones}, {Juvela}, {Keih{\"a}nen}, {Keskitalo}, {Kisner}, {Knoche}, {Kunz},
  {Kurki-Suonio}, {Lagache}, {Lamarre}, {Lasenby}, {Lattanzi}, {Lawrence},
  {Leonardi}, {Lesgourgues}, {Levrier}, {Liguori}, {Lilje}, {Linden-V{\o}rnle},
  {L{\'o}pez-Caniego}, {Lubin}, {Mac{\'\i}as-P{\'e}rez}, {Maggio}, {Maino},
  {Mand olesi}, {Mangilli}, {Marshall}, {Martin}, {Mart{\'\i}nez-Gonz{\'a}lez},
  {Masi}, {Matarrese}, {Mazzotta}, {McGehee}, {Melchiorri}, {Mendes},
  {Mennella}, {Migliaccio}, {Mitra}, {Miville-Desch{\^e}nes}, {Moneti},
  {Montier}, {Morgante}, {Mortlock}, {Moss}, {Munshi}, {Murphy}, {Naselsky},
  {Nati}, {Natoli}, {Netterfield}, {N{\o}rgaard-Nielsen}, {Noviello},
  {Novikov}, {Novikov}, {Oxborrow}, {Paci}, {Pagano}, {Pajot}, {Paladini},
  {Paoletti}, {Pasian}, {Patanchon}, {Pearson}, {Pelkonen}, {Perdereau},
  {Perotto}, {Perrotta}, {Pettorino}, {Piacentini}, {Piat}, {Pierpaoli},
  {Pietrobon}, {Plaszczynski}, {Pointecouteau}, {Polenta}, {Pratt},
  {Pr{\'e}zeau}, {Prunet}, {Puget}, {Rachen}, {Reach}, {Rebolo}, {Reinecke},
  {Remazeilles}, {Renault}, {Renzi}, {Ristorcelli}, {Rocha}, {Rosset},
  {Rossetti}, {Roudier}, {Rubi{\~n}o-Mart{\'\i}n}, {Rusholme}, {Sandri},
  {Santos}, {Savelainen}, {Savini}, {Scott}, {Seiffert}, {Shellard}, {Spencer},
  {Stolyarov}, {Sudiwala}, {Sunyaev}, {Sutton}, {Suur-Uski}, {Sygnet},
  {Tauber}, {Terenzi}, {Toffolatti}, {Tomasi}, {Tristram}, {Tucci}, {Tuovinen},
  {Umana}, {Valenziano}, {Valiviita}, {Van Tent}, {Vielva}, {Villa}, {Wade},
  {Wandelt}, {Wehus}, {Yvon}, {Zacchei}, \& {Zonca}}]{Planck2015}
{Planck Collaboration}, {Ade}, P.~A.~R., {Aghanim}, N., {et~al.} 2016, \aap,
  594, A28

\bibitem[{{Poglitsch} {et~al.}(2010){Poglitsch}, {Waelkens}, {Geis},
  {Feuchtgruber}, {Vandenbussche}, {Rodriguez}, {Krause}, {Renotte}, {van
  Hoof}, {Saraceno}, {Cepa}, {Kerschbaum}, {Agn{\`e}se}, {Ali}, {Altieri},
  {Andreani}, {Augueres}, {Balog}, {Barl}, {Bauer}, {Belbachir}, {Benedettini},
  {Billot}, {Boulade}, {Bischof}, {Blommaert}, {Callut}, {Cara}, {Cerulli},
  {Cesarsky}, {Contursi}, {Creten}, {De Meester}, {Doublier}, {Doumayrou},
  {Duband}, {Exter}, {Genzel}, {Gillis}, {Gr{\"o}zinger}, {Henning},
  {Herreros}, {Huygen}, {Inguscio}, {Jakob}, {Jamar}, {Jean}, {de Jong},
  {Katterloher}, {Kiss}, {Klaas}, {Lemke}, {Lutz}, {Madden}, {Marquet},
  {Martignac}, {Mazy}, {Merken}, {Montfort}, {Morbidelli}, {M{\"u}ller},
  {Nielbock}, {Okumura}, {Orfei}, {Ottensamer}, {Pezzuto}, {Popesso},
  {Putzeys}, {Regibo}, {Reveret}, {Royer}, {Sauvage}, {Schreiber}, {Stegmaier},
  {Schmitt}, {Schubert}, {Sturm}, {Thiel}, {Tofani}, {Vavrek}, {Wetzstein},
  {Wieprecht}, \& {Wiezorrek}}]{Poglitsch2010}
{Poglitsch}, A., {Waelkens}, C., {Geis}, N., {et~al.} 2010, \aap, 518, L2

\bibitem[{{Qiu} \& {Zhang}(2009)}]{Qiu2009}
{Qiu}, K. \& {Zhang}, Q. 2009, \apjl, 702, L66

\bibitem[{{Ragan} {et~al.}(2012){Ragan}, {Henning}, {Krause}, {Pitann},
  {Beuther}, {Linz}, {Tackenberg}, {Balog}, {Hennemann}, {Launhardt}, {Lippok},
  {Nielbock}, {Schmiedeke}, {Schuller}, {Steinacker}, {Stutz}, \&
  {Vasyunina}}]{Ragan2012}
{Ragan}, S., {Henning}, T., {Krause}, O., {et~al.} 2012, \aap, 547, A49

\bibitem[{{Reach} {et~al.}(2005){Reach}, {Megeath}, {Cohen}, {Hora}, {Carey},
  {Surace}, {Willner}, {Barmby}, {Wilson}, {Glaccum}, {Lowrance}, {Marengo}, \&
  {Fazio}}]{Reach2005}
{Reach}, W.~T., {Megeath}, S.~T., {Cohen}, M., {et~al.} 2005, \pasp, 117, 978

\bibitem[{{Reid} {et~al.}(2009){Reid}, {Menten}, {Zheng}, {Brunthaler},
  {Moscadelli}, {Xu}, {Zhang}, {Sato}, {Honma}, {Hirota}, {Hachisuka}, {Choi},
  {Moellenbrock}, \& {Bartkiewicz}}]{Reid2009}
{Reid}, M.~J., {Menten}, K.~M., {Zheng}, X.~W., {et~al.} 2009, \apj, 700, 137

\bibitem[{{Rieke} {et~al.}(2004){Rieke}, {Young}, {Engelbracht}, {Kelly},
  {Low}, {Haller}, {Beeman}, {Gordon}, {Stansberry}, {Misselt}, {Cadien},
  {Morrison}, {Rivlis}, {Latter}, {Noriega-Crespo}, {Padgett}, {Stapelfeldt},
  {Hines}, {Egami}, {Muzerolle}, {Alonso-Herrero}, {Blaylock}, {Dole}, {Hinz},
  {Le Floc'h}, {Papovich}, {P{\'e}rez-Gonz{\'a}lez}, {Smith}, {Su}, {Bennett},
  {Frayer}, {Henderson}, {Lu}, {Masci}, {Pesenson}, {Rebull}, {Rho}, {Keene},
  {Stolovy}, {Wachter}, {Wheaton}, {Werner}, \& {Richards}}]{Rieke2004}
{Rieke}, G.~H., {Young}, E.~T., {Engelbracht}, C.~W., {et~al.} 2004, \apjs,
  154, 25

\bibitem[{{Robitaille} {et~al.}(2007){Robitaille}, {Whitney}, {Indebetouw}, \&
  {Wood}}]{Robitaille2007}
{Robitaille}, T.~P., {Whitney}, B.~A., {Indebetouw}, R., \& {Wood}, K. 2007,
  \apjs, 169, 328

\bibitem[{{Rodr{\'\i}guez} {et~al.}(2021){Rodr{\'\i}guez}, {Hofner}, {Araya},
  {Zhang}, {Linz}, {Kurtz}, {G{\'o}mez}, {Carrasco-Gonz{\'a}lez}, \&
  {Rosero}}]{Rodriguez2021b}
{Rodr{\'\i}guez}, T.~M., {Hofner}, P., {Araya}, E.~D., {et~al.} 2021, arXiv
  e-prints, arXiv:2109.01243

\bibitem[{{Roman-Duval} {et~al.}(2010){Roman-Duval}, {Jackson}, {Heyer},
  {Rathborne}, \& {Simon}}]{RomanDuval2010}
{Roman-Duval}, J., {Jackson}, J.~M., {Heyer}, M., {Rathborne}, J., \& {Simon},
  R. 2010, \apj, 723, 492

\bibitem[{{Rosen} {et~al.}(2020){Rosen}, {Offner}, {Sadavoy}, {Bhandare},
  {V{\'a}zquez-Semadeni}, \& {Ginsburg}}]{Rosen2020}
{Rosen}, A.~L., {Offner}, S. S.~R., {Sadavoy}, S.~I., {et~al.} 2020, \ssr, 216,
  62

\bibitem[{{Rosolowsky} {et~al.}(2010){Rosolowsky}, {Dunham}, {Ginsburg},
  {Bradley}, {Aguirre}, {Bally}, {Battersby}, {Cyganowski}, {Dowell},
  {Drosback}, {Evans}, {Glenn}, {Harvey}, {Stringfellow}, {Walawender}, \&
  {Williams}}]{Rosolowsky2010}
{Rosolowsky}, E., {Dunham}, M.~K., {Ginsburg}, A., {et~al.} 2010, \apjs, 188,
  123

\bibitem[{{S{\'a}nchez-Monge} {et~al.}(2010){S{\'a}nchez-Monge}, {Palau},
  {Estalella}, {Kurtz}, {Zhang}, {Di Francesco}, \&
  {Shepherd}}]{SanchezMonge2010}
{S{\'a}nchez-Monge}, {\'A}., {Palau}, A., {Estalella}, R., {et~al.} 2010,
  \apjl, 721, L107

\bibitem[{{Sault} {et~al.}(1995){Sault}, {Teuben}, \& {Wright}}]{miriad}
{Sault}, R.~J., {Teuben}, P.~J., \& {Wright}, M.~C.~H. 1995, in Astronomical
  Society of the Pacific Conference Series, Vol.~77, Astronomical Data Analysis
  Software and Systems IV, ed. R.~A. {Shaw}, H.~E. {Payne}, \& J.~J.~E.
  {Hayes}, 433

\bibitem[{{Schilke}(2015)}]{Schilke2015}
{Schilke}, P. 2015, in EAS Publications Series, Vol.~75, EAS Publications
  Series, 227--235

\bibitem[{{Schilke} {et~al.}(1997){Schilke}, {Walmsley}, {Pineau des Forets},
  \& {Flower}}]{Schilke1997}
{Schilke}, P., {Walmsley}, C.~M., {Pineau des Forets}, G., \& {Flower}, D.~R.
  1997, \aap, 321, 293

\bibitem[{{Schlingman} {et~al.}(2011){Schlingman}, {Shirley}, {Schenk},
  {Rosolowsky}, {Bally}, {Battersby}, {Dunham}, {Ellsworth-Bowers}, {Evans},
  {Ginsburg}, \& {Stringfellow}}]{Schlingman2011}
{Schlingman}, W.~M., {Shirley}, Y.~L., {Schenk}, D.~E., {et~al.} 2011, \apjs,
  195, 14

\bibitem[{{Schneider} {et~al.}(2012){Schneider}, {Csengeri}, {Hennemann},
  {Motte}, {Didelon}, {Federrath}, {Bontemps}, {Di Francesco}, {Arzoumanian},
  {Minier}, {Andr{\'e}}, {Hill}, {Zavagno}, {Nguyen-Luong}, {Attard},
  {Bernard}, {Elia}, {Fallscheer}, {Griffin}, {Kirk}, {Klessen}, {K{\"o}nyves},
  {Martin}, {Men'shchikov}, {Palmeirim}, {Peretto}, {Pestalozzi}, {Russeil},
  {Sadavoy}, {Sousbie}, {Testi}, {Tremblin}, {Ward-Thompson}, \&
  {White}}]{Schneider2012}
{Schneider}, N., {Csengeri}, T., {Hennemann}, M., {et~al.} 2012, \aap, 540, L11

\bibitem[{{Schw{\"o}rer} {et~al.}(2019){Schw{\"o}rer}, {S{\'a}nchez-Monge},
  {Schilke}, {M{\"o}ller}, {Ginsburg}, {Meng}, {Schmiedeke}, {M{\"u}ller},
  {Lis}, \& {Qin}}]{Schwoerer2019}
{Schw{\"o}rer}, A., {S{\'a}nchez-Monge}, {\'A}., {Schilke}, P., {et~al.} 2019,
  \aap, 628, A6

\bibitem[{{Semenov} {et~al.}(2010){Semenov}, {Hersant}, {Wakelam}, {Dutrey},
  {Chapillon}, {Guilloteau}, {Henning}, {Launhardt}, {Pi{\'e}tu}, \&
  {Schreyer}}]{Semenov2010}
{Semenov}, D., {Hersant}, F., {Wakelam}, V., {et~al.} 2010, \aap, 522, A42

\bibitem[{{Shimajiri} {et~al.}(2015){Shimajiri}, {Sakai}, {Kitamura},
  {Tsukagoshi}, {Saito}, {Nakamura}, {Momose}, {Takakuwa}, {Yamaguchi},
  {Sakai}, {Yamamoto}, \& {Kawabe}}]{Shimajiri2015}
{Shimajiri}, Y., {Sakai}, T., {Kitamura}, Y., {et~al.} 2015, \apjs, 221, 31

\bibitem[{{Shirley} {et~al.}(2013){Shirley}, {Ellsworth-Bowers}, {Svoboda},
  {Schlingman}, {Ginsburg}, {Rosolowsky}, {Gerner}, {Mairs}, {Battersby},
  {Stringfellow}, {Dunham}, {Glenn}, \& {Bally}}]{Shirley2013}
{Shirley}, Y.~L., {Ellsworth-Bowers}, T.~P., {Svoboda}, B., {et~al.} 2013,
  \apjs, 209, 2

\bibitem[{{Shirley} {et~al.}(2000){Shirley}, {Evans}, {Rawlings}, \&
  {Gregersen}}]{Shirley2000}
{Shirley}, Y.~L., {Evans}, Neal~J., I., {Rawlings}, J. M.~C., \& {Gregersen},
  E.~M. 2000, \apjs, 131, 249

\bibitem[{{Slysh} {et~al.}(1999){Slysh}, {Val'tts}, {Kalenskii}, {Voronkov},
  {Palagi}, {Tofani}, \& {Catarzi}}]{Slysh1999}
{Slysh}, V.~I., {Val'tts}, I.~E., {Kalenskii}, S.~V., {et~al.} 1999, \aaps,
  134, 115

\bibitem[{{Smith} {et~al.}(2009){Smith}, {Longmore}, \& {Bonnell}}]{Smith2009}
{Smith}, R.~J., {Longmore}, S., \& {Bonnell}, I. 2009, \mnras, 400, 1775

\bibitem[{{Soler} {et~al.}(2017){Soler}, {Ade}, {Angil{\`e}}, {Ashton},
  {Benton}, {Devlin}, {Dober}, {Fissel}, {Fukui}, {Galitzki}, {Gandilo},
  {Hennebelle}, {Klein}, {Li}, {Korotkov}, {Martin}, {Matthews}, {Moncelsi},
  {Netterfield}, {Novak}, {Pascale}, {Poidevin}, {Santos}, {Savini}, {Scott},
  {Shariff}, {Thomas}, {Tucker}, {Tucker}, \& {Ward-Thompson}}]{Soler2017}
{Soler}, J.~D., {Ade}, P.~A.~R., {Angil{\`e}}, F.~E., {et~al.} 2017, \aap, 603,
  A64

\bibitem[{{Soler} {et~al.}(2019){Soler}, {Beuther}, {Rugel}, {Wang}, {Clark},
  {Glover}, {Goldsmith}, {Heyer}, {Anderson}, {Goodman}, {Henning},
  {Kainulainen}, {Klessen}, {Longmore}, {McClure-Griffiths}, {Menten},
  {Mottram}, {Ott}, {Ragan}, {Smith}, {Urquhart}, {Bigiel}, {Hennebelle},
  {Roy}, \& {Schilke}}]{Soler2019}
{Soler}, J.~D., {Beuther}, H., {Rugel}, M., {et~al.} 2019, \aap, 622, A166

\bibitem[{{Sridharan} {et~al.}(2002){Sridharan}, {Beuther}, {Schilke},
  {Menten}, \& {Wyrowski}}]{Sridharan2002}
{Sridharan}, T.~K., {Beuther}, H., {Schilke}, P., {Menten}, K.~M., \&
  {Wyrowski}, F. 2002, \apj, 566, 931

\bibitem[{{Steer} {et~al.}(1984){Steer}, {Dewdney}, \& {Ito}}]{Steer1984}
{Steer}, D.~G., {Dewdney}, P.~E., \& {Ito}, M.~R. 1984, \aap, 137, 159

\bibitem[{{Stetson}(1987)}]{Stetson1987}
{Stetson}, P.~B. 1987, \pasp, 99, 191

\bibitem[{{Sunada} {et~al.}(2007){Sunada}, {Nakazato}, {Ikeda}, {Hongo},
  {Kitamura}, \& {Yang}}]{Sunada2007}
{Sunada}, K., {Nakazato}, T., {Ikeda}, N., {et~al.} 2007, \pasj, 59, 1185

\bibitem[{{Tackenberg} {et~al.}(2012){Tackenberg}, {Beuther}, {Henning},
  {Schuller}, {Wienen}, {Motte}, {Wyrowski}, {Bontemps}, {Bronfman}, {Menten},
  {Testi}, \& {Lefloch}}]{Tackenberg2012}
{Tackenberg}, J., {Beuther}, H., {Henning}, T., {et~al.} 2012, \aap, 540, A113

\bibitem[{{Tan} {et~al.}(2014){Tan}, {Beltr{\'a}n}, {Caselli}, {Fontani},
  {Fuente}, {Krumholz}, {McKee}, \& {Stolte}}]{Tan2014}
{Tan}, J.~C., {Beltr{\'a}n}, M.~T., {Caselli}, P., {et~al.} 2014, Protostars
  and Planets VI, 149

\bibitem[{{Taquet} {et~al.}(2020){Taquet}, {Codella}, {De Simone},
  {L{\'o}pez-Sepulcre}, {Pineda}, {Segura-Cox}, {Ceccarelli}, {Caselli},
  {Gusdorf}, {Persson}, {Alves}, {Caux}, {Favre}, {Fontani}, {Neri}, {Oya},
  {Sakai}, {Vastel}, {Yamamoto}, {Bachiller}, {Balucani}, {Bianchi},
  {Bizzocchi}, {Chac{\'o}n-Tanarro}, {Dulieu}, {Enrique-Romero}, {Feng},
  {Holdship}, {Lefloch}, {Jaber Al-Edhari}, {Jim{\'e}nez-Serra}, {Kahane},
  {Lattanzi}, {Ospina-Zamudio}, {Podio}, {Punanova}, {Rimola}, {Sims},
  {Spezzano}, {Testi}, {Theul{\'e}}, {Ugliengo}, {Vasyunin}, {Vazart}, {Viti},
  \& {Witzel}}]{Taquet2020}
{Taquet}, V., {Codella}, C., {De Simone}, M., {et~al.} 2020, \aap, 637, A63

\bibitem[{{Tig{\'e}} {et~al.}(2017){Tig{\'e}}, {Motte}, {Russeil}, {Zavagno},
  {Hennemann}, {Schneider}, {Hill}, {Nguyen Luong}, {Di Francesco}, {Bontemps},
  {Louvet}, {Didelon}, {K{\"o}nyves}, {Andr{\'e}}, {Leuleu}, {Bardagi},
  {Anderson}, {Arzoumanian}, {Benedettini}, {Bernard}, {Elia}, {Figueira},
  {Kirk}, {Martin}, {Minier}, {Molinari}, {Nony}, {Persi}, {Pezzuto},
  {Polychroni}, {Rayner}, {Rivera-Ingraham}, {Roussel}, {Rygl}, {Spinoglio}, \&
  {White}}]{Tige2017}
{Tig{\'e}}, J., {Motte}, F., {Russeil}, D., {et~al.} 2017, \aap, 602, A77

\bibitem[{{Tychoniec} {et~al.}(2019){Tychoniec}, {Hull}, {Kristensen}, {Tobin},
  {Le Gouellec}, \& {van Dishoeck}}]{Tychoniec2019}
{Tychoniec}, {\L}., {Hull}, C. L.~H., {Kristensen}, L.~E., {et~al.} 2019, \aap,
  632, A101

\bibitem[{{Urquhart} {et~al.}(2018){Urquhart}, {K{\"o}nig}, {Giannetti},
  {Leurini}, {Moore}, {Eden}, {Pillai}, {Thompson}, {Braiding}, {Burton},
  {Csengeri}, {Dempsey}, {Figura}, {Froebrich}, {Menten}, {Schuller}, {Smith},
  \& {Wyrowski}}]{Urquhart2018}
{Urquhart}, J.~S., {K{\"o}nig}, C., {Giannetti}, A., {et~al.} 2018, \mnras,
  473, 1059

\bibitem[{{Valdettaro} {et~al.}(2001){Valdettaro}, {Palla}, {Brand},
  {Cesaroni}, {Comoretto}, {Di Franco}, {Felli}, {Natale}, {Palagi}, {Panella},
  \& {Tofani}}]{Valdettaro2001}
{Valdettaro}, R., {Palla}, F., {Brand}, J., {et~al.} 2001, \aap, 368, 845

\bibitem[{{van der Tak} {et~al.}(2000){van der Tak}, {van Dishoeck}, {Evans},
  \& {Blake}}]{vanderTak2000}
{van der Tak}, F. F.~S., {van Dishoeck}, E.~F., {Evans}, Neal~J., I., \&
  {Blake}, G.~A. 2000, \apj, 537, 283

\bibitem[{{Vasyunina} {et~al.}(2014){Vasyunina}, {Vasyunin}, {Herbst}, {Linz},
  {Voronkov}, {Britton}, {Zinchenko}, \& {Schuller}}]{Vasyunina2014}
{Vasyunina}, T., {Vasyunin}, A.~I., {Herbst}, E., {et~al.} 2014, \apj, 780, 85

\bibitem[{{Walcher} {et~al.}(2011){Walcher}, {Groves}, {Budav{\'a}ri}, \&
  {Dale}}]{Walcher2011}
{Walcher}, J., {Groves}, B., {Budav{\'a}ri}, T., \& {Dale}, D. 2011, \apss,
  331, 1

\bibitem[{{Wang} {et~al.}(2011){Wang}, {Zhang}, {Wu}, \& {Zhang}}]{Wang2011}
{Wang}, K., {Zhang}, Q., {Wu}, Y., \& {Zhang}, H. 2011, \apj, 735, 64

\bibitem[{{Wang} {et~al.}(2016){Wang}, {Audard}, {Fontani},
  {S{\'a}nchez-Monge}, {Busquet}, {Palau}, {Beuther}, {Tan}, {Estalella},
  {Isella}, {Gueth}, \& {Jim{\'e}nez-Serra}}]{Wang2016}
{Wang}, Y., {Audard}, M., {Fontani}, F., {et~al.} 2016, \aap, 587, A69

\bibitem[{{Williams} {et~al.}(1994){Williams}, {de Geus}, \&
  {Blitz}}]{Williams1994}
{Williams}, J.~P., {de Geus}, E.~J., \& {Blitz}, L. 1994, \apj, 428, 693

\bibitem[{{Wilson} \& {Rood}(1994)}]{Wilson1994}
{Wilson}, T.~L. \& {Rood}, R. 1994, \araa, 32, 191

\bibitem[{{Wouterloot} \& {Brand}(1989)}]{Wouterloot1989b}
{Wouterloot}, J.~G.~A. \& {Brand}, J. 1989, \aaps, 80, 149

\bibitem[{{Wouterloot} {et~al.}(1993){Wouterloot}, {Brand}, \&
  {Fiegle}}]{Wouterloot1993}
{Wouterloot}, J.~G.~A., {Brand}, J., \& {Fiegle}, K. 1993, \aaps, 98, 589

\bibitem[{{Wouterloot} {et~al.}(1989){Wouterloot}, {Henkel}, \&
  {Walmsley}}]{Wouterloot1989a}
{Wouterloot}, J.~G.~A., {Henkel}, C., \& {Walmsley}, C.~M. 1989, \aap, 215, 131

\bibitem[{{Wouterloot} \& {Walmsley}(1986)}]{Wouterloot1986}
{Wouterloot}, J.~G.~A. \& {Walmsley}, C.~M. 1986, \aap, 168, 237

\bibitem[{{Wouterloot} {et~al.}(1988){Wouterloot}, {Walmsley}, \&
  {Henkel}}]{Wouterloot1988}
{Wouterloot}, J.~G.~A., {Walmsley}, C.~M., \& {Henkel}, C. 1988, \aap, 203, 367

\bibitem[{{Wu} {et~al.}(2004){Wu}, {Wei}, {Zhao}, {Shi}, {Yu}, {Qin}, \&
  {Huang}}]{Wu2004}
{Wu}, Y., {Wei}, Y., {Zhao}, M., {et~al.} 2004, \aap, 426, 503

\bibitem[{{Wu} {et~al.}(2005){Wu}, {Zhang}, {Chen}, {Yang}, {Wei}, \&
  {Ho}}]{Wu2005}
{Wu}, Y., {Zhang}, Q., {Chen}, H., {et~al.} 2005, \aj, 129, 330

\bibitem[{{Wyrowski} {et~al.}(1999){Wyrowski}, {Schilke}, {Walmsley}, \&
  {Menten}}]{Wyrowski1999}
{Wyrowski}, F., {Schilke}, P., {Walmsley}, C.~M., \& {Menten}, K.~M. 1999,
  \apjl, 514, L43

\bibitem[{{Yang} {et~al.}(2002){Yang}, {Jiang}, {Wang}, {Ju}, \&
  {Wang}}]{Yang2002}
{Yang}, J., {Jiang}, Z., {Wang}, M., {Ju}, B., \& {Wang}, H. 2002, \apjs, 141,
  157

\bibitem[{{Zeng} {et~al.}(2020){Zeng}, {Zhang}, {Jim{\'e}nez-Serra}, {Tercero},
  {Lu}, {Mart{\'\i}n-Pintado}, {de Vicente}, {Rivilla}, \& {Li}}]{Zeng2020}
{Zeng}, S., {Zhang}, Q., {Jim{\'e}nez-Serra}, I., {et~al.} 2020, \mnras, 497,
  4896

\bibitem[{{Zhang} {et~al.}(2007){Zhang}, {Hunter}, {Beuther}, {Sridharan},
  {Liu}, {Su}, {Chen}, \& {Chen}}]{Zhang2007}
{Zhang}, Q., {Hunter}, T.~R., {Beuther}, H., {et~al.} 2007, \apj, 658, 1152

\bibitem[{{Zhang} {et~al.}(2005){Zhang}, {Hunter}, {Brand}, {Sridharan},
  {Cesaroni}, {Molinari}, {Wang}, \& {Kramer}}]{Zhang2005}
{Zhang}, Q., {Hunter}, T.~R., {Brand}, J., {et~al.} 2005, \apj, 625, 864

\bibitem[{{Zhang} {et~al.}(2015){Zhang}, {Wang}, {Lu}, \&
  {Jim{\'e}nez-Serra}}]{Zhang2015}
{Zhang}, Q., {Wang}, K., {Lu}, X., \& {Jim{\'e}nez-Serra}, I. 2015, \apj, 804,
  141

\bibitem[{{Zhang} {et~al.}(2009){Zhang}, {Wang}, {Pillai}, \&
  {Rathborne}}]{Zhang2009}
{Zhang}, Q., {Wang}, Y., {Pillai}, T., \& {Rathborne}, J. 2009, \apj, 696, 268

\bibitem[{{Zinnecker} \& {Yorke}(2007)}]{Zinnecker2007}
{Zinnecker}, H. \& {Yorke}, H.~W. 2007, \araa, 45, 481

\end{thebibliography}

\begin{appendix}
\section{Moment maps}\label{app:moment0maps}

\begin{figure*}
\centering
\includegraphics[]{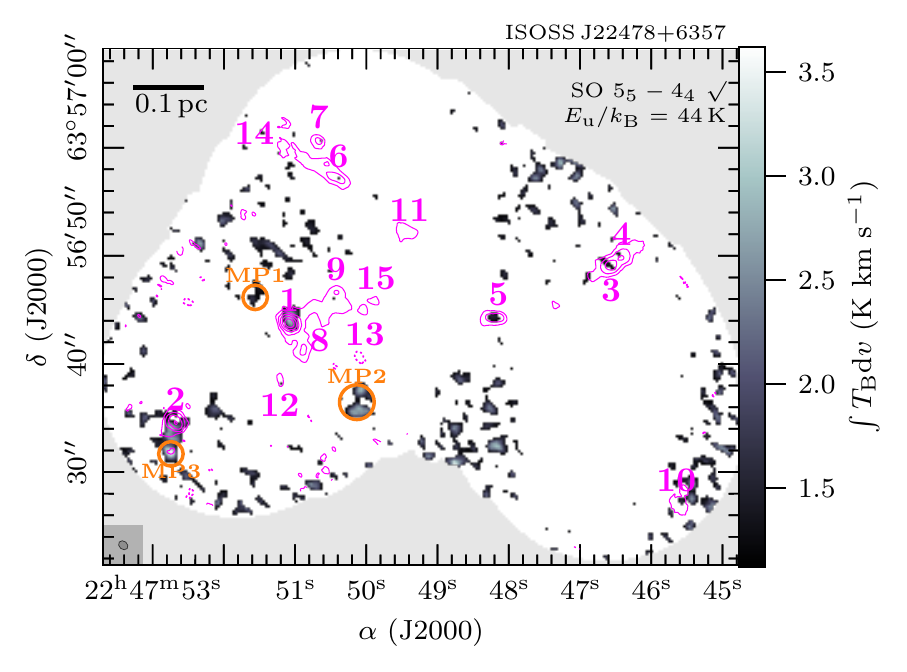}
\includegraphics[]{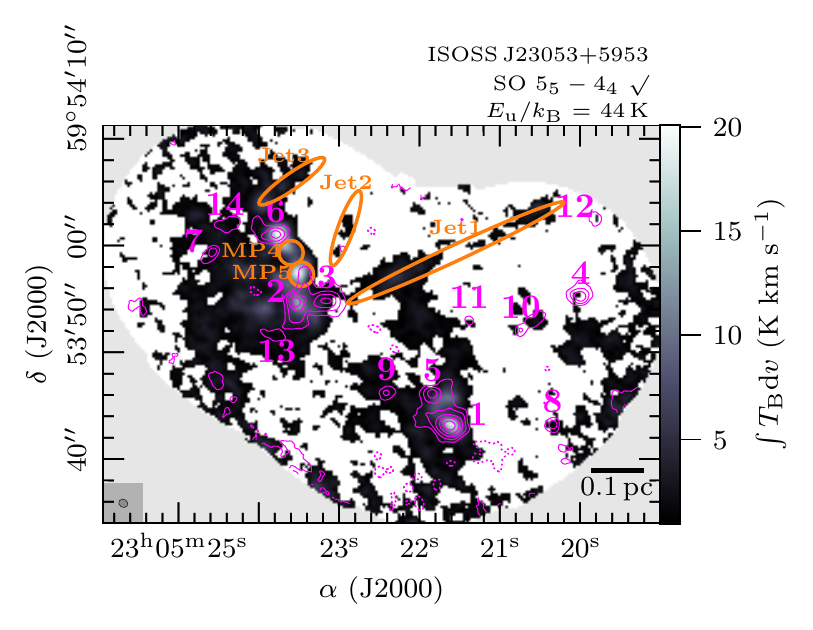}
\caption{The same as Fig. \ref{fig:moment0_H2CO_3_03_2_02}, but for SO $5_{5}-4_{4}$.}
\label{fig:moment0_SO_5_5_4_4}
\end{figure*}

\begin{figure*}
\centering
\includegraphics[]{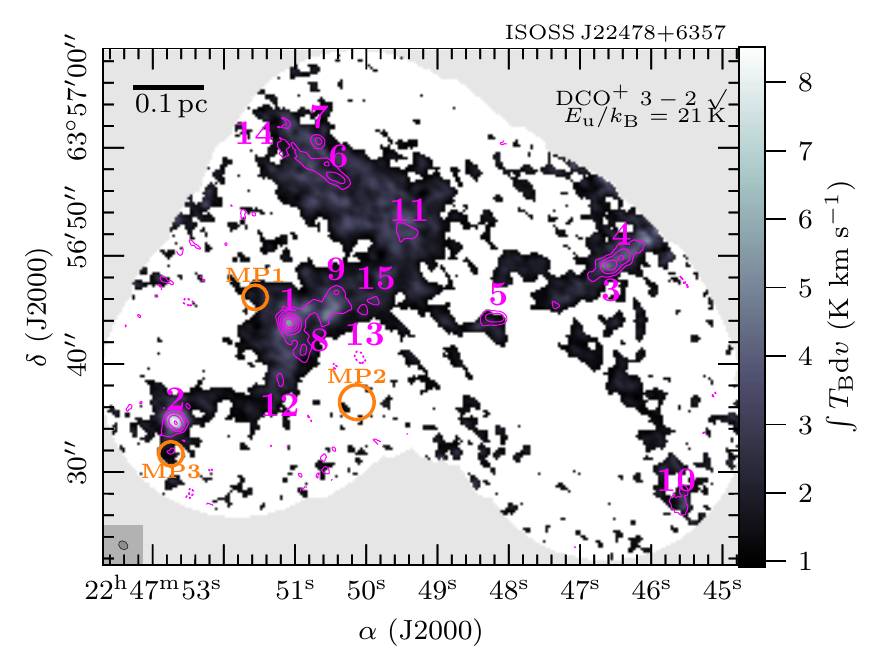}
\includegraphics[]{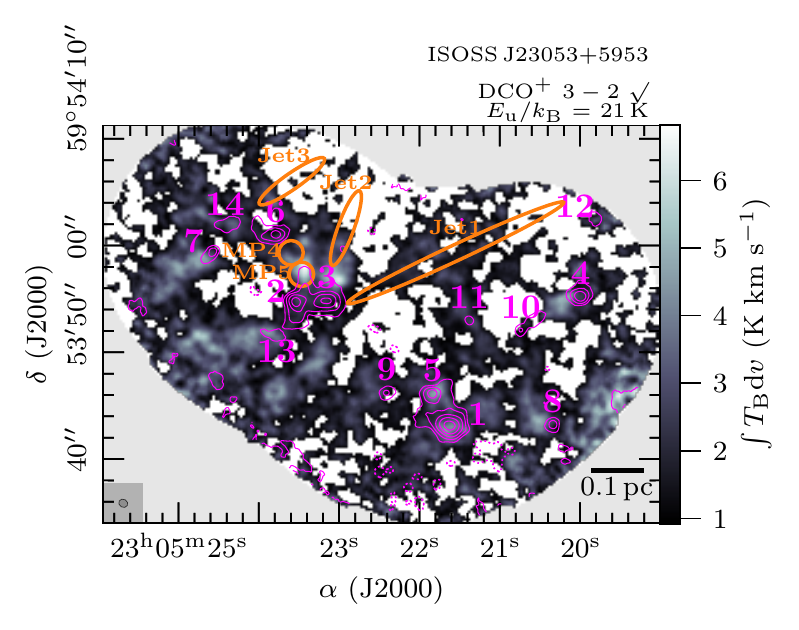}
\caption{The same as Fig. \ref{fig:moment0_H2CO_3_03_2_02}, but for DCO$^{+}$ $3-2$.}
\label{fig:moment0_DCO+_3_2}
\end{figure*}

\begin{figure*}
\centering
\includegraphics[]{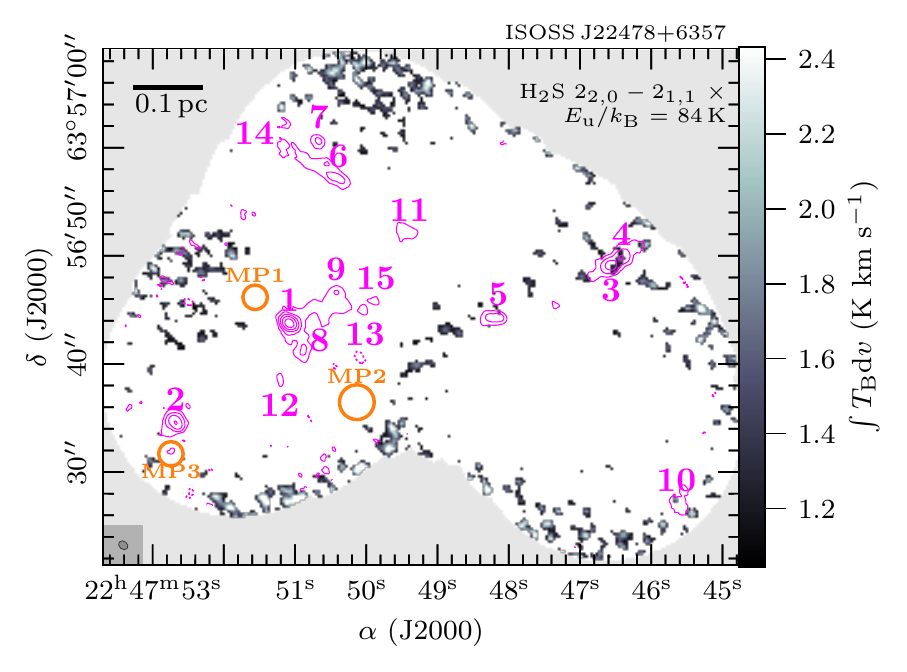}
\includegraphics[]{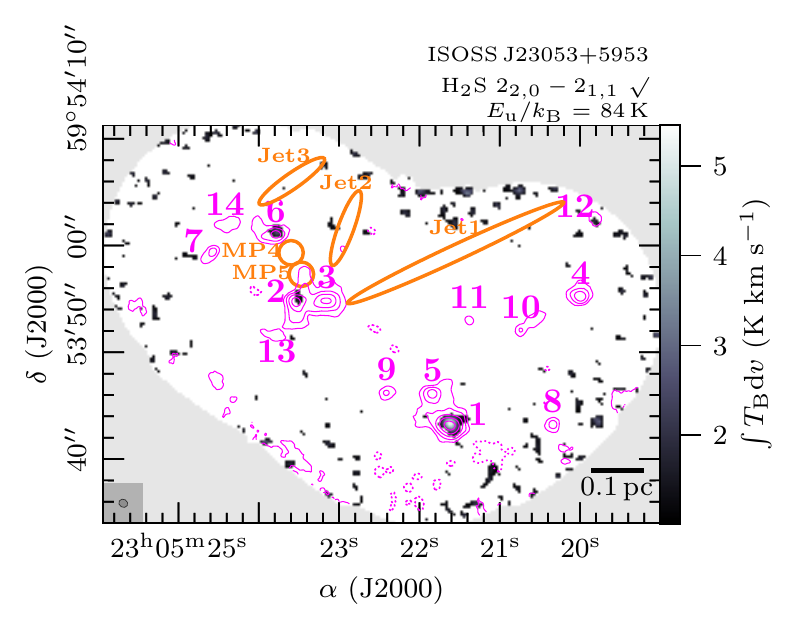}
\caption{The same as Fig. \ref{fig:moment0_H2CO_3_03_2_02}, but for H$_{2}$S $2_{2,0}-2_{1,1}$.}
\label{fig:moment0_H2S_2_20_2_11}
\end{figure*}

\begin{figure*}
\centering
\includegraphics[]{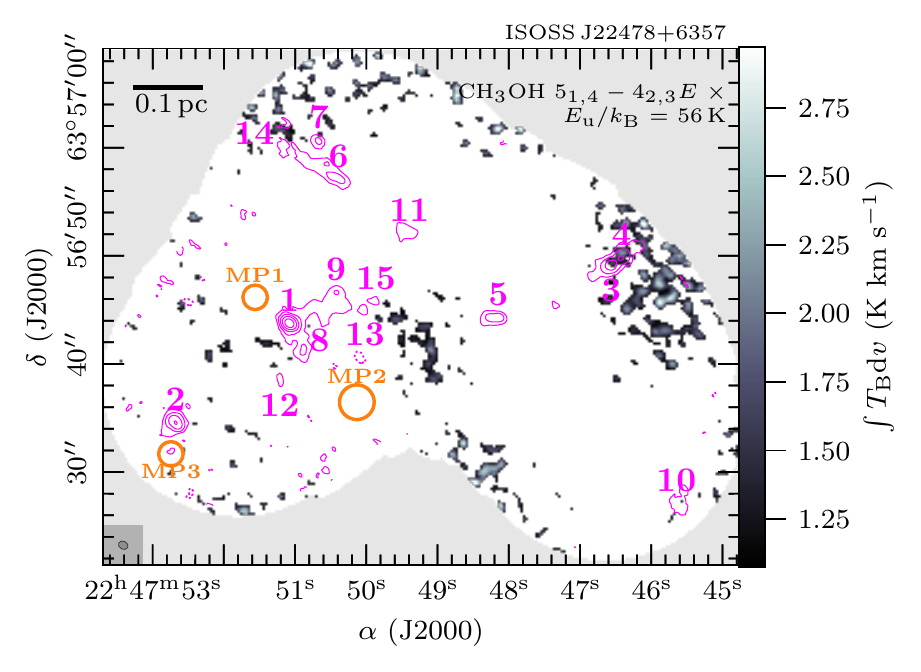}
\includegraphics[]{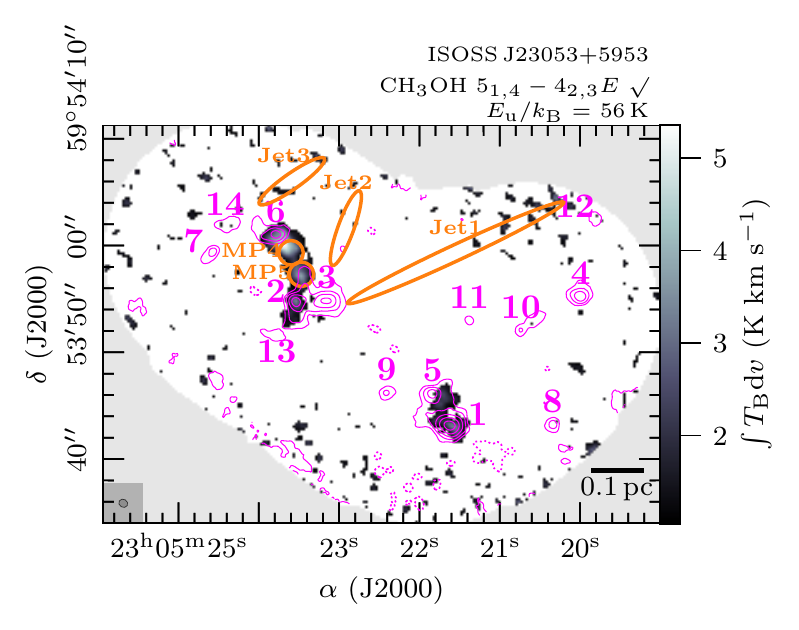}
\caption{The same as Fig. \ref{fig:moment0_H2CO_3_03_2_02}, but for CH$_{3}$OH $5_{1,4}-4_{2,3}E$.}
\label{fig:moment0_CH3OH_5_1_4_2}
\end{figure*}

\begin{figure*}
\centering
\includegraphics[]{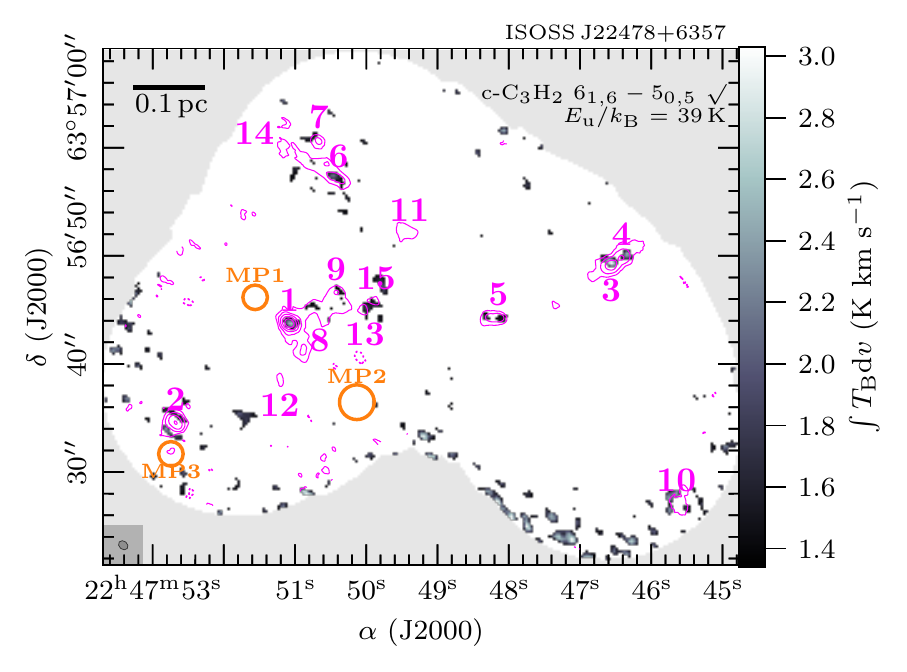}
\includegraphics[]{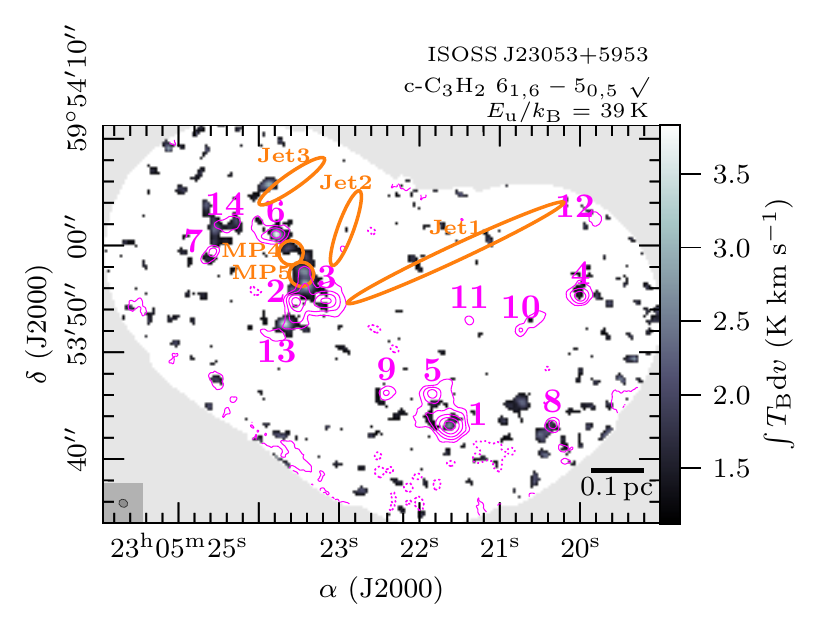}
\caption{The same as Fig. \ref{fig:moment0_H2CO_3_03_2_02}, but for c-C$_{3}$H$_{2}$ $6_{1,6}-5_{0,5}$.}
\label{fig:moment0_c-C3H2_6_16_5_05}
\end{figure*}

\begin{figure*}
\centering
\includegraphics[]{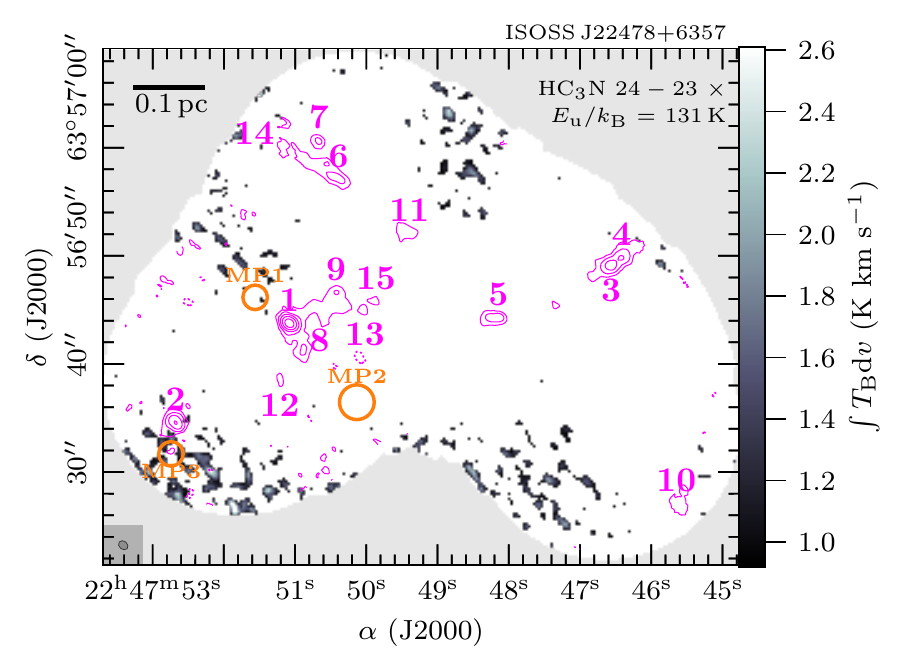}
\includegraphics[]{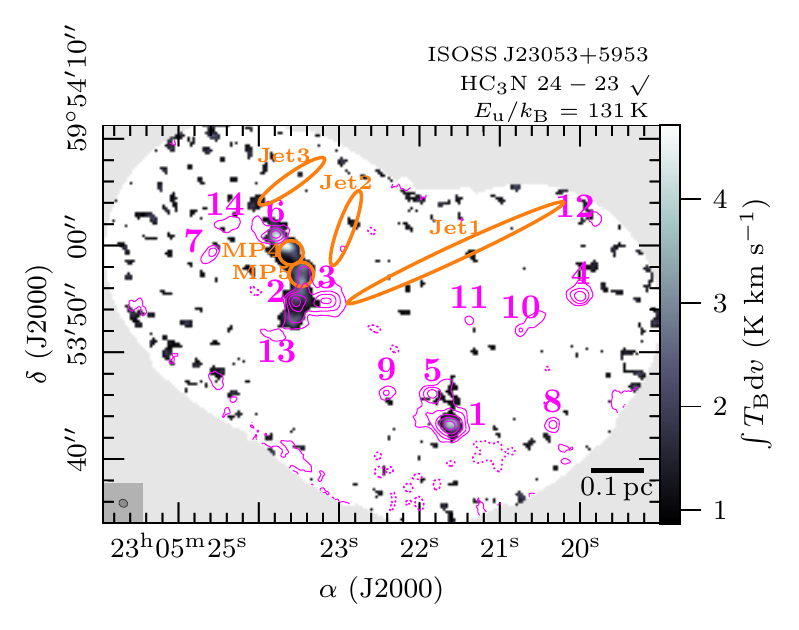}
\caption{The same as Fig. \ref{fig:moment0_H2CO_3_03_2_02}, but for HC$_{3}$N $24-23$.}
\label{fig:moment0_HC3N_24_23}
\end{figure*}

\begin{figure*}
\centering
\includegraphics[]{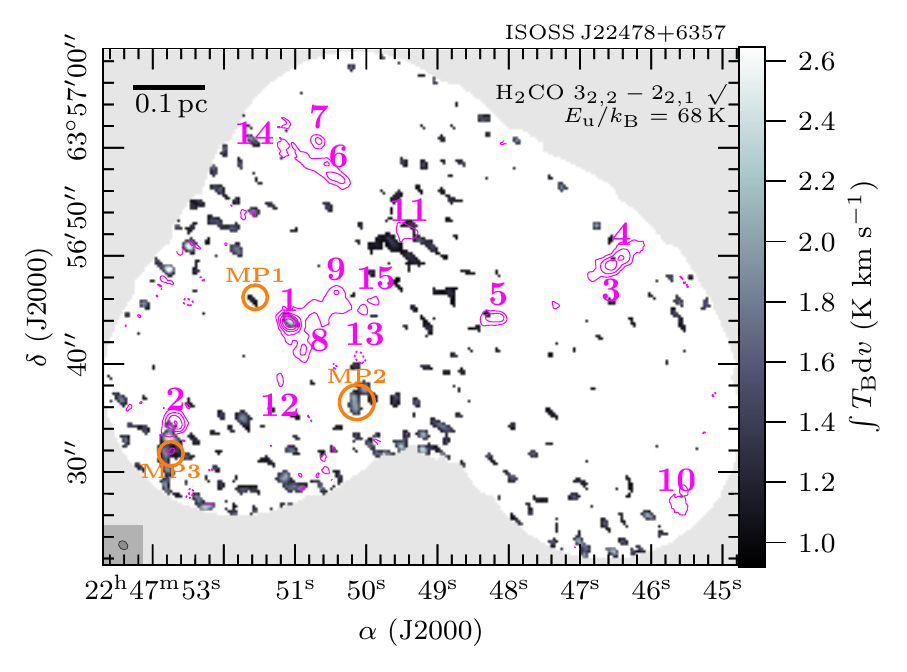}
\includegraphics[]{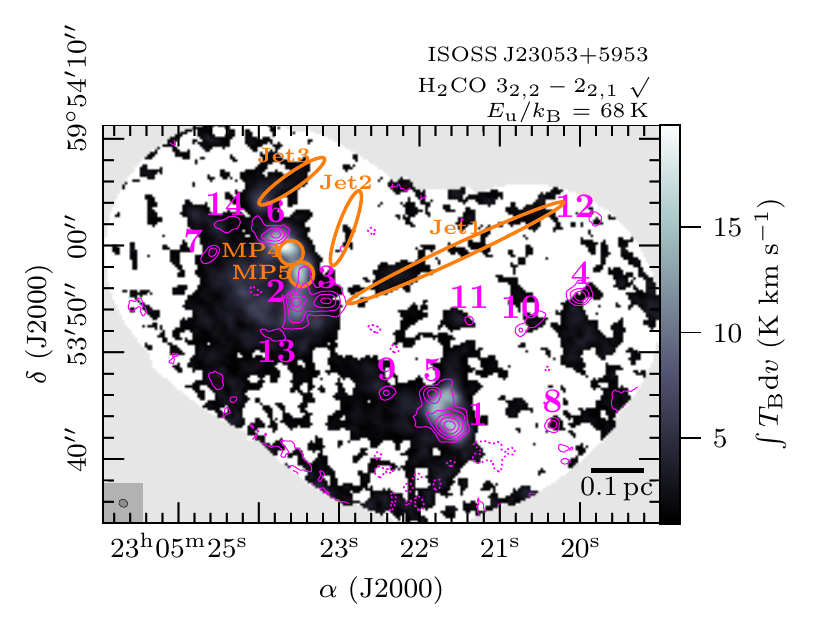}
\caption{The same as Fig. \ref{fig:moment0_H2CO_3_03_2_02}, but for H$_{2}$CO $3_{2,2}-2_{2,1}$.}
\label{fig:moment0_H2CO_3_22_2_21}
\end{figure*}

\begin{figure*}
\centering
\includegraphics[]{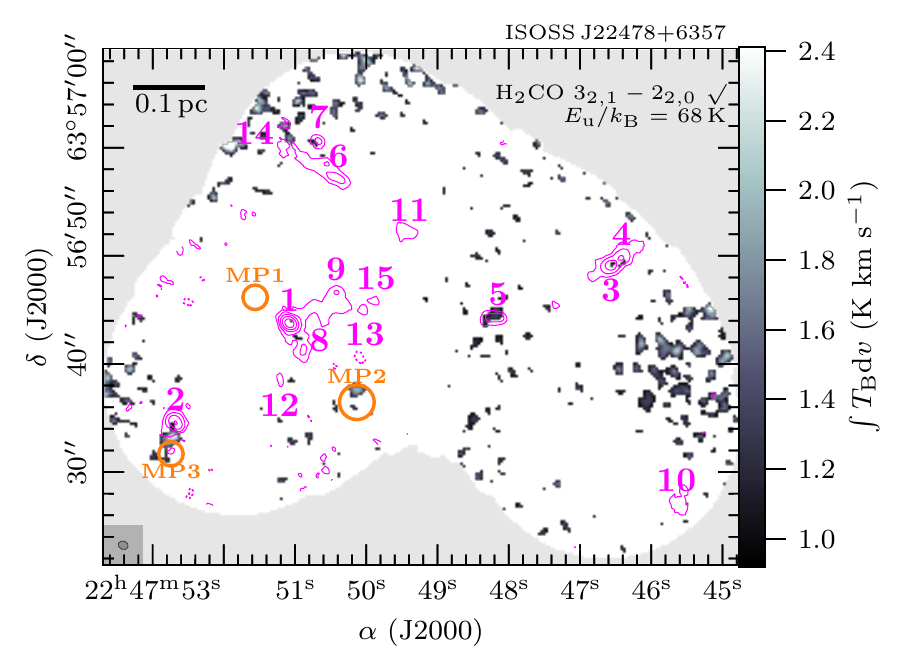}
\includegraphics[]{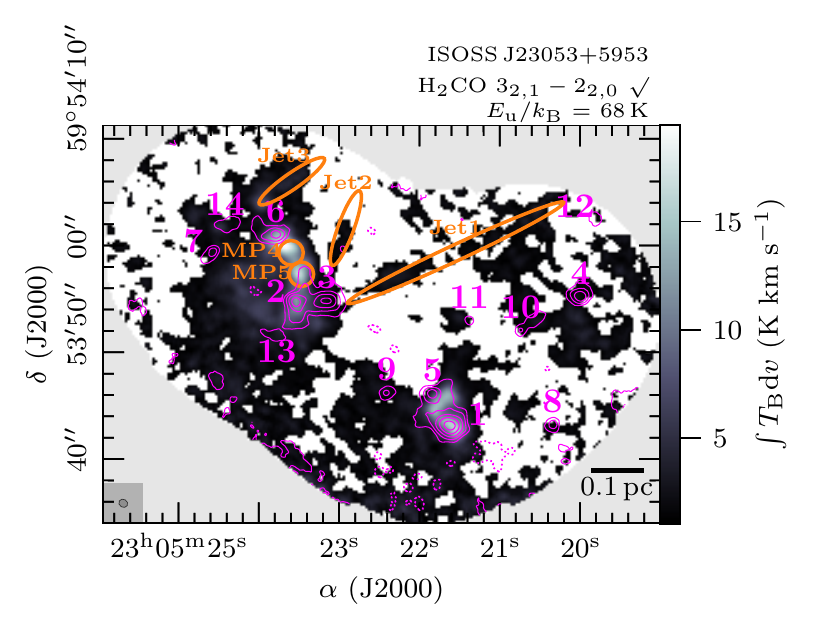}
\caption{The same as Fig. \ref{fig:moment0_H2CO_3_03_2_02}, but for H$_{2}$CO $3_{2,1}-2_{2,0}$.}
\label{fig:moment0_H2CO_3_21_2_20}
\end{figure*}

\begin{figure*}
\centering
\includegraphics[]{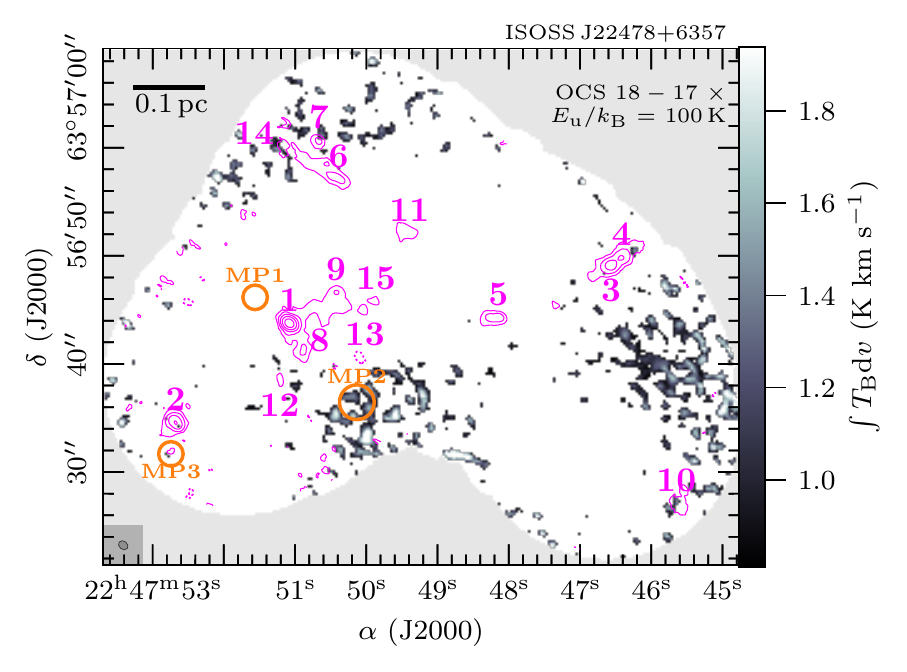}
\includegraphics[]{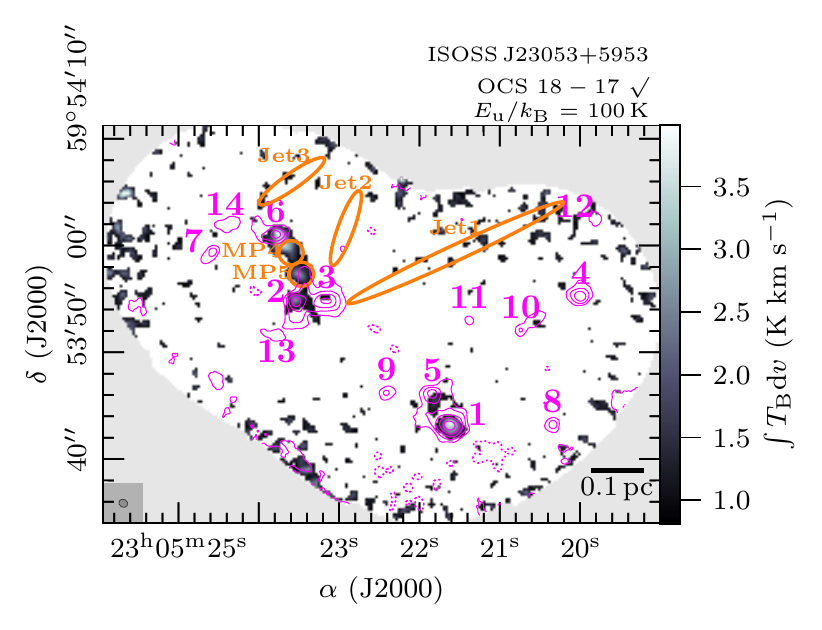}
\caption{The same as Fig. \ref{fig:moment0_H2CO_3_03_2_02}, but for OCS $18-17$.}
\label{fig:moment0_OCS_18_17}
\end{figure*}

\begin{figure*}
\centering
\includegraphics[]{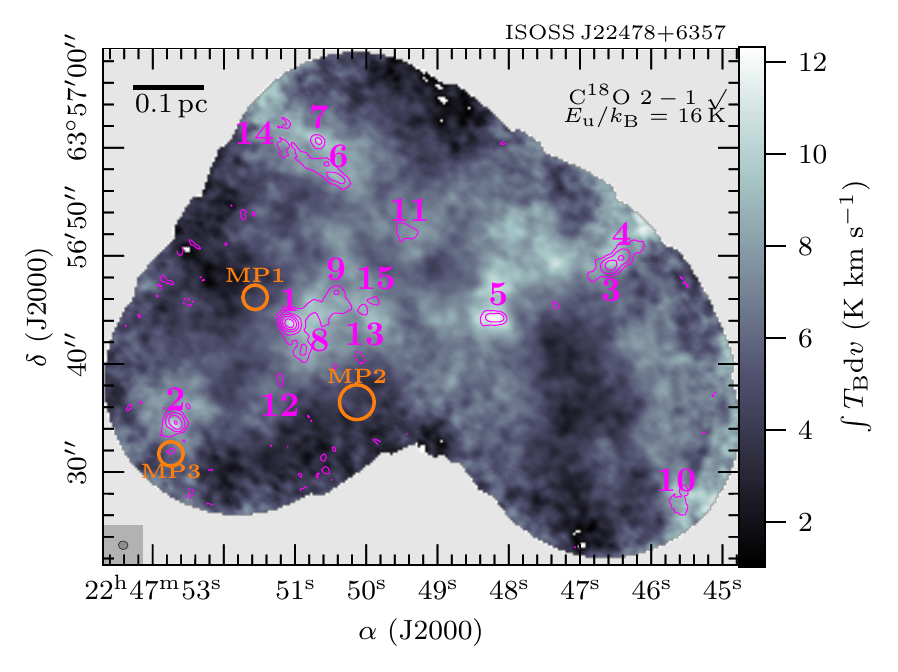}
\includegraphics[]{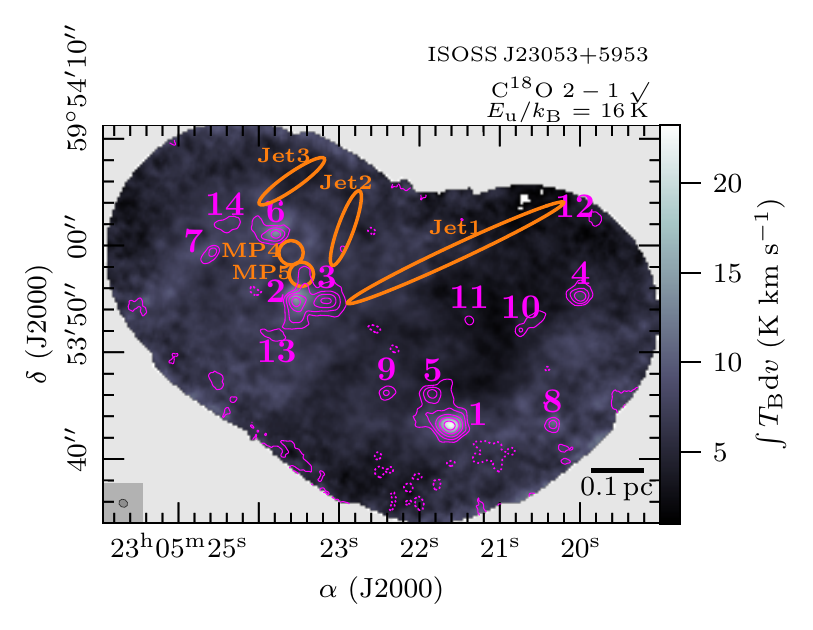}
\caption{The same as Fig. \ref{fig:moment0_H2CO_3_03_2_02}, but for C$^{18}$O $2-1$.}
\label{fig:moment0_C18O_2_1}
\end{figure*}

\begin{figure*}
\centering
\includegraphics[]{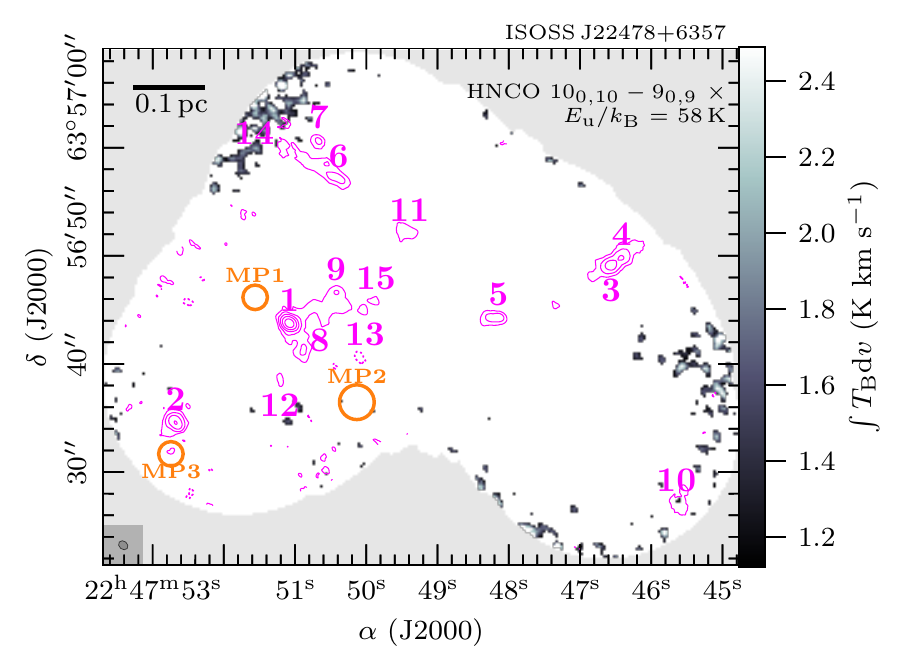}
\includegraphics[]{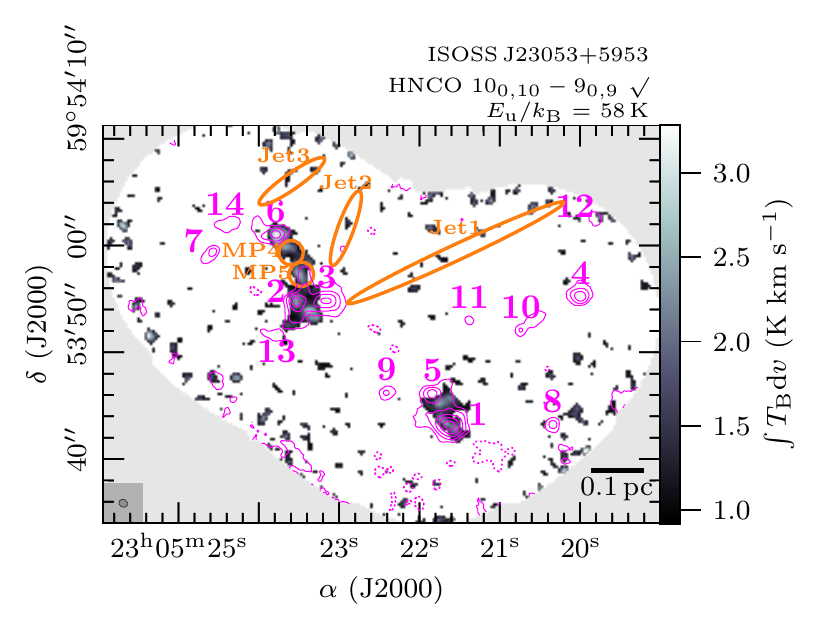}
\caption{The same as Fig. \ref{fig:moment0_H2CO_3_03_2_02}, but for HNCO $10_{0,10}-9_{0,9}$.}
\label{fig:moment0_HNCO_10_010_9_09}
\end{figure*}

\begin{figure*}
\centering
\includegraphics[]{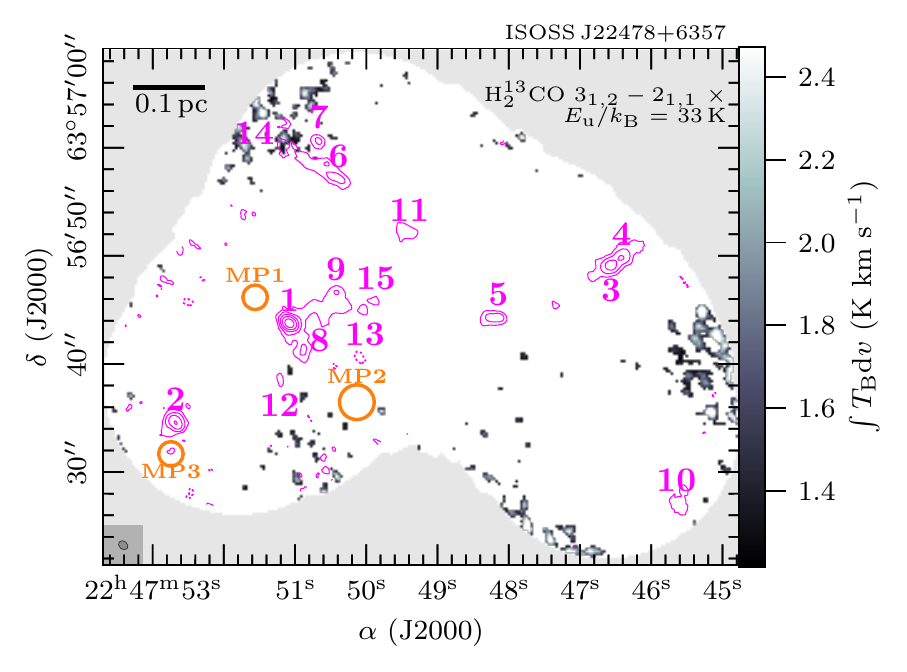}
\includegraphics[]{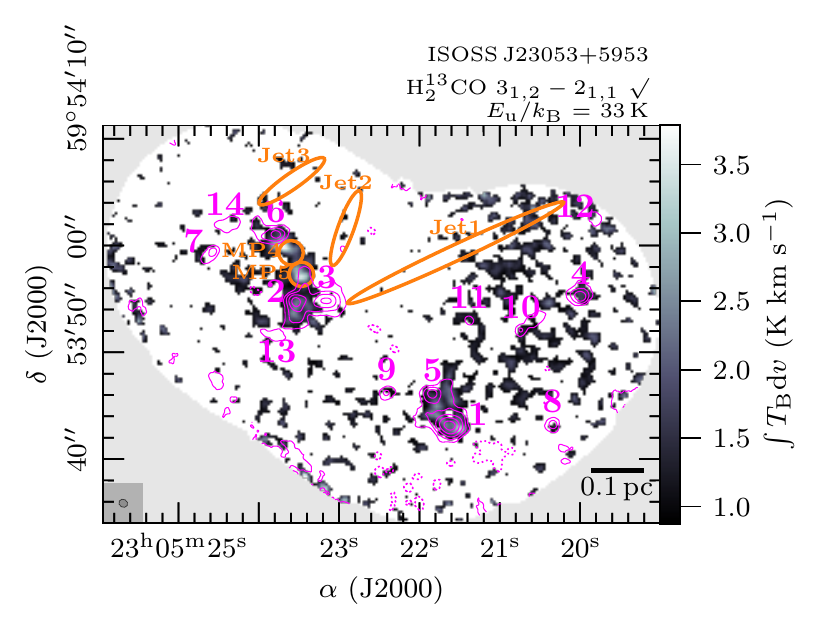}
\caption{The same as Fig. \ref{fig:moment0_H2CO_3_03_2_02}, but for H$_{2}^{13}$CO $3_{1,2}-2_{1,1}$.}
\label{fig:moment0_H213CO_3_12_2_11}
\end{figure*}

\begin{figure*}
\centering
\includegraphics[]{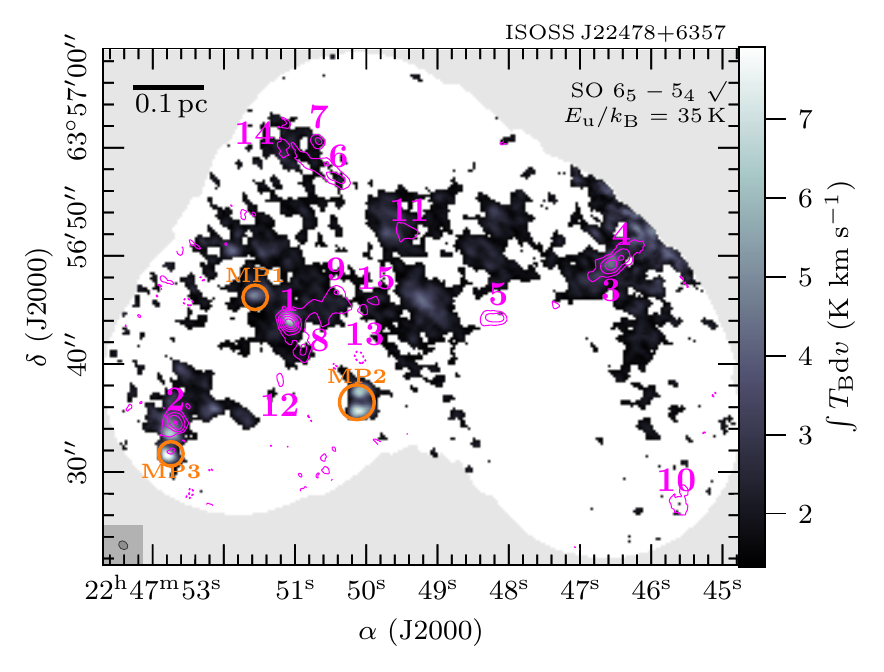}
\includegraphics[]{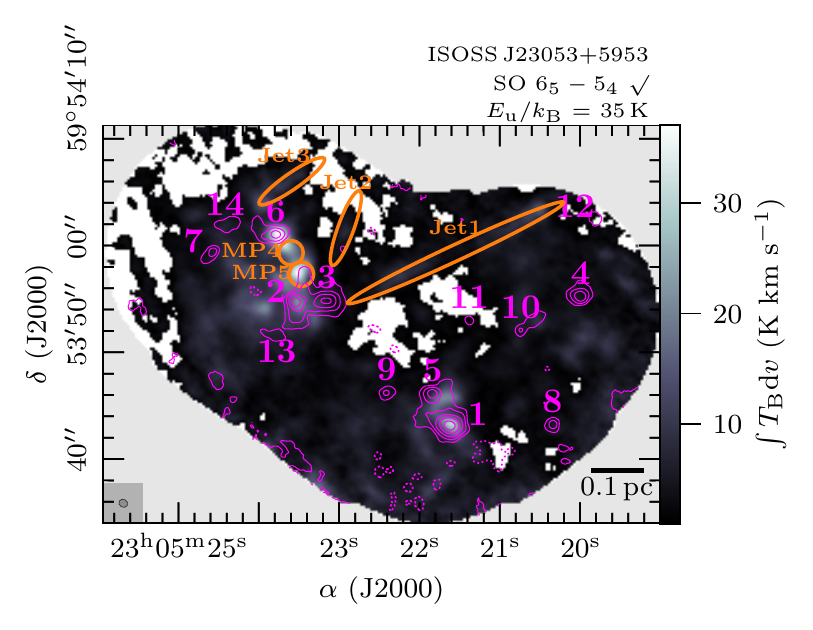}
\caption{The same as Fig. \ref{fig:moment0_H2CO_3_03_2_02}, but for SO $6_{5}-5_{4}$.}
\label{fig:moment0_SO_6_5_5_4}
\end{figure*}

\begin{figure*}
\centering
\includegraphics[]{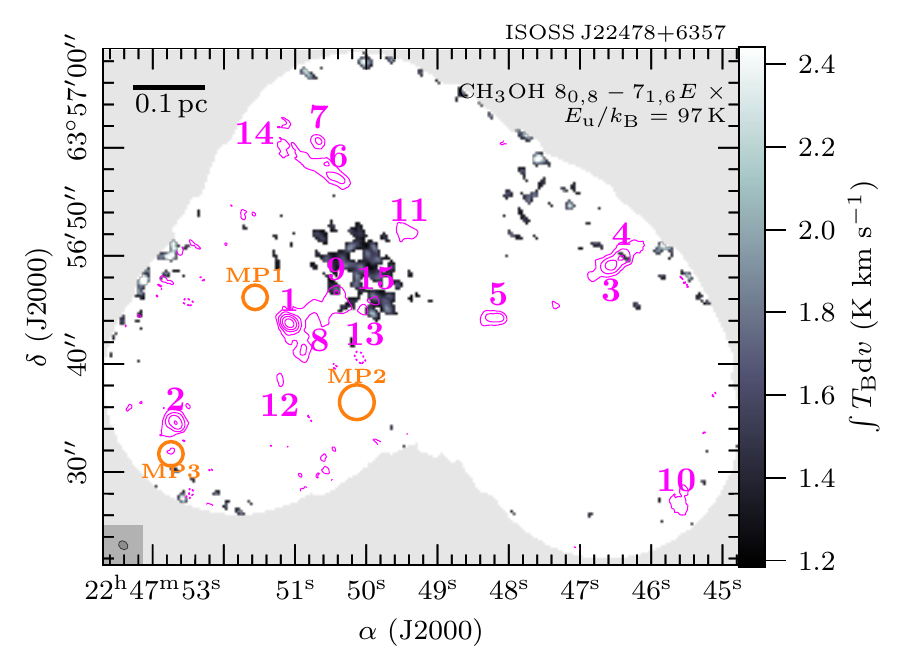}
\includegraphics[]{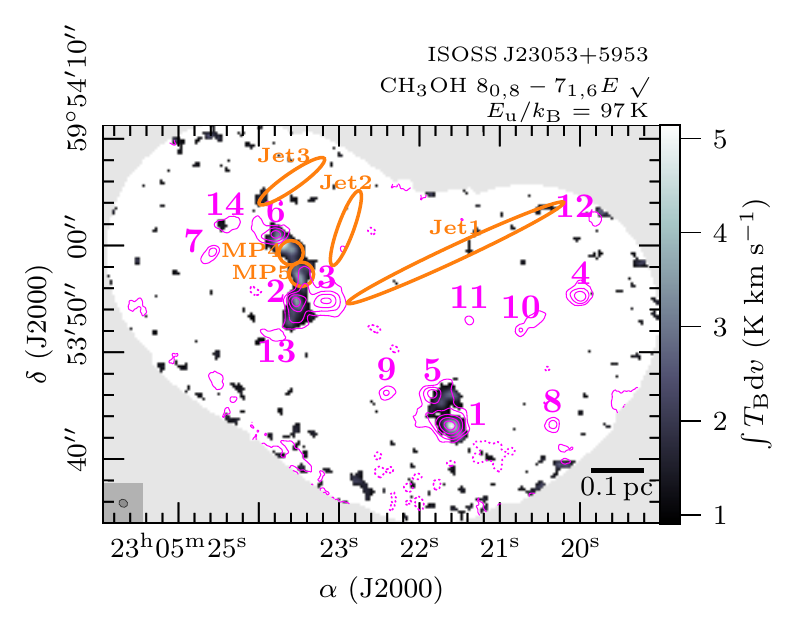}
\caption{The same as Fig. \ref{fig:moment0_H2CO_3_03_2_02}, but for CH$_{3}$OH $8_{0,8}-7_{1,6}E$.}
\label{fig:moment0_CH3OH_8_0_7_1}
\end{figure*}

\begin{figure*}
\centering
\includegraphics[]{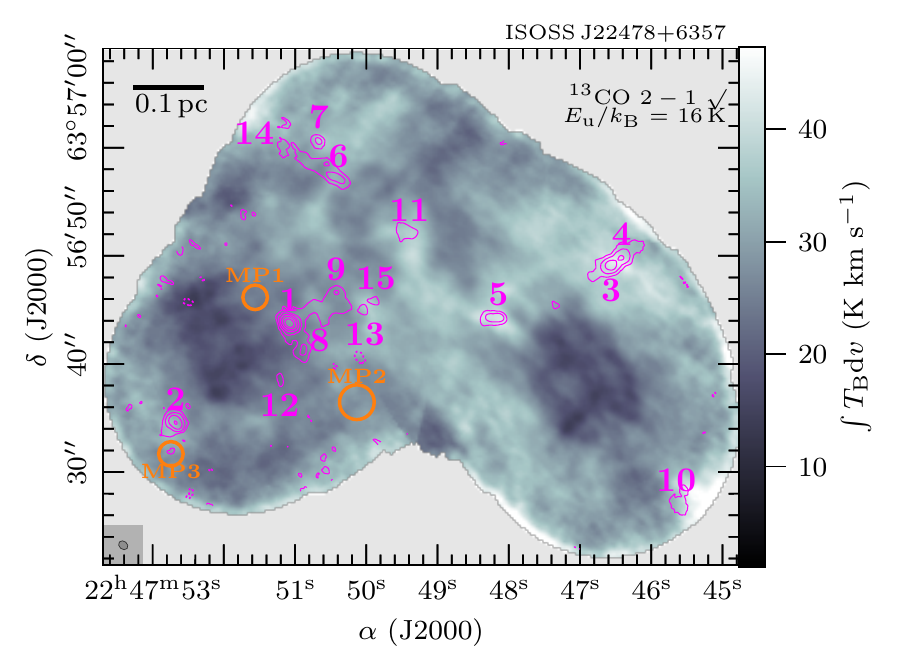}
\includegraphics[]{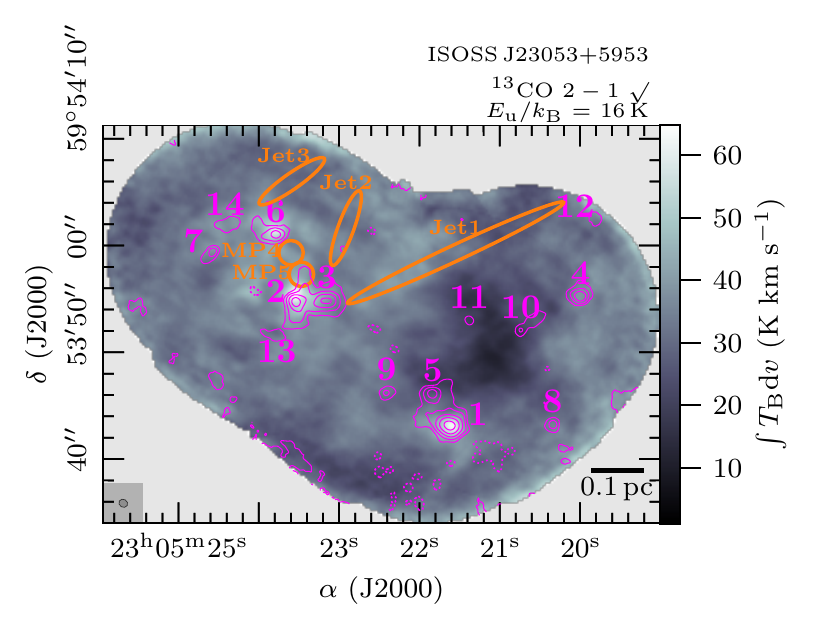}
\caption{The same as Fig. \ref{fig:moment0_H2CO_3_03_2_02}, but for $^{13}$CO $2-1$.}
\label{fig:moment0_13CO_2_1}
\end{figure*}

\begin{figure*}
\centering
\includegraphics[]{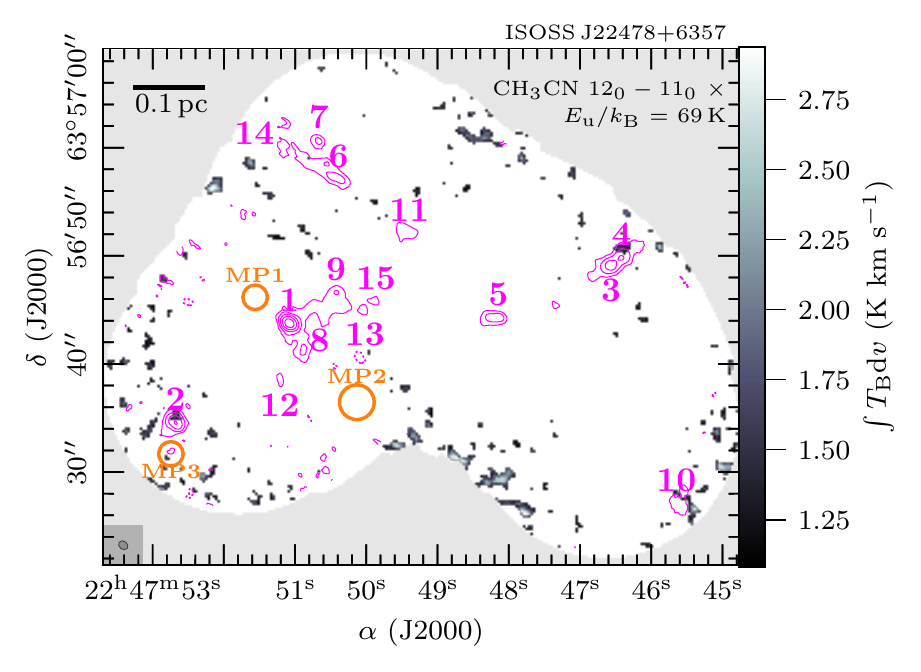}
\includegraphics[]{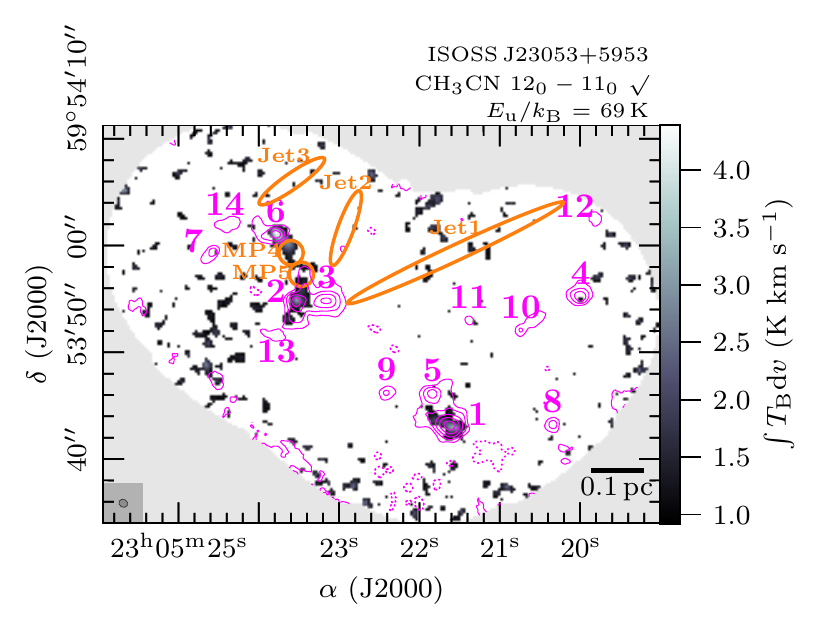}
\caption{The same as Fig. \ref{fig:moment0_H2CO_3_03_2_02}, but for CH$_{3}$CN $12_{0}-11_{0}$.}
\label{fig:moment0_CH3CN_12_0_11_0}
\end{figure*}

\begin{figure*}
\centering
\includegraphics[]{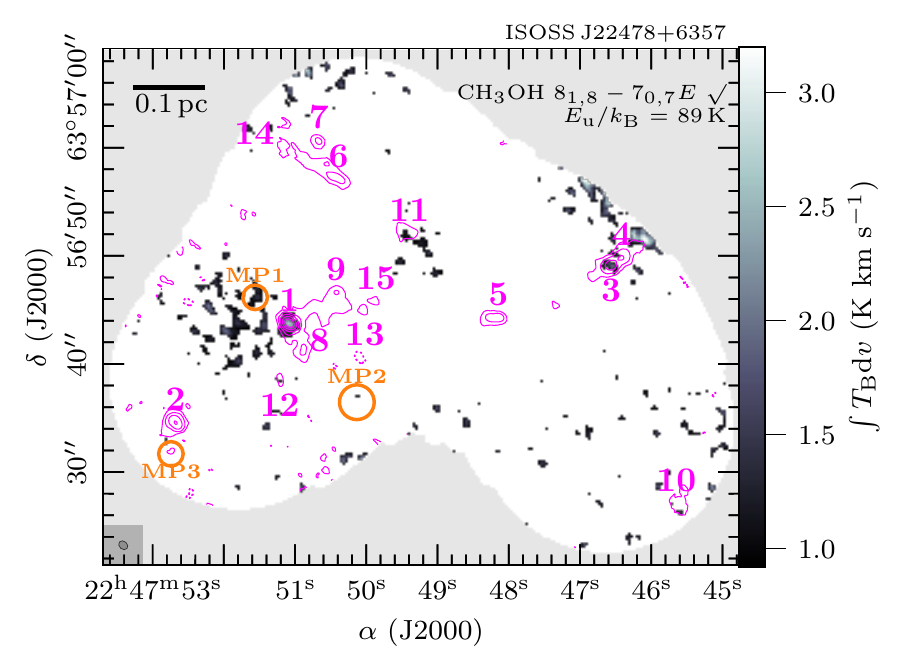}
\includegraphics[]{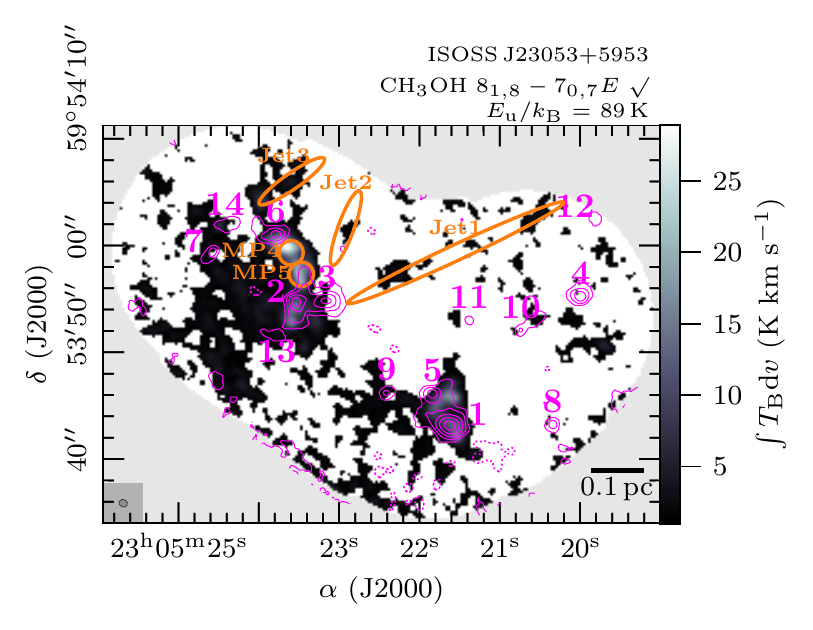}
\caption{The same as Fig. \ref{fig:moment0_H2CO_3_03_2_02}, but for CH$_{3}$OH $8_{1,8}-7_{0,7}E$.}
\label{fig:moment0_CH3OH_8_-1_7_0}
\end{figure*}

\begin{figure*}
\centering
\includegraphics[]{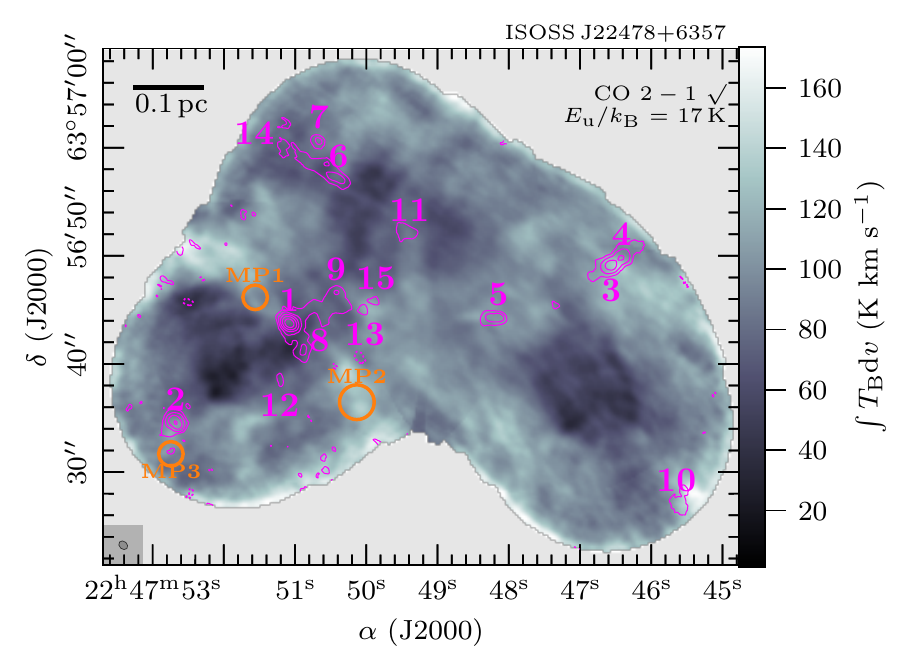}
\includegraphics[]{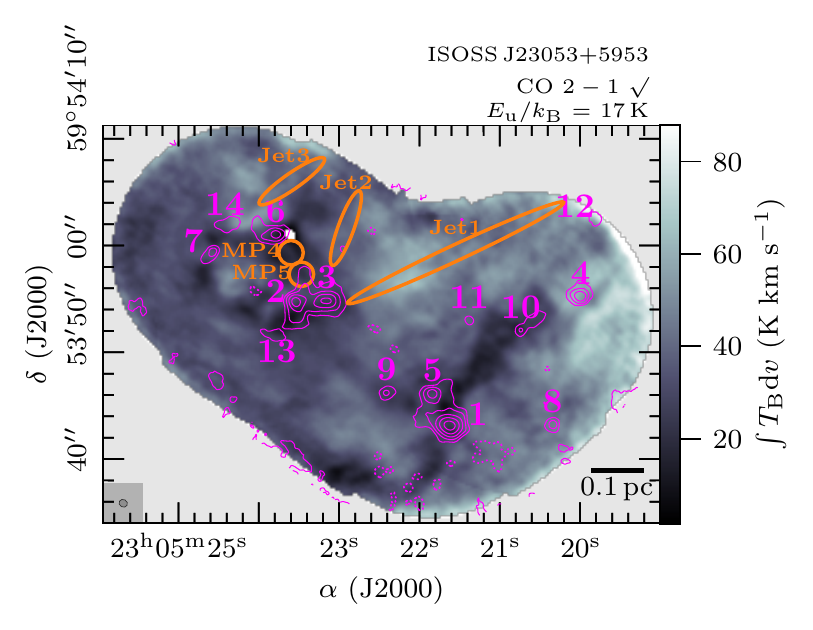}
\caption{The same as Fig. \ref{fig:moment0_H2CO_3_03_2_02}, but for CO $2-1$.}
\label{fig:moment0_CO_2_1}
\end{figure*}

\begin{figure*}
\centering
\includegraphics[]{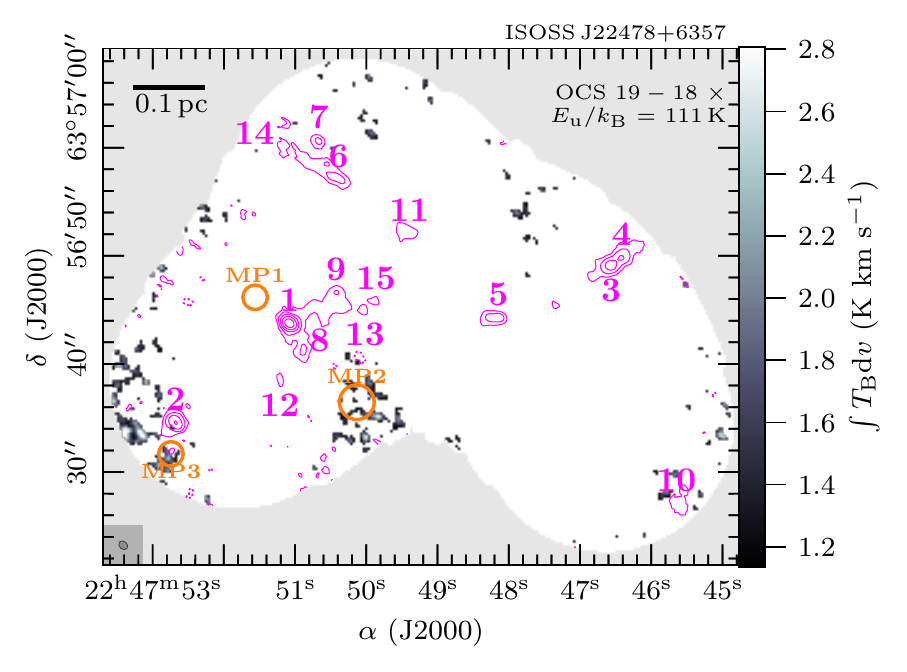}
\includegraphics[]{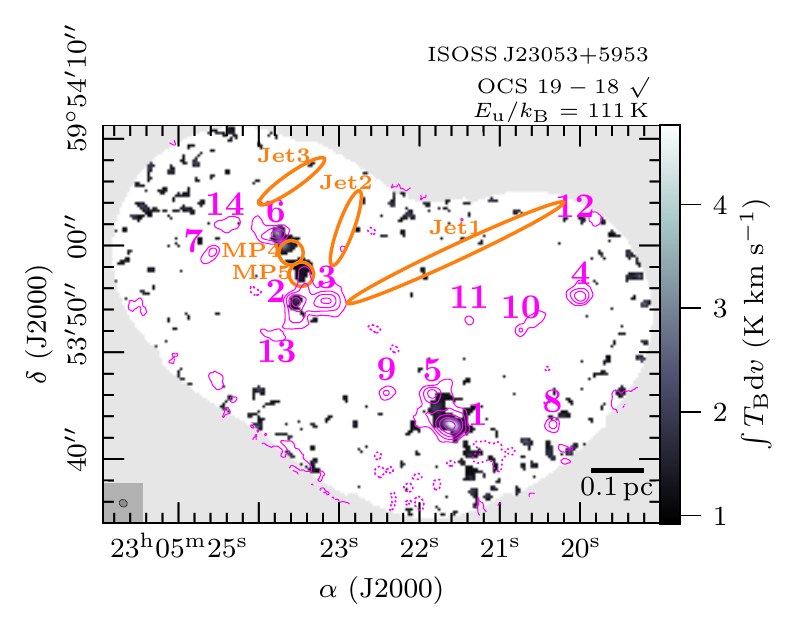}
\caption{The same as Fig. \ref{fig:moment0_H2CO_3_03_2_02}, but for OCS $19-18$.}
\label{fig:moment0_OCS_19_18}
\end{figure*}

\begin{figure*}
\centering
\includegraphics[]{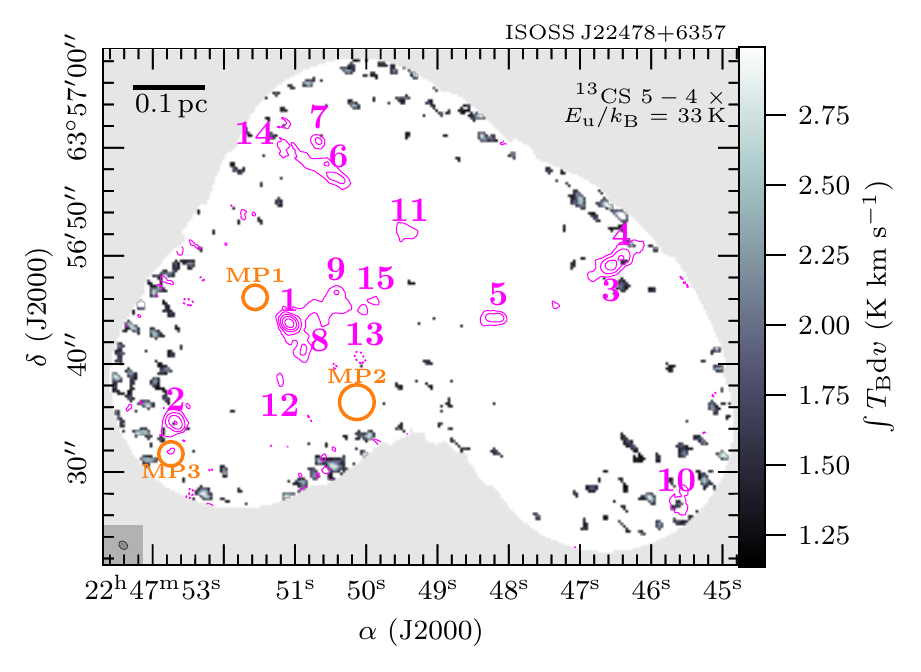}
\includegraphics[]{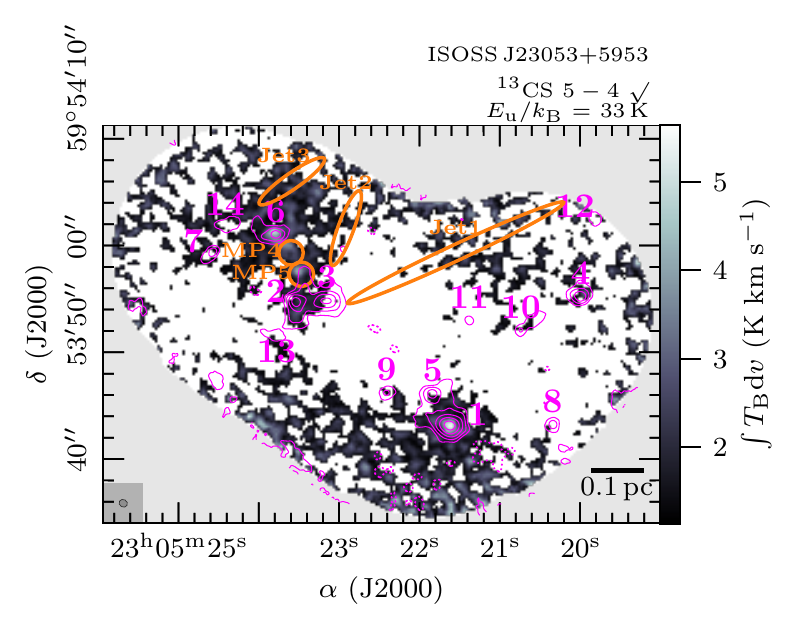}
\caption{The same as Fig. \ref{fig:moment0_H2CO_3_03_2_02}, but for $^{13}$CS $5-4$.}
\label{fig:moment0_13CS_5_4}
\end{figure*}

\begin{figure*}
\centering
\includegraphics[]{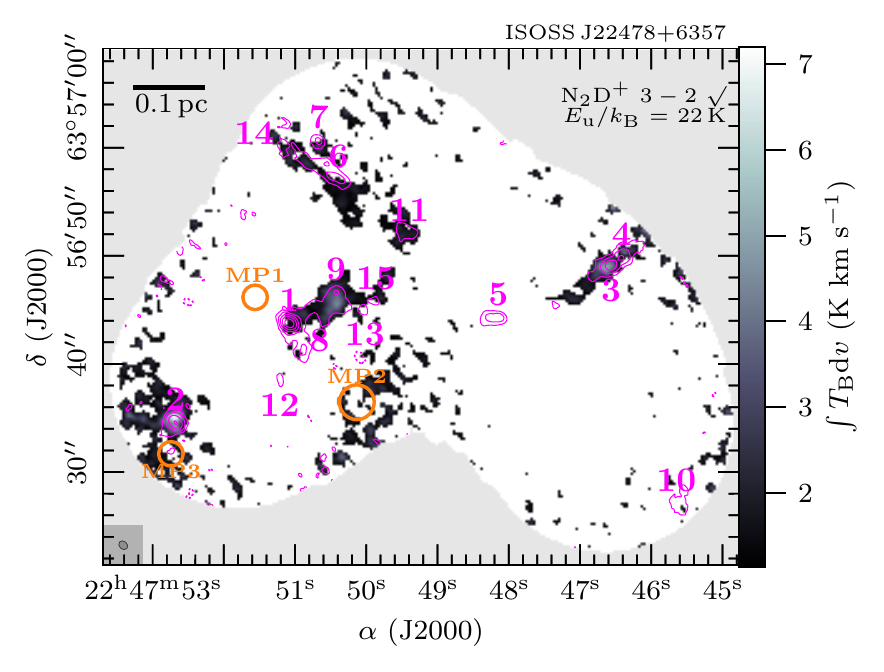}
\includegraphics[]{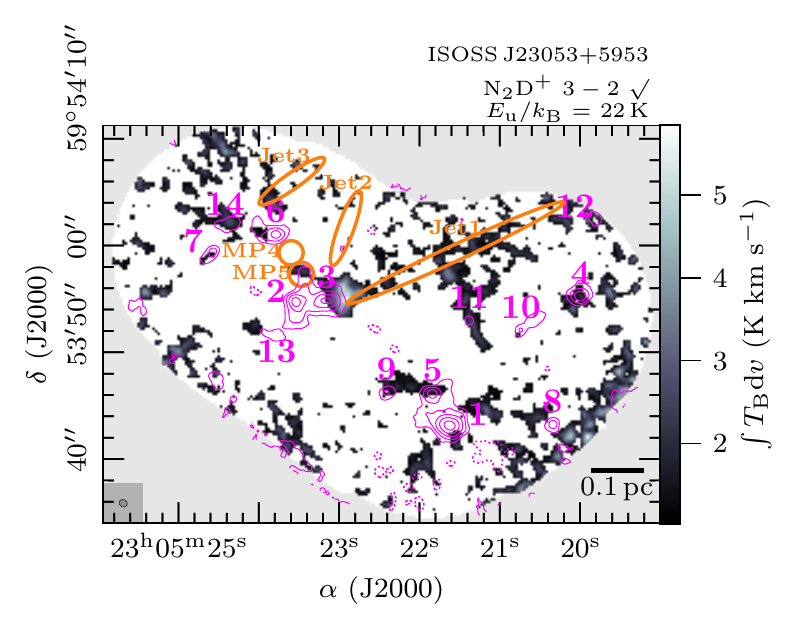}
\caption{The same as Fig. \ref{fig:moment0_H2CO_3_03_2_02}, but for N$_{2}$D$^{+}$ $3-2$.}
\label{fig:moment0_N2D+_3_2}
\end{figure*}

\begin{figure*}
\centering
\includegraphics[]{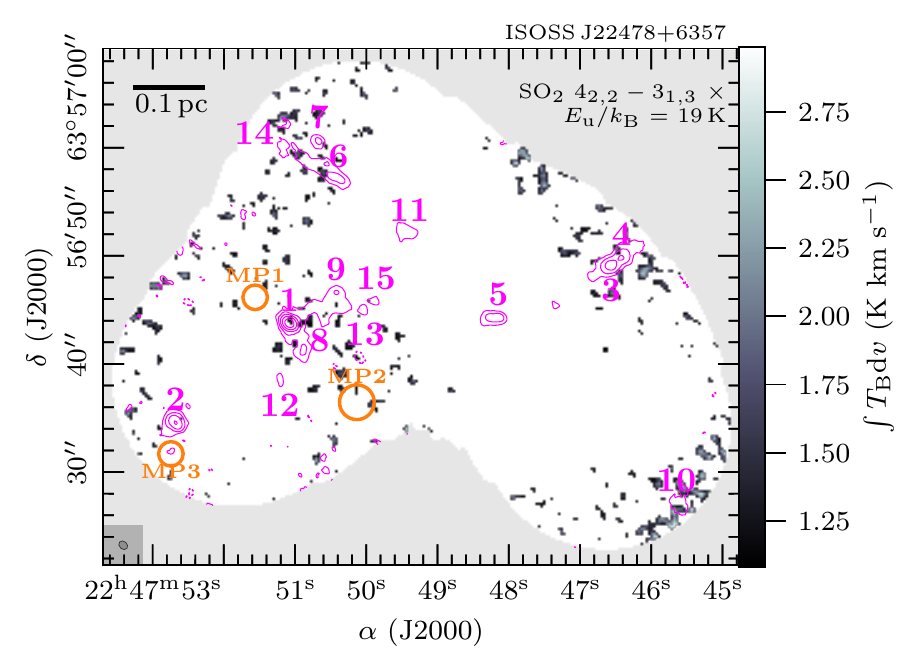}
\includegraphics[]{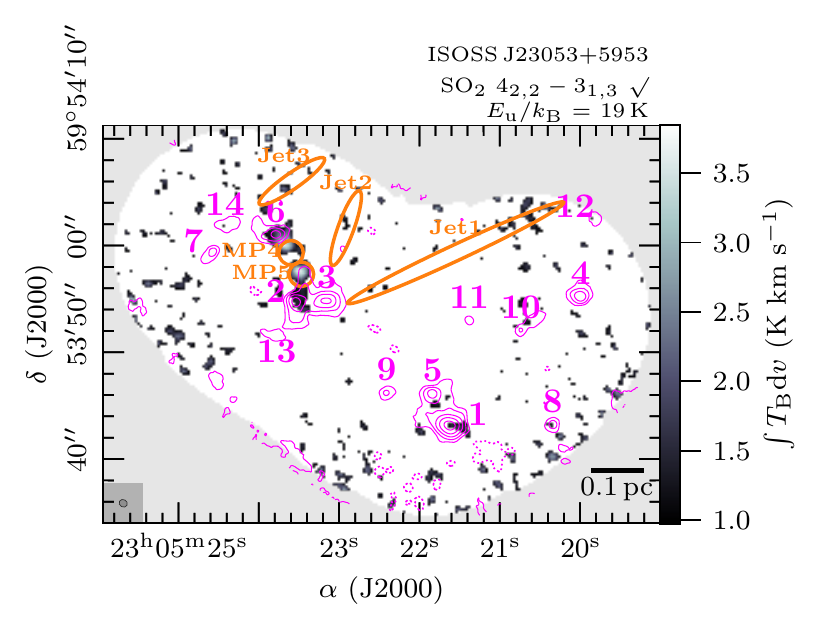}
\caption{The same as Fig. \ref{fig:moment0_H2CO_3_03_2_02}, but for SO$_{2}$ $4_{2,2}-3_{1,3}$.}
\label{fig:moment0_SO2_4_22_3_13}
\end{figure*}

\begin{figure*}
\centering
\includegraphics[]{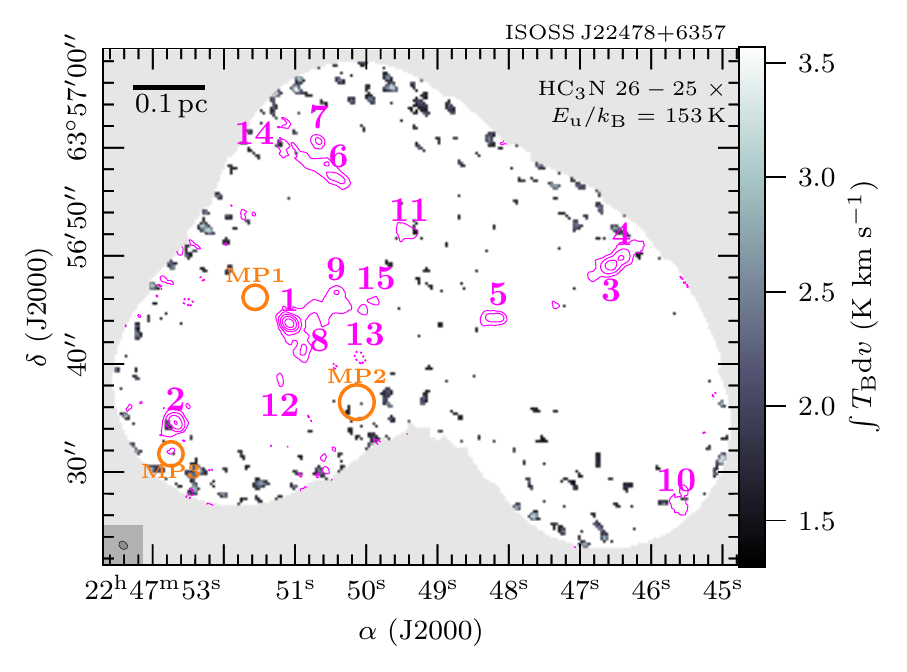}
\includegraphics[]{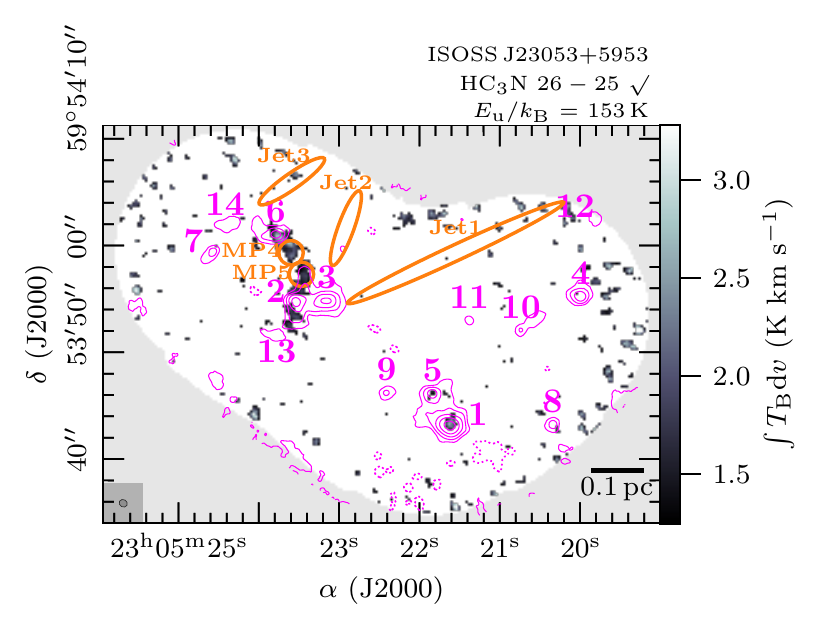}
\caption{The same as Fig. \ref{fig:moment0_H2CO_3_03_2_02}, but for HC$_{3}$N $26-25$.}
\label{fig:moment0_HC3N_26_25}
\end{figure*}

\begin{figure*}
\centering
\includegraphics[]{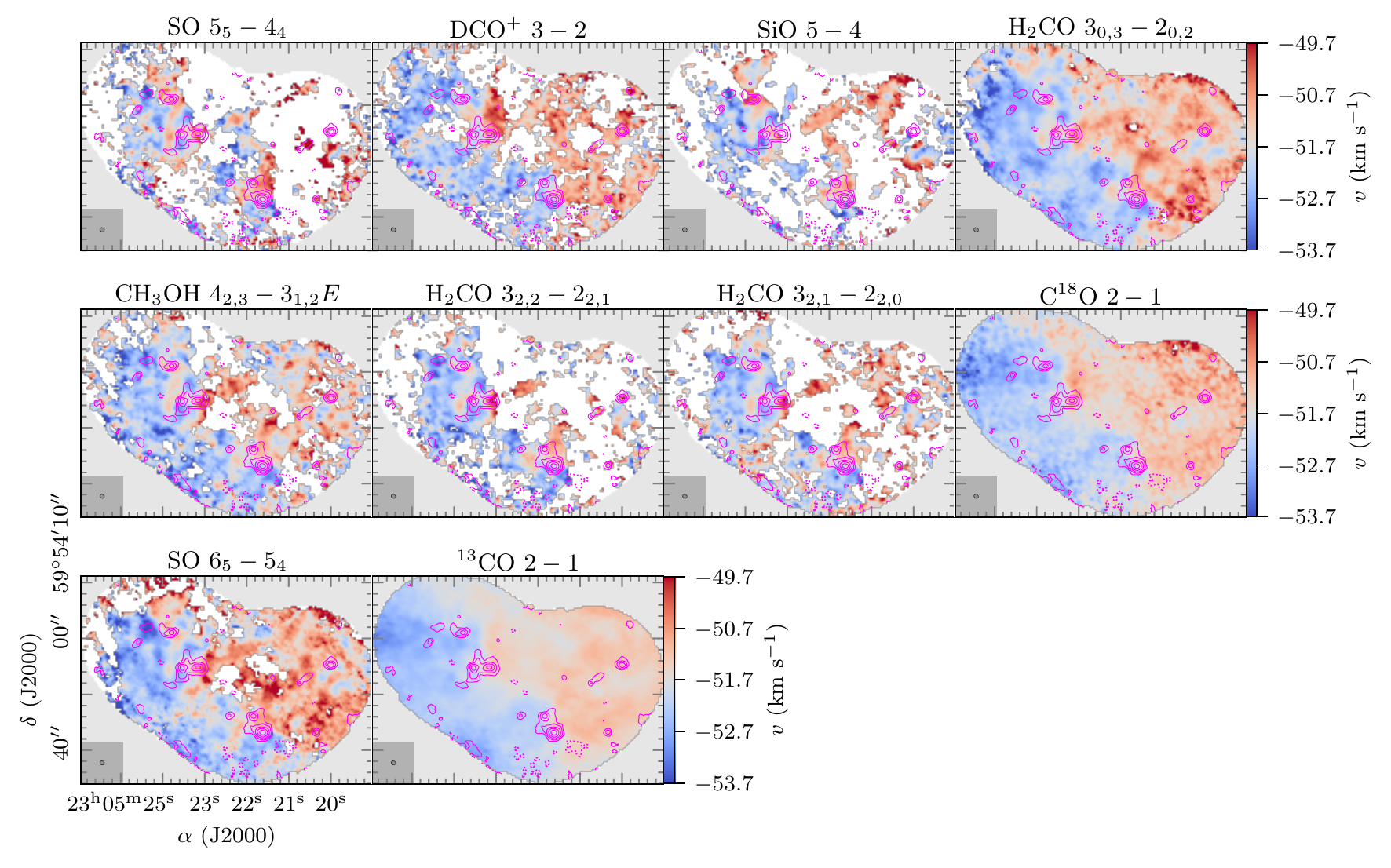}
\caption{Intensity-weighted peak velocity (moment 1) maps of molecular lines with extended emission in ISOSS\,J23053+5953. The moment 1 map is presented in color scale (using a threshold of S/N $\geq 3$ of the line integrated intensity). The NOEMA 1.3\,mm continuum data are shown in pink contours. The dotted pink contour marks the $-5\sigma_\mathrm{cont}$ level. Solid pink contour levels are 5, 10, 20, 40, and 80$\sigma_\mathrm{cont}$. The synthesized beam of the continuum data is shown in the bottom left corner. The synthesized beam of the spectral line data is similar.}
\label{fig:moment1}
\end{figure*}

	In Sect. \ref{sec:moment0} we used the line integrated intensity in order to study and compare the spatial morphology of the molecular emission. The moment 0 map of H$_{2}$CO $3_{0,3}-2_{0,2}$, SiO $5-4$, and CH$_{3}$OH $4_{2,3}-3_{1,2}E$ are shown in Sect. \ref{sec:moment0} in Figs. \ref{fig:moment0_H2CO_3_03_2_02}, \ref{fig:moment0_SiO_5_4}, and \ref{fig:moment0_CH3OH_4_2_3_1}, respectively. Here, the line integrated intensity maps of the remaining transitions (Table \ref{tab:lineprops}) are presented in Figs. \ref{fig:moment0_SO_5_5_4_4} $-$ \ref{fig:moment0_HC3N_26_25}. The line emission was integrated from $\varv_{\mathrm{LSR}} - 3$\,km\,s$^{-1}$ to $\varv_{\mathrm{LSR}} + 3$\,km\,s$^{-1}$. The intensity-weighted peak velocity (moment 1) map of molecular lines with spatially extended emission in ISOSS\,J23053+5953 is shown in Fig. \ref{fig:moment1}.

\section{Molecular column densities derived with \texttt{XCLASS}}\label{app:XCLASSfits}

\begin{table*}
\caption{Molecular column densities derived with \texttt{XCLASS} (Sect. \ref{sec:molecularcolumndensities}).}
\label{tab:XCLASSresults1}
\centering
\begin{tabular}{lccccccc}
\hline\hline
 & C$^{18}$O & SO & $^{13}$CS & OCS & DCO$^{+}$ & N$_{2}$D$^{+}$ & H$_{2}$S\\
Core & $N$ & $N$ & $N$ & $N$ & $N$ & $N$ & $N$\\
 & (cm$^{-2}$) & (cm$^{-2}$) & (cm$^{-2}$) & (cm$^{-2}$) & (cm$^{-2}$) & (cm$^{-2}$) & (cm$^{-2}$) \\
\hline
ISOSS\,J22478+6357 1 & 1.8(16)$^{+2.1(16)}_{-5.0(15)}$ & 1.7(14)$^{+1.4(14)}_{-1.6(14)}$ & $<$5.8(12) & $<$5.6(15) & 2.3(13)$^{+3.3(13)}_{-5.9(12)}$ & 2.3(12)$^{+2.1(12)}_{-2.3(12)}$ & $<$1.2(14)\\ 
ISOSS\,J22478+6357 2 & 1.2(16)$^{+4.6(15)}_{-5.4(15)}$ & $<$1.3(14) & $<$5.1(12) & $<$9.2(15) & 1.6(13)$^{+3.5(12)}_{-8.7(12)}$ & 3.0(12)$^{+2.3(12)}_{-9.5(11)}$ & $<$7.1(13)\\ 
ISOSS\,J22478+6357 3 & 2.0(16)$^{+1.7(16)}_{-8.0(15)}$ & 1.7(14)$^{+5.1(14)}_{-1.7(14)}$ & $<$1.4(12) & $<$9.6(15) & 3.9(13)$^{+4.8(12)}_{-3.0(13)}$ & 7.9(12)$^{+5.6(12)}_{-3.3(12)}$ & 2.7(14)$^{+5.4(14)}_{-2.2(14)}$\\ 
ISOSS\,J22478+6357 4 & 4.4(16)$^{+2.7(16)}_{-1.4(16)}$ & $<$2.2(14) & $<$1.6(13) & $<$9.6(15) & 3.4(12)$^{+2.7(12)}_{-1.3(12)}$ & 4.0(12)$^{+9.8(12)}_{-3.4(12)}$ & 1.6(14)$^{+2.6(14)}_{-1.3(14)}$\\ 
ISOSS\,J22478+6357 5 & 4.1(16)$^{+1.8(16)}_{-1.7(16)}$ & 4.8(13)$^{+4.7(13)}_{-4.5(13)}$ & $<$4.8(12) & $<$8.3(15) & 6.0(12)$^{+9.6(12)}_{-5.4(12)}$ & $<$3.6(11) & $<$3.0(14)\\ 
ISOSS\,J22478+6357 6 & 9.0(15)$^{+1.0(16)}_{-3.5(14)}$ & $<$2.1(14) & $<$6.6(12) & $<$2.6(14) & 3.3(12)$^{+2.0(12)}_{-3.1(12)}$ & $<$2.1(12) & $<$8.6(13)\\ 
ISOSS\,J22478+6357 7 & 5.2(16)$^{+3.2(16)}_{-5.9(15)}$ & $<$1.7(14) & $<$1.3(13) & $<$1.0(16) & 4.1(12)$^{+5.9(12)}_{-1.5(12)}$ & $<$2.2(12) & $<$6.1(13)\\ 
ISOSS\,J22478+6357 8 & 2.4(16)$^{+6.2(15)}_{-8.3(15)}$ & 3.7(14)$^{+1.7(15)}_{-3.4(14)}$ & $<$4.6(13) & $<$3.3(14) & 3.9(12)$^{+1.3(13)}_{-2.1(12)}$ & 1.6(12)$^{+9.7(11)}_{-1.6(12)}$ & $<$1.0(15)\\ 
ISOSS\,J22478+6357 9 & 3.0(16)$^{+2.7(16)}_{-1.0(16)}$ & $<$3.5(13) & $<$1.3(12) & $<$1.0(16) & 5.8(12)$^{+3.2(12)}_{-3.7(12)}$ & 7.9(12)$^{+2.1(13)}_{-3.4(12)}$ & $<$2.9(13)\\ 
ISOSS\,J22478+6357 10 & 7.0(15)$^{+2.6(15)}_{-1.9(15)}$ & $<$5.3(13) & 1.5(13)$^{+1.6(13)}_{-1.2(13)}$ & $<$8.8(15) & 8.1(13)$^{+5.9(12)}_{-3.8(13)}$ & $<$1.7(12) & 1.5(14)$^{+2.3(13)}_{-1.4(14)}$\\ 
ISOSS\,J22478+6357 11 & 9.5(15)$^{+1.8(15)}_{-6.0(15)}$ & $<$5.6(14) & $<$5.3(12) & $<$6.7(14) & 2.8(12)$^{+6.1(12)}_{-1.1(12)}$ & 1.1(13)$^{+3.4(12)}_{-1.1(13)}$ & $<$5.6(13)\\ 
ISOSS\,J22478+6357 12 & 1.0(16)$^{+3.5(15)}_{-6.0(15)}$ & $<$7.8(13) & $<$1.7(12) & $<$1.0(16) & 4.3(12)$^{+5.6(12)}_{-3.1(12)}$ & 3.5(12)$^{+5.3(12)}_{-3.3(12)}$ & $<$9.2(14)\\ 
ISOSS\,J22478+6357 13 & 6.9(15)$^{+3.0(15)}_{-1.2(15)}$ & 9.1(14)$^{+1.9(15)}_{-9.0(14)}$ & $<$3.2(12) & $<$9.9(15) & 6.7(12)$^{+1.0(13)}_{-5.9(12)}$ & 3.4(12)$^{+4.1(12)}_{-3.3(12)}$ & $<$8.9(13)\\ 
ISOSS\,J22478+6357 14 & 8.4(15)$^{+6.8(15)}_{-2.3(15)}$ & $<$8.7(13) & $<$5.3(12) & $<$1.0(16) & 1.0(13)$^{+1.5(13)}_{-9.7(12)}$ & $<$5.0(12) & $<$2.1(14)\\ 
ISOSS\,J22478+6357 15 & 1.5(16)$^{+7.1(15)}_{-7.6(15)}$ & $<$2.8(14) & $<$1.8(13) & $<$1.5(14) & 8.3(12)$^{+7.6(12)}_{-5.8(12)}$ & 3.6(12)$^{+3.7(12)}_{-3.5(12)}$ & $<$4.2(13)\\ 
\hline 
ISOSS\,J23053+5953 1 & 3.1(16)$^{+1.5(15)}_{-1.8(16)}$ & 7.1(14)$^{+4.6(14)}_{-1.4(14)}$ & 3.4(13)$^{+3.8(13)}_{-9.4(12)}$ & 9.1(14)$^{+5.6(14)}_{-3.5(14)}$ & 9.8(12)$^{+6.8(12)}_{-5.4(12)}$ & $<$2.6(11) & 5.0(14)$^{+7.8(14)}_{-2.7(14)}$\\ 
ISOSS\,J23053+5953 2 & 2.4(16)$^{+4.0(16)}_{-3.8(15)}$ & 4.3(14)$^{+5.6(14)}_{-5.9(13)}$ & $<$3.6(13) & $<$4.9(14) & 9.8(12)$^{+7.1(13)}_{-6.7(12)}$ & $<$1.0(14) & 3.9(14)$^{+1.6(15)}_{-1.3(14)}$\\ 
ISOSS\,J23053+5953 3 & 2.2(16)$^{+1.5(16)}_{-1.0(16)}$ & 2.6(14)$^{+1.5(14)}_{-2.6(14)}$ & 6.9(12)$^{+3.9(13)}_{-5.9(12)}$ & $<$2.7(14) & 7.4(12)$^{+9.0(12)}_{-6.2(12)}$ & 2.4(12)$^{+1.1(13)}_{-1.8(12)}$ & $<$6.2(13)\\ 
ISOSS\,J23053+5953 4 & 1.5(16)$^{+4.0(15)}_{-6.3(15)}$ & 5.6(14)$^{+1.7(15)}_{-5.5(14)}$ & $<$2.4(13) & $<$2.3(14) & 6.7(12)$^{+2.1(13)}_{-1.5(12)}$ & 4.2(12)$^{+2.0(13)}_{-3.7(12)}$ & $<$1.0(14)\\ 
ISOSS\,J23053+5953 5 & 1.2(16)$^{+1.9(16)}_{-5.7(15)}$ & 3.6(14)$^{+1.1(15)}_{-3.2(14)}$ & $<$4.4(12) & $<$4.3(14) & 1.2(13)$^{+8.4(12)}_{-5.7(12)}$ & 3.8(12)$^{+4.5(12)}_{-3.5(12)}$ & $<$7.7(13)\\ 
ISOSS\,J23053+5953 6 & 6.1(16)$^{+1.2(16)}_{-2.6(16)}$ & 9.1(14)$^{+1.0(15)}_{-1.1(14)}$ & 5.8(13)$^{+4.4(13)}_{-2.5(13)}$ & 5.6(14)$^{+3.5(14)}_{-2.2(14)}$ & $<$3.0(12) & $<$3.9(11) & 4.6(14)$^{+3.1(14)}_{-2.8(14)}$\\ 
ISOSS\,J23053+5953 7 & 2.5(16)$^{+1.4(16)}_{-7.7(15)}$ & 2.5(14)$^{+5.2(14)}_{-2.4(14)}$ & $<$8.8(12) & $<$1.3(15) & 4.8(12)$^{+1.3(13)}_{-4.1(12)}$ & $<$1.0(12) & $<$1.0(15)\\ 
ISOSS\,J23053+5953 8 & 3.2(16)$^{+1.9(16)}_{-4.5(15)}$ & $<$2.0(14) & $<$1.1(13) & $<$1.6(14) & 1.1(13)$^{+2.6(13)}_{-4.5(12)}$ & $<$6.4(11) & $<$8.5(13)\\ 
ISOSS\,J23053+5953 9 & 2.1(16)$^{+9.7(15)}_{-5.7(15)}$ & $<$1.7(14) & $<$1.0(13) & $<$1.6(15) & 1.7(12)$^{+9.5(11)}_{-1.5(12)}$ & 5.0(11)$^{+1.5(12)}_{-4.4(11)}$ & $<$8.9(13)\\ 
ISOSS\,J23053+5953 10 & 1.2(16)$^{+6.6(15)}_{-5.0(15)}$ & 3.6(14)$^{+9.2(13)}_{-3.6(14)}$ & $<$8.0(12) & $<$9.8(15) & 6.5(12)$^{+2.5(12)}_{-6.1(12)}$ & $<$1.0(12) & $<$4.8(13)\\ 
ISOSS\,J23053+5953 11 & 6.6(15)$^{+1.5(16)}_{-2.6(15)}$ & $<$5.5(13) & $<$5.8(12) & $<$1.0(16) & 8.9(12)$^{+3.0(13)}_{-6.2(12)}$ & 7.2(12)$^{+1.9(13)}_{-6.9(12)}$ & $<$5.0(13)\\ 
ISOSS\,J23053+5953 12 & 2.3(16)$^{+3.0(16)}_{-1.3(16)}$ & 1.1(14)$^{+2.4(13)}_{-1.1(14)}$ & $<$1.3(13) & $<$2.1(14) & 4.4(12)$^{+2.0(12)}_{-2.8(12)}$ & 1.2(13)$^{+2.6(13)}_{-7.3(12)}$ & $<$3.3(13)\\ 
ISOSS\,J23053+5953 13 & 1.4(16)$^{+1.1(16)}_{-6.9(15)}$ & 2.0(14)$^{+1.8(14)}_{-2.0(14)}$ & $<$4.1(12) & $<$2.7(15) & 2.5(12)$^{+1.9(12)}_{-8.9(11)}$ & $<$3.7(11) & $<$1.0(15)\\ 
ISOSS\,J23053+5953 14 & 4.5(15)$^{+2.7(15)}_{-6.5(14)}$ & $<$3.6(14) & $<$7.8(12) & $<$2.6(14) & 1.1(13)$^{+4.2(13)}_{-7.1(12)}$ & 4.2(12)$^{+3.4(13)}_{-4.0(12)}$ & $<$1.0(15)\\ 
\hline 
\end{tabular}
\tablefoot{a(b) = a $\times 10^{\mathrm{b}}$. Uncertainties are estimated with the MCMC error estimation algorithm. Fits that were discarded are shown as upper limits. The constraints are explained in Sect. \ref{sec:molecularcolumndensities}.}
\end{table*}

\begin{table*}
\caption{Molecular column densities derived with \texttt{XCLASS} (Sect. \ref{sec:molecularcolumndensities}).}
\label{tab:XCLASSresults2}
\centering
\begin{tabular}{lcccccc}
\hline\hline
Core & H$_{2}$CO & HNCO & HC$_{3}$N & c-C$_{3}$H$_{2}$ & CH$_{3}$OH & CH$_{3}$CN\\
 & $N$ & $N$ & $N$ & $N$ & $N$ & $N$\\
 & (cm$^{-2}$) & (cm$^{-2}$) & (cm$^{-2}$) & (cm$^{-2}$) & (cm$^{-2}$) & (cm$^{-2}$) \\
\hline
ISOSS\,J22478+6357 1 & 1.8(14)$^{+4.9(14)}_{-6.5(13)}$ & $<$6.1(13) & $<$8.6(12) & 3.8(13)$^{+1.0(14)}_{-3.3(13)}$ & 2.8(15)$^{+1.7(16)}_{-1.9(15)}$ & $<$2.4(14)\\ 
ISOSS\,J22478+6357 2 & 1.0(14)$^{+1.5(14)}_{-6.1(13)}$ & $<$1.0(15) & $<$3.5(12) & $<$4.4(13) & $<$1.1(16) & $<$3.5(14)\\ 
ISOSS\,J22478+6357 3 & 9.1(13)$^{+3.8(14)}_{-3.8(13)}$ & $<$1.0(15) & $<$1.1(13) & $<$3.3(13) & $<$1.1(15) & $<$5.0(14)\\ 
ISOSS\,J22478+6357 4 & $<$1.6(14) & $<$8.0(12) & $<$7.1(12) & 1.5(13)$^{+2.7(12)}_{-1.5(13)}$ & $<$1.6(16) & $<$9.9(12)\\ 
ISOSS\,J22478+6357 5 & 1.0(14)$^{+8.7(14)}_{-9.3(13)}$ & $<$6.3(12) & $<$7.3(12) & 3.5(13)$^{+3.5(13)}_{-3.3(13)}$ & $<$3.3(14) & $<$5.0(14)\\ 
ISOSS\,J22478+6357 6 & 1.1(14)$^{+1.0(15)}_{-1.1(14)}$ & $<$1.7(12) & $<$5.4(12) & 5.1(13)$^{+1.4(14)}_{-4.4(13)}$ & $<$2.4(14) & $<$1.9(13)\\ 
ISOSS\,J22478+6357 7 & 1.3(14)$^{+5.4(14)}_{-1.3(14)}$ & $<$4.3(13) & $<$1.8(12) & 2.4(13)$^{+1.2(13)}_{-2.3(13)}$ & $<$1.4(15) & $<$9.1(12)\\ 
ISOSS\,J22478+6357 8 & 6.9(13)$^{+1.3(14)}_{-6.9(13)}$ & $<$3.4(12) & $<$1.5(13) & $<$2.5(13) & $<$6.7(14) & $<$6.2(12)\\ 
ISOSS\,J22478+6357 9 & $<$3.0(14) & $<$9.7(14) & $<$2.6(12) & 3.3(13)$^{+1.6(13)}_{-3.2(13)}$ & $<$4.4(15) & $<$1.5(13)\\ 
ISOSS\,J22478+6357 10 & $<$4.8(13) & 5.3(13)$^{+1.1(13)}_{-5.1(13)}$ & $<$1.3(13) & $<$2.2(13) & $<$5.6(14) & $<$1.3(14)\\ 
ISOSS\,J22478+6357 11 & 9.4(14)$^{+4.4(14)}_{-9.4(14)}$ & $<$9.8(14) & $<$6.0(12) & 2.7(13)$^{+6.5(13)}_{-2.4(13)}$ & $<$1.3(14) & $<$7.0(12)\\ 
ISOSS\,J22478+6357 12 & $<$9.2(13) & $<$6.8(13) & $<$2.4(13) & 5.1(13)$^{+9.1(13)}_{-4.9(13)}$ & $<$2.4(15) & $<$4.5(14)\\ 
ISOSS\,J22478+6357 13 & $<$1.8(13) & $<$1.0(15) & $<$4.2(13) & 7.0(13)$^{+1.0(14)}_{-6.1(13)}$ & $<$1.7(14) & $<$5.0(14)\\ 
ISOSS\,J22478+6357 14 & 4.3(13)$^{+6.8(13)}_{-4.3(13)}$ & $<$5.4(13) & $<$8.7(12) & $<$4.7(13) & $<$1.1(15) & $<$1.2(13)\\ 
ISOSS\,J22478+6357 15 & 5.6(13)$^{+9.6(13)}_{-5.6(13)}$ & $<$9.7(14) & $<$1.3(12) & 4.3(13)$^{+2.0(14)}_{-3.8(13)}$ & $<$1.8(15) & $<$5.7(13)\\ 
\hline 
ISOSS\,J23053+5953 1 & 4.2(15)$^{+4.0(14)}_{-3.7(14)}$ & 1.3(14)$^{+5.7(14)}_{-1.2(14)}$ & 3.6(13)$^{+3.4(13)}_{-1.5(13)}$ & 1.3(14)$^{+9.7(13)}_{-1.2(14)}$ & 5.9(15)$^{+8.5(14)}_{-1.1(15)}$ & 5.9(13)$^{+5.2(13)}_{-4.1(13)}$\\ 
ISOSS\,J23053+5953 2 & 1.1(15)$^{+7.0(14)}_{-9.1(13)}$ & 1.5(14)$^{+2.6(14)}_{-1.4(14)}$ & $<$2.2(13) & $<$3.6(13) & 4.8(15)$^{+9.8(14)}_{-5.6(14)}$ & 3.0(13)$^{+8.8(13)}_{-6.4(12)}$\\ 
ISOSS\,J23053+5953 3 & 2.8(14)$^{+3.6(14)}_{-1.0(14)}$ & $<$3.9(13) & $<$3.9(12) & $<$5.3(13) & 8.8(14)$^{+4.3(15)}_{-5.4(14)}$ & $<$5.7(12)\\ 
ISOSS\,J23053+5953 4 & 2.0(14)$^{+1.2(14)}_{-5.3(13)}$ & $<$1.8(13) & $<$3.2(12) & 2.0(13)$^{+5.3(13)}_{-1.9(13)}$ & $<$6.6(14) & $<$5.0(14)\\ 
ISOSS\,J23053+5953 5 & 4.2(14)$^{+5.1(14)}_{-9.9(13)}$ & $<$4.1(13) & $<$1.2(13) & $<$3.4(13) & 1.8(15)$^{+1.3(15)}_{-5.5(14)}$ & $<$1.2(13)\\ 
ISOSS\,J23053+5953 6 & 2.3(15)$^{+8.9(14)}_{-6.4(14)}$ & 9.4(13)$^{+5.9(14)}_{-8.3(13)}$ & 3.0(13)$^{+3.1(13)}_{-2.0(13)}$ & 8.4(13)$^{+1.6(14)}_{-7.3(13)}$ & 4.1(15)$^{+3.1(14)}_{-6.3(14)}$ & 5.2(13)$^{+8.7(13)}_{-2.5(13)}$\\ 
ISOSS\,J23053+5953 7 & 2.1(14)$^{+2.3(14)}_{-5.7(13)}$ & $<$2.5(13) & $<$4.7(12) & $<$3.1(13) & 1.0(15)$^{+3.8(15)}_{-6.5(14)}$ & $<$8.9(12)\\ 
ISOSS\,J23053+5953 8 & 1.5(14)$^{+1.2(14)}_{-4.5(13)}$ & $<$9.9(14) & $<$6.7(12) & 3.6(13)$^{+4.8(13)}_{-3.1(13)}$ & 1.4(15)$^{+2.8(15)}_{-1.4(15)}$ & $<$5.0(14)\\ 
ISOSS\,J23053+5953 9 & 3.2(14)$^{+1.4(14)}_{-1.8(14)}$ & $<$1.4(14) & $<$5.7(13) & $<$1.3(13) & 6.7(14)$^{+4.4(14)}_{-3.1(14)}$ & $<$9.0(12)\\ 
ISOSS\,J23053+5953 10 & 1.1(14)$^{+2.8(14)}_{-4.5(13)}$ & $<$4.1(13) & $<$1.8(12) & 3.7(13)$^{+1.4(14)}_{-3.4(13)}$ & 7.1(14)$^{+1.0(15)}_{-6.6(14)}$ & $<$7.5(12)\\ 
ISOSS\,J23053+5953 11 & 1.2(14)$^{+2.6(14)}_{-4.7(13)}$ & $<$5.0(13) & $<$1.0(12) & $<$1.5(13) & $<$7.4(14) & $<$6.6(12)\\ 
ISOSS\,J23053+5953 12 & 1.1(14)$^{+4.2(14)}_{-7.2(13)}$ & $<$1.0(14) & $<$7.9(12) & $<$7.8(12) & 1.5(14)$^{+8.7(14)}_{-8.2(13)}$ & $<$4.6(14)\\ 
ISOSS\,J23053+5953 13 & 4.1(14)$^{+5.3(14)}_{-1.1(14)}$ & $<$4.7(13) & $<$2.5(12) & 6.9(13)$^{+2.6(14)}_{-6.0(13)}$ & 8.9(14)$^{+4.9(14)}_{-5.7(14)}$ & $<$1.6(13)\\ 
ISOSS\,J23053+5953 14 & 2.1(15)$^{+2.3(15)}_{-1.1(15)}$ & $<$1.2(13) & $<$3.8(12) & 5.8(13)$^{+5.4(14)}_{-4.9(13)}$ & 2.4(15)$^{+1.5(15)}_{-6.6(14)}$ & $<$8.8(13)\\ 
\hline 
\end{tabular}
\tablefoot{a(b) = a $\times 10^{\mathrm{b}}$. Uncertainties are estimated with the MCMC error estimation algorithm. Fits that were discarded are shown as upper limits. The constraints are explained in Sect. \ref{sec:molecularcolumndensities}.}
\end{table*}

\begin{figure*}
\includegraphics[]{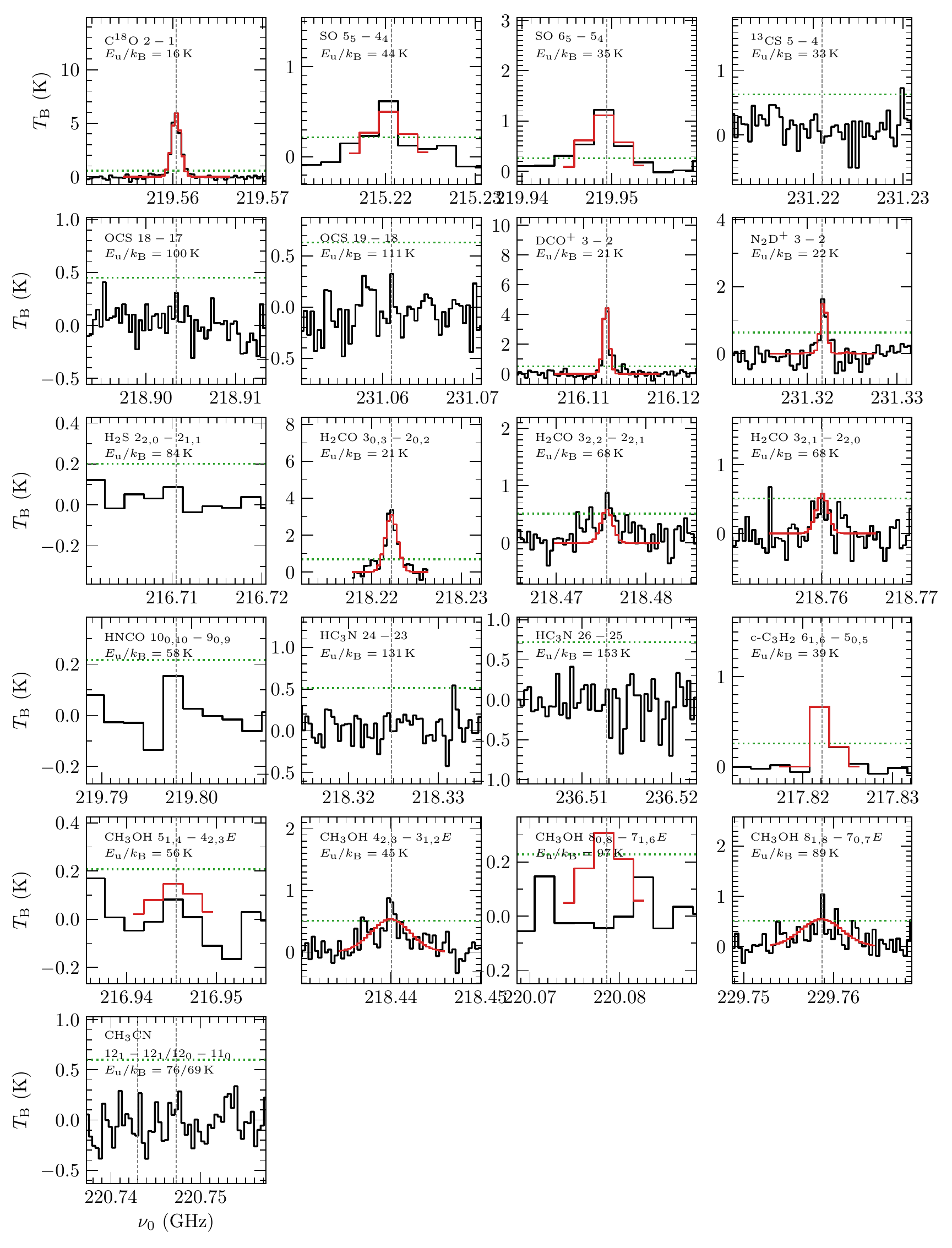}
\caption{Example spectrum of all molecular transitions fitted with \texttt{XCLASS} toward core 1 in ISOSS\,J22478+6357. In each panel, the observed spectrum of the emission line and the corresponding \texttt{XCLASS} fit are presented by the black and red lines, respectively. The green dotted line shows the $3\sigma_\mathrm{line}$ level of the transition (Table \ref{tab:lineobs}). The gray dashed line indicates the rest frequency of the transition (Table \ref{tab:lineprops}). If \texttt{XCLASS} fits were discarded no fit is shown (the constraints are explained in Sect. \ref{sec:molecularcolumndensities}). An example spectrum of core 1 in ISOSS\,J23053+5953 is shown in Fig. \ref{fig:spectrum_fit_23053}.}
\label{fig:spectrum_fit_22478}
\end{figure*}

\begin{figure*}
\includegraphics[]{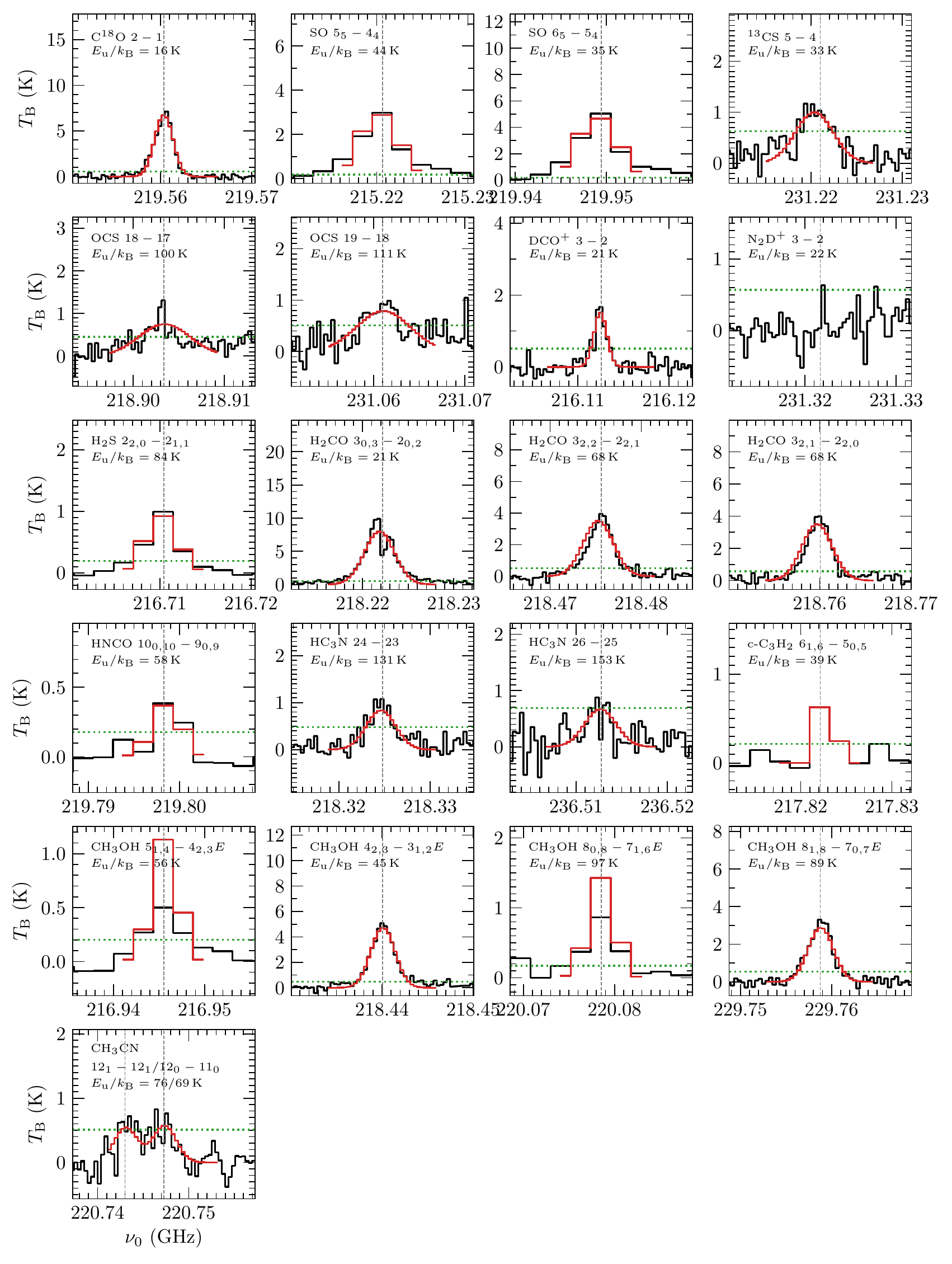}
\caption{The same as Fig. \ref{fig:spectrum_fit_22478}, but for core 1 in ISOSS\,J23053+5953.}
\label{fig:spectrum_fit_23053}
\end{figure*}

	In Sect. \ref{sec:XCLASSMapFitting} we present H$_{2}$CO parameter maps derived with \texttt{XCLASS}. The calculation of the modeled spectral lines in \texttt{XCLASS} follows the assumption of LTE and a 1D isothermal source. In the dense parts of star-forming regions the LTE conditions are valid. Using the \texttt{myXCLASSMapFit} function, we fitted in each pixel the observed H$_{2}$CO spectral line data with one emission component within the full FOV if the peak intensity is $>$1\,K. This threshold corresponds to a minimum S/N of $\sim$5 in the high-resolution data (Table \ref{tab:lineobs}). In both regions, toward the edges of the mosaic the noise increases and therefore the fit results toward the edges of the mosaic are not reliable.
	
	In \texttt{XCLASS}, the best-fit parameters can be obtained using several algorithms with the implemented model optimizer package \texttt{MAGIX} \citep{MAGIX}. We used an algorithm chain utilizing the Genetic algorithm, optimizing global minima, and the Levenberg-Marquart algorithm, optimizing local minima, with 50 iterations each to derive the best-fit parameter set for each species. The transition properties of the fitted lines are taken from the Virtual Atomic and Molecular Data Centre \citep[VAMDC,][]{Endres2016} with entries from the Cologne Database for Molecular Spectroscopy \citep[CDMS,]{CDMS} and the Jet Propulsion Laboratory \citep[JPL,][]{JPL} catalog. The line properties of the three fitted H$_{2}$CO transitions are summarized in Table \ref{tab:lineprops}. We varied each fit parameter within the following ranges: $\theta_{\mathrm{source}} = 0.1 - 2''$, $T_{\mathrm{rot}} = 1 - 200$\,K, $N = 10^{12} - 10^{17}$\,cm$^{-2}$, $\Delta \varv = 0.5 - 10$\,km\,s$^{-1}$, and $\varv_{\mathrm{off}} = -10 - +10$\,km\,s$^{-1}$. A further input is the beam size of the interferometric observations $\theta_{\mathrm{beam}}$ and as in \texttt{XCLASS} only a single value can be given as an input, we computed the mean of the major and minor axis of the synthesized beam: $\theta_{\mathrm{beam}} = \frac{\theta_{\mathrm{maj}} + \theta_{\mathrm{min}}}{2}$ (the synthesized beam of each spectral line data product is listed in Table \ref{tab:lineobs}).

	In order to quantify the molecular content of all mm cores (listed in Table \ref{tab:positions}), we fitted the spectral line data with \texttt{XCLASS} in order to derive molecular column densities (Sect. \ref{sec:molecularcolumndensities}). As we analyze the spectrum at the position of the core and not the full FOV of both regions, minor changes were employed in the \texttt{XCLASS} fitting routine in order to make the fit results as reliable as possible: In order to determine a robust velocity $\varv_\mathrm{LSR}$ for each core, we first fitted the C$^{18}$O $2-1$ transition with one emission component. We then corrected all spectra with the obtained velocity offset. The results for the systemic velocity of each core are summarized in Table \ref{tab:positions}. The fit parameter ranges for the source size and rotation temperature were $\theta_{\mathrm{source}} = 0.1 - 10''$, $T_{\mathrm{rot}} = 1 - 200$\,K, respectively. Setting 1\,K as the lower rotation temperature limit is in no opposition to the fact that the gas temperatures should be higher than the cosmic microwave background (CMB), $T_\mathrm{CMB} = 2.73$\,K, as toward all species and cores the derived rotation temperature is $T_{\mathrm{rot}} > 3$\,K. The fit parameter ranges for the column density $N$ and line width $\Delta \varv$ were specified individually for each molecule. The emission of each line was fitted around $\nu_{0} \pm 6$\,MHz (the rest frequency $\nu_{0}$ of all transitions is listed in Table \ref{tab:lineprops}). With both regions being line-poor in general, most of the detected lines do not suffer from line blending, except for CH$_{3}$CN. The transition of CH$_{3}$CN $12_{0}-11_{0}$ is blended with $12_{1}-11_{1}$. Both transitions were considered when fitting the molecular emission in \texttt{XCLASS}.
	
	The same algorithm chain as used for the H$_{2}$CO parameter maps for the was used. In order to estimate the uncertainties of the fit parameters, the Markov Chain Monte Carlo (MCMC) error estimation algorithm was used afterward with 50 iterations.
	
	The column densities of all fitted species and 29 cores are listed in Tables \ref{tab:XCLASSresults1} and \ref{tab:XCLASSresults2}. The uncertainties were estimated using the MCMC error estimation algorithm in \texttt{XCLASS}. Upper limits are listed for fits that were discarded. The constraints are explained in Sect. \ref{sec:molecularcolumndensities}. An example of the observed and \texttt{XCLASS} modeled spectrum for all fitted lines of core 1 in ISOSS\,J22478+6357 and ISOSS\,J23053+5953 is shown in Fig. \ref{fig:spectrum_fit_22478} and Fig. \ref{fig:spectrum_fit_23053}, respectively.

\section{MUSCLE results}

	For four cores in our sample, which have enough species detected in the 1.3\,mm setup and for which the density and temperature profiles could be derived, we applied the physical-chemical model \texttt{MUSCLE} in Sect. \ref{sec:MUSCLE} to estimate the chemical timescale $\tau_\mathrm{chem}$. As the initial chemical composition of the gas is unknown and the physical structure is static in \texttt{MUSCLE}, we modeled each core with three initial chemical conditions, referred to as the HMPO, HMC, and UCH{\sc ii} model. The initial chemical conditions are taken from a study of 59 HMSFRs by \citet{Gerner2014,Gerner2015}. 

	In each time step, the modeled column densities, computed with \texttt{ALCHEMIC}, are compared with the observed column densities. The time step with the lowest $\chi^{2}$ is taken as the best fit. Table \ref{tab:MUSCLEresults} shows for each of the four modeled cores and each initial condition model, the best-fit chemical timescale $\tau_\mathrm{chem}$, the corresponding $\chi^{2}$ value, and the percentage of well modeled species $\Upsilon$. A molecule is considered as ``well modeled'' when the modeled and observed column density agree within a factor of ten. A weighted $\bar \chi^{2}$ was computed taking into account the percentage of well modeled species $\Upsilon$: $\bar \chi^{2} = \frac{\chi^2}{\Upsilon^2}$. For each core, the initial condition model with the lowest $\bar \chi^{2}$ was taken as the best-fit model.

\clearpage
\begin{table*}
\caption{\texttt{MUSCLE} results for all initial condition models (HMPO, HMC, UCH{\sc ii}).}
\label{tab:MUSCLEresults}
\centering
\begin{tabular}{lcccc|cccc|cccc}
\hline\hline
Core & \multicolumn{4}{c}{HMPO model} & \multicolumn{4}{c}{HMC model} & \multicolumn{4}{c}{UCH{\sc ii} model} \\
\cline{2-5} \cline{6-9} \cline{10-13}
 & $\tau_\mathrm{chem}$ & $\chi^{2}$ & $\Upsilon$ & $\bar \chi^{2}$ & $\tau_\mathrm{chem}$ & $\chi^{2}$ & $\Upsilon$ & $\bar \chi^{2}$ & $\tau_\mathrm{chem}$ & $\chi^{2}$ & $\Upsilon$ & $\bar \chi^{2}$\\
 & (yr) & & (\%) & & (yr) & & (\%) & & (yr) & & (\%) & \\
\hline
ISOSS\,J22478+6357 1 & 3.1e+04 & 0.293 & 85.7 & 0.399 & 6.0e+04 & 0.339 & 78.6 & 0.548 & $\ldots$ & $\ldots$ & $\ldots$ & $\ldots$\\ 
ISOSS\,J23053+5953 1 & 1.9e+04 & 0.458 & 71.4 & 0.898 & 4.9e+04 & 0.556 & 57.1 & 1.702 & 8.4e+04 & 0.457 & 64.3 & 1.105\\ 
ISOSS\,J23053+5953 2 & 1.9e+04 & 0.365 & 71.4 & 0.715 & 6.0e+04 & 0.497 & 64.3 & 1.203 & 8.4e+04 & 0.469 & 64.3 & 1.135\\ 
ISOSS\,J23053+5953 6 & 1.8e+04 & 0.494 & 71.4 & 0.969 & 6.7e+04 & 0.500 & 71.4 & 0.980 & 8.4e+04 & 0.448 & 64.3 & 1.084\\ 
\hline 
\end{tabular}
\tablefoot{For each initial condition model, the best-fit chemical timescale $\tau_\mathrm{chem}$, the $\chi^{2}$ value obtained from comparing the modeled and observed column densities, the percentage of well-modeled molecules $\Upsilon$, and the weighted $\bar \chi^2 = \frac{\chi^2}{\Upsilon^2}$ are shown. Models that have a higher initial temperature as the measured core temperature ($T_{\mathrm{init,500}} > T_{\mathrm{500}}$) are not considered and indicated by an ellipsis.}
\end{table*}

\end{appendix}

\end{document}